\documentstyle[12pt]{report}

\renewcommand{\r}[1]{$(\!\!~\ref{#1})$}                   %Referanser
\newcommand{\BEA}{\begin{eqnarray}}
\newcommand{\EEA}{\end{eqnarray}}
\newcommand{\BA}{\begin{array}}
\newcommand{\EA}{\end{array}}
\newcommand{\BE}{\begin{equation}}
\newcommand{\EE}{\end{equation}}
\newcommand{\BT}{\begin{tabular}}
\newcommand{\ET}{\end{tabular}}
\newcommand{\BTA}{\begin{tabbing}}
\newcommand{\ETA}{\end{tabbing}}
\newcommand{\BC}{\begin{center}}
\newcommand{\EC}{\end{center}}

\newcommand{\SL}{\\[1mm]}
\newcommand{\NN}{\nonumber\\[4mm]}
\newcommand{\nn}{\nonumber\\}
\newcommand{\LP}{\left}
\newcommand{\RP}{\right}
\newcommand{\EQ}{\equiv}

\newcommand{\MB}{& & \mbox{}}
\newcommand{\nmb}{\nn & & \mbox{}}

\newcommand{\ABS}[1]{\left| #1 \right|}                   %Absoluttverdi
\newcommand{\R}{\mbox{{\sc I}\hspace{-0.25 em}{\bf R}}}   %Euklidrom
\newcommand{\PD}[2]{\frac{\partial #1}{\partial #2}}      %Partiell derivert
\newcommand{\HA}{\frac{1}{2}}                             %En halv
\def\f{\frac}                                             %Br¢k
\def\P{\partial}                                          %Partial

\def\L{{\cal L}}                                          %Calligraphic L
\def\H{{\cal H}}                                          %Calligraphic H

\newcommand{\US}[2]{\SS\SS #1^{#2}}
\newcommand{\LS}[2]{#1_{#2}}
\newcommand{\DUS}[2]{\bar{#1}^{\SS\dot{#2}}}
\newcommand{\DLS}[2]{\bar{#1}_{\SS\dot{#2}}}

\newcommand{\SA}{\s^{\mu}_{\a\dot{\a}}}

\newcommand{\TSTB}{\t\s^{\mu}\bar{\t}}

\newcommand{\tb}{\bar{\t}}

\newcommand{\SWA}{\sin\theta_{\mbox{w}}}
\newcommand{\CWA}{\cos\theta_{\mbox{w}}}
\newcommand{\SWAS}{\sin^{2}\theta_{\mbox{w}}}
\newcommand{\CWAS}{\cos^{2}\theta_{\mbox{w}}}

\def\a{\alpha}
\def\b{\beta}
\def\g{\gamma}

\def\D{\Delta}
\def\d{\delta}
\def\e{\varepsilon}
\def\t{\theta}

\def\la{\lambda}
\def\La{\Lambda}
\def\s{\sigma}

  \newenvironment{PROOF}
    {\begin{quote}\small{\bf Proof :\ }}{\normalsize\end{quote}}

\newcommand{\HS}{\hspace{1.5mm}}                            %Et lite hopp
\renewcommand{\SS}{\hspace{0.3mm}}

\begin{document}

\hfill{hep-ph/9506369}

\bigskip
\vspace*{4cm}
\begin{center}
{\Large  A Review of Minimal Supersymmetric Electro Weak Theory}\\*[2cm]
\large
 Ingve Simonsen\footnote{Email: ingves@phys.unit.no}\\*[4mm]
        University of Trondheim\\
        Department of Physics\\
        N-7055 Dragvoll\\
        NORWAY
\end{center}

\abstract{ In this review article we study the Minimal
           Supersymmetric Electro-Weak theory.
           The Lagrangian is constructed step by step in great detail,
           both in the superfield and component field formalism
           --- both on and off shell.
	   Furthermore the Lagrangian is written in the more familiar
           four component formalism.
           Electro weak symmetry breaking is discussed,
           and the physical chargino- and neutralino states
           are introduced and discussed.}

\newpage

\tableofcontents

\chapter*{Preface}

When I first started to work on supersymmetry,
my interest fell upon the minimal supersymmetric electro-weak theory,
or if you like, on supersymmetric quantum flavour dynamics (S-QFD).
To my disappointment,
as a person at that time with no background in supersymmetry,
I was not able to find any good {\em detailed} review article
on this subject.

In this report I try to present such a review article
with the hope that it may be useful to others.
The material is presented in great detail, and somebody may rightly
say that the presentation is too comprehensive.
For that reason most of the detailed calculations are
reserved for the appendices.
However, my personal motivation for including so much details was to
easy the chance of following the calculations step by step for a
person not familiar with the Minimal Supersymmetric Standard Model.

In this report I have mostly followed the notation
used by the authors of ref.~\cite{KAKU,WESS,MK} and I give
some useful formulae and comments about notation in appendix A.
These references together with ref.~\cite{SUSY} are also good introductions
to the necessary background of supersymmetry needed for this report.
\BEA
    --  \star --
       \nonumber
\EEA
I would like to take the opportunity  to express my deep appreciation to
Prof. Dr.tech. Haakon~A.~Olsen at the University of Trondheim, Norway.
He has been very helpful, and I in particular thank him for
great many  stimulating and  clarifying discussions and for
fruitful suggestions during the course of this work.

\vspace*{2cm}
\noindent
Trondheim, April 1995\\*[1cm]
%\vspace*{1cm}
\noindent
Ingve Simonsen

%\end{document}

\cleardoublepage

\chapter{Supersymmetric Extension of QFD.}
   \label{Chapter Supersymmetric Extension of QFD.}

We will now start the construction of a supersymmetric extension of
QFD of leptons.
In this chapter the Lagrangian, in the superfield
formalism, will be derived.

However, before we do so, we will say a few words about
possible extensions of the Standard Model (SM).

\section{Possible Extensions of the Standard Model.}

In a supersymmetric theory,
any fermionic state has to be accompanied by a bosonic one, and vice
versa. In the early days of SUSY, one had hoped that some of the states
required by SUSY, could be identified with some of the known particle states.
For instance one tried to identify the spin-$0$ fields associated with the
neutrino- and the electron-fields, as the photon and Higgs-field
respectively~\cite{FAY75}. Unfortunately, this idea runs into difficulties.
Firstly, if one of the spin-$0$ neutrino states is associated with the photon,
what happens to the lepton-quark symmetry?
Secondly, and more
convincing, is the observation that the spin-$0$ states, associated with
the leptons and quarks, carry lepton number and colour respectively.
By demanding a theory with unbroken colour and electromagnetism, only the
scalar neutrino can acquire a vacuum expectation value. This results in a
theory with the unwanted possibility of lepton number violation.
However, this scenario can not be completely ruled out~\cite{ELL85}, but no
realistic model, with such properties, exists.
Thus, in consequence, one is forced to introduce a complete Higgs
(SUSY) multiplet in addition to the multiplets of leptons and quarks.

In the SM, it is sufficient with only one Higgs doublet (and its
charge conjugated) in order to generate
masses for the leptons and charge-$\f{1}{3}$ and -$\f{2}{3}$ quarks.
In SUSY, however, one has to have at least two
Higgs doublets if suitable mass terms shall be
generated~\cite{SAK81,INO82a,INO82b,INO84}.
The reason is rather technical and relies
on the fact that SUSY do not allow for charge conjugation\footnote{
Two Higgs doublets are also needed in order to avoid gauge anomalies
originating from the spin-$\HA$ higgsinos.}.

\subsection{The Minimal Supersymmetric Standard Model.}

The different supersymmetric extensions of the SM are naturally
divided into two main classes.
The first one, is the
Minimal Supersymmetric Standard Model (MSSM)~[\ref{INO82a}---\ref{LI89}]
containing the minimal number of fields and parameters required to
construct a realistic model of leptons and quarks.
The second class, goes under the name of
Non-Minimal Supersymmetric Standard Models (NMSSM)~\cite{NMSSM}.
Several such models can also be constructed,
but they typically increase the number of
parameters (and fields) without any corresponding
increase in predictive power and physical motivation.

The MSSM has a high degree of predictivity, and within this model
all masses and coupling constants of the Higgs boson sector, can be
calculated at tree level.

Since the MSSM is the most attractive one from a practical point of view,
and since no theoretical aspects (at present) seem to discredit it,
we will be considering this model in the present work.
It is also interesting to note that the MSSM has survived
{\em all} the stringent
phenomenological tests coming from resent
LEP-experiments, and that in most of its parameter space the (relevant)~MSSM
predictions  are impressively close
to the SM values (calculated for a relative
light SM Higgs)~\cite{BAR90}.

\subsubsection{Model Ingredients.}

In a more complete way, the central ingredients of the MSSM can be
defined by the following points:
\begin{itemize}
  \item
      The minimal gauge group: $SU(3)\times SU(2)\times U(1)$.
  \item
      The minimal particle content, holding three generations of
       leptons and quarks, twelve gauge bosons (defined in the usual way),
       two Higgs doublets and, of course, all these particles superpartners.
   \item SUSY breaking parametrized by {\em soft} breaking terms.
   \item An exact discrete R-parity.
\end{itemize}

The three first points need no further comments at this early stage.
However, the same can not be said about the fourth point.
If we construct
a theory based on the three first points only, a theory possessing
baryon- and lepton-number violation will emerge~\cite{DIM81}.
The terms responsible for this,
give unacceptable physics (fast rates of nucleon decay).
Thus, these terms must somehow be avoided, and
it is believed that this can only be done in a satisfactory manner by
introducing additional symmetries, e.g. gauge- or discrete-symmetries.
The last possibility is used in the MSSM.
Here an unbroken R-symmetry~\cite{FAY75,FAY77,SAL75}
with a corresponding R-parity, or equivalently matter-parity,
is introduced in order to eliminate
the offending terms. The R-parity of a state is related to its
spin~(S), baryon-number~(B), and lepton-number~(L) according to
\BEA
    R_{p} &=& \LP(-1\RP)^{2J+3B+L}.
\EEA
Note that the assumption of baryon- and lepton-number conservation
implies the conservation of R-parity.

Furthermore,
an immediate consequence of the above expression is that all SM particles
(including the Higgs bosons) are R-even, while their superpartners
are R-odd. As a result the ``new" supersymmetric particles
can only be pair-produced, and any of their decay products have to contain
an odd number of supersymmetric particles. This implies that the
lightest supersymmetric particle (LSP) has to be stable, since it has no
allowed decay channels.

\section{The Lagrangian for Supersymmetric QFD.}

In this section, we shall construct
a (minimal)~supersymmetric extension of QFD.
We have chosen to work within the framework of the MSSM,
and consider supersymmetric QFD to be a part of this
more fundamental theory\footnote{An
alternative contemplation could be to consider the
MSSM for leptons only.
Hence the $SU(3)$-gauge invariance becomes trivial as in the SM~(of leptons),
where all fields except the quark- and gluon-fields are
$SU(3)$-singlets,
and a non-trivial $SU(2)\times U(1)$ theory remains.
This resulting theory may be considered,
as is correct,
to be a supersymmetric extension of
QFD (or equivalently the Glasow-Weinberg-Salam theory).}.
Thus the content of the Higgs-sector is
defined to contain two Higgs-doublets,
as we discussed in the previous section.
% Note that the supersymmetric extension of QFD
% coming from the above program, is a minimal theory.

In order to construct the Lagrangian of supersymmetric QFD (S-QFD),
we will assume that the theory can be viewed as a low-energy limit of
a SUGRAV-theory.
Thus the Lagrangian of S-QFD
has to have the form
\BEA
   \L_{S-QFD} &=& \L_{SUSY} + \L_{soft}.
      \label{The Lagrangian for Supersymmetric QFD prop 1}
\EEA
Here $\L_{SUSY}$ is a supersymmetric piece,
while $\L_{Soft}$ explicitly breaks SUSY.

The ultimate aim of this section, will be
to specify the different terms of $\L_{S-QFD}$.
However before we do so, we have to define
the different fields which are present in S-QFD.

The first version of the MSSM was constructed in the early eighties
by the authors of
refs.~\ref{KAK82} and \ref{FLO83} and later discussed in
refs.~\ref{INO84} and \ref{MAJ84}.
They promoted all the lepton fields of the SM
to chiral superfields, one for each generation. The same we will do, and
denote these superfields by $\hat{l}(x,\t,\tb)$ and
$\hat{\nu}_{l}(x,\t,\tb)$. Here the former contains
the charged leptons (like the electron) and the latter the
corresponding neutrinos.
Here the generational indices have been suppressed\footnote{Summation
over the generational indices will be
understood everywhere, if
nothing else is said to indicate otherwise.}.

It is useful, and we will henceforward use it,
to assume, as for ``ordinary" QFD, that the
neutrinos are completely left-handed. Hence the left-handed
lepton superfields~(for each generation) can be arranged in an $SU(2)$-doublet
and the right-handed in an $SU(2)$-singlet according to\footnote{Here the
subscripts L and R mean left- and right-handed respectively.}~\footnote{From
now on we will use hats~($\hat{}$) on the
superfield quanteties of our S-QFD model.}
\BEA
   \hat{L}(x,\t,\tb) &=&
      \LP( \BA{c} \hat{\nu}_{l}(x,\t,\tb) \\
                  \hat{l}(x,\t,\tb)          \EA \RP)_{L},
             \label{The Lagrangian for Supersymmetric QFD prop 2} \SL
   \hat{R} &=& \hat{l}_{R}(x,\t,\tb).
        \label{The Lagrangian for Supersymmetric QFD prop 3}
\EEA

{}From the previous section, we already know that the MSSM, and hence S-QFD,
contains two doublets of (chiral) Higgs superfields,
which we will defined as
\BEA
   \hat{H}_{1}(x,\t,\tb) &=&
       \LP( \BA{c} \hat{H}_{1}^{1}(x,\t,\tb) \\
                   \hat{H}_{1}^{2}(x,\t,\tb)     \EA \RP),
               \label{The Lagrangian for Supersymmetric QFD prop 4}
\EEA
and
\BEA
   \hat{H}_{2}(x,\t,\tb)  &=&
         \LP( \BA{c} \hat{H}_{2}^{1}(x,\t,\tb) \\
                     \hat{H}_{2}^{2}(x,\t,\tb)    \EA \RP).
                \label{The Lagrangian for Supersymmetric QFD prop 5}
\EEA
Note that the upper index on these superfields,
say $\hat{H}_{1}^{2}(x,\t,\tb)$,
is an $SU(2)$ index taking values in the set $\{1,2\}$.
The same applies to $\hat{L}(x,\t,\tb)$.

As for non-supersymmetric QFD, S-QFD possesses an
$SU(2) \times U(1)$-gauge invariance.
This means that the theory contains four different gauge vector superfields
--- $\hat{V}'(x,\t,\tb)$ for the $U(1)$-gauge group and
$\hat{V}^{a}(x,\t,\tb)$ ($a=1,2,3$) for $SU(2)$.
\begin{table}[tbh]
 \BC
 \begin{tabular}{|l|c|c|r|} \hline
  Multiplet        &  Superfields
       & \multicolumn{2}{|c|}{Quantum Numbers } \\ \cline{3-4}
  type             &                               &  $SU(2)$    & $U(1)$
                 \\ \hline \hline

  Matter           &  $\hat{L}(x,\t,\tb)$          &   doublet   &  $-1$ \\
                   &  $\hat{R}(x,\t,\tb)$          &   singlet   &  $2$ \\
                   &  $\hat{H}_{1}(x,\t,\tb) $     &   doublet   &  $-1$  \\
                   &  $\hat{H}_{2}(x,\t,\tb) $     &   doublet   &  $1$
                 \\  \hline
  Gauge            &  $\hat{V}'(x,\t,\tb)$         &  singlet    &  0 \\
                   &  $\hat{V}^{a}(x,\t,\tb)$      &  triplet    &  0
                  \\ \hline
  \end{tabular}
     \caption{The notation and quantum numbers used
            for the superfields in S-QFD (for leptons).
            The index a  labels $SU(2)$ triplets of gauge bosons.
            All the generational indices are suppressed.}
     \label{Table Superfields}
 \EC
\end{table}
As usual we
will take the gauge vector superfields to be Lie algebra valued, i.e.
\BEA
   \hat{V}'(x,\t,\tb)       &=& {\bf Y} \hat{v}'(x,\t,\tb), \SL
           \label{The Lagrangian for Supersymmetric QFD prop 6}
   \hat{V}(x,\t,\tb)    &=& {\bf T}^{a} \hat{V}^{a}(x,\t,\tb),
                  \hspace{1.5cm}       a = 1,2,3.
           \label{The Lagrangian for Supersymmetric QFD prop 7}
\EEA
Here ${\bf Y}$ and ${\bf T}^{a}$ are the generators of
$U(1)$ and $SU(2)$ respectively.

In table~\ref{Table Superfields} the above definitions, together with
the quantum numbers, are summerized.

\subsection{The Supersymmetric Term $\L_{SUSY}$.}

The term $\L_{SUSY}$, is obtained
by ``supersymmetrizing" the Lagrangian of ordinary QFD.
In this generalizing procedure, the Yang-Mills Lagrangian~\cite{KAKU,WESS,MK}
is useful.
However the S-QFD Lagrangian becomes slightly more complicated due to the fact
that we have a larger gauge group, a richer
particle spectrum with both left- and right-handed states, and in addition
a Higgs-sector as well to take into account.

With the identifications we made in the previous chapter for
the kinetic terms of chiral-  and vector-superfields,
the S-QFD Lagrangian reads
\BEA
   \L_{SUSY} &=&   \L_{Lepton}
                  + \L_{Gauge}
                  + \L_{Higgs},
        \label{The Supersymmetric Term prop 1}
\EEA
where
\BEA
  \L_{Lepton}
     &=& \int d^{4}\t\;\LP[\,\hat{L}^{\dagger}e^{2g\hat{V}+g'\hat{V}'}
                           \hat{L} +
                           \hat{R}^{\dagger}e^{2g\hat{V}+g'\hat{V}'}
                           \hat{R}\,\RP],
        \label{The Supersymmetric Term prop 2}  \SL
  \L_{Gauge}
     &=&  \f{1}{4} \int  d^{4}\t\;
         \LP[\,
                \US{W}{a\,\a}\LS{W}{\a}^{a}
             +  \US{W}{'\,\a}\LS{W}{\a}' \,
         \RP]\d^{2}(\tb)   + h.c.\,,
        \label{The Supersymmetric Term prop 3}
\EEA
and finally
\BEA
  \L_{Higgs}
     &=&  \int d^{4}\t\;\LP[\,\hat{H}_{1}^{\dagger}e^{2g\hat{V}+g'\hat{V}'}
                           \hat{H}_{1}
                         + \hat{H}_{2}^{\dagger}e^{2g\hat{V}+g'\hat{V}'}
                           \hat{H}_{2}
               +  W\,\d^{2}(\tb) + \bar{W}\,\d^{2}(\t) \RP]\!.\hspace{2mm}
        \label{The Supersymmetric Term prop 4}
\EEA
Here $g$ and $g'$ are the (gauge) coupling constants for $SU(2)$ and $U(1)$
respectively and
$\LS{W}{\a}$ and $\LS{W}{\a}'$ are the $SU(2)$- and
$U(1)$-fieldstrengths defined by
\BEA
   \LS{W}{\a}     &=&  -\f{1}{8g}\,\bar{D}\bar{D}e^{-2g\hat{V}}\LS{D}{\a}
 e^{2g\hat{V}},
           \label{The Supersymmetric Term prop 5}\SL
   \LS{W}{\a}'    &=&  -\f{1}{4}\,D D\DLS{D}{\a} \hat{V}'.
           \label{The Supersymmetric Term prop 6}
\EEA
Furthermore,
$W \equiv W[\hat{L},\hat{R},\hat{H}_{1},\hat{H}_{2}]$
is the superpotential of the theory which we will discuss in a
moment\footnote{We will not write the fieldstrengths without spinor indices
so confusion between the symbols for the superpotential and the fieldstengths
will arise.}.

The factors of 2 appearing in eqs.~\r{The Supersymmetric Term prop 2},
\r{The Supersymmetric Term prop 4} and \r{The Supersymmetric Term prop 5}
in connection with the $SU(2)$-coupling constant g, are inserted
for convenience. With this choice the (non-SUSY) fieldstrength $V^{a}_{\mu\nu}$
contained in $\LS{W}{\a}$ correspondes to that of the SM.

\subsubsection{The Superpotential.}

In order to give a complete expression for $\L_{SUSY}$,
the superpotential $W[\hat{L},\hat{R},\hat{H}_{1},\hat{H}_{2}]$ has to
be specified.
%Since W is not a trivial generalization
%of the corresponding expression from
%sect.~\ref{Supersymmetric Yang-Mills Theory.},
%particular care has to be taken.
The superpotential can at maximum be cubic in
the superfields in order to guarantee a renormalizable theory.

In the MSSM the superpotential takes on the form
\BEA
    W &=&  W_{H} + W_{Y}, \nonumber
              \label{The Supersymmetric Term prop 7}
\EEA
with the ``Higgs-part" given by
\BEA
   W_{H} &=& \mu\; \e^{ij}\hat{H}_{1}^{i}\hat{H}_{2}^{j}, \nonumber
          \label{The Supersymmetric Term prop 7a}
\EEA
and the corresponding
``Yukawa-part" by\footnote{Here $\hat{Q}$ is a quark $SU(2)$-doublet
while $\hat{U}$ and $\hat{D}$ are quark $SU(2)$-singlets.}
\BEA
     W_{Y}[\hat{L},\hat{R},\hat{H}_{1}\hat{H}_{2}]
             &=& \e^{ij}\LP[\,f\hat{H}^{i}_{1}\hat{L}^{j}\hat{R}
               +f_{1}\hat{H}^{i}_{1}\hat{Q}^{j}\hat{D}
               +f_{2}\hat{H}^{j}_{2}\hat{Q}^{i}\hat{U}\,\RP]. \nonumber
               \label{The Supersymmetric Term prop 8}
\EEA
Here  $\mu$ is a mass parameter and
$\e^{ij}$ is an anti-symmetric tensor defined by
\BEA
  \e &=& \LP( \BA{rr}   0 & 1 \\
                       -1 & 0     \EA \RP).
                \label{The Supersymmetric Term prop 7b}
\EEA
Furthermore, $f$, $f_{1}$ and $f_{2}$ are all
(Yukawa)~coupling constants containing
one generational index which has been suppressed. It is often the case
that only the largest Yukawa couplings (for the third generation)
are of importance. However, we will not in  particular take a stand on
this point.

As alluded to earlier, we will not be concerned about the
quark-sector of S-QFD. Hence the superpotential reduces
to
\BEA
    W &=& W_{H} + W_{Y} \nn
      &=& \mu\; \e^{ij}\hat{H}_{1}^{i}\hat{H}_{2}^{j} +
          f\;\e^{ij}\hat{H}^{i}_{1}\hat{L}^{j}\hat{R}.
          \label{The Supersymmetric Term prop 9}
\EEA
The first term of the above superpotential needs some further comments.
If this term is missing (i.e. $\mu = 0$), the theory has
an additional Peccei-Quinn symmetry~\cite{PEC77}.
Under this symmetry the Higgs superfield $\hat{H}_{1}$
undergoes a phase transformation.
In cases where the bosonic component of $\hat{H}_{1}^{1}$
gets a non-vanishing vacuum expectation value,
this symmetry is spontaneously broken.
The result of such a breaking is an experimentally
unacceptable Weinberg-Wilczek axion~\cite{WIL78}.
Hence, $\mu \neq 0$ is required in order to get a physically
acceptable theory.

\subsection{The Soft SUSY-Breaking Term $\L_{Soft}$.}

The most general soft SUSY breaking terms where described
by Giraedello and Grisaru~\cite{soft}. They found that
the allowed terms can be categorized as follows;
scalar mass tems, gaugino mass tems and finally trilinear
scalar interaction terms.
However, S-QFD, as the MSSM, has to possess R-invariance, as referred to
in the previous section. This implicates that trilinear terms contained in
$\LP.W \RP|_{\t=0}$, have to be disregarded (and we do
it from now) since they are not R-invariant.
The actual proof of this fact will be given in
subsect.~\ref{subsect: The R-Invariance}.

By adjusting the remaining allowed soft terms to our notation of S-QFD,
one gets the following Lagrangian (appropriate to Fermi scale)
in terms of superfields:
\BEA
   \L_{Soft} &=& \L_{SMT} + \L_{GMT},
        \label{The Soft SUSY-Breaking Term prop 2aaa}
\EEA
where the scalar mass term~(SMT) piece reads
\BEA
   \L_{SMT} &=& -\int d^{4}\t\; \LP[\,
            M_{L}^{2}\;\hat{L}^{\dagger}\hat{L}
          + m_{R}^{2}\hat{R}^{\dagger}\hat{R}
          + m_{1}^{2} \hat{H}_{1}^{\dagger}\hat{H}_{1} \RP.
\nmb  \hspace{1.7cm} \LP.
          + m_{2}^{2} \hat{H}_{2}^{\dagger}\hat{H}_{2}
          - m_{3}^{2}\e^{ij}\LP(\hat{H}_{1}^{i}\hat{H}_{2}^{j}+h.c.\RP)
             \RP]
          \d^{4}(\t,\tb),
          \label{The Soft SUSY-Breaking Term prop 2}
\EEA
and the gauge mass term~(GMT) is
\BEA
   \L_{GMT} &=&   \HA \int d^{4}\t\; \LP[
              \LP(\,M\; W^{a\;\a}W^{a}_{\a}
            + M' \;   W^{'\,\a}W'_{\a}\,\RP)
            + h.c. \RP]
            \d^{4}(\t,\tb).
            \label{The Soft SUSY-Breaking Term prop 3}
\EEA
Here
\BEA
  M_{L}^{2}\;\hat{L}^{\dagger}\hat{L} &=&
      m_{\tilde{\nu}}^{2}\;\hat{\nu}^{\dagger}\hat{\nu}+
      m_{L}^{2}\;\hat{l}^{\dagger}_{L}\hat{l}_{L}, \nonumber
\EEA
while the (soft)~mass-parameters M and $M'$
are corresponding to the SU(2)- and U(1)- gauge group
respectively.
The factor of $\HA$ in front of $\L_{GMT}$
is inserted for later convenience.

Within the framework of MSSM, the different
couplings and mass-terms, appearing in the above Lagrangian, are all
undetermined both in origin and magnitude.
However
%, as we hinted at in subsect.~\ref{sec: Soft SUSY Breaking},
they are usually interpreted as remnants of
a more fundamental spontaneously broken ($N=1$) SUGRAV-theory.
Keep in mind that at the Fermi scale, which we are working at,
one deals with renormalized parameters which are connected to their
values at the Planck scale via the renormalization
group equations.

\subsection{Conclusion.}

To conclude this section, we collect our results for the
Lagrangian $\L_{S-QFD}$, in terms of superfields,
for later reference.
It reads:
\BEA
  \L_{S-QFD} &=&   \L_{SUSY} + \L_{Soft} \NN
      &=& \int d^{4}\t\;\LP\{\,\hat{L}^{\dagger}e^{2g\hat{V}+g'\hat{V}'}
                           \hat{L} +
                           \hat{R}^{\dagger}e^{2g\hat{V}+g'\hat{V}'}
                           \hat{R} \RP.   \nn
     & & \mbox{} \hspace{1.3cm}
          + \f{1}{4}
         \LP[\LP(\,
                \US{W}{a\,\a}\LS{W}{\a}^{a}
             +  \US{W}{'\,\a}\LS{W}{\a}' \,
              \RP)\d^{2}(\tb)   + h.c.\,\RP]  \nn
      & & \mbox{}\hspace{1.3cm}
          +\hat{H}_{1}^{\dagger}e^{2g\hat{V}+g'\hat{V}'}
                           \hat{H}_{1}
                         + \hat{H}_{2}^{\dagger}e^{2g\hat{V}+g'\hat{V}'}
                           \hat{H}_{2}\nn
      & & \mbox{}\hspace{1.3cm}
               + W\,\d^{2}(\tb) + \bar{W}\,\d^{2}(\t) \nn
      & & \mbox{}\hspace{1.3cm}
          -\LP[\,
            M_{L}^{2}\,\hat{L}^{\dagger}\hat{L}
          + m_{R}^{2}\,\hat{R}^{\dagger}\hat{R}
          + m_{1}^{2}\, \hat{H}_{1}^{\dagger}\hat{H}_{1} \RP.
      \nmb \hspace{2.1cm}  \LP.
          + m_{2}^{2} \hat{H}_{2}^{\dagger}\hat{H}_{2}
          - m_{3}^{2}\e^{ij}\LP(\hat{H}_{1}^{i}\hat{H}_{2}^{j}+h.c.\RP)
             \RP]
          \d^{4}(\t,\tb)
       \nmb\hspace{1.3cm} \LP.
       + \HA\LP[\LP(\,M\, W^{a\;\a}W^{a}_{\a}
            + M' \,   W^{'\,\a}W'_{\a}\,\RP)
            + h.c.\,\RP] \d^{4}(\t,\tb) \RP\}.
\EEA

\section{Invariances of the Lagrangian $\L_{S-QFD}$.}

In this section, we will establish some of the symmetries of
$\L_{S-QFD}$, and we start by demonstrating the SUSY invariance of
$\L_{SUSY}$.

\subsection{The SUSY Invariance of $\L_{SUSY}$.}
     \label{SUBSECT: The SUSY Invariance}

It is well known that the
highest (mass) dimensional component of any superfield combination
is always supersymmetric (up to a total derivative)~\cite{KAKU,WESS,MK}.
With this in mind, the SUSY-invariance of $\L_{SUSY}$ is easy to
verify, due to its possible formulation in terms of superfields\footnote{Later
on, when the component-form of $\L_{SUSY}$ is obtained, we will
also verify the SUSY invariance explicitly without any reference to
the superfield formalism. As we will see then, this line of action is much more
demanding then the approach made here.}.

With eq.~\r{The Berezin Integral prop 20} we have that
a four dimensional integration with respect to Grassmann variables
projects out the $\t\t\,\tb\tb$-component of the integrand. This
is the highest, non-vanishing dimensional component possible,
because of the anti-commuting property of the Grassmann variables.
Hence, we may on this ground conclude that
$\L_{Lepton}$ and the two first terms of $\L_{Higgs}$ are
supersymmetric.

%From eqs.~\r{Chiral Superfields prop 6a}  and
%\r{Chiral Superfields prop 6b} and in view
%of%eq.~\r{Chiral Superfields prop 10}, we have that.
The highest component
of a product of two or three left-handed (right-handed) chiral superfields
is a $\t\t$-component ($\tb\tb$-component).
Hence, since $\hat{L}$,  $\hat{R}$,  $\hat{H}_{1}$, $\hat{H}_{2}$
and the fieldstrenghts $W_{\a}$ and $W_{\a}'$
are all left-handed chiral superfields,
 while
their hermitian conjugated are right-handed,
$\L_{Gauge}$ and the remaining terms of $\L_{Higgs}$
are SUSY-invariant. Note that the two-dimensional
delta functions over a Grassmann algebra, are inserted in
order to adopt with the four-dimensional Grassmann integration.

Hence $\L_{SUSY}$ is proven to be SUSY-invariant.

As have been stated up to several time, $\L_{Soft}$ breaks
SUSY. To see this, it is enough to note that
\BEA
    \int d^{4}\t\;  \hat{S}\,\, \d^{4}(\t,\tb) &=&
       \LP.\hat{S}\RP|_{\t=\tb=0}
\EEA
is a (mass)~dimensional zero term, with
 $\hat{S}$ being any superfield (or superfield combination).
Then according to our earlier discussion
%{sect: Transformations properties of Component Fields},
$\L_{Soft}$ is notoriously not SUSY-invariant.

\subsection{The Gauge Invariance of $\L_{S-QFD}$.}

The gauge transformations on  chiral- and vector-superfields
are  defined by
\BEA \LP. \BA{lclcl}
  \Phi'(x,\t,\tb) &=& e^{-ig\Lambda(x,\t,\tb)}\Phi(x,\t,\tb),
  \hspace{0.5cm} & &\DLS{D}{\a}\Lambda = 0  \\*[2mm]
  \Phi^{'\dagger}(x,\t,\tb) &=& \Phi^{\dagger}(x,\t,\tb)e^{ig
   \Lambda^{\dagger}(x,\t,\tb)}
   \hspace{0.5cm}& & \LS{D}{\a}\Lambda^{\dagger} = 0 \\*[2mm]
   e^{gV'} &=& e^{-ig\Lambda^{\dagger}}\,e^{gV}\,
         e^{ig\Lambda}  & & \EA \RP\}.
         \label{Non-Abelian Gauge Transformations prop 6}
\EEA
and that of the fieldstrength $\LS{W}{\a}^{a}$ by
\BEA
   \LS{W}{\a} \rightarrow \LS{W}{\a}'
    &=&  e^{-ig\Lambda}\,\LS{W}{\a}\,e^{ig\Lambda}.
       \label{Supersymmetric Field Strengths prop 8}
\EEA
These transformations will be extensively used in this subsection.

We start by showing the $SU(2)$ invariance of the theory.

\subsubsection{The $SU(2)$-Invariance.}

Since $[\hat{V},\hat{V}']=[\hat{\La},\hat{V}']=0$,  the
term $\hat{L}^{\dagger}e^{2g\hat{V}+g'\hat{V}'}\hat{L}$ is shown to be
$SU(2)$-gauge invariant as follows
\BEA
  \hat{L}^{\dagger}e^{2g\hat{V}+g'\hat{V}'}\hat{L}
    =  \hat{L}^{\dagger}e^{2g\hat{V}}e^{g'\hat{V}'}\hat{L}
 &\longrightarrow&
    \hat{L}^{\dagger}e^{2ig\hat{\La}^{\dagger}}
    e^{-2ig\hat{\La}^{\dagger}}
    e^{2g\hat{V}} e^{2ig\hat{\La}}
    e^{g'\hat{V}'}  e^{-2ig\hat{\La}}
    \hat{L}  \nn
   &=& \hat{L}^{\dagger}e^{2g\hat{V}+g'\hat{V}'}\hat{L}.
\EEA
The invariance of the corresponding kinetic terms of $\hat{R}$,
$\hat{H}_{1}$ or $\hat{H}_{2}$  are shown in the same
manner\footnote{Note that the invariance of the term containing  $\hat{R}$
is trivial since  $\hat{R}$ transform like a singlet under $SU(2)$.}.

If we can show that $W^{\a a}W_{\a}^{a}$, $W^{'\,\a}W_{\a}'$,
and the superpotential
$W \equiv W[\hat{L},\hat{R},\hat{H}_{1} ,\hat{H}_{2}]$ are gauge invariant,
then we have established the SU(2)-invariance of $\L_{SUSY}$.
This is so because
the invariance of the other terms can be obtained by hermitian conjugation.
{}From eq.~\r{Supersymmetric Field Strengths prop 8} we have
\BEA
   W^{\a\;a}W_{\a}^{a} = \f{1}{k}\,Tr \LP(W^{\a}W_{\a}\RP)
      &\longrightarrow& \f{1}{k}\,Tr \LP(
          e^{-2ig\hat{\La}} W^{\a} e^{2ig\hat{\La}}
          e^{-2ig\hat{\La}} W_{\a} e^{2ig\hat{\La}} \RP)\nn
      &=& \f{1}{k}\,Tr \LP(W^{\a}W_{\a}\RP)\nn
      &=& W^{\a\;a}W_{\a}^{a}
\EEA
Here we have used the cyclic property of the trace.
The SU(2)-invariance of $W^{'\,\a}W_{\a}'$ is trivial since
$W_{\a}'$ is a singlet under this group.

Now we shall demonstrate the invariance of the superpotential W, and we
start by $W_{H}$,
\BEA
  W_{H} = \mu\e^{ij}\;\hat{H}_{1}^{i}\hat{H}_{2}^{j}
     &\longrightarrow&
       \mu \e^{ij}\; \LP[e^{-2ig\hat{\La}}\hat{H}_{1}\RP]^{i}\,
                     \LP[e^{-2ig\hat{\La}}\hat{H}_{2}\RP]^{j},
                     \hspace{0.4cm} i,j = 1,2 \nn
%     &=& \mu \e^{ij}\;{\cal U}^{ik} \hat{H}_{1}^{k}\,
%                     {\cal U}^{jl} \hat{H}_{2}^{l}
%             \hspace{1.5cm} {\cal U} = e^{-2ig\hat{\La}}, \nn
     &=& \mu \e^{ij}\,{\cal U}^{ik}{\cal U}^{jl} \;
              \hat{H}_{1}^{k}\hat{H}_{2}^{l},
              \hspace{2.4cm} {\cal U} = e^{-2ig\hat{\La}}.
\EEA
In order for $W_{H}$ to be invariant we must have
\BEA
      \e^{kl} &=& \e^{ij}\,{\cal U}^{ik}{\cal U}^{jl}.
\EEA
This relation is in fact satisfied as we now will show.
The matrix ${\cal U}= e^{-2ig\hat{\La}}$ is obviously a
$2\times 2$-matrix, and its determinant is
\BEA
   \det {\cal U} =  e^{-2igTr\LP(\hat{\La}\RP)} = 1,
\EEA
since $Tr\LP(\hat{\La}\RP)\equiv Tr\LP(T^{a}\hat{\La}^{a}\RP) = 0$.
Hence ${\cal U}$ is  an $SU(2)$-matrix.
Then ${\cal U}$, as any $SU(2)$-matrix, can be written as
\BEA
    {\cal U} &=& \LP( \BA{cc}
          \hat{A}               &   \hat{B} \\
          - \hat{B}^{\dagger}   &   \hat{A}^{\dagger} \EA \RP),
\EEA
with
\BEA
    \hat{A}^{\dagger}\hat{A}+\hat{B}^{\dagger}\hat{B} &=& 1.
\EEA
Here $\hat{A}$ and $\hat{B}$ are functionals of the
chiral superfields $\hat{\La}^{a}$. Their actual  dependence on
these superfields are  of no importance to us, so we will
not worry about them.

Hence
\BEA
   \e^{ij}\,{\cal U}^{ik}{\cal U}^{jl}
        &=& \LP[{\cal U}^{T}\e\,{\cal U}\RP]^{kl} \nn
        &=& \LP(\BA{cc}
               0       & \hat{A}^{\dagger}\hat{A}+\hat{B}^{\dagger}\hat{B} \\
               -\LP(\hat{A}^{\dagger}\hat{A}+\hat{B}^{\dagger}\hat{B}\RP) & 0
             \EA \RP)^{kl} \nn
        &=& \LP( \BA{rr}
              0  & 1 \\
              -1 & 0     \EA \RP)^{kl}\nn
        &=& \e^{kl},
\EEA
and $W_{H}$ is (gauge) invariant under $SU(2)$.

The invariance of $W_{Y}$ is showed as above since
$\hat{H}_{1}$ and $\hat{L}$ are both doublets under SU(2),
while $\hat{R}$ is a singlet under this group.
Thus the superpotential
$W= W_{H}+W_{Y}$ is $SU(2)$-gauge invariant.

We now would like to draw the attention towards the SUSY-breaking
term $\L_{Soft}$. Because of the particular form of $\L_{GMT}$
(cf. eqs.~\r{The Soft SUSY-Breaking Term prop 3})
the only invariance which has not been checked yet, is that of $\L_{SMT}$.
Since
\BEA
  \hat{L}^{\dagger}\hat{L}
   &\longrightarrow&
      \hat{L}^{\dagger}e^{2ig\hat{\La}}e^{-2ig\hat{\La}}\hat{L}
    = \hat{L}^{\dagger}\hat{L},
\EEA
is invariant, and the same applies for the corresponding terms
of $\hat{R}$, $\hat{H}_{1}$ and $\hat{H}_{2}$,
we may conclue that $\L_{SMT}$, and thus $\L_{Soft}$,
are SU(2)-invariant\footnote{Note
that the last two terms of $\L_{SMT}$ are invariant
for the same reason as for instance $W_{H}$.}.

Thus the total Lagrangian $\L_{S-QFD}$ is SU(2)-gauge invariant as it should.

\subsubsection{The $U(1)$-Invariance.}

Many of the invariances showed above easily generalize to U(1) with the
substitutions $2g\rightarrow g'$,
$T^{a}\hat{\La}^{a}\rightarrow Y\hat{\la}'=\hat{\La}'$.
This applies to all terms containing only vector superfields,
and terms built out of vector superfields and only one type of
chiral superfields\footnote{The different types in our
model are $\hat{L}$, $\hat{R}$,  $\hat{H}_{1}$ and  $\hat{H}_{2}$.}.

The remaining U(1)-invariance to check, is that of terms holding
two, or more, types of chiral superfields. Such terms are only
contained in the superpotential W, and the invariance is proved as
follows
\BEA
   W_{H} = \mu\e^{ij}\;\hat{H}_{1}^{i}\hat{H}_{2}^{j}
         &\longrightarrow &
%          \mu\e^{ij}\;e^{-ig'Y_{H_{1}}\hat{\la}'}
%                e^{-ig'Y_{H_{2}}\hat{\la}'}\, \hat{H}_{1}^{i}
%                      \hat{H}_{2}^{j} \nn
%         &=&
                 \mu\e^{ij}\; e^{-ig'\LP(Y_{H_{1}}+Y_{H_{2}}\RP) \hat{\la}'}\,
                 \hat{H}_{1}^{i}\hat{H}_{2}^{j} \nn
          &=&    W_{H}
\EEA
and
\BEA
   W_{Y} =
        f\e^{ij}\;\hat{H}_{1}^{i}\hat{L}^{j}\hat{R}
      &\longrightarrow &
           f\e^{ij}\; e^{-ig'\LP(Y_{H_{1}}+Y_{L}+Y_{R}\RP)\hat{\la}'}
            \hat{H}_{1}^{i}\hat{L}^{j}\hat{R} \nn
      &=&   W_{Y}
\EEA
since $Y_{H_{1}}+Y_{H_{2}} = 0$ and $Y_{H_{1}}+Y_{L}+Y_{R} = 0$
according to table~\ref{Table Superfields}. Hence the theory is U(1)-invariant
as well.

This completes the proof of the full SU(2)$\times$U(1) gauge invariance of the
theory.

\subsection{The R-Invariance.}
     \label{subsect: The R-Invariance}

The definition of R-symmetry, generated by the operator ${\bf R}$,
was introduced by the authors of
refs.~\ref{FAY75} and \ref{SAL75}. It acts on left-handed chiral superfields
$\Phi(x,\t,\tb)$, and its (right-handed) hermitian conjugated, as follows
\BEA
   {\bf R} \Phi(x,\t,\tb) &=& e^{2 i n_{\Phi}\a}\Phi(x, e^{-i\a}\t,e^{i\a}\tb )
          \label{The R-Invariance prop 1} \SL
   {\bf R} \Phi^{\dagger}(x,\t,\tb) &=&
          e^{-2 i n_{\Phi}\a}\Phi^{\dagger}(x, e^{-i\a}\t, e^{i\a}\tb ),
          \label{The R-Invariance prop 2}
\EEA
and on vector multiplets according to
\BEA
   {\bf R} V(x,\t,\tb) &=& V(x, e^{-i\a}\t, e^{i\a}\tb ).
         \label{The R-Invariance prop 3}
\EEA
Here $\a$ is a continuous real parameter, while $n_{\Phi}$ is
called the {\em R-character} of the chiral superfield $\Phi(x,\t,\tb)$.

In terms of component fields, the above transformations read
for the chiral multiplet
\BEA
   \LP.  \BA{lcr}
          A       &\longrightarrow&       e^{2in_{\Phi}\a} A \\
          \psi    &\longrightarrow&       e^{2i\LP(n_{\Phi}-\HA\RP)\a} \psi\\
          F       &\longrightarrow&       e^{2i\LP(n_{\Phi}-1\RP)\a} F
             \EA \RP\},
           \label{The R-Invariance prop 4}
\EEA
and for the vector multiplet
\BEA
   \LP.  \BA{lcr}
          f       &\longrightarrow&       f \\
          \psi    &\longrightarrow&       e^{-i\a} \psi\\
          m       &\longrightarrow&       e^{-2i\a} m\\
          V_{\mu} &\longrightarrow&       V_{\mu} \\
          \la     &\longrightarrow&       e^{i\a} \la\\
          d       &\longrightarrow&       d
          \EA \RP\}.
          \label{The R-Invariance prop 5}
\EEA
Here the transformations for the remaining components are
given by hermitian conjugation.

For products of left-handed chiral superfields we have~\cite{FAYET}
\BEA
  {\bf R} \prod_{a}\;\Phi_{a}(x,\t,\tb)
    &=& e^{2i\sum_{a}n_{a}\a}\,\prod_{a}\Phi(x, e^{-i\a}\t,e^{i\a}\tb ),
      \label{The R-Invariance prop 6}
\EEA
and the following general superfield terms are all R-invariant:
\BEA
   & &   \int d^{4}\t\; \Phi^{\dagger}(x,\t,\tb) \Phi(x,\t,\tb),
              \label{The R-Invariance prop 8}\SL
   & &   \int d^{4}\t\; \Phi^{\dagger}(x,\t,\tb) e^{V(x,\t,\tb)}
                   \Phi(x,\t,\tb),
                   \label{The R-Invariance prop 7} \SL
   & &   \int d^{4}\t\; \prod_{a}\Phi_{a}(x,\t,\tb)\,\d^{2}(\tb),
              \hspace{2cm} \mbox{if  } \sum_{a} n_{a} = 1,
              \label{The R-Invariance prop 9}\SL
   & &   \int d^{4}\t\; \prod_{a}\Phi_{a}(x,\t,\tb)\,\d^{4}(\t,\tb),
              \hspace{1.55cm} \mbox{if  } \sum_{a} n_{a} = 0,
               \label{The R-Invariance prop 10}
\EEA
%  \begin{PROOF}
%   Since the proofs of eqs.~\r{The R-Invariance prop 6} to
%   \r{The R-Invariance prop 7} are trivial in view of the
%   definitions~\r{The R-Invariance prop 1} and \r{The R-Invariance prop 2},
%   we will only prove eqs.~\r{The R-Invariance prop 9} and
%   \r{The R-Invariance prop 10}.
%
%   From  eq.~\r{The R-Invariance prop 6} it follows
%   \BEA
%       \int d^{4}\t\; \prod_{a}\Phi_{a}(x,\t,\tb)\,\d^{2}(\tb)
%        &\longrightarrow &
%          e^{2i\sum_{a}n_{a}\a} \;\int d^{4}\t\;
%              \prod_{a}\Phi_{a}(x,e^{-i\a} \t,e^{i\a} \tb)\,\d^{2}(\tb)\nn
%        &=&  e^{2i\LP(\sum_{a}n_{a}-1\RP)\a} \;\int d^{4}\t\;
%              \prod_{a}\Phi_{a}(x,\t, \tb)\,\d^{2}(\tb). \nonumber
%    \EEA
%    Here we have used the fact that a $\t$-integration is equivalent
%    to a projection as discussed in appendix~\ref{App: Grassmann Variables},
%%    and that ``$d^{4}\t$" and the Grassmann $\d$-functions
%    are not affected by the R-transformation. The operator ${\bf R}$
%    operates only on superfields.
%    In order
%    for the above expression to be invariant,
%    one must have $\sum_{a} n_{a} = 1$
%    which is in agreement with eq.~\r{The R-Invariance prop 9}.
%
% \end{PROOF}

Now returning to S-QFD, we have at once, from the above results,
that $\L_{S-QFD}$ is R-invariant if and only if
\BEA
    n_{1} + n_{2} &=& 1, \SL
    n_{1}+n_{L}+n_{R} &=& 1.
\EEA
Here we have used obvious notation, and we have chosen to give the
superfields arranged in doublets, the same R-character for convenience.
Since the R-characters of the superfields in question are  somewhat
ambiguous, we will in addition take up the convention of all n's
being positive. With the choices made in table~\ref{table: R-character},
$\L_{S-QFD}$ is R-invariant, as it should.

Before we close this chapter, we will make one concluding remark.
{}From eqs.~\r{The R-Invariance prop 9} and \r{The R-Invariance prop 10}
it is obvious that both $\int\! d^{4}\t \;W\,\d^{2}(\t)$ (from $\L_{SUSY}$)
and the soft term $\int\! d^{4}\t \;W\,\d^{4}(\t,\tb)$
can not be R-invariant at
the same time. On the other hand, R-invariance alone does not
favour one from the other. However, the unbroken S-QFD theory,
described by $\L_{SUSY}$, must have appropriate Yukawa-terms.
This implies that $\int\! d^{4}\t \;W\,\d^{2}(\t)$ must be
included in  $\L_{SUSY}$, while the soft term
$\int\! d^{4}\t \;W\,\d^{4}(\t,\tb)$
has to be excluded from $\L_{Soft}$~(due to R-invariance)
as mentioned earlier in this chapter.

%This concludes this chapter.
\begin{table}[tbh]
 \BC
 \begin{tabular}{|l|c|} \hline
  Superfields                  & {\bf R}-character \\ \hline\hline

 $\hat{L}(x,\t,\tb)$          &     $1/4$    \\
 $\hat{R}(x,\t,\tb)$          &     $1/4$    \\
 $\hat{H}_{1}(x,\t,\tb) $     &     $1/2$         \\
 $\hat{H}_{2}(x,\t,\tb) $     &     $1/2$         \\ \hline
  \end{tabular}
  \caption{The R-character of the different chiral
           superfields of S-QFD.}
        \label{table: R-character}
 \EC
\end{table}

\cleardoublepage

\chapter{Component Field Expansion of $\L_{S-QFD}$.}
   \label{CHAPTER: Component Field Expansion}

In this chapter, the component expansion of the full Lagrangian $\L_{S-QFD}$
will be developed, even if it is $\L_{SUSY}$ which will be our main consurn.
Much of the explicit calculations
are pretty lengthy and are performed in the
appendices.
The time-consuming possedure of explicitely proving the SUSY-invariance
of $\L_{SUSY}$ will also be given in this chapter.
Finally we will transform the full Lagrangian into four-component
notation.

\section{Component Expansion of $\L_{SUSY}$.}
    \label{SECT: Component Expansion}

Before we go into the component expansion of $\L_{SUSY}$, the component
form of the different superfields of the model have to be given.
This we will do now.

In the previous chapter, we arranged for one of the lepton superfields
to be an SU(2)-doublet ($\hat{L}$) and the other an singlet ($\hat{R}$).
These chiral superfields will
%in accordance with eq.~\r{Chiral Superfields prop 7}
be given the following component
expansions\footnote{These component expansions, and coming, would
been simpler in the $(y,\t)$-basis. However, this basis will not
often be used, so we have decided to work in the
$(x,\t,\tb)$-basis form the very beginning.}
\BEA
  \hat{L}(x,\t,\tb) &=&  \LP( \BA{l}   \hat{\nu}_{l}(x,\t,\tb) \\
                                       \hat{l}(x,\t,\tb)   \EA \RP)_{L} \nn
             &=&    \tilde{L}(x)
                 +  i\;\TSTB \;\P_{\mu}\tilde{L}(x)
                 -  \f{1}{4}\;\t\t\;\tb\tb\;\P^{\mu}\P_{\mu}\tilde{L}(x) \nn
             & & \mbox{}
                 +  \sqrt{2}\;\t L^{(2)}(x)
                 +  \f{i}{\sqrt{2}}\;\t\t\;\tb\bar{\s}^{\mu}\P_{\mu}L^{(2)}(x)
                 +  \t\t\;F_{L}(x),
                 \label{Component Field Expansion prop 1} \SL
  \hat{R}(x,\t,\tb)
      &=&     \hat{l}_{R}(x) \nn
       &=&    \tilde{R}(x)
            + i\,\TSTB\;\P_{\mu} \tilde{R}(x)
            - \f{1}{4} \; \t\t\;\tb\tb\;\P^{\mu}\P_{\mu} \tilde{R}(x)\nn
       & & \mbox{}
            + \sqrt{2}\;\t R^{(2)}(x)
            + \f{i}{\sqrt{2}}\;\t\t\,\tb\bar{\s}^{\mu}\P_{\mu}R^{(2)}(x)
            + \t\t\;F_{R}(x).
            \label{Component Field Expansion prop 2}
\EEA
\begin{table}[tbh]
 \BC
 \begin{tabular}{|l|c|r|r|} \hline
  Field name       & Symbol              &  Spin   &  Charge  \\ \hline \hline

  Leptons         & $L^{(2)\,1}   $     &  $1/2$  &    0      \\
                  & $L^{(2)\,2}   $     &  $1/2$  &  $-1$     \\
                  & $R^{(2)}      $     &  $1/2$  &   $1$     \\
  Sleptons        & $\tilde{L}^{1}$     &    $0$  &   $0$     \\
                  & $\tilde{L}^{2}$     &    $0$  &  $-1$     \\
                  & $\tilde{R}$         &    $0$  &  $ 1$     \\
  Higgs bosons    & $H^{1}_{1}$         &    $0$  &  $ 0$     \\
                  & $H^{1}_{2}$         &    $0$  &  $-1$     \\
                  & $H^{2}_{1}$         &    $0$  &   $1$     \\
                  & $H^{2}_{2}$         &    $0$  &   $0$     \\
  Higgsinos       & $\psi_{H_{1}}^{1}$  &  $1/2$  &   $0$     \\
                  & $\psi_{H_{1}}^{2}$  &  $1/2$  &  $-1$     \\
                  & $\psi_{H_{2}}^{1}$  &  $1/2$  &   $1$     \\
                  & $\psi_{H_{2}}^{2}$  &  $1/2$  &   $0$     \\
  Gauge bosons    & $V^{a}_{\mu}$       &    $1$  &    --     \\
                  & $V'_{\mu}$          &    $1$  &    --     \\
  Gauginos        & $\la^{a}$           &  $1/2$  &    --     \\
                  & $\la'$              &  $1/2$  &    --     \\ \hline
\end{tabular}
\EC
  \caption{A summary of the SM-fields and their superpartners present in
             the S-QFD model.
             The quantum numbers of the various fields are also summarized.
             All fermion fields are given in terms of two-component
             (Weyl) spinors.  }
          \label{Table:  Total Component fields}
\end{table}
\vspace{0.5cm}

Here the component fields are defined by
\BEA
  \BA{lllll}
     \tilde{L}(x) = \LP( \BA{c}  \tilde{\nu}_{l}(x) \\ \tilde{l}_{L}(x)\EA \RP)
            & \hspace{0.4cm} &
     L^{(2)}(x)   = \LP( \BA{c}  \nu_{l}^{(2)}(x) \\ l^{(2)}(x) \EA \RP)_{L}
            &   \hspace{0.4cm} &
     F_{L}(x) = \LP( \BA{c}  f^{\nu}(x) \\ f^{l}_{L}(x) \EA \RP),
  \EA
  \label{Component Field Expansion prop 3}
\EEA
and\footnote{The relation $\tilde{R}=\tilde{l}^{\dagger}_{R}$
(with a dagger on only one side) may seem a little bit strange at first sight.
It is introduced for convenience, and in particular to let
$\tilde{L}^{\dagger}$ and $\tilde{R}^{\dagger}$ both create negatively
charged sleptons. If we have identified $\tilde{R}= \tilde{l}_{R}$, then
$\tilde{R}^{\dagger}$ would have
created positively charged sleptons\cite{HABER}.}
\BEA
  \BA{lllll}
      \tilde{R}(x) = \tilde{l}_{R}^{\dagger}(x)
            & \hspace{0.6cm} &
     R^{(2)}(x)   = l^{(2)}_{R}(x)
            &   \hspace{0.6cm} &
     F_{R}(x) = f^{l}_{R}(x).
  \EA
  \label{Component Field Expansion prop 4}
\EEA

\begin{table}[tbh]
 \BC
 \begin{tabular}{|l|c|r|r|} \hline
  Field name       & Symbol              &  Spin   &  Charge  \\ \hline \hline

Auxiliary Lepton Fields  & $f^{\nu}$    &    $0$  &   $0$     \\
                         & $f^{l}_{L}$  &    $0$  &  $-1$     \\
                         & $f^{l}_{R}$  &    $0$  &   $1$     \\
Auxiliary Higgs Fields   & $f_{1}^{1}$  &    $0$  &   $0$     \\
                         & $f_{1}^{2}$  &    $0$  &  $-1$     \\
                         & $f_{2}^{1}$  &    $0$  &   $1$     \\
                         & $f_{2}^{2}$  &    $0$  &   $0$     \\
Auxiliary Gauge Fields   & $D^{a}$      &    $1$  &   --     \\
                         & $D'$         &    $1$  &   --     \\  \hline
\end{tabular}
\EC
  \caption{A summary of the auxiliary fields of the S-QFD model
             and their quantum numbers.}
          \label{Table:  Auxiliary Component fields}
\end{table}
\vspace{0.7cm}
In the same way, we have for the two Higgs (doublet) superfields
\BEA
  \hat{H}_{1}(x,\t,\tb) &=& \LP( \BA{c}   \hat{H}_{1}^{1}(x,\t,\tb) \\
                                          \hat{H}_{1}^{2}(x,\t,\tb) \EA \RP)\nn
%                                           \hspace{5cm}
%   \hspace{3cm}
             &=&    H_{1}(x)
                 +  i\;\TSTB \;\P_{\mu}H_{1}(x)
                 -  \f{1}{4}\;\t\t\;\tb\tb\;\P^{\mu}\P_{\mu}H_{1}(x) \nn
             & & \mbox{}
                 +  \sqrt{2}\;\t \tilde{H}_{1}^{(2)}(x)
                 +  \f{i}{\sqrt{2}}\;\t\t\;\tb\bar{\s}^{\mu}
                           \P_{\mu}\tilde{H}_{1}^{(2)}(x)
                 +  \t\t\;F_{1}(x),
                 \label{Component Field Expansion prop 5}\SL
  \hat{H}_{2}(x,\t,\tb) &=& \LP( \BA{c}   \hat{H}_{2}^{1}(x,\t,\tb) \\
                                          \hat{H}_{2}^{2}(x,\t,\tb)\EA \RP) \nn
             &=&    H_{2}(x)
                 +  i\;\TSTB \;\P_{\mu}H_{2}(x)
                 -  \f{1}{4}\;\t\t\;\tb\tb\;\P^{\mu}\P_{\mu}H_{2}(x) \nn
             & & \mbox{}
                 +  \sqrt{2}\;\t \tilde{H}_{2}^{(2)}(x)
                 +  \f{i}{\sqrt{2}}\;\t\t\;\tb\bar{\s}^{\mu}\P_{\mu}
                           \tilde{H}_{2}^{(2)}(x)
                 +  \t\t\;F_{2}(x),
                 \label{Component Field Expansion prop 6}
\EEA
where the component fields read
\BEA
 \BA{lllll}
    H_{1}(x)     = \LP( \BA{c}  H_{1}^{1}(x) \\ H_{1}^{2}(x) \EA \RP)
         & \hspace{0.4cm} &
   \tilde{H}_{1}^{(2)}(x) =
        \LP( \BA{c}  \psi^{1}_{H_{1}}(x) \\ \psi_{H_{1}}^{2}(x) \EA \RP)
         & \hspace{0.4cm} &
   F_{1}(x) = \LP( \BA{c}  f_{1}^{1}(x) \\ f_{1}^{2}(x) \EA \RP),
 \EA
    \label{Component Field Expansion prop 7}
\EEA
and
\BEA
  \BA{lllll}
    H_{2}(x)     = \LP( \BA{c}  H_{2}^{1}(x) \\ H_{2}^{2}(x) \EA \RP)
         & \hspace{0.4cm} &
    \tilde{H}_{2}^{(2)}(x) =
            \LP( \BA{c}  \psi^{1}_{H_{2}}(x) \\ \psi_{H_{2}}^{2}(x) \EA \RP)
         & \hspace{0.4cm} &
   F_{2}(x) = \LP( \BA{c}  f_{2}^{1}(x) \\ f_{2}^{2}(x) \EA \RP).
 \EA
     \label{Component Field Expansion prop 8}
\EEA
Note that {\em all} the F-fields are auxiliary fields, which later
on, when constructing the on-shell Lagrangian, will be removed
through the Euler-Lagrange equations.

Here hats ($\hat{\, \,}$), as in the previous chapter, indicate
superfields while tildes ($\tilde{\, \,}$) denote
supersymmetric partners
of the SM particles. The subscripts L and R on fermionic-fields,
mean as usual, left- and right-handed fields\footnote{When those
subscripts occur on bosonic-fields, say on $\tilde{L}_{L}$, it
only denotes a particular field and has nothing to do with
left-and right-handed fields (which are not defined for bosonic-fields).},
while the superscript ``$(2)$" means that we are dealing
with two-component (Weyl) spinors. The same goes for the
SU(2) components of the higgsino doublets $\tilde{H}_{1}^{(2)}$ and
$\tilde{H}_{1}^{(2)}$, even if the above mentioned superscript
is missing on the $\psi$'s.

The (minimal) S-QFD model also contains vector multiplets.
As a matter of convenience, we choose to work in the WZ-gauge.
In this gauge the component expansions of the SU(2)- and
U(1)-gauge superfields $\hat{V}={\bf T}^{a}\hat{V}^{a}$ and
$\hat{V}'={\bf Y} \hat{v}'$, are given by
\BEA
 \hat{V}^{a}(x,\t,\tb)
     &=& - \;\TSTB\;V^{a}_{\mu}(x)
         + i\;\t\t\;\tb\bar{\lambda}^{a}(x)
         - i\;\tb\tb\;\t\lambda^{a}(x)
         + \HA\;\t\t\,\tb\tb\;D^{a}(x),
         \label{Component Field Expansion prop 9}
\EEA
and
\BEA
\hat{v}'(x,\t,\tb)
     &=& - \;\TSTB\;V'_{\mu}(x)
         + i\;\t\t\;\tb\bar{\lambda}'(x)
         - i\;\tb\tb\;\t\lambda'(x)
         + \HA\;\t\t\,\tb\tb\;D'(x).
         \label{Component Field Expansion prop 10}
\EEA
Here $\la^{a}(x)$ and $\la'(x)$ are the two-component (Weyl) gaugino fields,
the superpartners of the (SM) gauge bosons, and the D-fields
are auxiliary fields.

With the above definitions, the Lagrangian $\L_{SUSY}$ can be
expanded in terms of  component fields.
In appendix~\ref{APP: YM-Lagrangian}, this calculation is performed
in detail, and the result is according to eq.~\r{app comp L_sub_SUSY}
\BEA
\L_{SUSY}
  &=&   \LP(D^{\mu}\tilde{L}\RP)^{\dagger}\LP(D_{\mu}\tilde{L}\RP)
      + \LP(D^{\mu}\tilde{R}\RP)^{\dagger}\LP(D_{\mu}\tilde{R}\RP)
%      \nn
%  & & \mbox{}
      - i\;\bar{L}^{(2)}\bar{\s}^{\mu}D_{\mu}L^{(2)}
      - i\;\bar{R}^{(2)}\bar{\s}^{\mu}D_{\mu}R^{(2)}  \nn
  & & \mbox{}
      +\tilde{L}^{\dagger}\LP(gT^{a}D^{a}-\HA g'D'\RP)\tilde{L}
      +\tilde{R}^{\dagger}g'D'\tilde{R}  \nn
  & & \mbox{}
      + \sqrt{2}i\;\tilde{L}^{\dagger}\LP(gT^{a}\lambda^{a}-\HA
                   g'\lambda'\RP)L^{(2)}
      - \sqrt{2}i\;\bar{L}^{(2)}\LP(gT^{a}\bar{\lambda}^{a}-
              \HA g'\bar{\lambda}'\RP)\tilde{L} \nn
  & & \mbox{}
      + \sqrt{2}i\;\tilde{R}^{\dagger}g'\lambda'R^{(2)}
      - \sqrt{2}i\;\bar{R}^{(2)}g'\bar{\lambda}'\tilde{R} \nn
  & & \mbox{}
      + F^{\dagger}_{L}F_{L}  +F^{\dagger}_{R}F_{R} \nn
  & & \mbox{}
      - i\;\bar{\lambda}^{a}\bar{\s}^{\mu}D_{\mu}\la^{a}
      - i\;\bar{\lambda}'\bar{\s}^{\mu}D_{\mu}\lambda' \nn
  & & \mbox{}
      - \f{1}{4}\;\LP(\;V^{a\;\mu\nu}V^{a}_{\mu\nu}
                       + V^{'\mu\nu}V'_{\mu\nu}\;\RP)
      + \HA\;\LP(\;D^{a}D^{a}+D'D'\;\RP) \nn
  & & \mbox{}
      + \LP(D^{\mu}H_{1}\RP)^{\dagger}\LP(D_{\mu}H_{1}\RP)
      + \LP(D^{\mu}H_{2}\RP)^{\dagger}\LP(D_{\mu}H_{2}\RP) \nn
  & & \mbox{}
      - i\;\bar{\tilde{H}}_{1}^{(2)}\bar{\s}^{\mu}D_{\mu}\tilde{H}_{1}^{(2)}
      - i\;\bar{\tilde{H}}_{2}^{(2)}\bar{\s}^{\mu}D_{\mu}\tilde{H}_{2}^{(2)}\nn
  & & \mbox{}
      + H^{\dagger}_{1}\LP(gT^{a}D^{a}-\HA g'D'\RP)H_{1}
      + H^{\dagger}_{2}\LP(gT^{a}D^{a}+\HA g'D'\RP)H_{2} \nn
  & & \mbox{}
      + \sqrt{2}i\;H^{\dagger}_{1}\LP(gT^{a}\lambda^{a}-\HA g'\lambda'\RP)
                     \tilde{H}_{1}^{(2)}
      - \sqrt{2}i\;\bar{\tilde{H}}_{1}^{(2)}\LP(gT^{a}\bar{\lambda}^{a}-
               \HA g'\bar{\lambda}'\RP) H_{1} \nn
  & & \mbox{}
      + \sqrt{2}i\;H^{\dagger}_{2}\LP(gT^{a}\lambda^{a}+\HA g'\lambda'\RP)
             \tilde{H}_{2}^{(2)}
      - \sqrt{2}i\;\bar{\tilde{H}}_{2}^{(2)}\LP(gT^{a}\bar{\lambda}^{a}
                +\HA g'\bar{\lambda}'\RP) H_{2} \nn
  & & \mbox{}
      + F^{\dagger}_{1}F_{1}  +F^{\dagger}_{2}F_{2} \nn
  & & \mbox{}
      + \mu \e^{ij}\;
           \LP[\;H^{i}_{1}F_{2}^{j}
              + H_{1}^{i\,\dagger}F_{2}^{j\,\dagger}
              + F_{1}^{i}H_{2}^{j}
              + F_{1}^{i\,\dagger}H_{2}^{j\,\dagger}
              - \tilde{H}_{1}^{(2)\,i}\tilde{H}_{2}^{(2)\,j}
              - \bar{\tilde{H}_{1}}^{(2)\,i}\bar{\tilde{H}_{2}}^{(2)\,j}
             \;\RP] \nn
  & & \mbox{}
       + f\e^{ij}\;
           \LP[\;F_{1}^{i}\tilde{L}^{j}\tilde{R}
              +  F_{1}^{i\,\dagger}\tilde{L}^{j\,\dagger}\tilde{R}^{\dagger}
              +  H_{1}^{i}F_{L}^{j}\tilde{R}
              +  H_{1}^{i\,\dagger}F_{L}^{j\,\dagger}\tilde{R}^{\dagger}\RP.\nn
  & & \mbox{} \hspace{1.8cm}
              +  H_{1}^{i}\tilde{L}^{j} F_{R}
              +  H_{1}^{i\,\dagger}\tilde{L}^{j\,\dagger}F_{R}^{\dagger}
              -  \tilde{H}_{1}^{(2)\,i}L^{(2)\,j}\tilde{R}
              -  \bar{\tilde{H}_{1}}^{(2)\,i}\bar{L}^{(2)\,j}
               \tilde{R}^{\dagger} \nn
  & & \mbox{} \hspace{1.8cm} \LP.
              -  H_{1}^{i}L^{(2)\,j}R^{(2)}
              -  H_{1}^{i\,\dagger}\bar{L}^{(2)\,j}\bar{R}^{(2)}
              -  R^{(2)}\tilde{H}_{1}^{(2)\,i}\tilde{L}^{j}
              -  \bar{R}^{(2)}\bar{\tilde{H}_{1}}^{(2)\,i}\tilde{L}^{j\,
             \dagger}\;\RP]
  \nmb
     +   t.d.
           \label{Off-Shall L sub SUSY}
\EEA
Here t.d. means a total derivative and
$D_{\mu}$ is the standard $SU(2)\times U(1)$-covariant derivative
defined by
\BEA
   D_{\mu} &=& \P_{\mu} + i g{\bf T}^{a} V^{a}_{\mu}
               +ig'\f{{\bf Y}}{2} V'_{\mu}, \hspace{1cm}
    a = 1,2,3.
\EEA
Note that when $D_{\mu}$ operates on e.g. the gauginos $\la^{a}$ and $\la'$,
which lay in the adjoint representation of
SU(2) and U(1) respectively, i.e.
\BEA
       \LP[T^{c}_{adj}\RP]^{ab} &=& -if^{cab}, \nn
         Y_{adj}&=& 0, \nonumber
\EEA
we have~(cf. eq.~\r{app: Component Field Expansion prop 10asa1})
\BEA
   D_{\mu} \la^{a}
              &=& \P_{\mu}\la^{a} - g f^{abc} V^{b}_{\mu}\la^{c},
                        \label{Component Field Expansion prop 10asa1} \SL
   D_{\mu} \la' &=& \P_{\mu} \la'.
                       \label{Component Field Expansion prop 10asa2}
\EEA

The various fields of the  Lagrangian~\r{Off-Shall L sub SUSY}
are summarized in tables~\ref{Table:  Total Component fields} and
\ref{Table:  Auxiliary Component fields}.
Note that this Lagrangian contains auxiliary fields, i.e. F- and D-fields,
and thus is off-shell.

\section{Elimination of the Auxiliary Fields.}
  \label{sect. Elimination of the Auxiliary Fields}

The aim of this section will be to construct the on-shell Lagrangian,
i.e. to eliminate the different auxiliary fields given in
table~\ref{Table:  Auxiliary Component fields}.
When we do so, we will see that mass terms for Higgs-bosons
and different interaction terms between Higgses, Leptons and Sleptons,
without any Lepton-Slepton interaction, will appear.

If we pick all the terms from the off-shell Lagrangian~\r{Off-Shall L sub SUSY}
containing Lepton-, Higgs- and Gauge-auxiliary fields
(F- and D-fields) we get
\BEA
  \L_{Aux} &=& \L_{Aux-F} + \L_{Aux-D},
     \label{Elimination of the Auxiliary Fields prop 1}
\EEA
with
\BEA
    \L_{Aux-F}  &=&
          F_{L}^{\dagger}F_{L}
       +  F_{R}^{\dagger}F_{R}
       +  F_{1}^{\dagger}F_{1}
       +  F_{2}^{\dagger}F_{2} \nn
   & & \mbox{}
      + \mu\;\e^{ij}\LP[\,
            H_{1}^{i}F_{2}^{j}
         +   H_{1}^{i\,\dagger}F_{2}^{j\,\dagger}
         +   F_{1}^{i}H_{2}^{j}
         +   F_{1}^{i\,\dagger}H_{2}^{j\,\dagger}\,\RP] \nn
    & & \mbox{}
    +   f\;\e^{ij}\LP[\,
            F_{1}^{i}\tilde{L}^{j}\tilde{R}
         +  F_{1}^{i\,\dagger}\tilde{L}^{j\,\dagger}\tilde{R}^{\dagger}
         +  H_{1}^{i}F_{L}^{j}\tilde{R}
         +  H_{1}^{i\,\dagger}F_{L}^{j\,\dagger}\tilde{R}^{\dagger}\RP. \nn
     & & \mbox{}  \LP. \hspace{1.2cm}
         +  H_{1}^{i}\tilde{L}^{j}F_{R}
         +  H_{1}^{i\,\dagger}\tilde{L}^{j\,\dagger}F_{R}^{\dagger} \, \RP],
         \label{Elimination of the Auxiliary Fields prop 2}
\EEA
and
\BEA
   \L_{Aux-D} &=&
       \HA\;\LP(\;D^{a}D^{a}+D'D'\;\RP) \nn
  & & \mbox{}
      +\tilde{L}^{\dagger}\LP(gT^{a}D^{a}-\HA g'D'\RP)\tilde{L}
      +\tilde{R}^{\dagger}g'D'\tilde{R}  \nn
  & & \mbox{}
      + H^{\dagger}_{1}\LP(gT^{a}D^{a}-\HA g'D'\RP)H_{1}
      + H^{\dagger}_{2}\LP(gT^{a}D^{a}+\HA g'D'\RP)H_{2}.
      \label{Elimination of the Auxiliary Fields prop 3}
\EEA

We will now show that these fields can be eliminated through
the Euler-Lagrange equations~\cite{HAT vol 75}
\BEA
     \PD{\L}{\phi} - \P_{\mu}\PD{\L}{(\P_{\mu}\phi)} &=& 0,  \nonumber
       \label{Euler-Lagrange Equation}
\EEA
where $\phi$ is {\em any} (also hermitian conjugated) Minkowski field.
Formally auxiliary fields are defined as fields having no
kinetic terms.
Thus, this definition immediately yields that the Euler-Lagrange equations
for auxiliary fields
simplify to $\PD{\L}{\phi}=0$.

Applying these simplified equations to various auxiliary F-fields
%to the Lagrangian~\r{Off-Shall L sub SUSY},
yields the following relations
\BEA
    F_{L}^{j\,\dagger}   &=& -f\;\e^{ij}H_{1}^{i}\tilde{R}, \SL
    F_{R}^{\dagger}      &=& -f\;\e^{ij}H_{1}^{i}\tilde{L}^{j}, \SL
    F_{1}^{i\,\dagger}   &=& -\mu\;\e^{ij}H_{2}^{j}
                             - f\;\e^{ij}\tilde{L}^{j}\tilde{R},\SL
    F_{2}^{j\,\dagger}   &=& -\mu\;\e^{ij}H_{1}^{i}.
\EEA
Expressions for, say $F_{L}^{j}$ and so on, are given by
hermitian conjugation of the above relations.
Substituting these expressions for the F-fields into
eq.~\r{Elimination of the Auxiliary Fields prop 2} yields according to
eq.~\r{aux prop 6aaaaaa}
\BEA
  \L_{Aux-F} &=&
        -\mu^{2}\, H_{1}^{\dagger}H_{1}
        -\mu^{2}\, H_{2}^{\dagger}H_{2}
        -\mu f \LP[\, H_{2}^{\dagger}\tilde{L}\,\tilde{R}
                  +\tilde{L}^{\dagger}H_{2}\,\tilde{R}^{\dagger}\,\RP]\nn
      & & \mbox{}
     - f^{2}\LP[ \,
       \tilde{L}^{\dagger}\tilde{L}\,\tilde{R}^{\dagger}\tilde{R}
     + H_{1}^{\dagger}H_{1}\LP( \tilde{L}^{\dagger}\tilde{L}
                                + \tilde{R}^{\dagger}\tilde{R} \RP)
    - H_{1}^{\dagger}\tilde{L}\LP(H_{1}^{\dagger}\tilde{L}\RP)^{\dagger}\,\RP].
\EEA
Note that mass terms for the Higgs bosons and Higgs-Lepton
and Lepton-Lepton interactions  have now been generated
as we clamed at the beginning of this section.

The same program for the D-fields gives
\BEA
  D^{a} &=& -g\LP[\,
            \tilde{L}^{\dagger}T^{a}\tilde{L}
          + H_{1}^{\dagger}T^{a}H_{1}
          + H_{2}^{\dagger}T^{a}H_{2} \RP] , \SL
  D' &=&   \f{g'}{2}\,\tilde{L}^{\dagger}\tilde{L}
         - g'\, \tilde{R}^{\dagger}\tilde{R}
         + \f{g'}{2}\, H_{1}^{\dagger}H_{1}
         - \f{g'}{2}\, H_{2}^{\dagger}H_{2},
\EEA
and according to eq.~\r{aux prop 9} this means for $\L_{Aux-D}$
\BEA
  \L_{Aux-D}
        &=& -\f{g^2}{2}
         \LP(\,\tilde{L}^{\dagger}T^{a}\tilde{L}
        +H_{1}^{\dagger}T^{a}H_{1}+H_{2}^{\dagger}T^{a}H_{2}\,\RP)
         \LP(\,\tilde{L}^{\dagger}T^{a}\tilde{L}
       +H_{1}^{\dagger}T^{a}H_{1}+H_{2}^{\dagger}T^{a}H_{2}\,\RP)
    \nmb
        -\f{g'^2}{8}
          \LP(\,\tilde{L}^{\dagger}\tilde{L}
              -2\tilde{R}^{\dagger}\tilde{R}
          +H_{1}^{\dagger}H_{1}-H_{2}^{\dagger}H_{2}\,\RP)^2.
\EEA
Now Higgs-Higgs, Higgs-Slepton and Slepton-Slepton interactions have
come into play.

By substituting the expression for $\L_{Aux}$ back
into $\L_{SUSY}$, the on-shell Lagrangian is obtained.
According to eq.~\r{apendix On-Shall L sub SUSY}
the result is
\BEA
\L_{SUSY}
  &=&   \LP(D^{\mu}\tilde{L}\RP)^{\dagger}\LP(D_{\mu}\tilde{L}\RP)
      + \LP(D^{\mu}\tilde{R}\RP)^{\dagger}\LP(D_{\mu}\tilde{R}\RP)
      - i\;\bar{L}^{(2)}\bar{\s}^{\mu}D_{\mu}L^{(2)}
      - i\;\bar{R}^{(2)}\bar{\s}^{\mu}D_{\mu}R^{(2)}  \nn
  & & \mbox{}
      + \sqrt{2}i\;\tilde{L}^{\dagger}
        \LP(gT^{a}\lambda^{a}-\HA g'\lambda'\RP)L^{(2)}
      - \sqrt{2}i\;\bar{L}^{(2)}\LP(gT^{a}\bar{\lambda}^{a}
        -\HA g'\bar{\lambda}'\RP)\tilde{L} \nn
  & & \mbox{}
      + \sqrt{2}i\;\tilde{R}^{\dagger}g'\lambda'R^{(2)}
      - \sqrt{2}i\;\bar{R}^{(2)}g'\bar{\lambda}'\tilde{R} \nn
  & & \mbox{}
      - i\;\bar{\lambda}^{a}\bar{\s}^{\mu} D_{\mu}\lambda^{a}
      - i\;\bar{\lambda}'\bar{\s}^{\mu} D_{\mu}\lambda'
      - \f{1}{4}\;\LP(\;V^{a\;\mu\nu}V^{a}_{\mu\nu}
                       + V^{'\mu\nu}V'_{\mu\nu}\;\RP) \nn
  & & \mbox{}
      + \LP(D^{\mu}H_{1}\RP)^{\dagger}\LP(D_{\mu}H_{1}\RP)
      + \LP(D^{\mu}H_{2}\RP)^{\dagger}\LP(D_{\mu}H_{2}\RP) \nn
  & & \mbox{}
      - i\;\bar{\tilde{H}}_{1}^{(2)}\bar{\s}^{\mu}D_{\mu}\tilde{H}_{1}^{(2)}
      - i\;\bar{\tilde{H}}_{2}^{(2)}\bar{\s}^{\mu}D_{\mu}\tilde{H}_{2}^{(2)}\nn
  & & \mbox{}
      + \sqrt{2}i\;H^{\dagger}_{1}\LP(gT^{a}\lambda^{a}-\HA g'\lambda'\RP)
             \tilde{H}_{1}^{(2)}
      - \sqrt{2}i\;\bar{\tilde{H}}_{1}^{(2)}\LP(gT^{a}\bar{\lambda}^{a}
       -\HA g'\bar{\lambda}'\RP) H_{1} \nn
  & & \mbox{}
      + \sqrt{2}i\;H^{\dagger}_{2}\LP(gT^{a}\lambda^{a}+\HA g'\lambda'\RP)
\tilde{H}_{2}^{(2)}
      - \sqrt{2}i\;\bar{\tilde{H}}_{2}^{(2)}
        \LP(gT^{a}\bar{\lambda}^{a}+\HA g'\bar{\lambda}'\RP) H_{2} \nn
  & & \mbox{}
      -  \e^{ij}\;
           \LP[\;\mu\LP(\,
               \tilde{H}_{1}^{(2)\,i}\tilde{H}_{2}^{(2)\,j}
              + \bar{\tilde{H}}_{1}^{(2)\,i}\bar{\tilde{H}}_{2}^{(2)\,j}\,\RP)
       + f\LP(\,
                \tilde{H}^{(2)\,i}_{1} L^{(2)\,j}\tilde{R}
              +  \bar{\tilde{H}}_{1}^{(2)\,i}
        \bar{L}^{(2)\,j}\tilde{R}^{\dagger}\,\RP) \RP. \nn
  & & \mbox{} \hspace{1.3cm} \LP.
      +  f\LP(
                H_{1}^{i}L^{(2)\,j}R^{(2)}
              +  H_{1}^{i\,\dagger}\bar{L}^{(2)\,j}\bar{R}^{(2)}
              +  R^{(2)}\tilde{H}_{1}^{(2)\,i}\tilde{L}^{j}
              +  \bar{R}^{(2)}\bar{\tilde{H}_{1}}^{(2)\,i}
         \tilde{L}^{j\,\dagger}\,\RP)\RP] \nn
  & & \mbox{}
        -\mu^{2}\, H_{1}^{\dagger}H_{1}
        -\mu^{2}\, H_{2}^{\dagger}H_{2}
        -\mu f \LP[\, H_{2}^{\dagger}\tilde{L}\,\tilde{R}
                  +\tilde{L}^{\dagger}H_{2}\,\tilde{R}^{\dagger}\,\RP]\nn
   & & \mbox{}
     - f^{2}\LP[ \,
       \tilde{L}^{\dagger}\tilde{L}\,\tilde{R}^{\dagger}\tilde{R}
     + H_{1}^{\dagger}H_{1}\LP( \tilde{L}^{\dagger}\tilde{L}
                                + \tilde{R}^{\dagger}\tilde{R} \RP)
     - H_{1}^{\dagger}\tilde{L}\LP(H_{1}^{\dagger}\tilde{L}\RP)^{\dagger}\,\RP]
     \nmb
     -\f{g^2}{2}
         \LP(\,\tilde{L}^{\dagger}T^{a}\tilde{L}
        +H_{1}^{\dagger}T^{a}H_{1}+H_{2}^{\dagger}T^{a}H_{2}\,\RP)
         \LP(\,\tilde{L}^{\dagger}T^{a}\tilde{L}
          +H_{1}^{\dagger}T^{a}H_{1}+H_{2}^{\dagger}T^{a}H_{2}\,\RP)
      \nmb
        - \f{g'^2}{8} \LP(\,\tilde{L}^{\dagger}\tilde{L}-2\tilde{R}
      ^{\dagger}\tilde{R}+H_{1}^{\dagger}H_{1}-H_{2}^{\dagger}H_{2}\,\RP)^2
      + t.d.
           \label{On-Shall L sub SUSY}
\EEA
This concludes this section.

\sloppy

\section{Introducing the Photon-, W- and Z-Gauge Boson Fields.}

\fussy

In order for our model to be realistic, the sector of the theory
containing the SM-particles has to coincide with non-supersymmetric
QFD. In particular this means that the photon and
heavy W- and Z-bosons have to be present. However, ``generation"  of
heavy gauge bosons requires some sort of gauge symmetry breaking as in the SM,
and this will be discussed in detail in the next chapter.
Nevertheless, it is practical at this stage to introduce the
W- and Z-gauge fields even if they before
gauge symmetry breaking are massless.

In analogy with the Standard Model
%ter~\ref{Chapter: Recapitulation of the Standard Model}
we define
\BEA
   A_{\mu}(x) &=&   \cos\theta_{\mbox{w}}\, V'_{\mu}(x)
               + \sin\theta_{\mbox{w}}\,V^{3}_{\mu}(x),
               \label{Introducing the Photon... prop 1}\SL
   Z_{\mu}(x)  &=&   - \sin\theta_{\mbox{w}}\,V'_{\mu}(x)
                  + \cos\theta_{\mbox{w}} \,V^{3}_{\mu}(x),
                  \label{Introducing the Photon... prop 2}\SL
   W^{\pm}_{\mu}(x) &=& \f{V^{1}_{\mu}(x) \mp i V^{2}_{\mu}(x)}{\sqrt{2}},
                  \label{Introducing the Photon... prop 3}
\EEA
and for the corresponding spin-$1/2$ gauginos
\BEA
   \la_{A}(x)   &=&     \cos\theta_{\mbox{w}} \,\la'(x)
                   + \sin\theta_{\mbox{w}} \,\la^{3}(x),
                   \label{Introducing the Photon... prop 4}\SL
   \la_{Z}(x)   &=&   - \sin\theta_{\mbox{w}} \,\la'(x)
                   + \cos\theta_{\mbox{w}} \,\la^{3}(x),
                   \label{Introducing the Photon... prop 5}\SL
   \la^{\pm}(x) &=& \f{\la^{1}(x) \mp i \la^{2}(x)}{\sqrt{2}}.
                   \label{Introducing the Photon... prop 6}
\EEA
With these definitions the $SU(2)\times U(1)$-covariant derivative becomes
(cf. eq.~\r{Cov. deriv.})
\BEA
  D_{\mu} &=& \P_{\mu}+ig{\bf T}^{a}V^{a}_{\mu}+ig'\f{{\bf Y}}{2} V'_{\mu} \nn
      &=&    \P_{\mu}
          + \f{ig}{\sqrt{2}}  {\bf T}^{+} W^{+}_{\mu}
          + \f{ig}{\sqrt{2}}  {\bf T}^{-} W^{-}_{\mu}
          + i e {\bf Q} A_{\mu}
          + \f{ig}{\cos\theta_{\mbox{w}}}
            \LP[\, {\bf T}^{3}-{\bf Q}\sin^{2}\theta_{\mbox{w}}\,\RP]
Z_{\mu},\hspace{1cm}
                  \label{Introducing the Photon... prop 7}
\EEA
where the charge operator ${\bf Q}$ (with eigenvalues in units of the
elementary charge ``e") is
\BEA
     {\bf Q}       &=&  {\bf T}^{3} + \f{{\bf Y}}{2},
     \label{Introducing the Photon... prop 8}
\EEA
and
\BEA
     {\bf T}^{\pm} &=&  {\bf T}^{1} \pm i {\bf T}^{2}.
     \label{Introducing the Photon... prop 9}
\EEA
It is important to note that {\bf Q} and  the ${\bf T}$'s  are
assumed to operate on the same field as $D_{\mu}$.
For instance, if $D_{\mu}$ operates on an
SU(2)-doublet, $T^{a} = \s^{a}/2$, and $D_{\mu}$ is a $2\times 2$-matrix,
while for an SU(2)-singlet $T^{a} = 0$, and $D_{\mu}$ is no matrix at all.

In terms of the new fields
\r{Introducing the Photon... prop 1}--\r{Introducing the Photon... prop 6},
the Lagrangian $\L_{SUSY}$, in two-component notation, can be obtained from
appendix~\ref{App: 4 comp notation}
by substituting for the various terms of eq.~\r{On-Shall L sub SUSY}
rewritten in this  appendix.
%~\ref{App: 4 comp notation}.

Nevertheless, the result reads
\BEA
\L_{SUSY}
  &=&   \LP(D^{\mu}\tilde{L}\RP)^{\dagger}\LP(D_{\mu}\tilde{L}\RP)
      + \LP(D^{\mu}\tilde{R}\RP)^{\dagger}\LP(D_{\mu}\tilde{R}\RP)
      - i\;\bar{L}^{(2)}\bar{\s}^{\mu}D_{\mu}L^{(2)}
      - i\;\bar{R}^{(2)}\bar{\s}^{\mu}D_{\mu}R^{(2)}  \nn
  & & \mbox{}
   + ig\LP(    \tilde{L}^{\dagger}T^{+}L^{(2)}\la^{+}
            -  \bar{\la}^{+}\bar{L}^{(2)}T^{-}\tilde{L}   \RP)
      + ig\LP(    \tilde{L}^{\dagger}T^{-}L^{(2)}\la^{-}
            -  \bar{\la}^{-}\bar{L}^{(2)}T^{+}\tilde{L}   \RP) \nn
  \MB
    + \sqrt{2}ie\,Q_{i}\LP(
            \tilde{L}^{\dagger\,i} L^{(2)\,i} \la_{A}
          - \bar{\la}_{A}\bar{L}^{(2)\,i}\tilde{L}^{i}   \RP) \nn
  \MB
     + \f{\sqrt{2}ig}{\CWA}\LP( {\cal T}_{i}^{3}- Q_{i}\SWAS \RP)
         \LP[   \tilde{L}^{\dagger\,i}L^{(2)\,i}\la_{Z}
              - \bar{\la}_{Z}\bar{L}^{(2)\,i}\tilde{L}^{i}  \RP] \nn
   \MB
      +   \sqrt{2}i e \LP(\tilde{R}^{\dagger}\;R^{(2)}\la_{A}
          -\bar{\la}_{A}\bar{R}^{(2)}\;\tilde{R}\RP)
         -\sqrt{2}i g \f{\SWAS}{\CWA}
            \LP(\tilde{R}^{\dagger}\;R^{(2)}\la_{Z}-
                \bar{\la}_{Z}\bar{R}^{(2)}\;\tilde{R}\RP)
  \nmb
     -i\,\bar{\la}^{+}\bar{\s}^{\mu}\P_{\mu}\la^{+}
         -i\,\bar{\la}^{-}\bar{\s}^{\mu}\P_{\mu}\la^{-}
         -i\,\bar{\la}_{A}\bar{\s}^{\mu}\P_{\mu}\la_{A}
         -i\,\bar{\la}_{Z}\bar{\s}^{\mu}\P_{\mu}\la_{Z}
     \nmb
         +  g\CWA \LP[
                 \LP(\bar{\la}_{Z}\bar{\s}^{\mu}\la^{-}
             -\bar{\la}^{+}\bar{\s}^{\mu}\la_{Z}\RP)W^{+}_{\mu}
                -\LP(\bar{\la}_{Z}\bar{\s}^{\mu}\la^{+}
            -\bar{\la}^{-}\bar{\s}^{\mu}\la_{Z}\RP)W^{-}_{\mu}
                \RP.
     \nmb \hspace{2cm} \LP.
                + \LP(\bar{\la}^{+}\bar{\s}^{\mu}\la^{+}
          -\bar{\la}^{-}\bar{\s}^{\mu}\la^{-}\RP)Z_{\mu}\RP]
     \nmb
          + e \LP[
                  \LP(\bar{\la}_{A}\bar{\s}^{\mu}\la^{-}
             -\bar{\la}^{+}\bar{\s}^{\mu}\la_{A}\RP)W^{+}_{\mu}
                 - \LP(\bar{\la}_{A}\bar{\s}^{\mu}\la^{+}
               -\bar{\la}^{-}\bar{\s}^{\mu}\la_{A}\RP)W^{-}_{\mu}\RP.
     \nmb \hspace{0.8cm} \LP.
                +  \LP(\bar{\la}^{+}\bar{\s}^{\mu}\la^{+}
             -\bar{\la}^{-}\bar{\s}^{\mu}\la^{-}\RP)A_{\mu} \RP]
    \nmb
      - \f{1}{4}\;\LP(
            {\cal W}^{+\,\mu\nu}{\cal W}^{-}_{\mu\nu}
        + {\cal W}^{-\,\mu\nu}{\cal W}^{+}_{\mu\nu}
        + {\cal A}^{\mu\nu}{\cal A}_{\mu\nu}
        + {\cal Z}^{\mu\nu}{\cal Z}_{\mu\nu} \RP)\nn
  \MB
      + \LP(D^{\mu}H_{1}\RP)^{\dagger}\LP(D_{\mu}H_{1}\RP)
      + \LP(D^{\mu}H_{2}\RP)^{\dagger}\LP(D_{\mu}H_{2}\RP) \nn
  \MB
      - i\;\bar{\tilde{H}}_{1}^{(2)}\bar{\s}^{\mu}D_{\mu}\tilde{H}_{1}^{(2)}
      - i\;\bar{\tilde{H}}_{2}^{(2)}\bar{\s}^{\mu}D_{\mu}\tilde{H}_{2}^{(2)}\nn
  \MB
      + ig\LP(    H_{1}^{\dagger}T^{+}\tilde{H}_{1}^{(2)}\la^{+}
            -  \bar{\la}^{+}\bar{\tilde{H}}_{1}^{(2)}T^{-}H_{1}   \RP)
      + ig\LP(    H_{1}^{\dagger}T^{-}\tilde{H}_{1}^{(2)}\la^{-}
            -  \bar{\la}^{-}\bar{\tilde{H}}_{1}^{(2)}T^{+}H_{1}   \RP) \nn
  \MB
    + \sqrt{2}ie\,Q_{i}\LP(
            H_{1}^{\dagger\,i} \tilde{H}_{1}^{(2)\,i} \la_{A}
          - \bar{\la}_{A}\bar{\tilde{H}}_{1}^{(2)\,i}H_{1}^{i}   \RP) \nn
  \MB
     + \f{\sqrt{2}ig}{\CWA}\LP( {\cal T}_{i}^{3}- Q_{i}\SWAS \RP)
         \LP[   H_{1}^{\dagger\,i}\tilde{H}_{1}^{(2)\,i}\la_{Z}
              - \bar{\la}_{Z}\bar{\tilde{H}}_{1}^{(2)\,i}H_{1}^{i}  \RP] \nn
  \MB
        + ig\LP(    H_{2}^{\dagger}T^{+}\tilde{H}_{2}^{(2)}\la^{+}
            -  \bar{\la}^{+}\bar{\tilde{H}}_{2}^{(2)}T^{-}H_{2}   \RP)
            + ig\LP(    H_{2}^{\dagger}T^{-}\tilde{H}_{2}^{(2)}\la^{-}
            -  \bar{\la}^{-}\bar{\tilde{H}}_{2}^{(2)}T^{+}H_{2}   \RP) \nn
  \MB
    + \sqrt{2}ie\,Q_{i}\LP(
            H_{2}^{\dagger\,i} \tilde{H}_{2}^{(2)\,i} \la_{A}
          - \bar{\la}_{A}\bar{\tilde{H}}_{2}^{(2)\,i}H_{2}^{i}   \RP) \nn
  \MB
     + \f{\sqrt{2}ig}{\CWA}\LP( {\cal T}_{i}^{3}- Q_{i}\SWAS \RP)
         \LP[   H_{2}^{\dagger\,i}\tilde{H}_{2}^{(2)\,i}\la_{Z}
              - \bar{\la}_{Z}\bar{\tilde{H}}_{2}^{(2)\,i}H_{2}^{i}  \RP] \nn
  \MB
 -  \e^{ij}\;
           \LP[\;\mu\LP(\,
               \tilde{H}_{1}^{(2)\,i}\tilde{H}_{2}^{(2)\,j}
              + \bar{\tilde{H}}_{1}^{(2)\,i}\bar{\tilde{H}}_{2}^{(2)\,j}\,\RP)
       + f\LP(\,
                \tilde{H}_{1}^{(2)\,i}L^{(2)\,j}\tilde{R}
              +  \bar{\tilde{H}_{1}}^{(2)\,i}\bar{L}^{(2)\,j}
             \tilde{R}^{\dagger}\,\RP) \RP. \nn
  & & \mbox{} \hspace{1.3cm} \LP.
      +  f\LP(
                H_{1}^{i}L^{(2)\,j}R^{(2)}
              +  H_{1}^{i\,\dagger}\bar{L}^{(2)\,j}\bar{R}^{(2)}
              +  R^{(2)}\tilde{H}_{1}^{(2)\,i}\tilde{L}^{j}
              +  \bar{R}^{(2)}\bar{\tilde{H}_{1}}^{(2)\,i}
             \tilde{L}^{j\,\dagger}\,\RP)\RP] \nn
  & & \mbox{}
        -\mu^{2}\, H_{1}^{\dagger}H_{1}
        -\mu^{2}\, H_{2}^{\dagger}H_{2}
        -\mu f \LP[\, H_{2}^{\dagger}\tilde{L}\,\tilde{R}
                  +\tilde{L}^{\dagger}H_{2}\,\tilde{R}^{\dagger}\,\RP]\nn
   & & \mbox{}
     - f^{2}\LP[ \,
       \tilde{L}^{\dagger}\tilde{L}\,\tilde{R}^{\dagger}\tilde{R}
     + H_{1}^{\dagger}H_{1}\LP( \tilde{L}^{\dagger}\tilde{L}
                                + \tilde{R}^{\dagger}\tilde{R} \RP)
     - H_{1}^{\dagger}\tilde{L}\LP(H_{1}^{\dagger}
           \tilde{L}\RP)^{\dagger}\,\RP] \nn
   & & \mbox{}
         -\f{g^2}{2}
         \LP(\,\tilde{L}^{\dagger}T^{a}\tilde{L}+H_{1}^{\dagger}
                 T^{a}H_{1}+H_{2}^{\dagger}T^{a}H_{2}\,\RP)
         \LP(\,\tilde{L}^{\dagger}T^{a}\tilde{L}+H_{1}^{\dagger}
            T^{a}H_{1}+H_{2}^{\dagger}T^{a}H_{2}\,\RP)
      \nmb
        - \f{g'^2}{8} \LP(\,\tilde{L}^{\dagger}\tilde{L}
             -2\tilde{R}^{\dagger}\tilde{R}+H_{1}^{\dagger}H_{1}
            -H_{2}^{\dagger}H_{2}\,\RP)^2
      + t.d.
     \label{Introducing the Photon... prop 10}
\EEA
Here, cf. eqs.~\r{KIN A}, \r{KIN Z} , \r{KIN W1} and \r{KIN W2},
\BEA
{\cal A}_{\mu\nu} &=&   \cos\theta_{\mbox{w}}\, V'_{\mu\nu}
               + \sin\theta_{\mbox{w}}\,V^3_{\mu\nu}   \nn
  &=&  A_{\mu\nu} + ie\LP(\, W^{+}_{\mu}W^{-}_{\nu}
                      -  W^{-}_{\mu}W^{+}_{\nu}\,\RP),
                      \label{Introducing the Photon... prop 11}\SL
{\cal Z}_{\mu\nu} &=&    - \sin\theta_{\mbox{w}}\,V'_{\mu\nu}
                  + \cos\theta_{\mbox{w}} \,V^3_{\mu\nu} \nn
 &=&   Z_{\mu\nu} + ig\cos\theta_{\mbox{w}}\LP(\, W^{+}_{\mu}W^{-}_{\nu}
                                           -  W^{-}_{\mu}W^{+}_{\nu}\,\RP),
                      \label{Introducing the Photon... prop 12}\SL
{\cal W}^{+}_{\mu\nu} &=& \f{V^{1}_{\mu\nu} - i V^{2}_{\mu\nu}}{\sqrt{2}}\nn
   &=&   W^{+}_{\mu\nu}
      + ie\LP(A_{\mu}W^{+}_{\nu} - W^{+}_{\mu}A_{\nu}\RP)
      +ig \cos\theta_{\mbox{w}}\LP( Z_{\mu}W^{+}_{\nu}-W^{+}_{\mu}Z_{\nu}\RP),
      \label{Introducing the Photon... prop 13}\SL
{\cal W}^{-}_{\mu\nu} &=& \f{V^{1}_{\mu\nu} + i V^{2}_{\mu\nu}}{\sqrt{2}}\nn
   &=&   W^{-}_{\mu\nu}
      - ie\LP(A_{\mu}W^{-}_{\nu}-W^{-}_{\mu}A_{\nu}\RP)
      -ig\cos\theta_{\mbox{w}}\LP(Z_{\mu}W^{-}_{\mu} - W^{-}_{\mu}Z_{\nu} \RP),
      \label{Introducing the Photon... prop 14}
\EEA
and
$A_{\mu\nu}$, $Z_{\mu\nu}$ and $W^{\pm}_{\mu\nu}$ are the usual
fieldstrengths given by
\BEA
  A_{\mu\nu} &=& \P_{\mu}A_{\nu}-\P_{\nu}A_{\mu},
  \label{Introducing the Photon... prop 15}\SL
  Z_{\mu\nu} &=& \P_{\mu}Z_{\nu}-\P_{\nu}Z_{\mu},
  \label{Introducing the Photon... prop 16}\SL
  W^{\pm}_{\mu\nu} &=& \P_{\mu}W^{\pm}_{\nu}-\P_{\nu}W^{\pm}_{\mu}.
  \label{Introducing the Photon... prop 17}
\EEA
Note that the ``scripted kinetic terms" are defined in complete analogy with
eqs.~\r{Introducing the Photon... prop 1}--\r{Introducing the Photon... prop 3}
and that they also contain interaction terms
for the gauge bosons.

\section{Introducing Four-Component Spinors.}

In order to make use of the Lagrangian~\r{Introducing the Photon... prop 10}
in field theoretical calculations, it is  practical to express it
in terms of four-component spinors. This will be done in this section.

The interactions of the gauge-fermions of
eq.~\r{Introducing the Photon... prop 10} suggest that we introduce the
Majorana spinors
\BEA
  \tilde{A}(x) &=& \LP( \BA{r}  -i \la_{A}(x) \\
                              i \bar{\la}_{A}(x)   \EA   \RP),
             \label{Introducing Four-Component Spinors prop 1} \SL
  \tilde{Z}(x) &=& \LP( \BA{r}   -i \la_{Z}(x) \\
                               i \bar{\la}_{Z}(x)   \EA   \RP),
             \label{Introducing Four-Component Spinors prop 2}
\EEA
and the Dirac spinors
\BEA
  \tilde{W}(x) &=& \LP( \BA{r}   -i  \la^{+}(x) \\
                               i  \bar{\la}^{-}(x)   \EA   \RP),
       \label{Introducing Four-Component Spinors prop 3}\SL
  \tilde{W}^{c}(x) &=& \LP( \BA{r}   -i  \la^{-}(x) \\
                               i  \bar{\la}^{+}(x)   \EA   \RP).
       \label{Introducing Four-Component Spinors prop 3aaa}
\EEA
Here the Photino $\tilde{A}(x)$ and the Zino $\tilde{Z}(x)$ are
neutral fields, while the Wino-field describes charged ($\pm e$) Winos.
The state $\tilde{W}^{c}$ is the charge conjugated  of the
Wino-state $\tilde{W}$ (cf. eq.~\r{Charge conjugation}).

In sect.~\ref{SECT: Component Expansion} we saw that the Higgs-sector contains
two charged and neutral states~(cf. table~\ref{Table:  Total Component
fields}).
Hence we introduce the weak interacting neutral Majorana Higgsino states
\BEA
  \tilde{H}_{1} &=& \LP( \BA{r}  \psi_{H_{1}}^{1} \\
                                 \bar{\psi}_{H{1}}^{1}   \EA   \RP) ,
         \label{Introducing Four-Component Spinors prop 4} \SL
  \tilde{H}_{2} &=& \LP( \BA{r}  \psi_{H_{2}}^{2} \\
                                 \bar{\psi}_{H{2}}^{2}   \EA   \RP),
         \label{Introducing Four-Component Spinors prop 5}
\EEA
and the charged Dirac Higgsino states
\BEA
  \tilde{H}  &=& \LP( \BA{r}  \psi_{H_{2}}^{1} \\
                              \bar{\psi}_{H_{1}}^{2}   \EA   \RP),
         \label{Introducing Four-Component Spinors prop 6} \SL
  \tilde{H}^{c}  &=& \LP( \BA{r}  \psi_{H_{1}}^{2} \\
                              \bar{\psi}_{H_{2}}^{1}   \EA   \RP).
         \label{Introducing Four-Component Spinors prop 6aaa}
\EEA
The (four-component) leptons are as usual Dirac spinors,
and they have according
to subsect.~\ref{Subsect: Connection Between Two- and Four-Component Spinors.},
the form
\BEA
  l &=&  \LP( \BA{r}  l^{(2)}_{L} \\
                      \bar{l}^{(2)}_{R}   \EA   \RP).
         \label{Introducing Four-Component Spinors prop 7}
\EEA

By working in the Weyl basis for the $\g$-matrices
(cf. eqs.~\r{WEYL basis prop 1} and \r{WEYL basis prop 2}),
we demonstrate in great detail in appendix~\ref{App: 4 comp notation},
eq.~\r{app L sub SUSY 4-comp}, that the four component
version of the two-component Lagrangian~\r{On-Shall L sub SUSY}
(or equivalently ~\r{Introducing the Photon... prop 10})
is
\BEA
  \L_{SUSY}
  &=&   \LP(D^{\mu}\tilde{L}\RP)^{\dagger}\LP(D_{\mu}\tilde{L}\RP)
      + \LP(D^{\mu}\tilde{R}\RP)^{\dagger}\LP(D_{\mu}\tilde{R}\RP)
      - i\;\bar{L}\g^{\mu}D_{\mu}L
      - i\;\bar{R}\g^{\mu}D_{\mu}R
  \nmb
       -g\LP[\LP\{\bar{L}^{1}\tilde{W}\;\tilde{L}^{2}
       +  \bar{L}^{2}\tilde{W}^{c}\;\tilde{L}^{1}  \RP\}+h.c.\RP]
       +\sqrt{2} e
        \LP[\LP\{ \bar{L}^{2}\tilde{A}\;\tilde{L}^{2}
                - \bar{\tilde{A}}R \;\tilde{R} \RP\} + h.c. \RP]
   \nmb
         -\f{\sqrt{2}g}{\cos \theta_{\mbox{w}}}
           \LP[\LP\{
              \LP({\cal T}^{3}_{i}- Q_{i}\sin^{2}\theta_{\mbox{w}}\RP)
                    \bar{L}^{i}\tilde{Z}\;\tilde{L}^{i}
                  - \sin^{2}\theta_{\mbox{w}}\,
                       \bar{\tilde{Z}}R\;\tilde{R} \RP\}
                  + h.c.\RP]
    \nmb
           -i\,\bar{\tilde{W}}\g^{\mu}\P_{\mu}\tilde{W}
           -\f{i}{2}\,\bar{\tilde{A}}\g^{\mu}\P_{\mu}\tilde{A}
           -\f{i}{2}\,\bar{\tilde{Z}}\g^{\mu}\P_{\mu}\tilde{Z}
    \nmb
      -g\CWA \LP[ \; \bar{\tilde{Z}}\g^{\mu}\tilde{W}\;W^{-}_{\mu}
                 +  \bar{\tilde{W}}\g^{\mu}\tilde{Z}\;W^{+}_{\mu}
                 -  \bar{\tilde{W}}\g^{\mu}\tilde{W}\;Z_{\mu}  \RP]
   \nmb
     -e \LP[ \; \bar{\tilde{A}}\g^{\mu}\tilde{W}\;W^{-}_{\mu}
             +  \bar{\tilde{W}}\g^{\mu}\tilde{A}\;W^{+}_{\mu}
             -  \bar{\tilde{W}}\g^{\mu}\tilde{W}\;A_{\mu}  \RP]
   \nmb
        - \f{1}{4} {\cal W}^{+\,\mu\nu} {\cal W}^{-}_{\mu\nu}
        - \f{1}{4} {\cal W}^{-\,\mu\nu} {\cal W}^{+}_{\mu\nu}
        - \f{1}{4} {\cal Z}^{\mu\nu} {\cal Z}_{\mu\nu}
        - \f{1}{4} {\cal A}^{\mu\nu} {\cal A}_{\mu\nu}
    \nmb
      + \LP( D^{\mu}H_{1}\RP)^{\dagger}\LP(D_{\mu}H_{1}\RP)
           - \mu^{2} H_{1}^{\dagger}H_{1}
      + \LP(D^{\mu}H_{2}\RP)^{\dagger}\LP(D_{\mu}H_{2}\RP)
        -\mu^{2} H_{2}^{\dagger}H_{2}
    \nmb
      - \bar{\tilde{H}}\LP(i\g^{\mu}\P_{\mu}-\mu\RP)\tilde{H}
      - \f{i}{2}\;\bar{\tilde{H}}_{1}\g^{\mu}\P_{\mu}\tilde{H}_{1}
      - \f{i}{2}\;\bar{\tilde{H}}_{2}\g^{\mu}\P_{\mu}\tilde{H}_{2}
      - \f{\mu}{2} \bar{\tilde{H}}_{1}\tilde{H}_{2}
      - \f{\mu}{2} \bar{\tilde{H}}_{2}\tilde{H}_{1}
    \nmb
      -\f{g}{\sqrt{2}}\LP[\,\,\LP(\bar{\tilde{H}}\g^{\mu} P_{R} \tilde{H}_{1}
                    - \bar{\tilde{H}}\g^{\mu} P_{L} \tilde{H}_{2} \RP)
                         W^{+}_{\mu}+ h.c.\RP]
      + e\;\bar{\tilde{H}}\g^{\mu} \tilde{H}\;A_{\mu}
   \nmb
      +\f{g}{2\CWA}\LP[\,\,\LP(1-2\SWAS\RP)\bar{\tilde{H}}\g^{\mu} \tilde{H}
          -\HA \LP(
           \bar{\tilde{H}}_{1}\g^{\mu}\g_{5}\tilde{H}_{1}
           -\bar{\tilde{H}}_{2}\g^{\mu}\g_{5}\tilde{H}_{2} \RP)\,\,\RP] Z_{\mu}
   \nmb
           -g \LP[\, \LP(\bar{\tilde{W}}P_{R}\tilde{H}\;H^{1}_{1}
                   + \bar{\tilde{H}}P_{R}\tilde{W}\;H^{2}_{2}
                   + \bar{\tilde{H}}_{1}P_{R}\tilde{W}\;H^{2}_{1}
                   + \bar{\tilde{W}}P_{R}\tilde{H}_{2}\;H^{1}_{2}\RP)+h.c.\RP]
    \nmb
      + \sqrt{2} e \LP[\,\LP(\bar{\tilde{A}}P_{R}\tilde{H}\;H^{2}_{1}
               -  \bar{\tilde{H}}P_{R}\tilde{A}\;H^{1}_{2}\RP)+h.c. \RP]
   \nmb
      -\f{g}{\sqrt{2}\CWA}\LP[\,\LP\{
           \bar{\tilde{Z}}P_{R}\tilde{H}_{1}\;H^{1}_{1}
          - \bar{\tilde{H}}_{2}P_{R}\tilde{Z}\;H^{2}_{2} \RP.\RP.
    \nmb \hspace{2.8cm} \LP.\LP.
       - \LP(1-2\SWAS\RP)\LP(
           \bar{\tilde{Z}}P_{R}\tilde{H}\;H^{2}_{1}
         - \bar{\tilde{H}}P_{R}\tilde{Z}\;H^{1}_{2}\RP) \RP\}+ h.c. \RP]
    \nmb
     + f \! \LP[ \LP\{
         \bar{\tilde{H}}L^{1}\,\tilde{R}
       - \bar{\tilde{H}}_{1}L^{2}\,\tilde{R}
       + \bar{R}L^{1}\,H_{1}^{2}
       - \bar{R}L^{2}\,H_{1}^{1}
       + \bar{R}\tilde{H}^{c}\,\tilde{L}^{1}
       - \bar{R}\tilde{H}_{1}\,\tilde{L}^{2} \RP\}\! +h.c. \RP]
    \nmb
     - \mu f\! \LP[H_{2}^{\dagger}\tilde{L}\,\tilde{R}
             + h.c. \RP]
%    \nmb
     - f^{2} \! \LP[
           \tilde{L}^{\dagger}\tilde{L}\,\tilde{R}^{\dagger}\tilde{R}
         + H_{1}^{\dagger}H_{1}\!\LP( \tilde{L}^{\dagger}\tilde{L}
                                + \tilde{R}^{\dagger}\tilde{R} \RP)
     - H_{1}^{\dagger}\tilde{L}\!\LP(H_{1}^{\dagger}\tilde{L}\RP)^{\dagger}\RP]
    \nmb
    -\f{g^2}{2}
         \LP(\,\tilde{L}^{\dagger}T^{a}\tilde{L}+H_{1}^{\dagger}
                T^{a}H_{1}+H_{2}^{\dagger}T^{a}H_{2}\,\RP)
         \LP(\,\tilde{L}^{\dagger}T^{a}\tilde{L}+H_{1}^{\dagger}
                T^{a}H_{1}+H_{2}^{\dagger}T^{a}H_{2}\,\RP)
     \nmb
        - \f{g^2 \tan^{2} \t_{\mbox{w}} }{8}
             \LP(\,\tilde{L}^{\dagger}\tilde{L}-2\tilde{R}^{\dagger}
             \tilde{R}+H_{1}^{\dagger}H_{1}-H_{2}^{\dagger}H_{2}\,\RP)^2
      + t.d.
         \label{L sub SUSY 4-comp}
\EEA
Here $P_{L}$ and $P_{R}$ are the left- and right-handed projection
operators given by eqs.~\r{left proje} and \r{right proje}.

This concludes this section, and after the long discussion of
the Lagrangian $\L_{SUSY}$ we will finally draw our attention towards the
soft-breaking piece $\L_{Soft}$.

\section{Component Field Expansion of $\L_{Soft}$.}
    \label{SECT: Component Field Expansion}

{}From chapter~\ref{Chapter Supersymmetric Extension of QFD.},
eq.~\r{The Soft SUSY-Breaking Term prop 2aaa},
we recall that
\BEA
  \L_{Soft} &=&    \L_{SMT} + \L_{GMT},
    \label{Component Field Expansion of L sub Soft prop 1}
\EEA
with
%\footnote{Recall that  $M_{L}^{2}\;\hat{L}^{\dagger}\hat{L}
%= m_{\tilde{\nu}}^{2}\;\hat{\nu}^{\dagger}
%\hat{\nu}+ m_{L}^{2}\; \hat{L}^{\dagger}
%\hat{L}.}
\BEA
   \L_{SMT} &=& -\int d^{4}\t\; \LP[\,
            M_{L}^{2}\;\hat{L}^{\dagger}\hat{L}
          + m_{R}^{2}\hat{R}^{\dagger}\hat{R}
          + m_{1}^{2} \hat{H}_{1}^{\dagger}\hat{H}_{1} \RP.
\nmb  \hspace{1.7cm} \LP.
          + m_{2}^{2} \hat{H}_{2}^{\dagger}\hat{H}_{2}
          - m_{3}^{2}\e^{ij}\LP(\hat{H}_{1}^{i}\hat{H}_{2}^{j}+h.c.\RP)
             \RP] \d^{4}(\t,\tb),
             \label{Component Field Expansion of L sub Soft prop 2}
\EEA
and
\BEA
   \L_{GMT} &=&   \HA \int d^{4}\t\; \LP[
              \LP(\,M\; W^{\a\;a}W^{a}_{\a}
            + M' \;   W^{'\,\a}W'_{\a}\,\RP)
            + h.c. \RP]
            \d^{4}(\t,\tb).
              \label{Component Field Expansion of L sub Soft prop 3}
\EEA

Now the component expansion of $\L_{Soft}$ will be calculated, and
we start with $\L_{SMT}$.
With the component expansions of
$\hat{L}$, $\hat{R}$, $\hat{H}_{1}$
and $\hat{H}_{2}$ from sect.~\ref{SECT: Component Expansion},
we have (cf. \cite{KAKU,WESS,MK})
%with eq.~\r{Vector superfield prop 3bb}
\BEA
   \L_{SMT} &=&
          - M_{L}^{2}\;\tilde{L}^{\dagger}\tilde{L}
          - m_{R}^{2}\tilde{R}^{\dagger}\tilde{R}
          - m_{1}^{2} H_{1}^{\dagger}H_{1}
          - m_{2}^{2} H_{2}^{\dagger}H_{2}
\nmb  \hspace{1cm}
          + m_{3}^{2}\e^{ij}\LP(H_{1}^{i}H_{2}^{j}+h.c.\RP),
          \label{Component Field Expansion of L sub Soft prop 4}
\EEA
and correspondingly for $\L_{GMT}$
%with eq.~\r{Supersymmetric Field Strengths prop 1}
\BEA
  \L_{GMT} &=& -\HA M\LP(\la^{a}\la^{a}+\bar{\la}^{a}\bar{\la}^{a}\RP)
               -\HA M'\LP(\la'\la'+\bar{\la}'\bar{\la}'\RP).
               \label{Component Field Expansion of L sub Soft prop 5}
\EEA
Here $M_{L}^{2}\;\tilde{L}^{\dagger}\tilde{L}$ is defined in analogy with the
corresponding superfield definition, i.e.
$M_{L}^{2}\;\tilde{L}^{\dagger}\tilde{L} =
m_{\tilde{\nu}}^{2}\;\tilde{\nu}^{\dagger}\tilde{\nu}
 + m_{L}^{2}\;\tilde{l}^{\dagger}_{L}\tilde{l}_{L}$.

Since  eq.~\r{Component Field Expansion of L sub Soft prop 5} contains
two-component Weyl-spinors, we will, as in the previous section,
introduce four-component notation.

Hence, with  eq.~\r{Introducing Four-Component Spinors prop 3}  we have
\BEA
\lefteqn{-\HA M\LP(\la^{1}\la^{1}+\bar{\la}^{1}\bar{\la}^{1}\RP)
   -\HA M\LP(\la^{2}\la^{2}+\bar{\la}^{2}\bar{\la}^{2}\RP)}\hspace{2.5cm}\nn
     &=&  - M \LP(\la^{-}\la^{+}+\bar{\la}^{-}\bar{\la}^{+}\RP) \nn
     &=&   M_{\tilde{W}}\,\bar{\tilde{W}}\tilde{W},
\EEA
where $M_{\tilde{W}}= M$. Similarly,
with eqs.~\r{Introducing Four-Component Spinors prop 1}
and \r{Introducing Four-Component Spinors prop 2}
\BEA
\lefteqn{-\HA M\LP(\la^{3}\la^{3}+\bar{\la}^{3}\bar{\la}^{3}\RP)
   -\HA M'\LP(\la'\la'+\bar{\la}'\bar{\la}'\RP)}\hspace{1.5cm}\nn
   &=& -\HA\LP( M \SWAS+ M' \CWAS \RP)
          \LP(\la_{A}\la_{A}+\bar{\la}_{A}\bar{\la}_{A}\RP)
   \nmb
       -\HA\LP( M \CWAS+ M' \SWAS \RP)
          \LP(\la_{Z}\la_{Z}+\bar{\la}_{Z}\bar{\la}_{Z}\RP)
  \nmb
       -\HA\LP( M -  M' \RP) \sin{2\t_{\mbox{w}}}\;
          \LP(\la_{A}\la_{Z}+\bar{\la}_{A}\bar{\la}_{Z}\RP) \nn
   &=& \HA\LP( M \SWAS+ M' \CWAS \RP)
           \;\bar{\!\!\tilde{A}}\tilde{A}
       +\HA\LP( M \CWAS+ M' \SWAS \RP)
           \bar{\tilde{Z}}\tilde{Z}
   \nmb
       +\HA\LP( M -  M' \RP) \sin{2\t_{\mbox{w}}}\;
           \;\,\bar{\!\!\tilde{A}}\tilde{Z} \nn
   &=&  \HA\,  M_{\tilde{A}}\;\,\bar{\!\!\tilde{A}}\tilde{A}
         + \HA\,M_{\tilde{Z}}\,\bar{\tilde{Z}}\tilde{Z}
        +\HA\LP( M_{\tilde{Z}} -  M_{\tilde{A}} \RP)
           \tan{2\t_{\mbox{w}}}\;\,\bar{\!\!\tilde{A}}\tilde{Z},
           \label{Component Field Expansion of L sub Soft prop 6}
\EEA
where we have introduced the notation
\BEA
    M_{\tilde{A}} &=&  M'\CWAS+ M \SWAS ,
     \label{Component Field Expansion of L sub Soft prop 7}\SL
    M_{\tilde{Z}} &=&   M'\SWAS+ M \CWAS .
       \label{Component Field Expansion of L sub Soft prop 8}
\EEA
Thus eq.~\r{Component Field Expansion of L sub Soft prop 5} reads
\BEA
  \L_{GMT} &=&
         M_{\tilde{W}}\,\bar{\tilde{W}}\tilde{W}
          +  \HA\,  M_{\tilde{A}}\;\,\bar{\!\!\tilde{A}}\tilde{A}
          + \HA\,M_{\tilde{Z}}\,\bar{\tilde{Z}}\tilde{Z}
           +  \HA\LP( M_{\tilde{Z}} -  M_{\tilde{A}} \RP)\tan{2\t_{\mbox{w}}}
%          +\HA\LP( M -  M' \RP)\sin{2\t_{\mbox{w}}}
          \;\,\bar{\!\!\tilde{A}}\tilde{Z}, \hspace{0.6cm}
              \label{Component Field Expansion of L sub Soft prop 9}
\EEA
and this section is concluded.

\section{Conclusion \,\, --- \,\,
The\, Full\, Four-Component\, Lagrangian $\L_{S-QFD}$.}

With the results from eqs.~\r{L sub SUSY 4-comp}
and \r{Component Field Expansion of L sub Soft prop 9}
we may conclude for $\L_{S-QFD} = \L_{SUSY} + \L_{Soft}$
\BEA
  \L_{S-QFD}
  &=&   \LP(D^{\mu}\tilde{L}\RP)^{\dagger}\LP(D_{\mu}\tilde{L}\RP)
               - M_{L}^{2}\;\tilde{L}^{\dagger}\tilde{L}
      + \LP(D^{\mu}\tilde{R}\RP)^{\dagger}\LP(D_{\mu}\tilde{R}\RP)
              - m_{R}^{2}\tilde{R}^{\dagger}\tilde{R}
  \nmb
      - i\;\bar{L}\g^{\mu}D_{\mu}L
      - i\;\bar{R}\g^{\mu}D_{\mu}R
  \nmb
       -g\LP[\LP\{\bar{L}^{1}\tilde{W}\;\tilde{L}^{2}
       +  \bar{L}^{2}\tilde{W}^{c}\;\tilde{L}^{1}  \RP\}+h.c.\RP]
       +\sqrt{2} e
        \LP[\LP\{ \bar{L}^{2}\tilde{A}\;\tilde{L}^{2}
                - \bar{\tilde{A}}R \;\tilde{R} \RP\} + h.c. \RP]
   \nmb
         -\f{\sqrt{2}g}{\cos \theta_{\mbox{w}}}
           \LP[\LP\{
              \LP({\cal T}^{3}_{i}- Q_{i}\sin^{2}\theta_{\mbox{w}}\RP)
                    \bar{L}^{i}\tilde{Z}\;\tilde{L}^{i}
                  - \sin^{2}\theta_{\mbox{w}}\,
                        \bar{\tilde{Z}}R\;\tilde{R} \RP\}
                  + h.c.\RP]
    \nmb
           -\bar{\tilde{W}}\LP(i\g^{\mu}\P_{\mu}-M_{\tilde{W}}\RP)\tilde{W}
           -\f{1}{2}\,\bar{\tilde{A}}\LP(i\g^{\mu}\P_{\mu}
              -M_{\tilde{A}}\RP)\tilde{A}
           -\f{1}{2}\,\bar{\tilde{Z}}\LP(i\g^{\mu}
              \P_{\mu}-M_{\tilde{Z}}\RP)\tilde{Z}
    \nmb
           +  \HA\LP( M_{\tilde{Z}} -  M_{\tilde{A}} \RP)
                \tan{2\t_{\mbox{w}}}\;\,\bar{\!\!\tilde{A}}\tilde{Z}
    \nmb
      -g\CWA \LP[ \; \bar{\tilde{Z}}\g^{\mu}\tilde{W}\;W^{-}_{\mu}
                 +  \bar{\tilde{W}}\g^{\mu}\tilde{Z}\;W^{+}_{\mu}
                 -  \bar{\tilde{W}}\g^{\mu}\tilde{W}\;Z_{\mu}  \RP]
   \nmb
     -e \LP[ \; \bar{\tilde{A}}\g^{\mu}\tilde{W}\;W^{-}_{\mu}
             +  \bar{\tilde{W}}\g^{\mu}\tilde{A}\;W^{+}_{\mu}
             -  \bar{\tilde{W}}\g^{\mu}\tilde{W}\;A_{\mu}  \RP]
   \nmb
        - \f{1}{4} {\cal W}^{+\,\mu\nu} {\cal W}^{-}_{\mu\nu}
        - \f{1}{4} {\cal W}^{-\,\mu\nu} {\cal W}^{+}_{\mu\nu}
        - \f{1}{4} {\cal Z}^{\mu\nu} {\cal Z}_{\mu\nu}
        - \f{1}{4} {\cal A}^{\mu\nu} {\cal A}_{\mu\nu}
    \nmb
      + \LP( D^{\mu}H_{1}\RP)^{\dagger}\LP(D_{\mu}H_{1}\RP)
           -\LP( m^2_1+ \mu^{2}\RP) H_{1}^{\dagger}H_{1}
    \nmb
      + \LP(D^{\mu}H_{2}\RP)^{\dagger}\LP(D_{\mu}H_{2}\RP)
        -\LP( m^2_2+\mu^{2}\RP) H_{2}^{\dagger}H_{2}
       + m_{3}^{2}\e^{ij}\LP(H_{1}^{i}H_{2}^{j}+h.c.\RP)
    \nmb
      - \bar{\tilde{H}}\LP(i\g^{\mu}\P_{\mu}-\mu\RP)\tilde{H}
      - \f{i}{2}\;\bar{\tilde{H}}_{1}\g^{\mu}\P_{\mu}\tilde{H}_{1}
      - \f{i}{2}\;\bar{\tilde{H}}_{2}\g^{\mu}\P_{\mu}\tilde{H}_{2}
      - \f{\mu}{2} \bar{\tilde{H}}_{1}\tilde{H}_{2}
      - \f{\mu}{2} \bar{\tilde{H}}_{2}\tilde{H}_{1}
    \nmb
      -\f{g}{\sqrt{2}}\LP[\,\,\LP(\bar{\tilde{H}}\g^{\mu} P_{R} \tilde{H}_{1}
                    - \bar{\tilde{H}}\g^{\mu} P_{L} \tilde{H}_{2} \RP)
                         W^{+}_{\mu}+ h.c.\RP]
      + e\;\bar{\tilde{H}}\g^{\mu} \tilde{H}\;A_{\mu}
   \nmb
      +\f{g}{2\CWA}\LP[\,\,\LP(1-2\SWAS\RP)\bar{\tilde{H}}\g^{\mu} \tilde{H}
          -\HA \LP(
           \bar{\tilde{H}}_{1}\g^{\mu}\g_{5}\tilde{H}_{1}
           -\bar{\tilde{H}}_{2}\g^{\mu}\g_{5}\tilde{H}_{2} \RP)\,\,\RP] Z_{\mu}
   \nmb
           -g \LP[\, \LP(\bar{\tilde{W}}P_{R}\tilde{H}\;H^{1}_{1}
                   + \bar{\tilde{H}}P_{R}\tilde{W}\;H^{2}_{2}
                   + \bar{\tilde{H}}_{1}P_{R}\tilde{W}\;H^{2}_{1}
                   + \bar{\tilde{W}}P_{R}\tilde{H}_{2}\;H^{1}_{2}\RP)+h.c.\RP]
    \nmb
      + \sqrt{2} e \LP[\,\LP(\bar{\tilde{A}}P_{R}\tilde{H}\;H^{2}_{1}
               -  \bar{\tilde{H}}P_{R}\tilde{A}\;H^{1}_{2}\RP)+h.c. \RP]
   \nmb
      -\f{g}{\sqrt{2}\CWA}\LP[\,\LP\{
           \bar{\tilde{Z}}P_{R}\tilde{H}_{1}\;H^{1}_{1}
          - \bar{\tilde{H}}_{2}P_{R}\tilde{Z}\;H^{2}_{2} \RP.\RP.
    \nmb \hspace{2.8cm} \LP.\LP.
       - \LP(1-2\SWAS\RP)\LP(
           \bar{\tilde{Z}}P_{R}\tilde{H}\;H^{2}_{1}
         - \bar{\tilde{H}}P_{R}\tilde{Z}\;H^{1}_{2}\RP) \RP\}+ h.c. \RP]
    \nmb
     + f \! \LP[ \LP\{
         \bar{\tilde{H}}L^{1}\,\tilde{R}
       - \bar{\tilde{H}}_{1}L^{2}\,\tilde{R}
       + \bar{R}L^{1}\,H_{1}^{2}
       - \bar{R}L^{2}\,H_{1}^{1}
       + \bar{R}\tilde{H}^{c}\,\tilde{L}^{1}
       - \bar{R}\tilde{H}_{1}\,\tilde{L}^{2} \RP\}\! +h.c. \RP]
    \nmb
     - \mu f\! \LP[H_{2}^{\dagger}\tilde{L}\,\tilde{R}
             + h.c. \RP]
%    \nmb
     - f^{2} \! \LP[
           \tilde{L}^{\dagger}\tilde{L}\,\tilde{R}^{\dagger}\tilde{R}
         + H_{1}^{\dagger}H_{1}\!\LP( \tilde{L}^{\dagger}\tilde{L}
                                + \tilde{R}^{\dagger}\tilde{R} \RP)
     - H_{1}^{\dagger}\tilde{L}\!\LP(H_{1}^{\dagger}\tilde{L}\RP)^{\dagger}\RP]
    \nmb
    -\f{g^2}{2}
         \LP(\,\tilde{L}^{\dagger}T^{a}\tilde{L}+H_{1}^{\dagger}
                 T^{a}H_{1}+H_{2}^{\dagger}T^{a}H_{2}\,\RP)
         \LP(\,\tilde{L}^{\dagger}T^{a}\tilde{L}+H_{1}^{\dagger}
                 T^{a}H_{1}+H_{2}^{\dagger}T^{a}H_{2}\,\RP)
     \nmb
        - \f{g^2 \tan^{2} \t_{\mbox{w}} }{8}
            \LP(\,\tilde{L}^{\dagger}\tilde{L}-2\tilde{R}^{\dagger}
              \tilde{R}+H_{1}^{\dagger}H_{1}-H_{2}^{\dagger}H_{2}\,\RP)^2
      + t.d.
         \label{Total L sub SUSY 4-comp}
\EEA
This Lagrangian is the final result for our S-QFD theory, but before
we close this chapter we will
make several observations about this Lagrangian.

Firstly, it contains the correct kinetic terms for the
bosons~(sleptons, photons, Z-bosons, higgs bosons $\ldots$) and
fermions~(leptons, photinos, zinos, winos, $\ldots$) of the theory.

Secondly, it holds the well known SM-interaction terms for the
SM-particles, and in addition interaction terms between SM-
and SUSY-particles and SUSY-particles alone.
Note the rich number of different interactions, both cubic and
quadratic, that are possible in this theory.

Thirdly, we observe that for the wino- and charged higgsino-fields,
their charge conjugated fields also appear in the Lagrangian.
Such a situation is unknown from the SM.
In part~2 we will see that this has the
strange consequence that the theory will
contain fermion-number violating vertices
and propagators.

After these remarks we close this chapter.

\cleardoublepage

\chapter{Symmetry Breaking and Physical Fields.}

In this chapter the breaking of electroweak
gauge symmetry and the introduction of
physical states  will be demonstrated.

As alluded to earlier, the breaking of gauge symmetry is in the MSSM
directly connected to the breaking of supersymmetry.
In fact this breaking --- called radiative breaking ---
is an effect of radiative corrections to the soft
mass-parameters as we now will discuss in detail.

\section{Radiative $SU(2) \times U(1)$ Breaking.}

Our model has the
attractive virtue of allowing for
the possibility of a phenomenologically acceptable
radiative breaking of the electroweak gauge
symmetry~[\ref{INO82a}--\ref{LI89}, \ref{KAP87}--\ref{GAT88}].
This is obtained through a generalization of the original
Coleman-Weinberg mechanism~\cite{COL73}.
Radiative breaking also has the advantage,
when combined with some additional plausible assumptions,
of being very powerful since it excludes large regions of parameterspace
as we will see.
This takes part in increasing the predictiveness of the model.
Now we will work out the Coleman-Weinberg scheme~\cite{COL73}
for our supersymmetric field theory.

In SUSY-theories, one has two kinds of potentials ---
superpotentials and scalar potentials.
Superpotentials have been discussed earlier in this thesis, so in consequence
we now consider the scalar potential, which has its analogy in the SM.

Contributions to the MSSM scalar potential, $V_{MSSM}$, arise from three
sources ---  the auxiliary F- and D-fields and
the soft terms. We write
\BEA
   V_{MSSM} &=& V_{D} + V_{F} + V_{Soft},
      \label{Radiative Breaking prop 1}
\EEA
where\footnote{Generally can $D'\rightarrow D' + \xi$, where $\xi$
is a Fayet-Iliopoulos term~\cite{FAY74},
but we will henceforth assume that this term is neglectable.}~\footnote{Note
that it is $-V_{MSSM}$ which appears in the Lagrangian.}
\BEA
     V_{D} &=& - \L_{Aux-D}\nn
           &=& \f{g^2}{2}
         \LP(\,\tilde{L}^{\dagger}T^{a}\tilde{L}
              +H_{1}^{\dagger}T^{a}H_{1}+H_{2}^{\dagger}T^{a}H_{2}\,\RP)
         \LP(\,\tilde{L}^{\dagger}T^{a}\tilde{L}
               +H_{1}^{\dagger}T^{a}H_{1}+H_{2}^{\dagger}T^{a}H_{2}\,\RP)
    \nmb
        +\f{g'^2}{8}
          \LP(\,\tilde{L}^{\dagger}\tilde{L}-2\tilde{R}^{\dagger}
               \tilde{R}+H_{1}^{\dagger}H_{1}-H_{2}^{\dagger}H_{2}\,\RP)^2 ,\SL
     V_{F} &=& -\L_{Aux-F}\nn
         &=&
       \mu^{2}\, H_{1}^{\dagger}H_{1}
        +\mu^{2}\, H_{2}^{\dagger}H_{2}
        +\mu f \LP[\, H_{2}^{\dagger}\tilde{L}\,\tilde{R}
                  +\tilde{L}^{\dagger}H_{2}\,\tilde{R}^{\dagger}\,\RP]\nn
   & & \mbox{}
      + f^{2}\LP[ \,
       \tilde{L}^{\dagger}\tilde{L}\,\tilde{R}^{\dagger}\tilde{R}
     + H_{1}^{\dagger}H_{1}\LP( \tilde{L}^{\dagger}\tilde{L}
                                + \tilde{R}^{\dagger}\tilde{R} \RP)
    - H_{1}^{\dagger}\tilde{L}\LP(H_{1}^{\dagger}\tilde{L}\RP)^{\dagger}\,\RP],
\EEA
and
\BEA
    V_{Soft} &=& - \L_{SMT}  \nn
         &=&
            M_{L}^{2}\;L^{\dagger}L
          + m_{R}^{2}R^{\dagger}R
          + m_{1}^{2} H_{1}^{\dagger}H_{1}
          + m_{2}^{2} H_{2}^{\dagger}H_{2}
\nmb  \hspace{1cm}
          - m_{3}^{2}\e^{ij}\LP(H_{1}^{i}H_{2}^{j}+h.c.\RP).
             \label{Radiative Breaking prop 1aaaa}
\EEA

Now we leave this general scalar potential, and instead consider the
pure scalar Higgs potential because it is this potential which is of
interest in the discussion of gauge symmetry breaking.

\subsection{The Scalar Higgs Potential.}
    \label{SUBsect: The Scalar Higgs Potential.}

Thus, for the pure Higgs sector of the theory, the (tree-level)
scalar Higgs potential $V\equiv V_{Higgs}$
reads\footnote{This potential
is a special case of the general two-Higgs doublet
potential~\cite{HAB79,GEO78}.} according to
eqs.~\r{Radiative Breaking prop 1}--\r{Radiative Breaking prop 1aaaa}
\BEA
    V &=&   \LP(m_{1}^{2}+\mu^{2}\RP) H_{1}^{\dagger}\!H_{1}
          + \LP(m_{2}^{2}+\mu^{2}\RP) H_{2}^{\dagger}\!H_{2}
          - m_{3}^{2}\,\e^{ij}\LP(H_{1}^{i}H_{2}^{j} +h.c.\RP)
      \nmb
          +\f{g^{2}}{2} \LP( H_{1}^{\dagger}T^{a}H_{1}
                 +H_{2}^{\dagger}T^{a}H_{2}\RP)
                        \LP( H_{1}^{\dagger}T^{a}H_{1}
                  +H_{2}^{\dagger}T^{a}H_{2}\RP)
      \nmb
          + \f{g'^{2}}{8} \LP( H_{1}^{\dagger}\!H_{1}
                   -  H_{2}^{\dagger}\!H_{2}   \RP)^{2}.
              \label{Radiative Breaking prop 1aaa}
\EEA
However, in appendix~\ref{APP: Transcription of the Scalar Higgs Potential.}
this potential is rewritten for later convenience, and the result is
(cf.  eq.~\r{Transcription of the Scalar Higgs Potential prop 2})
\BEA
    V &=&   m_{1}^{2}\,H_{1}^{\dagger}H_{1}
          + m_{2}^{2}\,H_{2}^{\dagger}H_{2}
          - m_{3}^{2}\,\e^{ij}\LP(H_{1}^{i}H_{2}^{j} +h.c.\RP)
      \nmb
          +\f{1}{8}\LP(g^{2}+g'^{2}\RP)
                \LP( H_{1}^{\dagger}H_{1}-H_{2}^{\dagger}H_{2}\RP)^{2}
          + \f{g^{2}}{2} \LP|H_{1}^{\dagger}H_{2}\RP|^{2}.
            \label{Tree-level scalar potential}
\EEA
Here we have taken advantage of the arbitrary nature of the
soft mass-parameters $m^2_1$ and $m^2_2$, and absorbed
$\mu^2$ into these, i.e.
\BEA
       m_{1}^{2}+\mu^{2} &\longrightarrow&  m_{1}^{2}, \nn
       m_{2}^{2}+\mu^{2} &\longrightarrow& m_{2}^{2}. \nonumber
\EEA

Without loss of generality, we may choose the phases of the
(scalar) Higgs fields in such a way that all
mass parameters $m_{i}^{2}$ ($i=1,2,3$) are real
and that the vacuum expectation values (v.e.v.'s)
of the Higgs fields are non-negative.
As in the Standard Model~(SM), the $SU(2)\times U(1)$ gauge symmetry
has to be broken down to $U(1)_{EM}$.
This means that electromagnetism is unbroken and hence the
charged components of the Higgs-doublets can not
develop non-vanishing v.e.v.'s.
Hence
\BEA
   \LP< H_{1}\RP> &=& \LP(\BA{c} v_{1}\\ 0 \EA \RP),
      \label{Radiative Breaking prop 2} \SL
   \LP< H_{2}\RP> &=& \LP(\BA{c} 0 \\ v_{2} \EA \RP),
      \label{Radiative Breaking prop 3}
\EEA
and the potential becomes at the vacuum
\BEA
    V   &=&   m_{1}^{2}\,v_{1}^{2}
          + m_{2}^{2}\,v_{2}^{2}
          - 2 m_{3}^{2}\,v_{1}v_{2}
          +\f{1}{8}\LP(g^{2}+g'^{2}\RP)\LP[v_{1}^{2}-v_{2}^{2}\RP]^{2}.
              \label{Radiative Breaking prop 4}
\EEA
For this potential to be bound from below, e.g. in the direction $v_{1}=v_{2}$,
one has to be careful and demand
\BEA
  {\cal B} \equiv m_{1}^{2}+m_{2}^{2}-2 m_{3}^{2} \geq 0.
       \label{Radiative Breaking prop 5}
\EEA
This relation will hereafter be referred to as the stability condition.

{}From the SM Higgs-mechanism, it is a well-known fact that when
the Higgs v.e.v. is non-vanishing this signals breaking of
the $SU(2)\times U(1)$-symmetry because origo is ``unstable".
This situation applies equivalently well to the two Higgs
doublet model~\cite{HAB79,GEO78}.
However, what demands do we have to make  in order to obtain
non-vanishing v.e.v.'s?
As long as $V_{min}$ is non-negative,
the minimum~($V_{min}=0$) lies at the origo, i.e. at $v_{1}=v_{2}=0$,
and the gauge symmetry is unbroken.
Thus $V_{min}$ has to be negative to obtain breaking of gauge symmetry.

Now we will derive a condition on the mass parameters for this to happen.
Rewriting eq.~\r{Radiative Breaking prop 4} yields
\BEA
  V &=& \mbox{v}^{T}{\cal M}^{2}\mbox{v} +
          \f{1}{8}\LP(g^{2}+g'^{2}\RP)\LP[v_{1}^{2}-v_{2}^{2}\RP]^{2},
                 \label{Radiative Breaking prop 6}
\EEA
where
\BEA
     \mbox{v}      &=&  \LP( \BA{c} v_{1}\\ -v_{2}\EA\RP),
                 \label{Radiative Breaking prop 7} \nn
     {\cal M}^{2}  &=&  \LP( \BA{cc}  m_{1}^{2} & m_{3}^{2}\\
                                      m_{3}^{2} & m_{2}^{2}   \EA  \RP).
                                       \label{Radiative Breaking prop 8}
\nonumber
\EEA
Since ${\cal M}^{2}$ is a symmetric matrix, the following
is true for the quadratic form
$\mbox{v}^{T}{\cal M}^{2}\mbox{v}$~\cite{ANTON}
\BEA
     \la_{-}\ABS{\mbox{v}}^{2}\;\, \leq \;\,
    \mbox{v}^{T}{\cal M}^{2}\mbox{v}\;\,  \leq \;\,  \la_{+}\ABS{\mbox{v}}^{2}.
         \label{Radiative Breaking prop 9}
\EEA
Here $\la_{\pm}$ are the eigenvalues of ${\cal M}^{2}$ given by
\BEA
   \la_{\pm} &=& \HA \LP(m_{1}^{2}+m_{2}^{2}
        \pm \sqrt{\LP( m_{1}^{2}+m_{2}^{2} \RP)^{2}
          -4\LP(m_{1}^{2}m_{2}^{2}-m_{3}^{4}\RP)}\RP),
         \label{Radiative Breaking prop 10}
\EEA
and the norm $\ABS{\mbox{v}}$ is taken relative to the
inner product space $\R^{2}$.

%The vacuum expectation values $v_{1}$ and $v_{2}$ are
%choosen in such a way that the scalar potential has a (local) minimum.
Since the last term of eq.~\r{Radiative Breaking prop 6} is
non-negative, the quadratic form $\mbox{v}^{T}{\cal M}^{2}\mbox{v}$
has to be at its minimum value in order to get a minimum of V, i.e.
\BEA
   \mbox{v}^{T}{\cal M}^{2}\mbox{v} &=& \la_{-}\ABS{\mbox{v}}^{2}. \nonumber
%     \label{Radiative Breaking prop 11}
\EEA
Hence in order for $V_{min}<0$ we must have $\la_{-}< 0$,
something which implies
%and thus  one has to demand
\BEA
   \det{{\cal M}^{2}} = m_{1}^{2}m_{2}^{2}-m_{3}^{4} < 0.
      \label{Radiative Breaking prop 12}
\EEA
Thus if eq.~\r{Radiative Breaking prop 12},
in addition to the stability condition~\r{Radiative Breaking prop 5},
are satisfied, this signals $SU(2)\times U(1)$-gauge symmetry breaking.
Later on this will be demonstrated explicitly.
\begin{figure}[tbh]
\unitlength=1.00mm
\special{em:linewidth 0.4pt}
\linethickness{0.4pt}
\begin{picture}(144.00,54.00)
\put(35.00,28.00){\makebox(0,0)[cc]{$m_{1}^{2}+m_{2}^{2}-2\ABS{m_{3}^{2}}$}}
\put(80.00,28.00){\makebox(0,0)[cc]{$m_{1}^{2}m_{2}^{2}-m_{3}^{4}$}}
\bezier{12}(130.00,49.00)(130.00,47.00)(130.00,46.00)
\bezier{12}(130.00,44.00)(130.00,42.00)(130.00,41.00)
\bezier{12}(130.00,39.00)(130.00,37.00)(130.00,36.00)
\bezier{12}(130.00,34.00)(130.00,32.00)(130.00,31.00)
\bezier{12}(130.00,29.00)(130.00,27.00)(130.00,26.00)
\bezier{12}(130.00,24.00)(130.00,22.00)(130.00,21.00)
\bezier{12}(130.00,19.00)(130.00,17.00)(130.00,16.00)
\bezier{12}(130.00,14.00)(130.00,12.00)(130.00,11.00)
\bezier{8}(130.00,4.00)(130.00,3.00)(130.00,2.00)
\put(130.00,-3.00){\makebox(0,0)[cb]{$M_{Pl}$}}
\put(106.00,7.00){\makebox(0,0)[cb]{\small Unbroken $SU(2) \times U(1)$}}
\put(58.00,7.00){\makebox(0,0)[cb]{\small $SU(2) \times U(1)$ breaking}}
\put(26.00,7.00){\makebox(0,0)[cb]{\small Instability}}
\put(34.00,5.00){\vector(-1,0){18.00}}
\put(16.00,5.00){\vector(1,0){18.00}}
\put(34.00,5.00){\vector(1,0){45.00}}
\put(79.00,5.00){\vector(-1,0){45.00}}
\put(116.00,5.00){\vector(-1,0){37.00}}
\put(16.00,-8.00){\vector(0,1){62.00}}
\put(80.00,-3.00){\makebox(0,0)[cb]{$Q_{c}$}}
\put(35.00,-3.00){\makebox(0,0)[cb]{$Q_{s}$}}
\put(16.00,2.00){\vector(1,0){128.00}}
\put(83.00,5.00){\vector(1,0){47.00}}
\bezier{340}(73.00,-8.00)(87.00,31.00)(130.00,36.00)
\bezier{564}(130.00,49.00)(42.00,43.00)(29.00,-8.00)
\bezier{12}(130.00,9.00)(130.00,7.00)(130.00,6.00)
\end{picture}
   \vspace{0.7cm}
   \caption{The figure shows ${\cal B}$ and $\det {\cal M}$ as functions of
             scale Q. The various sectors where $SU(2)\times U(1)$
             is broken in a satisfactory/unsatisfactory
 	      manner are also indicated.}
    \label{FIG: rad_breaking}
    \vspace{0.4cm}
\end{figure}
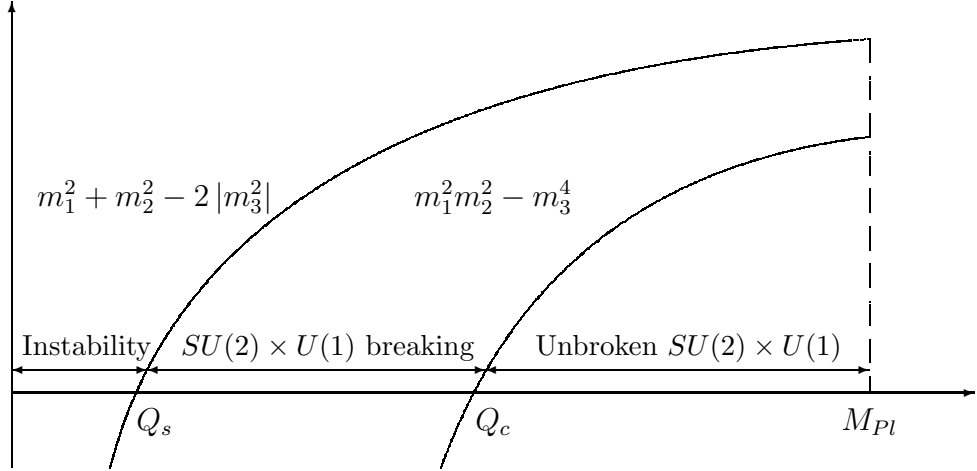

When condition~\r{Radiative Breaking prop 12} is satisfied, the
neutral components of $H_{1}$ and $H_{2}$ start to develop
non-vanishing v.e.v.'s~($v_{1}, v_{2} \neq 0$).
Now we will derive some useful relations, and
an expression for the potential at its minimum.
At $V_{\min}$, the potential  has to fulfil the
equations $\PD{V_{min}}{v_{1}}=\PD{V_{min}}{v_{2}}=0$ and
$\f{\P^{2}V_{min}}{\P v_{1}\P v_{2}}>0$.
This yields the following relations
\BEA
  m_{1}^{2}v_{1}-m_{3}^{2}v_{2}
            +\f{1}{4}\LP(g^{2}+g'^{2}\RP)\LP[v_{1}^{2}-v_{2}^{2}\RP]
             v_{1} &=& 0,
            \label{Radiative Breaking prop 12a}\SL
  m_{2}^{2}v_{2}-m_{3}^{2}v_{1}
            -\f{1}{4}\LP(g^{2}+g'^{2}\RP)\LP[v_{1}^{2}-v_{2}^{2}\RP]
              v_{2} &=& 0,
            \label{Radiative Breaking prop 12b} \SL
     -2m_{3}^{2}-\LP(g^{2}+g'^2\RP)\,v_1 v_2 & > & 0.
         \label{Radiative Breaking prop 12c}
\EEA
By multiplying eqs.~\r{Radiative Breaking prop 12a} and
\r{Radiative Breaking prop 12b} by $v_{1}^{-1}$ and $v_{2}^{-1}$
respectively, and then adding and subtracting
the resulting equations, we obtain
\BEA
  m_{1}^{2}+m_{2}^{2} &=& m_{3}^{2}\LP(\tan{\b}+\cot{\b}\RP),
     \label{Radiative Breaking prop 12d}\SL
  v_{1}^{2}-v_{2}^{2} &=& \f{-2}{g^2+g'^2}\LP[m_1^2-m_2^2
               -\LP(m_1^2+m_2^2\RP)\f{\tan{\b}
           -\cot{\b}}{\tan{\b}+\cot{\b}}\RP]\nn
      &=&  \f{-2}{g^2+g'^2}\LP[m_1^2-m_2^2
               +\LP(m_1^2+m_2^2\RP)\cos{2\b}\RP],
           \label{Radiative Breaking prop 12e}
\EEA
where
\BEA
    \tan\beta &=& \f{v_{2}}{v_{1}}.
      \label{Radiative Breaking prop 12f}
\EEA
Here the angle $\b$ is a new parameter of the model
and since $v_{1},v_{2} \geq 0$ we have
\BEA
       0 \;\;\leq\;\; \b \;\;\leq\;\;  \f{\pi}{2}.
\EEA

With eqs.~\r{Radiative Breaking prop 12a}, \r{Radiative Breaking prop 12b}
and \r{Radiative Breaking prop 12e} the minimum of the potential can
be written as
\BEA
  V_{min} &=& \f{-1}{2(g^{2}+g^{'\,2})}
    \LP[\LP(m_{1}^{2}-m_{2}^{2}\RP)+
     \LP(m_1^2+m_2^2\RP)\cos{2\b}\,\RP]^{2},
        \label{Radiative Breaking prop 13}
\EEA

All the parameters
of the model have a functional dependence on the renormalization
point\footnote{This Q-dependence may for instance come from the
renormalization-group-improved tree-level potential
which incorporates the large logarithmic corrections
proportional to  $\a\log \LP( M_{GUT}/Q\RP)$. Here $M_{GUT}$ is
a grand unification scale.} Q.
This in particular applies to the mass parameters $m_{i}^{2}$
($i=1,2,3$) and thus to $\det{{\cal M}^{2}(Q)}$.
To proceed, one has to take the complicated (coupled) renormalization
group equations (RGE's) into account.
This we will not do here,
but only refer the interested reader to the
literature~\cite{SAV82}.
The rest of the discussion of this section will
be kept on a qualitative level.

At the Planck scale, $M_{Pl}$, condition~\r{Radiative Breaking prop 12} is not
fulfilled, and hence the critical scale reads
$\det{{\cal M}^{2}(Q_{c})} = 0$, where $Q_{c}<M_{Pl}$.
Below $Q_{c}$, non-vanishing
Higgs v.e.v.'s start to develop, signalling $SU(2)\times U(1)$-breaking
as discussed earlier, but only as long as ${\cal B}(Q)\geq 0$.
However, for some particular scale $Q_{s}<Q_{c}$, ${\cal B}(Q_{s})<0$
is driven negative and for $Q<Q_{s}$ one is in an instability region
where $SU(2)\times U(1)$ is broken in an unsatisfactory manner.
Our picture is recapitulated in figure~\ref{FIG: rad_breaking}
for various scales Q.

Note that in the supersymmetric limit, where {\em all} (soft) mass
parameters of $\L_{Soft}$ are set equal to zero,
$\det{{\cal M}^{2}} = 0$ and no electroweak breaking is possible in
view of condition~\r{Radiative Breaking prop 12}. So, in our model
the gauge symmetry breaking is connected to the breaking
of supersymmetry, as we already have  noted several times\footnote{It is
possible to construct non-minimal models~\cite{NMSSM} where
the Higgs-sector is enlarged by
an $SU(2)\times U(1)$-gauge singlet and where the gauge-symmetry
and SUSY can be broken separately.}.

Before we close this section, we will make one final comment.
Instead of our naive use of the tree-level scalar
potential~\r{Tree-level scalar potential},
we should have used the full one-loop corrected effective
potential
\BEA
   V_{1}(Q) &=&   V(Q)+\D V(Q).  \nonumber
\EEA
Here $\D V$ is the one-loop radiative correction to the scalar potential and
 in the leading logarithm approximation it  reads
\BEA
   \D V(Q) &\sim &  m_{t}^{4}\log{\LP(\f{M_{GUT}}{Q}\RP)^{2}},  \nonumber
\EEA
where $m_{t}$ is the top-quark mass and $M_{GUT}$ some
super-high unification scale.
By choosing a low renormalization scale, one
gets substantial contributions from $\D V$.
Until recently, it was believed that
the large logarithmic terms could be
reabsorbed into the soft  parameters\footnote{Recall that the soft parameters
in our theory are arbitrary.}
of $V(Q)$, and in
consequence, $\D V$ only contained small logarithmic corrections.
However, this only applies to so-called field
independent radiative corrections.
For the field dependent corrections we still can get substantial
contributions as explained e.g. in ref.~\ref{ELL91}.

Even though the scalar potential can receive large corrections from $\D V$,
the use of the tree-level potential $V(Q)$ is adequate for our discussion.
Furthermore, it simplifies the discussion enormously.

%\include{start}

%\clearpage

\section{The Physical Higgs Boson Spectrum.}

In the previous section we derived the condition for electroweak
symmetry breaking. Henceforth we will assume that these conditions, i.e.
eqs.~\r{Radiative Breaking prop 5} and
\r{Radiative Breaking prop 12},  are fulfilled, and show that this implies the
correct symmetry breaking pattern.

In the SM one starts by expanding around the Higgs v.e.v.'s
and identify the new state as the physical state.
However, by performing the same scheme for the MSSM,
these new weak interacting eigenstates
do not represent physical (mass) eigenstates, as we will see.
So, before we proceed, we will
work out the physical Higgs boson states.

The physical eigenstates are obtained by diagonalizing the
Higgs boson mass-square matrix. This is most easily done in a real basis where
\BEA
   H_1 &=& \LP(\BA{r} h_1+i h_2  \\
                      h_3+i h_4  \EA \RP),
                       \label{The Physical Higgs Boson Spectrum prop 3}\SL
   H_2 &=& \LP(\BA{r} h_5+i h_6  \\
                      h_7+i h_8  \EA \RP).
                       \label{The Physical Higgs Boson Spectrum prop 4}
\EEA
In this basis the scalar (Higgs) potential~\r{Tree-level scalar potential}
 reads
\BEA
  V(h_i) &=&    m_1^2\sum_{i=1}^{4}h_i^2
              + m_2^2\sum_{i=5}^8h_i^2
       % \nmb
              - 2 m_3^2\LP(h_1h_7+h_4h_6-h_3h_5-h_2h_8\RP)
         \nmb
              + \f{1}{8}\LP(g^2+g'^2\RP)
           \LP[\sum_{i=1}^4 h_i^2-\sum_{i=5}^{8}h_i^2\RP]^2
         \nmb
              + \f{g^2}{2}\LP(h_1h_5+h_2h_6+h_3h_7+h_4h_8\RP)^2
         \nmb
              +\f{g^2}{2}\LP(h_1h_6+h_3h_8-h_2h_5-h_4h_7\RP)^2 .
              \label{The Physical Higgs Boson Spectrum prop 4a}
\EEA
{}From this potential it is apparent that the Higgs field basis
that we are working in can not be a physical basis since it
contains off-diagonal mass terms.
Thus we are forced to transform to a mass-eigenstate basis, and
the method which we will apply, is described in detail in ref.~\ref{HAB79}
for a general two doublet model.

The physical Higgs boson states are obtained by
diagonalizing the Higgs boson mass-square matrix\footnote{The
factor of $\HA$ in front of
definition~\r{The Physical Higgs Boson Spectrum prop 5}
stems from the normalization of the scalar
fields in eqs.~\r{The Physical Higgs Boson Spectrum prop 3}
and \r{The Physical Higgs Boson Spectrum prop 4}.}
given by~\cite{HAB79}
\BEA
   M^2_{ij} &=& \LP. \HA\;\f{\P^2 V}{\P h_i\; \P h_j}\RP|_{\;\mbox{min}}.
            \label{The Physical Higgs Boson Spectrum prop 5}
\EEA
Here the term ``min" means setting $\LP<h_1\RP>=v_1$, $\LP<h_7\RP>=v_2$
and $\LP<h_i\RP> =0$ for all other i's. Note from
eq.~\r{The Physical Higgs Boson Spectrum prop 4a} that the
``mixed" second order partial derivatives of $V(h_i)$ are continuous
and thus equal (i.e. $\f{\P^2 V}{\P\,h_i\P\,h_j}= \f{\P^2 V}{\P\,h_j\P\,h_i}$),
implying a symmetric mass-matrix, i.e. $M^2_{ij} = M^2_{ji}$.

Now, the different parts of the Higgs sector will be analyzed in detail,
and this will be the aim of the next three subsections.

\subsection{The Charged Higgs Sector; Indices 3, 4, 5 and 6.}

With eqs.~\r{The Physical Higgs Boson Spectrum prop 4a} and
\r{The Physical Higgs Boson Spectrum prop 5}
the Higgs boson mass-square matrix is easily calculated.
Observe that the real and imaginary sector decouple i.e.
\BEA
  M^2_{56}=M^2_{54}=M^2_{36}=M^2_{34} = 0.  \nonumber
\EEA
The remaining mass-square matrix components read
\BEA
  M^2_{55} &=& m_2^2-\f{1}{4}\LP(g^2+g'^2\RP)
         \LP(v_1^2-v_2^2\RP)+\HA g^2v_1^2\nn
           &=& \HA\LP(g^2+\f{2m_3^2}{v_1v_2}\RP)v_1^2,\nn
  M^2_{53} &=& m_3^2+\HA g^2 v_1v_2\nn
           &=& \HA\LP(g^2+\f{2m_3^2}{v_1v_2}\RP)v_1v_2,\nn
  M^2_{33} &=& m_1^2+\f{1}{4}\LP(g^2+g'^2\RP)
          \LP(v_1^2-v_2^2\RP)+\HA g^2v_2^2\nn
           &=& \HA\LP(g^2+\f{2m_3^2}{v_1v_2}\RP) v_2^2, \nonumber
\EEA
and
\BEA
  M^2_{66} &=& M^2_{55}, \nn
  M^2_{44} &=& M^2_{33}, \nn
  M^2_{64} &=& - M^2_{53}.
      \label{The Physical Higgs Boson Spectrum prop 6}
\EEA
Here eqs.~\r{Radiative Breaking prop 12a} and \r{Radiative Breaking prop 12b}
have been taken advantage of in eliminating the mass parameters
$m_1^2$ and $m_2^2$.
Hence in the basis's $(h_5,h_3)$ and $(-h_6,h_4)$, the charged Higgs
mass-square matrix
reads\footnote{The particular sign of the basis $(-h_6,h_4)$,
owing to the appearance of the sign in
eq.~\r{The Physical Higgs Boson Spectrum prop 6},
is chosen such that the two mass matrices coincide with each other.}
\BEA
  M_{\pm}^2 &=& \HA \LP(g^2+\f{2m_3^2}{v_1v_2}\RP)
                   \LP(\BA{cc} v_1^2    &  v_1 v_2 \\
                               v_1v_2   &  v_2^2     \EA \RP).
                   \label{The Physical Higgs Boson Spectrum prop 7}
\EEA
To obtain the physical charged Higgs states and their masses,
one has to orthogonal diagonalize
\footnote{Recall that
an orthogonal diagonalizable $n\times n$-matrix A always can be written
in the form  $A~=~PDP^{-1}$,
where D is given by
$D = \mbox{diag}  \LP( \BA{cccc}  \la_{1}& \la_{2} &\ldots &\la_{n} \EA \RP)$,
and P is the orthogonal matrix containing the eigenvectors in the following way
$P =  \LP( \BA{cccc} {\bf v}_1 & {\bf v}_2 & \ldots & {\bf v}_n \EA \RP)$.
Here $(\la_i, {\bf v}_i)$ are
corresponding sets of eigenvalues and eigenvectors.}
the matrix $M^2_{\pm}$  since
physical states always are orthogonal to each other.
Note that $M^2_{\pm}$ always will  be orthogonal diagonalizable because it is
symmetric~\cite{ANTON}.

By calculating the eigenvalues and the corresponding set of
orthonormal eigenvectors, the
charged mass matrix
$M^2_{\pm}$ can be written in the form~($tan\b=v_{2}/v_{1}$)
\BEA
  M^2_{\pm} &=&
     \LP( \BA{cc}   -\sin{\b}  & \cos{\b}\\
                     \cos{\b}  & \sin{\b}   \EA \RP)
     \LP( \BA{cc}     0  & 0\\
                     0   & m^2_{H^{\pm}}   \EA \RP)
     \LP( \BA{cc}   -\sin{\b}  & \cos{\b}\\
                     \cos{\b}  & \sin{\b}   \EA \RP),
                     \label{The Physical Higgs Boson Spectrum prop 8}
\EEA
where
\BEA
  m^2_{H^{\pm}} &=& \HA\LP(g^2+\f{2m_3^2}{v_1v_2}\RP)\LP(v_1^2+v_2^2\RP),
     \label{The Physical Higgs Boson Spectrum prop 9}
\EEA
is the mass-square of the physical charged Higgs-bosons.
Note that by this diagonalization procedure, two massless and two massive
states have appeared. The mass-zero states will be associated with Goldstone
bosons, as we will see in a moment.

After completing the diagonalizing procedure, the mass terms
for the charged Higgs-bosons in the Lagrangian can be written in the
following way
\BEA
\lefteqn{ \LP( \BA{cc} h_5 & h_3 \EA \RP)
M^2_{\pm} \LP( \BA{r} h_5 \\ h_3 \EA \RP)
 +  \LP( \BA{cc} -h_6 & h_4 \EA \RP) M^2_{\pm}
\LP( \BA{r} -h_6 \\ h_4 \EA \RP)} \hspace{1.5cm} \nn
  &=&
    \LP( \BA{cc} h_5+i h_6 & h_3-ih_4 \EA \RP)
M^2_{\pm} \LP( \BA{r} h_5-ih_6 \\ h_3+i h_4 \EA \RP)\nn
  &=&
     \LP( \BA{cc} H_2^1 & H_1^{2\;\dagger} \EA \RP) M^2_{\pm}
      \LP( \BA{l} H_2^{1\;\dagger} \\ H_1^2  \EA \RP) \nn
  &=&  \LP( \BA{r} -H_2^1 \sin{\b}+H_1^{2\;\dagger}\cos{\b} \\
                    H_2^1\cos{\b}+H_1^{2\;\dagger}\sin{\b} \EA \RP)^{T}
       \LP( \BA{cc}     0  & 0\\
                        0   & m^2_{H^{\pm}}   \EA \RP)
       \LP( \BA{r} -H_2^{1\;\dagger } \sin{\b}+H_1^{2}\cos{\b} \\
                    H_2^{1\;\dagger}\cos{\b}+H_1^{2}\sin{\b} \EA \RP)\nn
   &=& \LP( \BA{cc} G^{+}& H^{+}\EA\RP)
       \LP( \BA{cc}     0  & 0\\
                        0   & m^2_{H^{\pm}}   \EA \RP)
       \LP(\BA{r}  G^{-}\\H^{-}  \EA\RP).         \nonumber
%                \label{The Physical Higgs Boson Spectrum prop 9}
\EEA
Here
\BEA
   G^{-} &=& H_1^{2}\cos{\b}-H_2^{1\;\dagger} \sin{\b},
         \label{The Physical Higgs Boson Spectrum prop 9a}\SL
   H^{-} &=& H_1^{2}\sin{\b}+H_2^{1\;\dagger}\cos{\b},
         \label{The Physical Higgs Boson Spectrum prop 9b}
\EEA
and
\BEA
   G^{+} &=& \LP( G^{-}\RP)^{\dagger},\nn
   H^{+} &=& \LP( H^{-}\RP)^{\dagger}, \nonumber
\EEA
where $G^{\pm}$ are the charged Goldstone bosons while
$H^{\pm}$ are the charged Higgs bosons.

This completes this subsection.

\subsection{The Neutral Higgs Sector; Indices 2 and 8.}

In the previous subsection we saw that the charged Higgs sector
decouples into a real and an imaginary part.
This is also the case for the neutral Higgs sector as
the reader may easily verify
by showing that $M^2_{ij}=0$ for $i=1,7$ and $j=2,8$.
This owing to the fact that our theory is CP-invariant.
We start the discussion with the imaginary (CP-odd) sector, and
consider the real~(CP-even) part in the next subsection\footnote{The
various CP-assignments can be obtained by e.g. studying the
interactions of (neutral) Higgs- and gauge-bosons.}.

Proceeding as in the previous subsection the mass-square matrix becomes
\BEA
   \f{m_3^2}{v_1v_2}\LP( \BA{cc}   v_1^2    & v_1 v_2   \\
                                   v_1 v_2  & v_2^2    \EA \RP) \nonumber
\EEA
in the basis $(h_8, h_2)$.
By diagonalizing this matrix, which is identical to that for the charged
sector, the physical mass eigenstates are obtained as follows
\BEA
\lefteqn{\f{m_3^2}{v_1v_2}\LP( \BA{cc} h_8 & h_2\EA \RP)
          \LP( \BA{cc} v_1^2  & v_1v_2 \\
                       v_1v_2 & v_2^2  \EA \RP)
          \LP(\BA{r} h_8 \\ h_2 \EA \RP)}\hspace{1.5cm}\nn
    &=&  \LP( \BA{r} -h_8\sin{\b}+h_2\cos{\b} \\
                      h_8\cos{\b}+h_2\sin{\b} \EA \RP)^{T}
       \LP( \BA{cc}     0  & 0\\
                        0   & m^2_{H^{0}_3}   \EA \RP)
       \LP( \BA{r} -h_8\sin{\b}+h_2\cos{\b} \\
                    h_8\cos{\b}+h_2\sin{\b} \EA \RP)\nn
   &=& \LP( \BA{cc} \f{G^{0}}{\sqrt{2}} & \f{H^{0}_3}{\sqrt{2}} \EA\RP)
       \LP( \BA{cc}     0  & 0\\
                        0   & m^2_{H^{0}_3}  \EA \RP)
       \LP(\BA{r}  \f{G^{0}}{\sqrt{2}} \\
      \f{H^{0}_3}{\sqrt{2}}  \EA\RP).  \nonumber
%                \label{The Physical Higgs Boson Spectrum prop 10}
\EEA
Here
\BEA
   G^{0} &=&  \sqrt{2} \LP(\, h_2\cos{\b}-h_8\sin{\b}\, \RP) \nn
         &=&  \sqrt{2} \LP(\, \mbox{Im\,}
               H_1^1\cos{\b}-\mbox{Im\,}H_2^2\sin{\b}\,\RP),
         \label{The Physical Higgs Boson Spectrum prop 10a}\SL
   H^0_3 &=&  \sqrt{2} \LP(\, h_2\sin{\b}+h_8\cos{\b}\,\RP) \nn
         &=&     \sqrt{2} \LP(\, \mbox{Im\,}
            H_1^1\sin{\b}+\mbox{Im\,}H_2^2\cos{\b}\,\RP),
         \label{The Physical Higgs Boson Spectrum prop 10b}
\EEA
where $G^{0}$ is a Goldstone boson~(in this case neutral), and $H^{0}_{3}$ is
a neutral Higgs boson.
The mass of the Higgs boson
is\footnote{In sect.~\ref{SECT: The W-, Z- and Lepton Mass.}
we will show that the W- and Z-mass
are respectively given by $m_{\mbox{w}}^2=\HA g^2\LP(v_1^2+v_2^2\RP)$
and $m_{\mbox{z}}^2=\HA \LP(g^2+g'^2\RP)\LP(v_1^2+v_2^2\RP)$.}
\BEA
     m_{H^0_3}^2 &=& \f{m_3^2}{v_1v_2}\LP( v_1^2+v_2^2\RP)   \nn
                 &=& m_{H^{\pm}}^2-m_{\mbox{w}}^2.
                 \label{The Physical Higgs Boson Spectrum prop 10dgdb}
\EEA
The factors of $\sqrt{2}$ are inserted in order for these fields to have the
conventional kinetic energy terms.
%Note that also here we have a massless Goldstone boson, but now it is neutral.

\subsection{The Neutral Higgs Sector; Indices 1 and 7.}

After completing the diagonalizing of the neutral imaginary
sector, we will now consider the corresponding real sector.
For this sector the mass-square matrix reads
\BEA
   M_0^2 &=& \LP( \BA{cc} A & B \\
                          B & C   \EA \RP) \nonumber
\EEA
relative to the basis $(h_1, h_7)$.
Here we have introduced the abbreviations
\BEA
   A &=& \HA \LP(g^2+g'^2\RP)v_1^2+m_3^2\f{v_2}{v_1},
       \label{index 17 prop 3a}\nn
   B &=& -\HA \LP(g^2+g'^2\RP)v_1 v_2-m_3^2,
       \label{index 17 prop 3b} \nn
   C &=& \HA \LP(g^2+g'^2\RP)v_2^2+m_3^2\f{v_1}{v_2},
       \label{index 17 prop 3c}    \nonumber
\EEA
and we notice that $A, C \geq 0$ and $B \leq 0$.
Also here eqs.~\r{Radiative Breaking prop 12a} and \r{Radiative Breaking prop
12b}
have been used to eliminating the mass parameters $m_1^2$ and $m_2^2$.

The orthogonal diagonalization scheme for this sector is not as
straightforward as above. Accordingly some more details will be given.
The eigenvalues of $M^2_0$
\addtocounter{footnote}{-1}
read\footnotemark
\BEA
   m^{2}_{H^{0}_{1},\,H^{0}_{2}}  &=&
      \HA \LP[\;A+C\pm \sqrt{\LP( A-C \RP)^2 +4B^2}\;\RP] \nn
      &=& \HA \LP[\; m^2_{H^0_3}+m_{\mbox{z}}^2\pm
            \sqrt{\LP(m^2_{H^0_3}+m^2_{\mbox{z}}\RP)^2
                    -4m^2_{\mbox{z}}m^2_{H^0_3}\cos^2{2\b} }\;\RP],
         \label{index 17 prop 4}
\EEA
where the positive (negative) sign is associated with $m^2_{H^0_1}$
($m^2_{H^0_2}$). The corresponding eigenvectors
are\footnote{Here ${\bf v}_1$ and ${\bf v}_2$ correspond to
the eigenvalues $m^2_{H^0_1}$ and $m^2_{H^0_2}$ respectively.}
\BEA
  {\bf v}_{1,2} &=& N_{1,\,2}\LP( \BA{c}   1 \\
      \f{-(A-C)\pm \sqrt{\LP( A-C \RP)^2 +4B^2}}{2B} \EA \RP).
      \label{index 17 prop 5}
\EEA
Here $N_{1,\,2}$ are normalization constants.

As will become clear soon,
it is useful to introduce the mixing angel $\a$~(not
to be confused with the fine structure constant) defined by
\BEA
  \sin{2\a} &=& \f{2B}{\sqrt{\LP(A-C\RP)^2+4B^2}}\nn
             &=& -\sin{2\b}\;\LP( \f{m^2_{H^0_1}+m^2_{H^0_2}}{m^2_{H^0_1}
                    -m^2_{H^0_2}}\RP)\nn
%    \label{index 17 prop 7}\SL
  \cos{2\a} &=& \f{A-C}{\sqrt{ \LP(A-C\RP)^2+4B^2}}\nn
               &=& -\cos{2\b}\;\LP( \f{m^2_{H^0_3}
                   -m^2_{\mbox{z}}}{m^2_{H^0_1}-m^2_{H^0_2}}\RP).  \nonumber
%      \label{index 17 prop 8}
\EEA
{}From the mathematical identities  $\sin{2\a}=2\sin\a\cos\a$ and
$\cos{2\a}= \cos^2\a-\sin^2\a$, one easily obtains the second
order equation
\BEA
      x^2+2\cot\LP(2\a\RP) x -1 &=& 0,  \nonumber
     % f{\cos{2\a}}{\sin{2\a}}x-1 = 0, \nonumber
\EEA
where $x=\tan\a$. This equation generally has two distinct
solutions. However, earlier we have chosen $v_1, v_2 \geq 0$ or equivalently
$0\leq \b \leq \f{\pi}{2}$, something which according
to ref.~\ref{GUN86} implies that $-\f{\pi}{2} \leq \a \leq 0$.
With this constraint in mind, one can uniquely solve for x, and the
result is~(remember that $B\leq 0$)
%\footnote{Note from eqs.~\r{index 17 prop 3a} to \r{index 17 prop 3c}
%that $A\geq 0$, $B\leq 0$ and $C\geq 0$, and that $-\f{\pi}{2}\leq \a \leq 0$
%implies that $\f{\sin{\a}}{\cos{\a}}$, $\,\f{\cos{\a}}{\sin{\a}} \leq 0$.}
\BEA
   \tan\a &=& \f{-\LP(A-C\RP)+\sqrt{\,\LP(A-C\RP)^2+4B^2}}{2B},
             \label{index 17 prop 9a}
\EEA
and by inversion~(and some algebra)
\BEA
     \cot\a &=& \f{\LP(A-C\RP)+\sqrt{\,\LP(A-C\RP)^2+4B^2}}{2B}.
               \label{index 17 prop 9b}
\EEA
By comparing eqs.~\r{index 17 prop 9a} and \r{index 17 prop 9b} with
eq.~\r{index 17 prop 5}, we see that the second component of
${\bf v}_{1}$~(${\bf v}_{2}$) can up to a sign be identified with $\tan\a$
($\cot\a$). The mixing angle, $\a$, was defined
in order to obtain this.

Thus we choose $ N_1 = \cos{\a}$ and $N_2 = -\sin{\a}$ in order to
obtain an orthonormal eigenvector set, and
the mass-square matrix of the real neutral sector takes on the form
\BEA
  M^2_0 &=&
       \LP( \BA{cc} \cos{\a}  & -\sin{\a} \\
                    \sin{\a}  &  \cos{\a}  \EA \RP)
       \LP(\BA{cc}   m^2_{H^0_1}  &  0 \\
                       0          &  m^2_{H^0_2}  \EA \RP)
       \LP( \BA{cc} \cos{\a}  & -\sin{\a} \\
                    \sin{\a}  &  \cos{\a}  \EA \RP)^{-1}.
\EEA

The corresponding mass terms of  the Lagrangian now become
\BEA
\lefteqn{ \LP( \BA{cc} h_1 & h_7 \EA \RP) M^2_0
%   \LP(\BA{cc}   A  &  B \\
%                 B  &  C  \EA \RP)
   \LP(\BA{r} h_1 \\ h_7 \EA \RP)}\hspace{0.3cm} \nn
     &=& \LP( \BA{r} h_1 \cos{\a}+ h_7\sin{\a} \\
                     -h_1\sin{\a}+h_7\cos{\a}   \EA \RP)^{T}
         \LP(\BA{cc}   m^2_{H^0_1}  &  0 \\
                       0          &  m^2_{H^0_2}  \EA \RP)
         \LP( \BA{r} h_1 \cos{\a}+ h_7\sin{\a} \\
                     -h_1\sin{\a}+h_7\cos{\a}  \EA \RP). \hspace{0.5cm}
\EEA
When we now proceed by identifying the physical Higgs states $H^0_1$ and
$H^0_2$, we have to be careful. The reason is that these states, as any
physical states, have to have zero vacuum expectation values.
Hence we make the following identifications
\BEA
 \f{H^0_1}{\sqrt{2}} + v_1\cos{\a}+v_2\sin{\a}
           &=& h_1\cos{\a}+h_7\sin{\a} , \nn
 \f{H^0_2}{\sqrt{2}} - v_1\sin{\a}+v_2\cos{\a}
           &=& -h_1\sin{\a}+h_7\cos{\a},\nonumber
\EEA
or equivalently
\BEA
  H^0_1 &=& \sqrt{2}\LP[ \,\LP(\mbox{Re\,}
          H_1^1-v_1\RP)\cos{\a}+\LP(\mbox{Re\,}H_2^2-v_2\RP)\sin{\a}\RP],
      \label{index 17 prop 18}\SL
  H^0_2 &=& \sqrt{2}\LP[ \,-\LP(\mbox{Re\,}
          H_1^1-v_1\RP)\sin{\a}+\LP(\mbox{Re\,}H_2^2-v_2\RP)\cos{\a}\RP].
      \label{index 17 prop 19}
\EEA

This concludes this section.

\subsection{Conclusion and Comments.}

In the three previous subsections the physical content of the
Higgs sector of the MSSM was obtained.
It is the charged Higgs bosons~($H^{\pm}$),
the neutral Higgs bosons\footnote{Some authors use the notation
$H^{0}$, $h^{0}$ and $A^{0}$ instead of our $H^{0}_{1}$,  $H^{0}_{2}$
and $H^{0}_{3}$.}~($H^{0}_{i}$,     $i=1,2,3$)
and finally the charged~($G^{\pm}$) and neutral Goldstone bosons~($G^{0}$).

The new fields in terms of the ``old" are given in
eqs.~\r{The Physical Higgs Boson Spectrum prop 9a},
 \r{The Physical Higgs Boson Spectrum prop 9b},
\r{The Physical Higgs Boson Spectrum prop 10a},
 \r{The Physical Higgs Boson Spectrum prop 10b},
\r{index 17 prop 18} and \r{index 17 prop 19}.
However, in order to give the Lagrangian in terms of the physical fields,
we have to invert the above relations.
The results, obtained by straightforward calculations, are
\BEA
   H_1  &=&
     \LP( \BA{l}
       v_1+\f{1}{\sqrt{2}}\LP[\,
           H_1^0\cos{\a}-H_2^0\sin{\a}+iH_3^0\sin{\b}+iG^0\cos{\b}\,\RP] \\
       H^-\sin{\b}+G^-\cos{\b}  \EA \RP),
       \label{The Z-, W- and Lepton Mass prop 1} \SL
   H_2  &=&
     \LP( \BA{l}
       H^+\cos{\b}-G^+\sin{\b} \\
       v_2+\f{1}{\sqrt{2}}\LP[\,H_1^0\sin{\a}
           +H^0_2\cos{\a}+i H^0_3\cos{\b}-iG^0\sin{\b}\,\RP] \EA\RP).
              \label{The Z-, W- and Lepton Mass prop 2}
\EEA
By inserting these expressions into the
   Lagrangian~\r{Total L sub SUSY 4-comp} the
interactions (and Feynman rules) of the physical Higgs bosons can
be obtained.

{}From the formulae for the Higgs-masses obtained earlier,
eqs.~\r{The Physical Higgs Boson Spectrum prop 7},
\r{The Physical Higgs Boson Spectrum prop 10dgdb} and
\r{index 17 prop 4}, it is interesting to note that in the limit
$m_{H^{0}_{3}}\rightarrow \infty$ (fixed $\tan \b$),
$H^{\pm}$, $H^{0}_{1}$ (and $H^{0}_{3}$) decouple from the theory,
and thus the Higgs-sector contains only $H^{0}_{2}$.
In this limit, it is possible to show that $H^{0}_{2}$ is identical to
the Higgs of the (minimal) Standard Model.

It should be noticed that the Higgs-masses obtained in the previous subsections
are tree-level formulae.
They fulfil the following relations
\BEA
   m_{H^{\pm}} \;\;\geq \;\; M_{\mbox{w}},\nn
   m_{H^{0}_{2}} \;\;\leq \;\; m_{\mbox{z}} \;\;\leq \;\; m_{H^{0}_{1}},\nn
   m_{H^{0}_{3}} \;\; \geq \;\; m_{H^{0}_{2}}. \nonumber
\EEA
Since $ m_{H^{0}_{2}} \leq m_{\mbox{z}}$ (at tree-level)
it is believed, due to the
interaction picture of $H^{0}_{2}$, that $H^{0}_{2}$
could be produced and hopefully detected at LEP.
No Higgs has ever been seen and this may seem like a problem.
Thus it came like a relief to many physicists when it recently
was reported~\cite{ELL91}
(see subsect.~\ref{SUBsect: The Scalar Higgs Potential.})
that the MSSM Higgses could get
radiative corrections as large as ${\cal O}\LP(100\RP)$ GeV.
This at once may push the mass of $H^{0}_{2}$ far above that of the Z-boson
(and outside the LEP~1 discovery range).
These large radiative corrections also
have the implications~\cite{ELL91 PL B257 p 91 ref 1},
due to the unsuccessful Higgs searches at LEP~1, that\footnote{It is usual
to let $\tan \b$ varies in the range $1\leq \tan \b \leq 50$.}
\BEA
   \tan \b & \geq & 1,
      \label{constraint tan-beta}
\EEA
in the context of the MSSM.

%This concludes this section.

%\end{document}

\section{The W-, Z- and Lepton Mass.}
   \label{SECT: The W-, Z- and Lepton Mass.}

In this section we will give an illustrative  demonstration
(and a control for the sceptic one)
of the fact that our gauge symmetry breaking scheme is capable
of ``producing" masses  of the W- and Z-bosons
and the (charged) leptons.

As for the SM case, we will make use of the gauge freedom of the
theory and transform to the unitary gauge. This consists of setting
the Goldstone fields of eqs.~\r{The Z-, W- and Lepton Mass prop 1}
and \r{The Z-, W- and Lepton Mass prop 2} to zero,
but it will have no practical
implication for our discussion.

\subsection{The W- and Z-Mass.}

{}From the Lagrangian~\r{Total L sub SUSY 4-comp}
(after symmetry breaking) we pick
the following terms
% The piece $\LP(D^\mu H_1\RP)^{\dagger}\LP(D_\mu H_1\RP)~
%+~\LP(D^\mu H_2\RP)^{\dagger}\LP(D_\mu H_2\RP)$.
% of the Lagrangian~\r{} contains the terms
\BEA
 \lefteqn{ \LP(D^\mu \mbox{v}_1\RP)^\dagger
  \LP(D_\mu \mbox{v}_1\RP) +
  \LP(D^\mu \mbox{v}_2\RP)^\dagger
  \LP(D_\mu \mbox{v}_2\RP)}\hspace{1.5cm} \nn
    &=&
      \LP( \BA{c}  \f{ig}{2\CWA}\; v_1 Z^\mu\\
      \f{i g}{\sqrt{2}}\;v_1 W^{-\,\mu}\EA\RP)^\dagger
      \LP( \BA{c}  \f{ig}{2\CWA}\; v_1 Z_\mu\\
      \f{i g}{\sqrt{2}}\;v_1 W^{-}_{\mu}\EA\RP)
  \nmb
    +
      \LP( \BA{c}
      \f{i g}{\sqrt{2}}\;v_2 W^{+\,\mu}\\
       -\,\f{ig}{2\CWA}\; v_2 Z^\mu
       \EA\RP)^\dagger
       \LP( \BA{c}
      \f{i g}{\sqrt{2}}\;v_2 W^{+}_{\mu}\\
       -\,\f{ig}{2\CWA}\; v_2 Z_\mu
       \EA\RP)  \nn
    &=&  \f{g^2}{4\CWAS}\LP(v_1^2+v_2^2\RP) Z^\mu Z_\mu
    \nmb
       + \f{g^2}{4}\LP(v_1^2+v_2^2\RP)W^{+\,\mu}W^{-}_{\mu}
       + \f{g^2}{4}\LP(v_1^2+v_2^2\RP)W^{-\,\mu}W^{+}_{\mu}.\nonumber
\EEA
where $\mbox{v}_1 = \LP(\BA{cc} v_1 & 0 \EA \RP)^T$
and $\mbox{v}_2 = \LP(\BA{cc} 0 & v_2 \EA \RP)^T$.

Hence the Z- and W-mass can be identified as
\BEA
   m^2_{\mbox{w}} &=& \f{g^2}{2}\LP(v_1^2+v^2_2\RP),\SL
   m^2_{\mbox{z}} &=& \HA\,\f{g^2}{\CWAS}\LP(v_1^2+v_2^2\RP)
                  \;\;=\;\; \HA \LP(g^2+g'^2\RP) \LP(v_1^2+v_2^2\RP),
\EEA
which is consistent with the results from the SM.

Note that with the above results  $v^2_1+v^2_2$ is fixed by the W-mass.

\subsection{The Lepton Mass.}

Now the lepton mass will be paid attention.
The piece $f\e^{ij}\,\bar{R}L^i H_1^j +h.c.$, stemming from the
Yukawa piece of the superpotential,
%(and its hermitian conjugated),
gives raise to the lepton mass as we now will see.

With $H_{1}$ given by eq.~\r{The Z-, W- and Lepton Mass prop 1},
$f\e^{ij}\,\bar{R}L^i H_1^j +h.c.$ contains the following terms
\BEA
 - f\,\bar{R} L^2 v_1  +h.c.
    &=& -f v_1 \LP( \bar{\l}_{R}l_{L}+\bar{l}_{L}l_{R}\RP) \nn
    &=& -f v_1 \LP( \bar{l}\, P_{L}\, l + \bar{l}\, P_{R}\, l \RP) \nn
    &=& -f v_1\; \bar{l}l.  \nonumber
\EEA
Hence, we can make the identification
\BEA
   m_{l} &=& f v_1, \nonumber
\EEA
and as in the SM, we notice that the lepton mass
is undetermined by the theory.

For later use we observe that the Yukawa coupling f can be written as
\BEA
    f &=& \f{m_{l}}{v_1}\;\;=\;\; \f{g m_{l}}{\sqrt{2}m_{\mbox{w}}\cos\b}.
      \label{The Z-, W- and Lepton Mass prop 4}
\EEA

%In order for this result to fit with the SM-result, the relation
%$f~=~\f{G_e}{v_1}\sqrt{2(v_1^2+v_2^2)}$ has to be fulfilled.

\section{The Physical Slepton States.}
    \label{SECT: Slepton Mass-Matrix}

The Lagrangian~\r{Total L sub SUSY 4-comp} contains off-diagonal mass
terms for the sleptons in the basis $(\tilde{l}_{L}, \tilde{l}_{R} )$.
So, also here we have to perform a diagonalizing procedure to obtain
the physical mass eigenstates, and hence we have
%With eqs.~\r{four comp lagr}
%and~\r{Component Field Expansion of L sub Soft prop 1}
%we have
\BEA
  \L_{slept.}^{mass} &=&
    -\mu f v_{2}\;\tilde{l}_{L}^{\dagger}\tilde{l}_{R}
    -\mu f v_{2}\;\tilde{l}_{R}^{\dagger}\tilde{l}_{L}
    -f^{2}v_{1}^{2}\LP(\tilde{l}_{L}^{\dagger}\tilde{l}_{L}
                      +\tilde{l}_{R}^{\dagger}\tilde{l}_{R} \RP)
  \nmb
    -m_{L}^{2}\;\tilde{l}_{L}^{\dagger}\tilde{l}_{L}
    -m_{R}^{2}\;\tilde{l}_{R}^{\dagger}\tilde{l}_{R} \NN
  &=&
    -\LP(\BA{cc} \tilde{l}_{L}^{\dagger} & \tilde{l}_{R}^{\dagger} \EA \RP)
      \LP(\BA{cc}  m_{L}^{2} +f^{2}v_{1}^{2} & \mu f v_{2} \\
                   \mu f v_{2}               & m_{R}^{2}
             +f^{2}v_{1}^{2} \EA \RP)
      \LP( \BA{c} \tilde{l}_{L} \\ \tilde{l}_{R} \EA \RP) .  \nonumber
\EEA
By diagonalizing, one obtains the mass eigenstates (in the usual way)
\BEA
  \tilde{l}_{1} &=& \tilde{l}_{L}\cos{\t} +\tilde{l}_{R}\sin{\t}, \nn
  \tilde{l}_{2} &=&  \tilde{l}_{L}\sin{\t} -\tilde{l}_{R}\cos{\t}, \nonumber
\EEA
with\footnote{Notice from eqs.~\r{Radiative Breaking prop 12f}
and \r{The Z-, W- and Lepton Mass prop 4} that
$fv_2 = fv_1\f{v_2}{v_1}=m_l\tan\b$.}
\BEA
   \tan{2\t} &=& \f{2\mu f v_{2}}{\LP(m^{2}_{L}-m^{2}_{R}\RP)}
            \;\;=\;\; \f{ 2 \mu m_l \tan\b }{ \LP( m^2_L - m^2_R \RP) },
  \nonumber
\EEA
and masses respectively given by
\BEA
   M_{\tilde{l}_{1},\tilde{l}_{2}}^2
      &=& f^{2}v_{1}^{2}+\HA\LP[\LP(m_{L}^{2}+m^{2}_{R}\RP)
       \pm \sqrt{\LP(m^{2}_{L}-m^{2}_{R}\RP)^{2}+4\mu^{2}
              f^{2}v_{2}^{2}}\;\RP]\nn
  &=&     m^2_l+\HA\LP[ \LP(m_{L}^{2}+m^{2}_{R}\RP)
       \pm \sqrt{\LP(m^{2}_{L}-m^{2}_{R}\RP)^{2}+4\mu^{2}
              m^2_l\tan^2\b }\;\RP].
\EEA

Unfortunately, there do not exist much
information about the parameters contained
in the slepton mass matrix. All the same, we will assume maximal mixing,
i.e. $\t = \pi/4$ or
\BEA
    m^{2}_{L} = m^{2}_{R}= \tilde{m}^{2}.
\EEA
A motivation for this choice can be taken from
supersymmetric QED where this choice is made in order
to keep parity unbroken.

Hence
\BEA
  \tilde{l}_{1} &=& \f{\tilde{l}_{L} +\tilde{l}_{R}}{\sqrt{2}},\SL
  \tilde{l}_{2} &=& \f{\tilde{l}_{L} -\tilde{l}_{R}}{\sqrt{2}},
\EEA
and
\BEA
  M_{\tilde{l}_{1},\tilde{l}_{2}}^2
      &=& \tilde{m}^2+m^2_l \pm \ABS{\mu}m_l \tan\b.
\EEA

This concludes this section.

\section{Chargino and Neutralino Mixing.}
   \label{SECT: Chargino and Neutralino mixing}

The gaugino-higgsino sector of the
theory also contains off-diagonal mass terms,
as easily seen from the
Lagrangian~\r{Total L sub SUSY 4-comp}.
To obtain mass-eigenstates the now familiar diagonalization procedure
has to be performed, and the  resulting mass-eigenstates are  called
charginos, $\tilde{\chi}^{\pm}$, and neutralinos, $\tilde{\chi}^{0}$.
The discussion of these states will be the aim of
the present section.

\subsection{Chargino Mixing.}
    \label{SUBSECT: Chargino Mixing}

Charginos $\tilde{\chi}^+_i$ ($i=1,2$), which arise due to mixing
of Winos, $\tilde{W}^{\pm}$, and charged Higgsinos, $\tilde{H}^{\pm}$, are
four component Dirac spinors. Since there in principle are two
independent mixings, i.e. ($\tilde{W}^-,\tilde{H}^-$) and
($\tilde{W}^+,\tilde{H}^+$), we will need two
unitary matrices in order to diagonalize the resulting
mass-matrix~\cite{GRI83}.

{}From the Lagrangian~\r{Total L sub SUSY 4-comp} we pick the terms
\BEA
\L_{\tilde{\chi}^{\pm}}^{mass}
    &=&  -gv_1\,\bar{\tilde{W}} P_{R}\tilde{H}
        -gv_1\,\bar{\tilde{H}} P_{L}\tilde{W}
        -gv_2\,\bar{\tilde{H}} P_{R}\tilde{W}
        -gv_2\,\bar{\tilde{W}} P_{L}\tilde{H}
   \nmb
   + \mu\, \bar{\tilde{H}}\tilde{H}
   + M_{\tilde{W}}\,  \bar{\tilde{W}}\tilde{W}, \nonumber
\EEA
which in two-component form reads
\BEA
\L_{\tilde{\chi}^{\pm}}^{mass}
   &=& ig \LP[
                  v_1 \psi^2_{H_1}\la^+
                  + v_2\la^-\psi^1_{H_2}
           \RP]
%    \nmb
          + \mu\; \psi^2_{H_1} \psi^1_{H_2}
        - M\,
            \la^{-}\la^{+}
             + h.c.
             \label{chargino prop 1}
\EEA
By introducing the notation
\BEA
   \psi^{+} \;\;=\;\; \LP( \BA{r} - i\la^{+} \\
     \psi^{1}_{H_{2}} \EA \RP), \hspace{1cm}
%   \label{chargino prop 2} \SL
   \psi^{-} \;\;=\;\; \LP( \BA{r} - i\la^{-} \\ \psi^{2}_{H_{1}} \EA \RP),
   \label{chargino prop 3}   \nonumber
\EEA
and
\BEA
     \Psi^{\pm} &=& \LP( \BA{c}   \psi^{+}  \\ \psi^{-}  \EA  \RP),  \nonumber
%          \label{chargino prop 4}
\EEA
eq.~\r{chargino prop 1} takes on the form
\BEA
  \L_{\tilde{\chi}^{\pm}}^{mass}
      &=&  \HA \LP( \Psi^{\pm} \RP)^{T} Y^{\pm} \Psi^{\pm} + h.c.  \nonumber
  %         \label{chargino prop 5}
\EEA
Here
\BEA
      Y^{\pm} &=& \LP(  \BA{cc} 0  & X^{T} \\
                                X  & 0      \EA  \RP),
                                \label{chargino prop 6}
\EEA
with
\BEA
   X &=&   \LP(  \BA{cc}  M  & -\sqrt{2} m_{\mbox{w}}\sin{\b}   \\
                          -\sqrt{2} m_{\mbox{w}}\cos{\b}  &   \mu   \EA  \RP).
                          \label{chargino prop 7}
\EEA
Now, two-component mass-eigenstates can be defined by ($i, j = 1,2$)
\BEA
     \chi^{+}_{i}  &=&  V_{ij} \psi^{+}_{j},
         \label{chargino prop 8} \SL
     \chi^{-}_{i}  &=&  U_{ij} \psi^{-}_{j},
         \label{chargino prop 9}
\EEA
where  U and V are unitary matrices, chosen in such a way that
\BEA
   U^{*} X V^{\dagger}  &=& M^{\pm}_{D}.
       \label{chargino prop 10}
\EEA
Here $M^{\pm}_{D}$ is the chargino mass-matrix.
Since we have assumed CP-invariance of our theory, this
in particular holds for the chargino sector. Thus
the chargino-masses will be real and non-negative\footnote{It
is possible to show that the masses read
\BEA
     M_{\tilde{\chi}_{1}}^{2} &=& A+\sqrt{B}, \nn
     M_{\tilde{\chi}_{2}}^{2} &=& A-\sqrt{B}, \nonumber
\EEA
where
\BEA
   A &=& \HA \LP( M^{2}+\mu^{2}  \RP) +  m^{2}_{\mbox{w}},\nn
   B &=& \f{1}{4} \LP(M^{2}-\mu^{2} \RP)^{2}
          + m^{4}_{\mbox{w}}\cos^{2}\LP( 2\b \RP)
          + m^{2}_{\mbox{w}}\LP( M^{2}+\mu^{2}+2\mu M\sin\LP( 2\b \RP)\RP)
     \nonumber.
\EEA  }.
Furthermore, the two-component spinors of eqs.~\r{chargino prop 8} and
\r{chargino prop 9} can be arranged in (four-component) Dirac-spinors
as follows\footnote{In what follows, we will use the abbreviation
$\tilde{\chi} \equiv \tilde{\chi}^{+}$.
Hence $\tilde{\chi}^{c} \equiv (\tilde{\chi}^{+})^{c}=\tilde{\chi}^-$ is a
negatively charged
chargino.}:
\BEA
     \tilde{\chi}_{i} &=& \LP(  \BA{c}  \chi^{+}_{i} \\
                                            \bar{\chi}^{-}_{i}  \EA  \RP)
                      ,\hspace{1.5cm}   i = 1,2.
            \label{chargino prop 11}
\EEA

The Lagrangian~\r{Total L sub SUSY 4-comp} is given in terms of the non
mass-eigenstates
$\tilde{W}$ and $\tilde{H}$, because it leads to simpler expressions for
the interaction terms. In converting to the (physical) charginos,
the following relations are useful
\BEA
P_{L}\tilde{W} &=& P_{L} V^{*}_{i1}\tilde{\chi}_{i},
         \label{chargino prop 12} \SL
P_{R}\tilde{W} &=& P_{R} U_{i1}\tilde{\chi}_{i},
         \label{chargino prop 13} \SL
P_{L}\tilde{H} &=& P_{L} V^{*}_{i2}\tilde{\chi}_{i},
         \label{chargino prop 14} \SL
P_{R}\tilde{H} &=& P_{R} U_{i2}\tilde{\chi}_{i}.
         \label{chargino prop 15}
\EEA
Here, repeated indices are summed from 1 to 2, and,
as usual, $P_{L}$ and $P_{R}$ are (projection) operators
projecting out the top two and bottom two components of a Dirac-spinor.
These relations are easy to prove with eqs.~\r{chargino prop 8},
\r{chargino prop 9}, \r{chargino prop 11} and the unitarity of
the matrices U and V. We will now demonstrate it for
eq.~\r{chargino prop 12}.

\begin{PROOF}
With eq.~\r{Introducing Four-Component Spinors prop 3},
the left-hand side of eq.~\r{chargino prop 12} reads
\BEA
  P_{L}\tilde{W} = P_{L}\LP( \BA{c} -i\la^{+} \\ i\bar{\la}^{-} \EA \RP)
                 = \LP( \BA{c} -i\la^{+} \\ 0 \EA \RP).
                                  \label{chargino prop 15aa}
\EEA
By pre-multiplying eq.~\r{chargino prop 8} with $V_{ik}^{*}$ ($i,k=1,2$), and
using the unitarity of V,  we obtain
\BEA
   V^{*}_{ik}\chi^{+}_{i} = V^{*}_{ik} V_{ij} \psi_{j}^{+}
                          = \d_{kj} \psi_{j}^{+} = \psi_{k}^{+}.  \nonumber
\EEA
Hence
\BEA
    \LP( \BA{c} -i\la^{+} \\ 0 \EA \RP) =
\LP( \BA{c} \psi^{+}_{1} \\ 0 \EA \RP)
           = V^{*}_{i1}  \LP( \BA{c} \chi^{+}_{i} \\ 0 \EA \RP)
           = P_{L} V^{*}_{i1} \tilde{\chi}_{i},  \nonumber
%               \label{chargino prop 15aaa}
\EEA
and by comparing this result with eq.~\r{chargino prop 15aa} the proof of
eq.~\r{chargino prop 12}   is competed.
\end{PROOF}

In the same way the following relations for the charge-conjugated
fields are obtained
\BEA
   P_{L}\tilde{W}^{c} &=& P_{L} U^{*}_{i1}\tilde{\chi}^{c}_{i},
    \label{chargino prop 12bb} \SL
   P_{R}\tilde{W}^{c} &=& P_{R} V_{i1}\tilde{\chi}^{c}_{i},
    \label{chargino prop 13bb} \SL
   P_{L}\tilde{H}^{c} &=& P_{L} U^{*}_{i2}\tilde{\chi}^{c}_{i},
    \label{chargino prop 14bb} \SL
   P_{R}\tilde{H}^{c} &=& P_{R} V_{i2}\tilde{\chi}^{c}_{i}.
    \label{chargino prop 15bb}
\EEA

Observe that by hermitian conjugation,
two corresponding sets of equations, like
for instance $ \bar{\tilde{W}}P_{R} =  V_{i1}\bar{\tilde{\chi}}_{i}\;P_{R}$,
can be obtained.

By this observation we conclude this subsection, and instead consider
neutralino mixing.

\subsection{Neutralino Mixing.}
    \label{SUBSECT: Neutralino Mixing}

Neutralinos~($\tilde{\chi}^{0}_{i}, i=1,\ldots, 4$)
are Majorana-spinors arising due to  mixing of
photino, zino and neutral higgsinos.

The appropriate mass-terms are
\BEA
\L_{\tilde{\chi}^{0}}^{mass}
    &=&  - \f{g}{\sqrt{2}\CWA}\LP[\,\LP\{
               v_{1}\,\bar{\tilde{Z}}P_{R}\tilde{H}_{1}
             - v_{2}\,\bar{\tilde{H}}_{2}P_{R}\tilde{Z}\RP\} + h.c. \RP]
         -\f{\mu}{2}\,\bar{\tilde{H}}_{1}\tilde{H}_{2}
         -\f{\mu}{2}\,\bar{\tilde{H}}_{2}\tilde{H}_{1}
   \nmb
         + \HA M_{\tilde{A}}\,\bar{\tilde{A}}\tilde{A}
         + \HA M_{\tilde{Z}}\,\bar{\tilde{Z}}\tilde{Z}
         + \HA \LP( M_{\tilde{Z}} -  M_{\tilde{A}}
\RP)\tan{2\t_{\mbox{w}}}\,\bar{\tilde{A}}\tilde{Z},  \nonumber
%             \label{Neutralino Mixing prop 1}
\EEA
and in two-component form they read
\BEA
 \L_{\tilde{\chi}^{0}}^{mass}
   &=&
         \f{ig}{\sqrt{2}\CWA}\LP\{
               v_{1}\,\la_{Z}\psi^{1}_{H_{1}}
             - v_{2}\,\la_{Z}\psi^{2}_{H_{2}} \RP\}
             -\mu\,\psi^{1}_{H_{1}}\psi^{2}_{H_{2}}
   \nmb
         - \HA M_{\tilde{A}}\,\la_{A}\la_{A}
         - \HA M_{\tilde{Z}}\,\la_{Z}\la_{Z}
         - \HA \LP( M_{\tilde{Z}} -
      M_{\tilde{A}} \RP)\tan{2\t_{\mbox{w}}}\,\la_{A}\la_{Z}.\hspace{1cm}
               \label{Neutralino Mixing prop 2}
\EEA
In the basis
\BEA
    \psi^{0} &=& \LP(  \BA{cccc}
         -i\la_{A} & -i\la_{Z} &
     \psi^{1}_{H_{1}} & \psi^{2}_{H_{2}}  \EA  \RP)^{T},
            \label{Neutralino Mixing prop 3}
\EEA
eq.~\r{Neutralino Mixing prop 2} can be written in the form
\BEA
    \L_{\tilde{\chi}^{0}}^{mass}
       &=& \HA \, \LP(\psi^{0}\RP)^{T}Y^{0}\psi^{0} + h.c.,
       \label{Neutralino Mixing prop 4}
\EEA
where $Y^{0}$ reads
\BEA
   Y^{0} &=&
     \LP( \BA{cccc}
         M_{\tilde{A}}         & \HA\LP(M_{\tilde{Z}}
       -M_{\tilde{A}}\RP)\tan{2\t_{\mbox{w}}} & 0 & 0 \\
         \HA\LP(M_{\tilde{Z}} -M_{\tilde{A}}\RP)\tan{2\t_{\mbox{w}}}
          & M_{\tilde{Z}} & -m_{\mbox{z}}\cos{\b} & m_{\mbox{z}}\sin{\b} \\
         0 & -m_{\mbox{z}}\cos{\b} & 0 & -\mu \\
         0 & m_{\mbox{z}}\sin{\b} & -\mu & 0 \EA \RP).\nn
           \label{Neutralino Mixing prop 5}
\EEA
Note that $Y^{0}$ is symmetric, something which has to do with the Majorana
nature of the neutralinos. In consequence, only one unitary
matrix N is required in order to diagonalize $Y^{0}$:
\BEA
     N^{*}Y^{0}N^{\dagger} &=&  M^{0}_{D}.
         \label{Neutralino Mixing prop 6}
\EEA
Here $M^{0}_{D}$ is the diagonal neutralino mass matrix\footnote{Also here
the matrix N may be chosen in such a way that the elements of $M^{0}_{D}$
are real and non-negative.}.

As in the previous subsection we define two-component mass-eigenstates by
\BEA
    \chi^{0}_{i} &=& N_{ij} \psi^{0}_{j}, \hspace{1cm} i,j =1,\ldots, 4,
       \label{Neutralino Mixing prop 7}
\EEA
but in this case we arrange them in (four-component) Majorana  spinors
defined by
\BEA
   \tilde{\chi}^{0}_{i}  &=& \LP(  \BA{c}  \chi^{0}_{i} \\
                                           \bar{\chi}^{0}_{i}  \EA  \RP),
                            \hspace{1cm} i= 1,\ldots, 4.
                            \label{Neutralino Mixing prop 8}
\EEA
The relations corresponding to
eqs.~\r{chargino prop 12}--\r{chargino prop 15} read
\BEA
   P_{L} \tilde{A}  &=& P_{L} \, N^{*}_{i1}\tilde{\chi}^{0}_{i},
                                  \label{Neutralino Mixing prop 9} \SL
   P_{R} \tilde{A}  &=& P_{R}\,  N_{i1}\tilde{\chi}^{0}_{i},
                                 \label{Neutralino Mixing prop 10}       \SL
   P_{L} \tilde{Z}  &=& P_{L} \, N^{*}_{i2}\tilde{\chi}^{0}_{i},
                                  \label{Neutralino Mixing prop 11} \SL
   P_{R} \tilde{Z}  &=& P_{R}\,  N_{i2}\tilde{\chi}^{0}_{i}
                                  \label{Neutralino Mixing prop 12}\SL
   P_{L} \tilde{H}_{j}  &=& P_{L} \,  N^{*}_{i,j+2}\tilde{\chi}^{0}_{i},
                             \hspace{1cm} j =1,2,
                                  \label{Neutralino Mixing prop 13} \SL
   P_{R} \tilde{H}_{j}  &=& P_{R}\, N_{i,j+2}\tilde{\chi}^{0}_{i},
                                  \label{Neutralino Mixing prop 14}
\EEA
and they are obtained in the same fashion.
Here repeated indices  are assumed to be summed from 1 to 4.

\section{Concluding Remarks.}

In the previous chapter the full four-component Lagrangian for our
supersymmetric electroweak theory was established.
Furthermore, we in this chapter introduced the physical states
and described the gauge symmetry breaking which gives masses to the
gauge bosons and the charged leptons.

With these elements at hand,
one can in principle calculate any  process contained within
this minimal electro-weak theory.

\cleardoublepage

\newcommand{\PRD}{Phys. Rev. }
\newcommand{\PRL}{Phys. Rev. Lett. }
\newcommand{\NP}{Nucl. Phys. }
\newcommand{\PL}{Phys. Lett. }
\newcommand{\NC}{Riv. Nuovo Cimento }

\newcommand{\BI}[1]{\bibitem{#1}}

\appendix

\chapter{Notation and Conventions.}
   \label{APP: Notation and Conventions.}

\renewcommand{\dag}{\dagger}

\section{Relativistic Notation.}

In this report we will adopt standard relativistic units, i.e.
\BEA
    \hbar = c = 1.
\EEA
A general contravariant   and covariant four-vector
will be denoted by
\BEA  \LP. \BA{lclcl}
   A^{\mu}  &=& (\SS A^{0} ; A^{1},A^{2},A^{3}\SS ) &=&
                 (\SS A^{0}; {\bf A} \SS )\\
   A_{\mu} &=&   (\SS A_{0} ; A_{1},A_{2},A_{3} \SS) &=&
                 (\SS A^{0};-{\bf A} \SS ) \EA \RP\}.
\EEA
The compact ``Feynman slash" notation
\BEA
      A\!\!\!/\;\; = \g^{\mu}A_{\mu},
\EEA
will be used.
The metric tensor, $g^{\mu\nu}$, which connects $A^{\mu}$ and $A_{\mu}$,
is defined by
\BEA
    g^{\mu\nu} &=& \mbox{diag}\,(1,-1,-1,-1).
     \label{metric tensor}
\EEA

Moreover, we will use the (relativistic) summation convention which states that
repeated Greek indices, $\mu,\nu,\rho,\s,\tau,$  are summed from 0 to 3
 and latin indices run from 1 to 3 unless specifically
indicated to the contrary.

The Minkowski product (the four-product) will be denoted by AB and defined as
\BE    AB \HS \EQ \HS A^{\mu}B_{\mu}
          \HS =   \HS A^{0}B^{0} - {\bf A}{\bf B}
\EE
Practical notation for the four-gradients, $\P^{\mu}$ and $\P_{\mu}$,
will be used
\BEA
    \P^{\mu} &\EQ&  \PD{}{x_{\mu}}
             =    (\SS \PD{}{t};\SS -\nabla \SS), \SL
       \P_{\mu} &\EQ&  \PD{}{x^{\mu}}
                = (\SS \PD{}{t};\SS \nabla \SS) .
\EEA

The totally antisymmetric Levi-Civita tensors in three and four dimensions
are respectively defined by
\BEA
       \e_{i j k} &=&
          \LP\{ \BT{rl} $ +1 $ &, \HS\HS for even permutations of 123 \SL
                        $ -1 $ &, \HS\HS for odd permutations \SL
                        $  0 $ &, \HS\HS otherwise,  \ET \RP.
                          \label{Levi-Civita3}
\EEA
\BEA
       \e_{\mu\nu\rho\s} &=&
          \LP\{ \BT{rl} $ +1 $ &, \HS\HS for even permutations of 0123 \SL
                        $ -1 $ &, \HS\HS for odd permutations \SL
                        $  0 $ &, \HS\HS otherwise,  \ET \RP.
                          \label{Levi-Civita4}
\EEA
where
\BEA   \e_{i j k} &=& \e^{i j k} \HS,   \SL
       \e_{\mu\nu\rho\s} &=& - \e^{\mu\nu\rho\s} \HS.   \EEA

\section{Pauli Matrices.}
       \label{sec Pauli Matrices}

The well known Pauli matrices are defined by
\BE
   \s^{1} = \LP( \BA{rr} 0          & 1\SL
                         1          & 0      \EA \RP) ,   \HS\HS
   \s^{2} = \LP( \BA{rr} 0          & -i   \SL
                         i          & 0      \EA \RP) ,   \HS\HS
   \s^{3} = \LP( \BA{rr} 1          & 0      \SL
                         0          & -1   \EA \RP),
\EE
and satisfy the commutator relation
\BEA
   [\s^{i},\s^{j}] &=& 2i\e^{ijk}\s^{k}, \hspace{1cm} i,j,k = 1,2,3 \nonumber.
\EEA
{}From this definition it is evident that
\BEA
       (\s^{i})^{\dag} &=& \s^{i},  \hspace{1cm} i=1,2,3,\SL
       (\s^{i})^{2}     &=& 1  ,\SL
       Tr(\s^{i})      &=& 0.
\EEA
For later use, we also introduce\footnote{Note that different signs
are used in the literature for the definition of this quantity.}
\BEA
      \s^{0} &=& \LP( \BA{cc}  1 & 0   \SL
                               0 & 1 \EA \RP),
\EEA
and a useful arrangement of these matrices is
\BEA
      \s^{\mu} = (\s^{0}\,;\,{\bf \s \!\!\!\! \s})
               = (\s^{0}\,;\,\s^{1}, \s^{2}, \s^{3}). \nonumber
\EEA
%In sect.~\ref{SECT: Connection between the Restricted},
%eq.~\r{connection SL(2,C) L prop 3}~(together with the results
%from sect.~\ref{Weyl Spinor Notation.}),
The index structure of the $\s$-matrices is given by
\BEA
      \s^{\mu} &=& [\s^{\mu}_{\a\dot{\a}}].
\EEA
We now introduce some ``Pauli related" matrices defined by
\BEA
    \bar{\s}^{\mu\;\dot{\a}\a} \equiv \s^{\mu\;\a\dot{\a}}
      = \e^{\dot{\a}\dot{\b}} \e^{\a\b} \s_{\b\dot{\b}}^{\mu},
\EEA
where the ``metrics" $\e$ and $\bar{\e}$
%from sect.~\ref{Weyl Spinor Notation.}
have been used. By direct computations one can establish the following
relations
\BEA
    \bar{\s}^{0} &=& \s^{0}  \label{Pauli Matrix prop 1}\SL
    \bar{\s}^{i} &=& -\,\s^{i} , \hspace{1cm}i=1,2,3
       \label{Pauli Matrix prop 2}.
\EEA
Moreover, the following relations are true
\BEA
    \s^{\mu}_{\a\dot{\a}}\bar{\s}_{\mu}^{\dot{\b}\b}
              &=& 2\,\d_{\a}^{\;\b} \d^{\;\dot{\b}}_{\dot{\a}}
       \label{Pauli Matrix prop 3}\SL
    Tr(\s^{\mu}\bar{\s}^{\nu}) &=& 2g^{\mu\nu}
       \label{Pauli Matrix prop 4}\SL
    (\s^{\mu}\bar{\s}^{\nu}+\s^{\nu}\bar{\s}^{\mu})_{\a}^{\;\b}
              &=& 2\,g^{\mu\nu}\d_{\a}^{\;\b}
       \label{Pauli Matrix prop 5}\SL
    (\bar{\s}^{\mu}\s^{\nu}+\bar{\s}^{\nu}\s^{\mu})_{\;\dot{\b}}^{\dot{\a}}
              &=& 2\,g^{\mu\nu} \d^{\dot{\a}}_{\;\dot{\b}}
       \label{Pauli Matrix prop 6}\SL
    (\s^{\mu}\bar{\s}^{\nu}\s^{\rho} + \s^{\rho}\bar{\s}^{\nu}\s^{\mu})
        &=& 2 \LP(g^{\mu\nu}\s^{\rho} + g^{\nu\rho}\s^{\mu}
           - g^{\mu\rho}\s^{\nu} \RP)
             \label{WESS A.16a}\SL
    (\bar{\s}^{\mu}\s^{\nu}\bar{\s}^{\rho}
    + \bar{\s}^{\rho}\s^{\nu}\bar{\s}^{\mu})
        &=& 2 \LP(g^{\mu\nu}\bar{\s}^{\rho}
        + g^{\nu\rho}\bar{\s}^{\mu} - g^{\mu\rho}\bar{\s}^{\nu} \RP)
             \label{WESS A.16a}\SL
    Tr(\s^{\mu}\bar{\s}^{\nu}\s^{\rho}\bar{\s}^{\s})
          &=& 2\,(g^{\mu\nu}g^{\rho\s}+g^{\mu\s}g^{\nu\rho}
               -g^{\mu\rho}g^{\nu\s} -i\e^{\mu\nu\rho\s}).
       \label{Pauli Matrix prop 7}
\EEA
Most of the above relations are easily proved by direct computations.
Besides, M\"{u}ller-Kirsten and Wiedemann~\cite[subsec. 1.3.5]{MK},
have proved most of them, and in particular
eq.~\r{Pauli Matrix prop 7} which is the most difficult one.

Anti-symmetric matrices $\s^{\mu\nu}$ and $\bar{\s}^{\mu\nu}$ are
defined by
\BEA
      \s^{\mu\nu} &=&  \f{i}{4}\,(\s^{\mu}\bar{\s}^{\nu}-
                   \s^{\nu}\bar{\s}^{\mu}),
                   \label{Notation prop 49}\SL
      \bar{\s}^{\mu\nu} &=& \f{i}{4}\,(\bar{\s}^{\mu}\s^{\nu}-
                  \bar{\s}^{\nu}\s^{\mu})
                   \label{Notation prop 49aa}.
\EEA
By utilizing the index structure of the $\s$-matrices, it is easily seen
that $\s^{\mu\nu}$ and $\bar{\s}^{\mu\nu}$ must have the
index structure $\s^{\mu\nu} = [(\s^{\mu\nu})_{\a}^{\;\b}]$ and
$\bar{\s}^{\mu\nu} = [(\bar{\s}^{\mu\nu})_{\;\;\dot{\a}}^{\dot{\b}}]$.
In fact are $\s^{\mu\nu}$  and $\bar{\s}^{\mu\nu}$ the generators of
$SL(2,C)$ in the spinor representations $(\HA,0)$ and $(0,\HA)$ respectively.
The proofs together with the establishment of the
below formulae can  be found in ref.~\ref{MK}:
\BEA
     \s^{\mu\nu\;\dag} &=& \bar{\s}^{\mu\nu}  \label{Pauli Matrix prop 8},\SL
     \s^{\mu\nu} &=& \f{1}{2i}\e^{\mu\nu\rho\s}\s_{\rho\s}
           \label{Pauli Matrix prop 9},\SL
     \bar{\s}^{\mu\nu} &=&-\, \f{1}{2i}\e^{\mu\nu\rho\s}\bar{\s}_{\rho\s}
             \label{Pauli Matrix prop 10},\SL
     Tr(\s^{\mu\nu}) &=& Tr(\bar{\s}^{\mu\nu}) \;=\; 0\SL
         \label{Pauli Matrix prop 11a}
     Tr(\s^{\mu\nu}\s^{\rho\s}) &=& \f{1}{2} \,(g^{\mu\rho}g^{\nu\s}
              -g^{\mu\s}g^{\nu\rho})+\f{i}{2}\e^{\mu\nu\rho\s}
             \label{Pauli Matrix prop 11},\SL
     Tr(\bar{\s}^{\mu\nu}\bar{\s}^{\rho\s}) &=& \f{1}{2}
            \,(g^{\mu\rho}g^{\nu\s}
              -g^{\mu\s}g^{\nu\rho})-\f{i}{2}\e^{\mu\nu\rho\s}
        \label{Pauli Matrix prop 12}.
\EEA

\section{Dirac Matrices.} \label{sec Dirac Matrices}

The Dirac $\g$-matices are defined by the anticommutation ~(Clifford) relations
\BEA
      \{\g^{\mu},\g^{\nu}\} & = & 2g^{\mu\nu}. \label{Dirac-matrices}
\EEA

{}From the four $\g$-matrices above, it is possible to define
a ``fifth $\g$-matrix" by
\BE
     \g_{5}  \equiv \g^{5}  \equiv i\g^{0}\g^{1}\g^{2}\g^{3}
%             = \f{i}{4!}\e^{\mu\nu\rho\s}\g_{\mu}\g_{\nu}\g_{\rho}\g_{\s}
               \label{Gamma5def}
\EE
It possesses the following properties
which follows easily from the definitions~\r{Dirac-matrices}
and \r{Gamma5def}
\BEA      \{ \g^{5},\g^{\mu}\} & = & 0, \label{g5gmu}     \SL
             (\g^{5})^{2} &=& 1.
\EEA

We will now state three explicit representations of the $\g$-matrices,
namely the so-called  Dirac representation, the
Majorana representation, and finally the Chiral representation.

\subsection{Representations}
   \label{Subsect: Representations}

The lowest non-trivial representation of these matrices is of dimension four.
and  we will concentrate on this represntation.
{}From now on, we will assume that a four dimensional representation is used.

\subsubsection{The Dirac Representation or Canonical Basis.}

In this particular representation the $\g$-matrices read
\BEA
  \g^{0} &=& \LP( \BA{rr}  1      & 0         \SL
                           0      & -1    \EA \RP) , \label{rep 1}\SL
  \g^{i} &=& \LP( \BA{rr}  0           & \s^{i}     \SL
                         \bar{\s}^{i}    & 0         \EA \RP) , \hspace{1cm}
                         i=1,2,3, \label{rep 2}\SL
  \g^{5} &=& \LP( \BA{rr}  0           & \s^{0}\SL
                         \bar{\s}^{0}  & 0         \EA \RP) , \label{rep 3}
\EEA
where 1 denotes the $2\times 2$ identity matrix and $\s^{\mu}$ and
$\bar{\s}^{\mu}$ are the Pauli matrices defined in the previous section.

\subsubsection{The Majorana Representation.}

In this representation  all $\g$-matricrs are pure imaginary and have
the explicit form:
\BEA
   \g^{0}  &=& \LP( \BA{rr}  0              &    \s^{2} \SL
                            - \bar{\s}^{2}  &    0        \EA  \RP), \SL
   \g^{1}  &=&  \LP( \BA{rr}  i\s^{3}     &  0 \SL
                              0          &  i\s^{3}  \EA \RP),  \SL
   \g^{2} &=&  \LP( \BA{rr}   0          &  -\s^{2}\SL
                              -\bar{\s}^{2} & 0  \EA \RP),\SL
   \g^{3} &=&  \LP( \BA{rr} -i\s^{1}     &  0 \SL
                              0          &  i\s^{1}  \EA \RP) ,
\EEA
and finally
\BEA
   \g^{5} &=&  \LP( \BA{rr}  \s^{2}      &  0 \SL
                              0          &  -\s^{2}  \EA \RP).
\EEA

\subsubsection{The Chiral representation or Weyl Basis.}
    \label{subsubsect: The Chiral representation}

This basis is of particular interest to persons doing SUSY.
In this representation the $\g$-matrices take on the explicite form
\BEA
  \g^{\mu} = \LP( \BA{rr} 0           & \s^{\mu}        \SL
                    \bar{\s}^{\mu}    & 0          \EA \RP) ,
                    \label{WEYL basis prop 1}\SL
  \g^{5} = \LP( \BA{rr} -1          & 0          \SL
                        0           & 1 \EA \RP) .
                   \label{WEYL basis prop 2}
\EEA

\section{Spinor Relations.}
    \label{SECT: Spinor Relations.}

In two-component notation we have the anti-symmetric $\e$-metric.
The tensor oby the following relations,
which are proven by stright forward calculations
\BEA
  \e^{\a\b}\;\e_{\g\d}  &=& \d^{\a}_{\;\;\d}\;\d^{\b}_{\;\;\g} -
       \d^{\a}_{\;\;\g}\;\d^{\b}_{\;\;\d}
         \label{Spinor Relations prop 2}, \\
  \e_{\dot{\a}\dot{\b}}\;\e^{\dot{\g}\dot{\d}}  &=&
       \d_{\dot{\a}}^{\;\;\dot{\d}}\;\d_{\dot{\b}}^{\;\;\dot{\g}} -
       \d_{\dot{\a}}^{\;\;\dot{\g}}\;\d_{\dot{\b}}^{\;\;\dot{\d}}
         \label{Spinor Relations prop 3}, \\
   \e^{\a\b}\;\e_{\b\g} &=& \d^{\a}_{\;\;\g}
         \label{Spinor Relations prop 4}, \\
   \e_{\dot{\a}\dot{\b}}\;\e^{\dot{\b}\dot{\g}}
          &=& \d_{\dot{\a}}^{\;\;\dot{\g}}
         \label{Spinor Relations prop 5}.
\EEA
%Here we have used the summazion convention which states that suppressed
%undotted spinor indices are summed from upper  left to lower right,
%while suppressed dotted indices are summed form lower left to upper right.
%In particular this means, for instance, that
%\BEA
%    \psi\chi             & \equiv & \US{\psi}{\a}\LS{\chi}{\a} \SL
%              \label{Useful Relations prop 6}
%    \bar{\psi}\bar{\chi} & \equiv & \DUS{\psi}{\a}\DLS{\chi}{\a} \SL
%              \label{Useful Relations prop 7}
%    \psi\s^{\mu}\chi     & \equiv & \US{\psi}{\a}\SA\LS{\chi}{\a} \SL
%              \label{Useful Relations prop 8}
%    etc.   \nonumber
%\EEA
 %With this summation convention the spinors could be viewed as matrices
 %with the  dimmensional assignments given in table~\ref{kklf}

%    **************************************

We start by postulating that the spinor components are
Grassmann numbers, i.e.
\BEA   \LP. \BA{lll}
 & &  \{\LS{\psi}{\a},\LS{\psi}{\b}\} = \{\US{\psi}{\a},\US{\psi}{\b}\}
     = \{\LS{\psi}{\a},\US{\psi}{\b}\} = 0 \\
 & & \{\DLS{\chi}{\a},\DLS{\chi}{\b}\} = \{\DUS{\chi}{\a},\DUS{\chi}{\b}\}
     = \{\DLS{\chi}{\a},\DUS{\chi}{\b}\} = 0
     \EA \RP\},
     \label{Spinor Relations prop 6}
\EEA
and also anti-commute with other Grassmann numbers (e.g. fermion
fields, spinor charges etc.).

With this postulate an expression like $\psi^{\a}\chi_{\a} =
 \psi_{2}\chi_{1}-\psi_{1}\chi_{2}$ do not vanish\footnote{This observation
can be taken as a motivation of the above postulate.}, and
in particular
 \BEA  \LP. \BA{lcl}
    \US{\psi}{\a}\LS{\chi}{\a} &=& - \LS{\chi}{\a}\US{\psi}{\a} \\
    \DLS{\psi}{\a}\DUS{\chi}{\a} &=& -\DUS{\chi}{\a}\DLS{\psi}{\a}
      \EA \RP\}.
      \label{Spinor Relations prop 7}
\EEA
Because of the signs in eq.~\r{Spinor Relations prop 7}, it is not
well-defined what we mean by $\psi\chi$ or $\bar{\psi}\bar{\chi}$.
To tackle this problem, we introduce the {\em summation convention} that
states that suppressed undotted spinor indices are summed from
upper left to downer right, while suppressed
dotted indices are summed from lower
left to upper right.
In particular this means, for instance, that
\BEA
       \psi\chi & \equiv & \US{\psi}{\a}\LS{\chi}{\a}
            \label{Spinor Relations prop 8} \\
       \bar{\psi}\bar{\chi} & \equiv & \DLS{\psi}{\a}\DUS{\chi}{\a}
            \label{Spinor Relations prop 9}\\
       \psi \s^{\mu}\bar{\chi} & \equiv & \US{\psi}{\a}\SA \DUS{\chi}{\a}
            \label{Spinor Relations prop 10} \\
       \textstyle{etc.} & & \nonumber
\EEA

We are now in position to establish some useful relations involving
spinors which will frequently be use in calculations.

Let $\psi$, $\t$ and $\chi$ be two-comonent (Weyl) spinors.
Then the following relations hold:
\BEA
     \psi\chi &=& \chi\psi
        \label{Spinor Relations prop 11}, \\
     \bar{\psi}\bar{\chi} &=& \bar{\chi}\bar{\psi}
         \label{Spinor Relations prop 12}, \\
     \LP(\psi\chi\RP)^{\dagger} &=& \bar{\chi}\bar{\psi}
         \label{Spinor Relations prop 13}, \\
     \psi\s^{\mu}\bar{\chi} &=& - \bar{\chi}\bar{\s}^{\mu}\psi
         \label{Spinor Relations prop 14}, \\
     \LP(\psi\s^{\mu}\bar{\chi}\RP)^{\dagger} &=& \chi\s^{\mu}\bar{\psi}
         \label{Spinor Relations prop 15}, \\
       \psi\s^{\mu\nu}\chi  &=&  - \chi \s^{\mu\nu}\psi
         \label{Spinor Relations prop 16}, \\
     \SA\DUS{\t}{\a}\;\t\s^{\nu}\tb &=&
          \tb\tb\;\LP[\;\HA \d_{\a}^{\;\;\b}g^{\mu\nu}-i\,\LP(\s^{\mu\nu}\RP)
             _{\a}^{\;\;\b}\;\RP]\;\LS{\t}{\b}
         \label{Spinor Relations prop 16a}, \\
      \psi \s^{\mu}\bar{\s}^{\nu}\chi &=&
         \psi \LP[\;-2i\,\s^{\mu\nu}+ g^{\mu\nu}\;\RP]\chi
         \label{Spinor Relations prop 16b}, \\
      \bar{\psi} \bar{\s}^{\mu}\s^{\nu}\bar{\chi}
       &=& \bar{\psi} \LP[\;-2i\,\bar{\s}^{\mu\nu} + g^{\mu\nu}\;\RP]\bar{\chi}
         \label{Spinor Relations prop 16c}, \\
     \US{\t}{\a}\US{\t}{\b} &=& -\f{1}{2}\,\e^{\a\b}\;\t\t
         \label{Spinor Relations prop 17}, \\
     \LS{\t}{\a}\LS{\t}{\b} &=& \f{1}{2}\,\e_{\a\b}\;\t\t
         \label{Spinor Relations prop 18}, \\
     \DUS{\t}{\a}\DUS{\t}{\b} &=& \f{1}{2}\,\e^{\dot{\a}\dot{\b}}\;\tb\tb
         \label{Spinor Relations prop 19}, \\
     \DLS{\t}{\a}\DLS{\t}{\b} &=& -\f{1}{2}\,\e_{\dot{\a}\dot{\b}}\;\tb\tb
         \label{Spinor Relations prop 20}.
\EEA
The final results are only stated here.
Most of the  explicite proofs are given in detail in
in ref.~\ref{MK}.

\subsection{Fierz Rearrangemant Formulae.}

Some other relations have proven useful. They go under the
name of Fierz Rearrangement formulae and read:
\BEA
   \t\psi \;\;\t\chi   &=& -\f{1}{2}\;\t\t \;\;\psi\chi
       \label{Fierz Rearrangemant Formula prop 1}, \\
   \tb\bar{\psi} \;\;\tb\bar{\chi}
              &=& -\f{1}{2}\;\tb\tb \;\;\bar{\psi}\bar{\chi}
       \label{Fierz Rearrangemant Formula prop 2}, \\
   \t\psi\;\;\DLS{\chi}{\a}
         &=& \f{1}{2}\;\t\s^{\mu}\bar{\chi}\;\US{\psi}{\a}\s_{\mu\;\a\dot{\a}}
       \label{Fierz Rearrangemant Formula prop 3}, \\
   \tb\bar{\psi}\;\;\US{\chi}{\a}
 &=& \f{1}{2}\;\tb\bar{\s}^{\mu}\chi\;\DLS{\psi}{\a}\bar{\s}_{\mu}^{\dot{\a}\a}
       \label{Fierz Rearrangemant Formula prop 4}, \\
%    \TSTB\;\t\s^{\nu}\tb &=& \f{1}{2}\;g^{\mu\nu}\;\t\t\;\tb\tb
%       \label{Fierz Rearrangemant Formula prop 5}, \\
    \psi_{1}\s^{\mu}\bar{\chi}_{1}\;\psi_{2}\s^{\nu}\bar{\chi}_{2}
     &=&  \HA g^{\mu\nu}\;\psi_{1}\psi_{2}\;\bar{\chi}_{1}\bar{\chi}_{2}
        \label{Fierz Rearrangemant Formula prop 5}, \\
   \t\chi\;\t\s^{\mu}\bar{\psi} &=& -\f{1}{2}\;\t\t\;\chi\s^{\mu}\bar{\psi}
       \label{Fierz Rearrangemant Formula prop 6}, \\
   \tb\bar{\chi}\;\tb\bar{\s}^{\mu}\psi
      &=& -\f{1}{2}\;\tb\tb\;\bar{\chi}\bar{\s}^{\mu}\psi
       \label{Fierz Rearrangemant Formula prop 7}, \\
   \psi\s^{\mu}\bar{\s}^{\nu}\chi &=& \chi\s^{\nu}\bar{\s}^{\mu}\psi
       \label{Fierz Rearrangemant Formula prop 8}.
\EEA
Neither these formulae we will prove explicitly.
The proofs can be found from the same source as above.

\section{Four Component notation.}
   \label{SECT: Four comp not}

\subsection{The Projections Operators.}

We start by defining the  projection operators, well known from SM,
\BEA
   P_{L} &=& \HA \LP(1-\g_5\RP),
      \label{left proje}\SL
   P_{R} &=& \HA \LP(1+\g_5\RP).
      \label{right proje}
\EEA

With the properties of the $\g$-matrices from sect.~\ref{sec Dirac Matrices},
it is straightforward to establish the relations
\BEA
  P_L + P_R &=& 1
    \label{pl plus pr equiv one},\SL
  P_L P_L &=& P_L,
     \label{PLPL}\SL
%  P_R P_R &=& P_R,\SL
  P_L P_R &=& P_R P_L \;\; = \;\; 0,
     \label{PL PR equiv zero}\SL
  P_L^\dagger &=& P_L ,\SL
  P_L \g^\mu &=& \g^\mu P_R,
     \label{PL PR equiv zero prop 1}
\EEA
and corresponding equations for $P_R$.

\subsection{Connection Between the Two- and Four-Component Spinors.}
    \label{Subsect: Connection Between Two- and Four-Component Spinors.}

Let us introduce the two
two-component Weyl spinors $\LS{\xi}{\a}$ and $\DUS{\eta}{\a}$
\BEA
   \LS{\xi}{\a} &\in & F, \nn
   \DUS{\eta}{\a} &\in & \dot{F}^{*}. \nonumber
\EEA
%where F and $\dot{F}^{*}$ are vector spaces.

The vector-spaces  $F$ and $\dot{F}^{*}$
are inequivalent representation spaces of SL(2,C).
Now we construct the direct sum space
\BEA
     D &=& F \oplus \dot{F}^{*}.
\EEA
This space is a four-dimensional representation space of SL(2,C).
The elements of D, are just the well-known four-component Dirac-spinors.

Thus a Dirac-spinor, $\Psi$, can be constructed from these Weyl spinors
according to
\BEA
   \Psi &=& \LP( \BA{c} \LS{\xi}{\a} \\ \DUS{\eta}{\a}  \EA \RP).
       \label{Dirac Spinor}
\EEA
Strictly speaking this is a Dirac-spinor in the Weyl-representation.
Thus we see that if we work in
the Weyl representation~(subsect.~\ref{Subsect: Representations})
we have a direct relation between two- and four-component spinors.
Throughout this subsection we will thus assume the Weyl-representation.

A Majorana spinor, $\la$, is a (four-component)
Dirac-spinor with the additional
condition
\BEA
     \la &=&  \la^{c} \;\; = \;\; C \bar{\la}^{T}.
     \label{Def: Majorana spinor}
\EEA
Here C is the charge conjugation matrix\footnote{For more
information on this matrix consider e.g. ref.~\ref{scadron}.}
while $\bar{\la}$ means, as usual, the
the Dirac adjoint spinor
$\bar{\la}= \la^{\dagger} \g_{0}$~(independent of represnetation).
In the Dirac-represntation C reads $C_{D} = i\g^{2}\g^{0}$,
and in the Weyl-representation~(with the
correct index structure)~\cite[p. 135]{MK}
\BEA
    C_{\mbox{w}} &=&  \LP( \BA{cc} \LP(i\s^{2}\bar{\s}^{0}\RP)_{\a}^{\;\;\b}
               &   0 \\

   0 & \LP(i\bar{\s}^{2}\s^{0}\RP)^{\dot{\a}}_{\;\;\dot{\b}} \EA \RP).
\EEA
Thus it is possible to show that~\cite[p. 140]{MK}
\BEA
   \Psi^{c} &=& C \bar{\Psi}^{T} \;\; = \;\;
    \LP( \BA{c} \LS{\eta}{\a} \\ \DUS{\xi}{\a}  \EA \RP),
       \label{Charge conjugation}
\EEA
i.e. the charge conjugation (in the Weyl representation) flips
$\xi$ and $\eta$.

Hence, we may conclude that a Majorana-spinor,$\la$, defined in
eq.~\r{Def: Majorana spinor}, can be written
\BEA
   \la &=& \LP( \BA{c} \LS{\xi}{\a} \\ \DUS{\xi}{\a}  \EA \RP).
      \label{Majorana Spinor}
\EEA

Furthermore, in the Weyl representation we have
\BEA
      P_{L} &=& \LP( \BA{cc}   1 & 0 \\ 0 & 0 \EA \RP),\nn
      P_{R} &=& \LP( \BA{cc}   0 & 0 \\ 0 & 1 \EA \RP),\nonumber
\EEA
and thus
\BEA
      \Psi_{L} &=& P_{L} \Psi \;\;=\;\;
           \LP( \BA{c} \LS{\xi}{\a} \\ 0  \EA \RP),
          \label{Left handed Dirac spinor} \nn
      \Psi_{R} &=& P_{R} \Psi \;\;=\;\;
          \LP( \BA{c} 0 \\ \DUS{\eta}{\a}  \EA \RP).
          \label{Right handed Dirac spinor} \nonumber
\EEA
The Dirac-adjoint spinor of $\Psi$, is
\BEA
   \bar{\Psi} &=& \Psi^{\dagger} \g_{0}
      \;\; = \;\;  \LP( \BA{cc}  \US{\eta}{\a} & \DLS{\xi}{\a} \EA \RP),
\EEA
as can be showed by straightforward calculations.

\subsubsection{Useful Relations Between Two- and Four-Component Spinors.}

Now we shall establish some relations, making the
transitions between two- and four-component spinors more
explicite and easy later on.
Let the Dirac- and Majorana-spinor, $\Psi(x)$ and $\la(x)$, be
defined as in eqs.~\r{Dirac Spinor} and \r{Majorana Spinor}.

Hence we have~(in the Weyl representation):
\BEA
   \bar{\Psi}_{1}\Psi_{2} &=& \eta_{1}\xi_{2} + \bar{\xi}_{1}\bar{\eta}_{2},
             \label{URBTFCS prop 0a} \SL
   \bar{\Psi}_{1}\g^{\mu}\Psi_{2}
           &=& \bar{\xi}_{1}\bar{\s}^{\mu}\xi_{2}
              - \bar{\eta}_{2}\bar{\s}^{\mu}\eta_{1},
              \label{URBTFCS prop 0b} \SL
   \bar{\Psi}_{1}\g_{5}\Psi_{2}
           &=& - \eta_{1}\xi_{2}+ \bar{\eta}_{2}\bar{\xi}_{1},
              \label{URBTFCS prop 0c} \SL
   \bar{\Psi}_{1}\g^{\mu}\g_{5}\Psi_{2}
           &=& - \bar{\xi}_{1}\bar{\s}^{\mu}\xi_{2}
          - \bar{\eta}_{2}\bar{\s}^{\mu}\eta_{1},
              \label{URBTFCS prop 0d} \SL
   \bar{\Psi}_{1}\g^{\mu}\P_{\mu}\Psi_{2}
           &=&  \eta_{1}\s^{\mu}\P_{\mu}\bar{\eta}_{2}
          + \bar{\xi}_{1}\bar{\s}^{\mu}\P_{\mu}\xi_{2} \nn
           &=&  \bar{\eta}_{2}\bar{\s}^{\mu}\P_{\mu}\eta_{1}
          + \bar{\xi}_{1}\bar{\s}^{\mu}\P_{\mu}\xi_{2}
              -\P_{\mu}\LP(\bar{\eta}_{2}\bar{\s}^{\mu}\eta_{1} \RP),
              \label{URBTFCS prop 0e} \SL
    \bar{\Psi}_1 P_{L} \Psi_2 &=& \eta_1\xi_2,
       \label{URBTFCS prop 1}  \SL
    \bar{\Psi}_1 P_{R} \Psi_2 &=& \bar{\xi}_1\bar{\eta}_2,
       \label{URBTFCS prop 2}\SL
    \bar{\Psi}_1\g^\mu P_{L} \Psi_2 &=& \bar{\xi}_1\bar{\s}^{\mu}\xi_2,
       \label{URBTFCS prop 3}\SL
    \bar{\Psi}_1\g^\mu P_{R} \Psi_2
                       %      &=&\eta_1{\s}^{\mu}\bar{\eta}_2, \nn
        &=& -\bar{\eta}_2\bar{\s}^{\mu}\eta_1  ,
        \label{URBTFCS prop 4}\SL
    \bar{\Psi}_1\g^\mu P_{L}\P_\mu \Psi_2
        &=& \bar{\xi}_1\bar{\s}^\mu\P_{\mu}\xi_2,
        \label{URBTFCS prop 5}\SL
    \bar{\Psi}_1\g^\mu P_{R}\P_\mu \Psi_2
        &=& \eta_1\s^\mu\P_{\mu}\bar{\eta}_2,\nn
        &=& \bar{\eta}_2\bar{\s}^\mu\P_{\mu}\eta_1
             -\P_{\mu}\LP(\bar{\eta}_2\bar{\s}^{\mu}\eta_1\RP).
        \label{URBTFCS prop 6}
\EEA

\section{Grassmann Variables.}
    \label{App: Grassmann Variables}

In this appendix a differentiation and integration calculus for Grassmann
variables will be established. The obtained results will be
extensively used in the text.

\sloppy

\subsection{Differentiation with respect to Grassmann Variables.}

\fussy

In supersymmetry the Grassmann variables,
which parametrize superspace, are important.
%\BEA
%  \{\LS{\t}{\a},\LS{\t}{\b}\} &=&  0, \SL
%  \{\DLS{\t}{\a},\DLS{\t}{\b}\} &=& 0, \SL
%   \{\LS{\t}{\a}, \DLS{\t}{\b} \} &=& 0.
%\EEA
Because of their anticommuting properties, they can not be continuos
varying variables. However,  they have to be  discrete  objects.
Hence, defining differentiation with respect to Grassmann variables in the
normal sense, as the ratio of two infinitesimal increments, has no
meaning. However, formally we can define differentiation, following common
practice, as
\BEA
   \PD{\LS{\t}{\a}}{\LS{\t}{\b}} &=& \d^{\b}_{\;\;\a},
       \label{Differentiation with respect to Grassmann Variables prop 1}\SL
   \PD{\US{\t}{\a}}{\US{\t}{\b}} &=& \d_{\b}^{\;\;\a},
   \label{Differentiation with respect to Grassmann Variables prop 2}\SL
   \PD{\DLS{\t}{\a}}{\DLS{\t}{\b}} &=& \d^{\dot{\b}}_{\;\;\dot{\a}},
     \label{Differentiation with respect to Grassmann Variables prop 3}\SL
   \PD{\DUS{\t}{\a}}{\DUS{\t}{\b}} &=& \d_{\dot{\b}}^{\;\;\dot{\a}}.
      \label{Differentiation with respect to Grassmann Variables prop 4}
\EEA
The $\e$-metric can be used to raise and lower indices of derivatives
according to\footnote{Take particular notice in the sign on the
right-hand side of these equations.}
\BEA
    \e^{\a\b}\PD{}{\US{\t}{\b}}  &=&  -\PD{}{\LS{\t}{\a}}
      \label{Differentiation with respect to Grassmann Variables prop 5}\SL
    \e_{\a\b}\PD{}{\LS{\t}{\b}}  &=&  -\PD{}{\US{\t}{\a}}
      \label{Differentiation with respect to Grassmann Variables prop 5a}
\EEA
and
\BEA
    \e^{\dot{\a}\dot{\b}}\PD{}{\DUS{\t}{\b}}  &=&  -\PD{}{\DLS{\t}{\a}}
      \label{Differentiation with respect to Grassmann Variables prop 6}\SL
    \e_{\dot{\a}\dot{\b}}\PD{}{\DLS{\t}{\b}}  &=&  -\PD{}{\DUS{\t}{\a}}
        \label{Differentiation with respect to Grassmann Variables prop 6a}
\EEA
\begin{PROOF}
  Let the eq.~\r{Differentiation with respect to Grassmann Variables prop 5}
  operates (from the left) on $\US{\t}{\g}$:
    \BEA
       \e^{\a\b}\PD{}{\US{\t}{\b}} \US{\t}{\g}
          &=& -\PD{}{\LS{\t}{\a}} \US{\t}{\g}.
          \nonumber
     \EEA
  Then by comparing each side of this equation we have
  \BEA
    \e^{\a\b}\PD{}{\US{\t}{\b}} \US{\t}{\g} &=&
           \e^{\a\b}\,\d_{\b}^{\;\;\g}\nn &=& \e^{\a\g}, \nonumber
  \EEA
  \BEA
     -\PD{}{\LS{\t}{\a}} \US{\t}{\g}
       &=& -\e^{\g\b}\,\PD{}{\LS{\t}{\a}}\;\LS{\t}{\b} \nn
       &=& - \e^{\g\b}\,\d^{\a}_{\;\;\b}              \nn
       &=& -\e^{\g\a}                                 \nn
       &=& \e^{\a\g}.                                 \nonumber
\EEA
Hence we can conclude
eq.~\r{Differentiation with respect to Grassmann Variables prop 5}
is fulfilled.
%\BEA
%    \e^{\dot{\a}\dot{\b}}\PD{}{\DUS{\t}{\b}}  &=&  -\PD{}{\DLS{\t}{\a}}.
%      \nonumber
%\EEA
The other relations in eqs.~\r{Differentiation with respect
to Grassmann Variables prop 5a}--\r{Differentiation with respect to
Grassmann Variables prop 6a}
are showed in a similar fashion.
\end{PROOF}

Due to the anticommuting character of $\t$ and $\tb$, we shall demand
that
%Furthermore, we shall demand, since $\t$ and $\tb$ are anticommuting
%variables, that
\BEA
  \LP\{\PD{}{\US{\t}{\a}},\PD{}{\US{\t}{\b}} \RP\} &=&
  \LP\{\PD{}{\DUS{\t}{\a}},\PD{}{\DUS{\t}{\b}} \RP\} \;\;=\;\;
  \LP\{\PD{}{\US{\t}{\a}},\PD{}{\DUS{\t}{\b}} \RP\} \;\;= \;\;0,
    \label{Differentiation with respect to Grassmann Variables prop 7}\SL
  \LP\{ \PD{}{\LS{\t}{\a}} , \LS{\t}{\b} \RP\} &=& \d^{\a}_{\;\;\b},
     \label{Differentiation with respect to Grassmann Variables prop 7a}\SL
  \LP\{ \PD{}{\DUS{\t}{\a}},
 \DUS{\t}{\b} \RP\} &=& \d_{\;\;\dot{\a}}^{\dot{\b}},
     \label{Differentiation with respect to Grassmann Variables prop 7b}
\EEA
and, since $\t$ and $\tb$ are concidered to be independent,
\BEA
  \LP\{\PD{}{\US{\t}{\a}},\DUS{\t}{\b} \RP\} =
  \LP\{\PD{}{\DUS{\t}{\a}},\US{\t}{\b} \RP\} = 0.
     \label{Differentiation with respect to Grassmann Variables prop 8}
\EEA
These equations yield directly that
\BEA
  \PD{\US{\t}{\b}}{\DUS{\t}{\a}} = \PD{\DUS{\t}{\b}}{\US{\t}{\a}} = 0,
      \nonumber
\EEA
and also an ``unusual" product rule~(with a minus sign) like e.g.
\BEA
   \PD{}{\US{\t}{\a}}\;(\LS{\t}{\b}\LS{\t}{\g}) &=&
     \LP(\PD{}{\US{\t}{\a}}\,\LS{\t}{\b}\RP)\;\LS{\t}{\g}
       - \LS{\t}{\b}\,\PD{}{\US{\t}{\a}}\,\LS{\t}{\g}
     = \d^{\a}_{\;\;\b}\LS{\t}{\g}-\LS{\t}{\b}\d^{\a}_{\;\;\g}.
       \label{Differentiation with respect to Grassmann Variables prop 8a}
\EEA
%An analogous formula holds
%for $\PD{}{\DUS{\t}{\a}}\;(\DLS{\t}{\b}\DLS{\t}{\g})$.

With the conventions for the differential operators established so far,
the following relations are true\footnote{When spinor indices are suppressed
on the differentiation symbols,
we will follow the convention
\BEA
  \PD{}{\t}\PD{}{\t}  &=&   \PD{}{\LS{\t}{\a}}\PD{}{\US{\t}{\a}},  \nn
  \PD{}{\tb}\PD{}{\tb}  &=&   \PD{}{\DUS{\t}{\a}}\PD{}{\DLS{\t}{\a}}. \nonumber
\EEA }
\BEA
   \PD{}{\US{\t}{\a}}\;\;\t\t &=& 2\,\LS{\t}{\a} ,
      \label{Differentiation with respect to Grassmann Variables prop 9} \SL
   \PD{}{\DUS{\t}{\a}}\;\;\tb\tb &=& -2\,\DLS{\t}{\a},
      \label{Differentiation with respect to Grassmann Variables prop 10} \SL
   \PD{}{\LS{\t}{\a}}\PD{}{\US{\t}{\a}}\;\, \t\t &=& 4 ,
      \label{Differentiation with respect to Grassmann Variables prop 11} \SL
   \PD{}{\DUS{\t}{\a}}\PD{}{\DLS{\t}{\a}}\;\; \tb\tb &=& 4.
      \label{Differentiation with respect to Grassmann Variables prop 12}
\EEA
\begin{PROOF}
We start by proving eqs.~\r{Differentiation with respect
to Grassmann Variables prop 9} and
\r{Differentiation with respect to Grassmann Variables prop 11}
\BEA
   \PD{}{\US{\t}{\a}}\;\;\t\t &=& \PD{}{\US{\t}{\a}} \US{\t}{\b}\LS{\t}{\b} \nn
     &=& \d_{\a}^{\;\;\b}\LS{\t}{\b}-\US{\t}{\b}\e_{\b\g}\d_{\a}^{\;\;\g} \nn
     &=& \LS{\t}{\a}+\LS{\t}{\g}\d_{\a}^{\;\;\g} \nn
     &=& 2\,\LS{\t}{\a}, \nonumber
\EEA
and
\BEA
   \PD{}{\LS{\t}{\a}}\PD{}{\US{\t}{\a}}\;\; \t\t &=&
      \PD{}{\LS{\t}{\a}}\PD{}{\US{\t}{\a}}\;\; \US{\t}{\b}\LS{\t}{\b} \nn
      &=& \PD{}{\LS{\t}{\a}}\LP(\PD{}{\US{\t}{\a}}
                \;\,\US{\t}{\b}\LS{\t}{\b}\RP)\nn
      &=& 2\;\PD{}{\LS{\t}{\a}}\;\;\LS{\t}{\a} \nn
      &=& 2\;\;\d_{\,\;\a}^{\a} \nn
      &=& 4. \nonumber
\EEA
Eqs.~\r{Differentiation with respect to Grassmann Variables prop 10}
and \r{Differentiation with respect to Grassmann Variables prop 12}
are proven in a similar way.
\end{PROOF}

\subsection{The Berezin Integral.}

In ordinary field theories, a translation invariant action is
constructed (assuming surface terms to vanish), by integrating a
Lagrangian density $\L(x)$ over $d^{4}x$. In a similar fashion,
SUSY invariant actions in superspace can be obtained
by an integration over the whole of superspace.

The aim of this section will be to define what we understand by
integration with respect to Grassmann variables, i.e. to define the
so-called Berezin integral~\cite{BER66}.

We will start by considering the simplest situation with only one
Grassmann variable $\zeta$. Since $\zeta^{n}=0,\;n \geq 2$, due to
the anticommuting property of $\zeta$, any function of $\zeta$,
$f(\zeta)$, has always the form
\BEA
   f(\zeta) &=& f(0)+\zeta f^{(1)}.
     \label{The Berezin Integral prop 1}
\EEA
Hence it is sufficient to define $\int\!\!d\zeta$ and
$\int\!\!d\zeta\; \zeta$ in order to let $\int\!\!d\zeta\; f(\zeta)$
be well-defined. Following F.A. Berezin~\cite{BER66} we define
\BEA
    \int\!\!d\zeta  &=& 0,
          \label{The Berezin Integral prop 2}\SL
    \int\!\!d\zeta \;\zeta &=& 1.
          \label{The Berezin Integral prop 3}
\EEA
Thus
\BEA
  \int\!\!d\zeta \;f(\zeta) &=& \int\!\!d\zeta
        \LP(f(0)+\zeta f^{(1)}\RP) =  f^{(1)} ,
\EEA
and formally differentiation and integration are the same, i.e.
\BEA
  \int\!\!d\zeta f(\zeta) &=& \PD{}{\zeta}\;f(\zeta).
          \label{The Berezin Integral prop 4}
\EEA
Two important properties, follow as a consequence of the
definitions~\r{The Berezin Integral prop 2},
\r{The Berezin Integral prop 3} and use of eq.~\r{The Berezin Integral prop 1},
should be noted
\BEA
   \int\!\!d\zeta\; f(\zeta+\kappa)  &=& \int\!\!d\zeta\; f(\zeta),
             \label{The Berezin Integral prop 5}\SL
   \int\!\!d\zeta \LP( a\, f(\zeta) +b\,  h(\zeta) \RP)
             &=& a \int\!\!d\zeta \; f(\zeta) +b \int\!\!d\zeta\; h(\zeta),
             \hspace{1cm}a,b \in C,
             \label{The Berezin Integral prop 6}
\EEA
i.e. translation invariance and complex linearity respectively.

Superspace is not parametrized in terms of only one Grassmann variable.
However, it ``contains" the Grassmann algebras $G_{2}=\{\t^{1},\t^{2}\}$
and $\bar{G}_{2}=\{\tb^{1},\tb^{2}\}$. To define integration on these
algebras, we have to generalize the above results. By demanding
\BEA
   \LP\{d\,\LS{\t}{\a},d\,\LS{\t}{\b}\RP\} =
     \LP\{d\,\LS{\t}{\a},\LS{\t}{\b}\RP\}  = 0,
       \label{The Berezin Integral prop 7} \SL
   \LP\{d\,\DLS{\t}{\a},d\,\DLS{\t}{\b}\RP\} =
     \LP\{d\,\DLS{\t}{\a},\DLS{\t}{\b}\RP\}  = 0,
       \label{The Berezin Integral prop 8}
\EEA
and using the definitions~\r{The Berezin Integral prop 2} and
\r{The Berezin Integral prop 3} we have
\BEA
  \int\!\!d\,\LS{\t}{1}d\,\LS{\t}{2} &=& 0,
      \label{The Berezin Integral prop 9}\SL
  \int\!\!d\,\LS{\t}{1}d\,\LS{\t}{2} \; \LS{\t}{1} &=&
  \int\!\!d\,\LS{\t}{1}d\,\LS{\t}{2} \; \LS{\t}{2} = 0,
      \label{The Berezin Integral prop 10}\SL
  \int\!\!d\,\LS{\t}{1}d\,\LS{\t}{2} \; \LS{\t}{1}\LS{\t}{2} &=& -1.
      \label{The Berezin Integral prop 11}
\EEA
Similar formulae hold for the algebra $\bar{G}_{2}$. Now the integral of
any function on $G_{2}$ and/or $\bar{G}_{2}$ can be obtained by
Taylor expansion and linearity.

We now define ``volume elements" of the anti-commuting part of superspace
\BEA
   d^{2}\t &=& -\f{1}{4}\; d\!\US{\t}{\a}\,d\LS{\t}{\a}
                 =  -\f{\e_{\a\b}}{4}\; d\!\US{\t}{\a}\,d\!\US{\t}{\b},
                 \label{The Berezin Integral prop 12}\SL
   d^{2}\tb &=& -\f{1}{4}\; d\DLS{\t}{\a}\,d\DUS{\t}{\a}
                 =  -\f{\e_{\dot{a}\dot{\b}}}{4}\; d\DLS{\t}{\a}\,d
                    \DLS{\t}{\b},
                    \label{The Berezin Integral prop 13}\SL
   d^{4}\tb &=& d^{2}\t\, d^{2}\tb.
                    \label{The Berezin Integral prop 14}
\EEA
With these definitions the following relations are true
\BEA
   \int\!\!d^{2}\t &=& \int\!\!d^{2}\tb = 0,
             \label{The Berezin Integral prop 14a} \SL
   \int \!\! d^{2}\t \; \US{\t}{\a}
        &=& \int\!\! d^{2} \tb \; \DLS{\t}{\a} \;\; = \;\; 0,
             \label{The Berezin Integral prop 14aa} \SL
   \int\!\!d^{2}\t\; \t\t &=& 1 ,
             \label{The Berezin Integral prop 15}  \SL
   \int\!\!d^{2}\tb\; \tb\tb &=& 1 ,
             \label{The Berezin Integral prop 16} \SL
   \int\!\!d^{4}\t\; \t\t\,\tb\tb &=& 1 .
             \label{The Berezin Integral prop 17}
\EEA
\begin{PROOF}
  Eq.~\r{The Berezin Integral prop 14a} and
  \r{The Berezin Integral prop 14aa} follow immediately from
  eq.~\r{The Berezin Integral prop 9} and
  \r{The Berezin Integral prop 10} and the corresponding equations for
  $\bar{G}_{2}$.

  By using the definition~\r{The Berezin Integral prop 11},
  \r{The Berezin Integral prop 12} and
  the fact that $\t\t= \e_{\a\b}\,\t^{\a}\t^{\b}=-2\,\t^{1}\t^{2}$,
  we have
  \BEA
     \int\!\!d^{2}\t\; \t\t &=& -\f{\e_{\a\b}}{4} \int\!\!d\US{\t}{\a}
       \,d\US{\t}{\b}\;\LP(-2\,\US{\t}{1}\US{\t}{2}\RP) \nn
       &=& \f{\e_{12}}{2}\int\!\!d\US{\t}{1} \,
               d\US{\t}{2}\;\US{\t}{1}\US{\t}{2}+
           \f{\e_{21}}{2}\int\!\!d\US{\t}{2}
               \,d\US{\t}{1}\;\US{\t}{1}\US{\t}{2} \nn
      &=& \e_{12}\int\!\!d\US{\t}{1}
               \,d\US{\t}{2}\;\US{\t}{1}\US{\t}{2} \nn
      &=& 1, \nonumber
      \label{The Berezin Integral prop 18}
\EEA
since $\e_{\a\b}$ is antisymmetric and $\e_{12}=-1$.
Eq.~\r{The Berezin Integral prop 16} is proved in the same way.
With eqs.~\r{The Berezin Integral prop 15} and
\r{The Berezin Integral prop 16} established, it is rather
straightforward to prove eq.~\r{The Berezin Integral prop 17}
\BEA
   \int\!\!d^{4}\t\; \t\t\,\tb\tb &=&  \int\!\!d^{2}\t\, d^{2}\tb \;
      \tb\tb\,\t\t  \nn
       &=& \int\!\!d^{2}\t \;\t\t \nn
       &=& 1. \nonumber
         \label{The Berezin Integral prop 19}
\EEA
\end{PROOF}

With the formulae obtained so far, the integral
$\int\!\!d^{4}\t\;\Phi(x,\t,\tb)$ of a general superfield
can be established.
Hence we have
\BEA
   \int\!\!d^{4}\t\;\Phi(x,\t,\tb) &=&
   \int\!\!d^{4}\t\;\LP( f(x)+\US{\t}{\a}\LS{\phi}{\a}(x) +
    \DLS{\t}{\a}\DUS{\psi}{\a}(x) +\t\t\,m(x)+\tb\tb\,n(x)\RP. \nn
    & & \mbox{}\hspace{1.2cm}\LP. +
    \TSTB \,V_{\mu}(x)+\t\t\,\DLS{\t}{\a}\DUS{\lambda}{\a}(x)+
    \tb\tb\,\US{\t}{\a}\LS{\psi}{\a}(x) + \t\t\,\tb\tb\,d(x)\RP)\nn
    &=& d(x).
     \label{The Berezin Integral prop 20}
\EEA
Thus, by integration with respect to Grassmann supercoordinates, the
$\t\t\;\tb\tb$-component of any integrand is always outprodjected.
This fact, as we will see, is rather useful when supersymmetric
Lagrangians are being constructed.

\subsection{Delta Functions on Grassmann Algebras.}

Delta functions on superspace
simplify the constructions of SUSY-invariant actions.
Let such delta functions  on $G_{2}$ and $\bar{G}_{2}$, both two and four
dimensional, be defined implicitly by
\BEA
  \int\!\!d^{2}\t\;f(\t)\,\d^{2}(\t) &=& f(0),
    \hspace{1cm}f(\t) \in G_{2},
     \label{Delta Functions on Grassmann Algebras prop 1}\SL
  \int\!\!d^{2}\tb\;g(\tb)\,\d^{2}(\tb) &=& g(0),
    \hspace{1cm}g(\tb) \in \bar{G}_{2},
    \label{Delta Functions on Grassmann Algebras prop 2}
\EEA
and
\BEA
    \int \!\! d^{4}\t \; h(,\t,\tb)\, \d^{4}(\t,\tb)
       &=& h(0,0),
       \hspace{1cm}h(\t,\tb) \in G_{2} \times \bar{G}_{2}.
       \label{Delta Functions on Grassmann Algebras prop 2a}
\EEA
This implies that
\BEA
    \d^{2}(\t) &=& \t\t,
    \label{Delta Functions on Grassmann Algebras prop 3} \SL
   \d^{2}(\tb) &=& \tb\tb,
   \label{Delta Functions on Grassmann Algebras prop 4}
\EEA
and
\BEA
    \d^{4} (\t,\tb) &=& \d^{2}(\t)\, \d^{2}(\tb) \;\;=\;\, \t\t\;\tb\tb,
    \label{Delta Functions on Grassmann Algebras prop 4a}
\EEA
as we now shall show.
\begin{PROOF}
   By using the anticommuting properties of the elements of $G_{2}$
   and eq.~\r{The Berezin Integral prop 15} we have
   \BEA
        \int\!\!d^{2}\t\;f(\t)\;\t\t &=&
          \int\!\!d^{2}\t\;\LP(f(0)+\US{\t}{\a}f^{(1)}_{\a}+
                \t\t \,f^{(2)}\RP)\t\t \nn
          &=& \int\!\!d^{2}\t\; \t\t\;f(0) \nn
          &=& f(0). \nonumber
\EEA
Hence, with the identification $\d^{2}(\t) = \t\t$,
eq.~\r{Delta Functions on Grassmann Algebras prop 1} is fulfilled,
something which shows that our identification is correct.

In a similar way eq.~\r{Delta Functions on Grassmann Algebras prop 4}
is seen to be consistent with
eq.~\r{Delta Functions on Grassmann Algebras prop 2}.

For the same reason as above we have
\BEA
     \int \!\! d^{4}\t \; h(,\t,\tb) \d^{4}(\t,\tb)
       &=& \int \!\! d^{4}\t \;
          \LP( h(0,0) + \US{\t}{\a} \PD{h}{\US{\t}{\a}}
                      +  \DLS{\t}{\a} \PD{h}{\DLS{\t}{\a}}
                      + \ldots \RP) \;\t\t\;\tb\tb \nn
        &=& h(0,0).
\EEA
Thus, the identification made in
eq.~\r{Delta Functions on Grassmann Algebras prop 4a}
is correct.
\end{PROOF}

%\cleardoublepage

\cleardoublepage

\chapter{The Two-Component Form of the Off-Shell Lagrangian $\L_{SUSY}$.}
   \label{APP: YM-Lagrangian}

In this appendix the expansion of $\L_{SUSY}$,
in the two-component formalism, will be
performed in detail.

However, before we address this problem, some general calculations
will be performed.
To be more specific, we will in sect.~\ref{The Non-Abelian Fieldstrength.}
calculate the component form of
the non-Abelian fieldstrength $W_{\a}$.
In sect.~\ref{SECT: Calculating TRACE}
this expansion will be used in obtaining the
component form of the kinetic term of vectorsuperfields.
Finally, in sect.~\ref{sect. left},
which concludes our general calculations
of this appendix, we derive the expansion of the
matter Lagrangian of a ${\cal G}\times U(1)$- gauge theory, where
${\cal G}$ is some non-Abelian gauge group.

\section{The Non-Abelian Fieldstrength.}
     \label{The Non-Abelian Fieldstrength.}

In this section we will calculate the component expansion of the
non-Abelian fieldstrengths, as defined\footnote{Here we have made the
substitution $g\rightarrow 2g$.}
by
\BEA
  \LS{W}{\a} &=& -\f{1}{8g}\,\bar{D}\bar{D}e^{-2gV}\LS{D}{\a}e^{2gV}, \\
  \DLS{W}{\a} &=& -\f{1}{8g}\,D D e^{-2gV}\DLS{D}{\a}e^{2gV}.
\EEA
We start by $\LS{W}{\a}$ and for simplicity we will work in the WZ-gauge.
By hermitian conjugation the corresponding
expression for $\DLS{W}{\a}$ is obtained.
It is practical to work in the basis $(y=x+i\t\s\tb,\t,\tb)$,  since then
the SUSY covariant derivatives take on a somewhat simpler form
%(cf.eqs.~\r{Chiral Superfields prop 4} and \r{Chiral Superfields prop 5})
\BEA
 \LS{D}{\a}(y,\t,\tb) &=& \PD{}{\US{\t}{\a}}+2i\SA
          \DUS{\t}{\a}\PD{}{y^{\mu}}, \\
          \label{The Non-Abelian Fieldstrength prop 3}
 \DLS{D}{\a}(y,\t,\tb) &=& -\PD{}{\DUS{\t}{\a}}.
          \label{The Non-Abelian Fieldstrength prop 4}
\EEA
Hence ($ \mbox{Im} A(x) = 0$)\footnote{When we work in the basis $(y,\t,\tb)$
the notation $\P_{\mu}$ will mean $\PD{}{y^{\mu}}$ if nothing else is
said to indicate otherwise. }
\BEA
  V^{a}(x,\t,\tb) &=& V^{a}(y-i\t\s\tb,\t,\tb) \nn
    &=& -\TSTB\,V^{a}_{\mu}(y)
        +  i\,\t\t\;\tb\bar{\lambda}^{a}(y) -i\,\tb\tb\;\t\lambda^{a}(y) \nn
    & & \mbox{}
       + \HA\;\t\t\;\tb\tb\;\LP[\;D^{a}(y)+i\,\P^{\mu}V^{a}_{\mu}(y)\;\RP],
          \label{The Non-Abelian Fieldstrength prop 5}
\EEA
where eq.~\r{Fierz Rearrangemant Formula prop 5} has been used.
{}From now on, the y-dependence of the component fields will be
understood and suppressed.

Our program will be to first calculate
$e^{-2gV}\LS{D}{\a}e^{2gV}$ and then the total
(non-Abelian) field strength. In these calculations, expressions for
$\LS{D}{\a} V^{a}$, $\LS{D}{\a}\LP(V^{a} V^{b}\RP)$,
and finally $V^{a}\LS{D}{\a} V^{b}$ are useful.
Hence we start by determine these expressions.

%Direct generalization of eqs.~\r{WZ-gauge prop 4} and
%\r{WZ_gauge prop 2} gives
%\BEA
%   e^{gV} &=& e^{gT^{a}V^{a}} \nn
%      &=& 1+gT^{a}V^{a} +\f{g^{2}}{2}T^{a}T^{b}
%              V^{a}V^{b},
%          \label{The Non-Abelian Fieldstrength prop 6}    \\*[1.5mm]
%      V^{a}\,V^{b} &=& \HA\;\t\t\;\tb\tb\;
%          V^{a\;\mu}(x)V^{b}_{\mu}(x) \nn
%      &=& \HA\;\t\t\;\tb\tb\;V^{a\;\mu}V^{b}_{\mu}.
%          \label{The Non-Abelian Fieldstrength prop 7}
%\EEA
Using the results from appendix~\ref{App: Grassmann Variables}, and in
particular eqs.~\r{Differentiation  with respect to Grassmann Variables prop 2}
and \r{Differentiation  with respect to Grassmann Variables prop 9}, yields
\BEA
   \LS{D}{\a} V^{a}
      &=& \LP(\PD{}{\US{\t}{\a}}+2i\SA\DUS{\t}{\a}\PD{}{y^{\mu}}\RP)
      \nmb   \times
            \LP(-\TSTB\;V^{a}_{\mu} +  i\,\t\t\;\tb\bar{\lambda}^{a}
                 -i\,\tb\tb\;\t\lambda^{a}
       + \HA\;\t\t\;\tb\tb\;\LP[\;D^{a}
        +i\,\P^{\mu}V^{a}_{\mu}\;\RP]\RP)\hspace{-0.4cm} \NN
    &=&   -\;\SA\DUS{\t}{\a}V^{a}_{\mu}
          +2i\;\LS{\t}{\a}\;\tb\bar{\lambda}^{a}
          - i\;\tb\tb\;\LS{\lambda}{\a}^{a}
          +\LS{\t}{\a}\;\tb\tb\;
               \LP[\;D^{a}+i\;\P^{\mu}V^{a}_{\mu}\;\RP]
     \nmb
          -2i\;\s^{\nu}_{\a\dot{\a}}\DUS{\t}{\a}\;\TSTB\;\P_{\nu}V^{a}_{\mu}
          -2 \;\SA\DUS{\t}{\a}\;\t\t\;
         \DLS{\t}{\b}\;\P_{\mu}\bar{\lambda}^{\dot{\b}\;a}\NN
    &=&   -\;\SA\DUS{\t}{\a}V^{a}_{\mu}
          +2i\;\LS{\t}{\a}\;\tb\bar{\lambda}^{a}
          - i\;\tb\tb\;\LS{\lambda}{\a}^{a}
    \nmb
          +\LS{\t}{\a}\;\tb\tb\;
               \LP[\;D^{a}+i\;\P^{\mu}V^{a}_{\mu}\;\RP]
    \nmb
          -2i\;\tb\tb\;\LP[\HA\d_{\a}^{\;\;\b}g^{\nu\mu}-i\;
                              \LP(\s^{\nu\mu}\RP)_{\a}^{\;\;\b}\;
                       \RP]\;\LS{\t}{\b}\;\P_{\nu}V^{a}_{\mu}
     \nmb
           + \t\t\;\tb\tb\; \e^{\dot{\a}\dot{\b}}
         \SA\;\P_{\mu}\bar{\la}^{a}_{\dot{\b}}  .
%          -2 \;\SA\DUS{\t}{\a}\;\t\t\;\DLS{\t}{\b}\;\
%                        \P_{\mu}\bar{\lambda}^{\dot{\b}\;a}.
             \nonumber
\EEA
Here eq.~\r{Spinor Relations prop 16a} has been used. By utilizing the
antisymmetry of $\s^{\mu\nu}$~(cf. sect.~\ref{sec Dirac Matrices})
and eq.~\r{Spinor Relations prop 17} (together
with a redefinition for the indices $\mu$ and $\nu$)
one obtains
\BEA
    \LS{D}{\a} V^{a}
    &=&   -\;\SA\DUS{\t}{\a}V^{a}_{\mu}
          +2i\;\LS{\t}{\a}\;\tb\bar{\lambda}^{a}
          - i\;\tb\tb\;\LS{\lambda}{\a}^{a} \nn
    & & \mbox{}
          +\tb\tb\;\LP\{ \; \LS{\t}{\a} D^{a}
                         -\LP( \s^{\mu\nu}\RP)_{\a}^{\;\;\b}\LS{\t}{\b}\;
                               \LP[\; \P_{\mu}V_{\nu}^{a}-\P_{\nu}
                                       V^{a}_{\mu}\;
                               \RP]\;
                  \RP\} \nn
    & & \mbox{}
          +\;\t\t\;\tb\tb\;\SA\,\P_{\mu}\bar{\lambda}^{\dot{\a}\;a}.
          \label{The Non-Abelian Fieldstrength prop 8}
\EEA
With eqs.~\r{Fierz Rearrangemant Formula prop 5}
and \r{Differentiation with respect to Grassmann Variables prop 9} we have
\BEA
   \LS{D}{\a}\;\LP(V^{a}V^{b}\RP)
      &=& \LS{D}{\a}\LP(\HA\,\t\t\,\tb\tb\;V^{a\,\mu}V^{b}_{\mu}\RP)\nn
      &=& \tb\tb\;\LS{\t}{\a}\;V^{a\;\mu}V^{b}_{\mu}.
       \label{The Non-Abelian Fieldstrength prop 9}
\EEA
Note that only the first term of $\LS{D}{\a}$
contributes to $D_{\a}\LP( V^{a} V^{b}\RP)$ due
to the anticommuting properties of the superspace parameters $\t$.

Furthermore with eq.~\r{Differentiation with respect
to Grassmann Variables prop 9}
\BEA
 V^{a}\LS{D}{\a}V^{b}
   &=&  \LP\{ -\TSTB\,V^{a}_{\mu}
+  i\,\t\t\;\tb\bar{\lambda}^{a}-i\,\tb\tb\;\t\lambda^{a}
       + \HA\;\t\t\;\tb\tb\;D^{a} \RP\}
   \nmb \times
       \LP\{  -\s_{\a\dot{\a}}^{\nu}\DUS{\t}{\a}V_{\nu}^{b}
              +2i\;\LS{\t}{\a}\,\tb\bar{\la}^{b}
              -i\;\tb\tb\,\la_{\a}^{b}   \RP.
   \nmb \hspace{1cm}  \LP.
              + \tb\tb\LP[  \LS{\t}{\a}D^{b}
                          -\LP(\s^{\s\nu}\RP)_{\a}^{\;\;\b}\LS{\t}{\b}
       \LP(\P_{\s}V^{b}_{\nu}-\P_{\nu}V_{\s}^{b}\RP)\RP]
                          + \t\t\;\tb\tb\;\s^{\nu}_{\a\dot{\a}}
             \,\P_{\nu}\bar{\la}^{\dot{\a}\,b}  \RP\}\nn
   &=&  \TSTB\;\s^{\nu}_{\a\dot{\a}}\DUS{\t}{\a}\;V^{a}_{\mu}V^{b}_{\nu}
        -2i\,\TSTB\;\LS{\t}{\a}\;\tb\bar{\lambda}^{b}\,V^{a}_{\mu}
        -i\;\t\t\;\tb\bar{\lambda}^{a}\;\s^{\nu}_{\a\dot{\a}}
                   \DUS{\t}{\a}V^{b}_{\nu}.   \nonumber
       \label{The Non-Abelian Fieldstrength prop 10}
\EEA
For later convenience, we rewrite this expression.
The first term is rewritten by eq.~\r{Spinor Relations prop 16a}, while
the two next terms are rewritten as follows:
\BEA
  -2i\,\TSTB\;\LS{\t}{\a}\;\tb\bar{\lambda}^{b}\;V^{a}_{\mu}
    &=& 2i\; \DLS{\t}{\g}\bar{\s}^{\mu\;\dot{\g}\g}\LS{\t}{\g}\;\LS{\t}{\a}
             \;\DLS{\t}{\a}\bar{\lambda}^{\dot{\a}\;b}\;V^{a}_{\mu} \nn
    &=& 2i\;(-\HA\e_{\dot{\g}\dot{\a}}\,\tb\tb)\;\bar{\s}^{\mu\;\dot{\g}\g}
              \;(\HA\e_{\g\a}\,\t\t)\;
           \bar{\lambda}^{\dot{\a}\;b}\;V^{a}_{\mu}\nn
    &=& -\;\f{i}{2}\;\t\t\;\tb\tb\;\SA\;
         \bar{\lambda}^{\dot{\a}\;b}\;V^{a}_{\mu} , \nonumber
\EEA
and similarly
\BEA
   -i\;\t\t\;\tb\bar{\lambda}^{a}\;\s^{\nu}_{\a\dot{\a}}
                   \DUS{\t}{\a}V^{b}_{\nu}
       &=& i\;\t\t\;\DUS{\t}{\a}\DUS{\t}{\b}\;\s^{\nu}_{\a\dot{\a}}\;
                \DLS{\lambda}{\b}^{a}V^{b}_{\nu}\nn
       &=&  \f{i}{2}\;\t\t\;\tb\tb\;\s^{\nu}_{\a\dot{\a}}\;
                \bar{\lambda}^{\dot{\a}\;a}V^{b}_{\nu} . \nonumber
\EEA
By collecting terms, eq.~\r{The Non-Abelian Fieldstrength prop 10}
reads
\BEA
   V^{a}\LS{D}{\a}V^{b}
     &=& i\;\tb\tb\;\LP(\s^{\mu\nu}\RP)_{\a}^{\;\;\b}\LS{\t}{\b}
                V^{a}_{\mu}V^{b}_{\nu}
         +\HA\; \tb\tb\;\LS{\t}{\a}\;V^{a\;\mu} V^{b}_{\mu}  \nn
     & & \mbox{}
         +\f{i}{2}\;\t\t\;\tb\tb\;\SA\;\LP[\;\bar{\lambda}^{\dot{\a}\;a}
                V^{b}_{\mu}-\bar{\lambda}^{\dot{\a}\;b}V^{a}_{\mu}\RP].
   \label{The Non-Abelian Fieldstrength prop 11}
\EEA
The reader should note that the first term of the above expression
is antisymmetric under the combined
index transformation $\mu \leftrightarrow \nu$ and $a \leftrightarrow b$,
while the second and third terms are symmetric and antisymmetric
respectively under $a \leftrightarrow b$.

Hence, by taking advantage of the fact that all powers of three (or higher)
of vector superfields in the WZ-gauge always vanish, we have
\BEA
 e^{-2gV}\LS{D}{\a}e^{2gV}
    &=& \LP( 1-2gT^{a}V^{a}+2g^{2}T^{a}T^{b}V^{a}V^{b}\RP)
         D_{\a} \LP( 1+ 2gT^{c}V^{c}+2g^{2}T^{c}T^{d}V^{c}V^{d}\RP) \nn
%    &=&   gT^{a}\LS{D}{\a}V^{a} \nn
%    & & \mbox {}
%        + \f{g^{2}}{2}\;T^{a}T^{b}\;\LP[\LS{D}{\a}
%              \LP(V^{a}V^{b} \RP)
%              -2\;V^{a}\LS{D}{\a}V^{b}\RP] \nn
%    & & \mbox {}
%        + \f{g^{3}}{2}\;T^{a}T^{b}T^{c}\;
%              \LP[\;V^{a}V^{b}\LS{D}{\a}V^{c}
%                     - V^{a}\LS{D}{\a}
%                         \LP(V^{b}V^{c}\RP)\;\RP]  \nn
%    & & \mbox {}
%        + \f{g^{4}}{4}T^{a}T^{b}T^{c}T^{d}\;
%               V^{a}V^{b}\LS{D}{\a}
%                      \LP(V^{c}V^{d}\RP)\nn
    &=&    2 gT^{a}\LS{D}{\a}V^{a}
        + 2g^{2}\;T^{a}T^{b}\;\LP[\LS{D}{\a}
              \LP(V^{a}V^{b} \RP)
              -2\;V^{a}\LS{D}{\a}V^{b}\RP].
            \label{The Non-Abelian Fieldstrength prop 12}
\EEA
We now rewrite the term in the square brackets, and with
eqs.~\r{The Non-Abelian Fieldstrength prop 9} and
\r{The Non-Abelian Fieldstrength prop 11} we obtain
\BEA
  \lefteqn{T^{a}T^{b}\;\LP[\LS{D}{\a}
              \LP(V^{a}V^{b} \RP)
              -2\;V^{a}\LS{D}{\a}V^{b}\RP]}\hspace*{1.5cm} \nn
     &=& T^{a}T^{b}\;\tb\tb\;
           \LP[\;-2i\;\LP(\s^{\mu\nu}\RP)_{\a}^{\;\;\b}
                 \LS{\t}{\b}V^{a}_{\mu}V^{b}_{\nu}
               - i\;\t\t\;\SA\;
                    \LP(\;\bar{\lambda}^{\dot{\a}\;a}V^{b}_{\mu}-
                          \bar{\lambda}^{\dot{\a}\;b}V^{a}_{\mu}\;
                    \RP)\;
           \RP].           \nonumber
\EEA
Since the terms in the square brackets are antisymmetric under the index
transformation $\mu \leftrightarrow \nu$ and $a \leftrightarrow b$
we have
\BEA
   \lefteqn{T^{a}T^{b}\;\LP[\LS{D}{\a}
              \LP(V^{a}V^{b} \RP)
              -2\;V^{a}\LS{D}{\a}V^{b}\RP]}\hspace*{1.5cm} \nn
     &=& \HA\;[T^{a},T^{b}]\;\tb\tb\;
            \LP[-2i\LP(\s^{\mu\nu}\RP)_{\a}^{\;\;\b}
                  \LS{\t}{\b}V^{a}_{\mu}V^{b}_{\nu}
                -i\;\t\t\;\SA
                   \LP(\;\bar{\lambda}^{\dot{\a}\;a}V^{b}_{\mu}-
                          \bar{\lambda}^{\dot{\a}\;b}V^{a}_{\mu}\;
                   \RP)\;
            \RP] \nn
     &=& f^{a b c}T^{c}\;\tb\tb\;
         \LP[\;\LP(\s^{\mu\nu}\RP)_{\a}^{\;\;\b}
                    \LS{\t}{\b}V^{a}_{\mu}V^{b}_{\nu}
                +\HA \;\t\t\;\SA
                   \LP(\;\bar{\lambda}^{\dot{\a}\;a}V^{b}_{\mu}-
                          \bar{\lambda}^{\dot{\a}\;b}V^{a}_{\mu}\;
                   \RP)\;
            \RP] \nn
      &=&  T^{a}\;\tb\tb\;
           \LP[\;f^{a b c}\LP(\s^{\mu\nu}\RP)_{\a}^{\;\;\b}
                    \LS{\t}{\b}V^{b}_{\mu}V^{c}_{\nu}
                   - f^{a b c}\;\t\t\;\SA
                       V^{b}_{\mu}\bar{\lambda}^{\dot{\a}\;c}
            \RP].
\EEA
Here we have used the antisymmetry of $f^{abc}$.
With this result, eq.~\r{The Non-Abelian Fieldstrength prop 12}
becomes
\BEA
   e^{-2gV}\LS{D}{\a}e^{2gV}
    &=& 2g\,T^{a}\LP[  -\;\SA\DUS{\t}{\a}V^{a}_{\mu}
          +2i\;\LS{\t}{\a}\;\tb\bar{\lambda}^{a}
          - i\;\tb\tb\;\LS{\lambda}{\a}^{a} \RP. \nn
    & & \mbox{} \hspace{1.3cm}
          +\tb\tb\;\LP\{ \; \LS{\t}{\a} D^{a}
                         -\LP( \s^{\mu\nu}\RP)_{\a}^{\;\;\b}
                 \LS{\t}{\b}\;V^{a}_{\mu\nu} \RP\} \nn
    & & \mbox{} \hspace{1.3cm} \LP.
          +\;\t\t\;\tb\tb\;\SA\LP\{
              \P_{\mu}\bar{\lambda}^{\dot{\a}\;a}
              - g f^{a b c}V^{b}_{\mu}
                      \bar{\la}^{\dot{\a}\;c} \RP\}  \RP],
\EEA
where
\BEA
  V_{\mu\nu}^{a} &=&
     \P_{\mu}V_{\nu}^{a}-\P_{\nu}V^{a}_{\mu}
             - g\,f^{a b c}V^{b}_{\mu}V^{c}_{\nu},
\EEA
is the non-Abelian, conventional fieldstength\footnote{The factor of two
in front of the coupling constant was inserted in order to make this
identification possible.}.

Hence, the total fieldstrength becomes with
eq.~\r{Differentiation with respect to Grassmann Variables prop 12}
\BEA
   W_{\a} &=&  -\f{1}{8g}\,\bar{D}\bar{D}e^{-2gV}\LS{D}{\a}e^{2gV} \nn
          &=&  -\f{1}{8g}\,2g\,T^{a}\LP[
               -4i\,\la^{a}_{\a}+4\LP(\LS{\t}{\a}D^{a}
         -\LP(\s^{\mu\nu}\RP)_{\a}^{\;\;\b}\LS{\t}{\b}\,V^{a}_{\mu\nu}\RP)\RP.
          \nmb  \hspace{2.5cm} \LP.
                +4\;\t\t\SA\LP(\P_{\mu}\bar{\la}^{\dot{\a}\;a}
            -g\,f^{abc}V^{b}_{\mu}\bar{\la}^{\dot{\a}\,c}\RP)\RP] \nn
          &=&   T^{a}\LP[    i\,\la^{a}_{\a} - \LS{\t}{\a}\,D^{a}
                           + \LP(\s^{\mu\nu}\RP)_{\a}^{\;\;\b}\LS{\t}{\b}
                                \,V^{a}_{\mu\nu}
                           - \t\t\,\SA\LP(\P_{\mu}\bar{\la}^{\dot{\a}\,a}
          - g\,f^{abc}V^{b}_{\mu}\bar{\la}^{\dot{\a}\,c}\RP)\RP].\hspace{2mm}
                           \label{W sup alpha}
\EEA
By hermitian conjugation the component expansion for $\bar{W}_{\dot{\a}}$
is obtained, and it reads
\BEA
   \bar{W}_{\dot{\a}}
          &=&   T^{a}\LP[  -i\,\bar{\la}^{a}_{\dot{\a}} - \DLS{\t}{\a}D^{a}
                           + \DLS{\t}{\b}\LP(\bar{\s}^{\mu\nu}\RP)_{\;\;
                    \dot{\a}}^{\dot{\b}}V^{a}_{\mu\nu}
                           - \tb\tb\,\SA\LP(\P_{\mu}\la^{\a\,a}
       - g\,f^{abc}V^{b}_{\mu}\la^{\a\,c}\RP)\RP].\hspace{9mm}
                           \label{Component form of the W-fieldstrength}
\EEA
The fieldstrengths with upper spinor indices are obtained in the usual
way by applying $\e^{\a\b}$ and $\e^{\dot{\a}\dot{\b}}$ to the above
expressions
\BEA
    \US{W}{\a} &=& \e^{\a\b} \LS{W}{\b},\SL
    \DUS{W}{\a} &=& \e^{\dot{\a}\dot{\b}}\DLS{W}{\b}.
\EEA

\section{Calculation $\int d^{4}\t\;\LP(1/4k\RP)
 \,Tr\LP(W^{\a}W_{\a}\RP)\,\d^{2}(\tb)+ h.c.$}
   \label{SECT: Calculating TRACE}

Since $W_{\a}$ is Lie-Algebra valued one has
\BEA
\lefteqn{\int d^{4}\t\;\f{1}{4k}\,
     Tr\LP(W^{\a}W_{\a}\RP)\,\d^{2}(\tb)}\hspace*{1.5cm} \nn
   &=&   \int d^{4}\t\;\f{1}{4k}\,Tr\LP(T^{a}
             T^{b}\RP)W^{\a\,a}W^{b}_{\a}\;\d^{2}(\tb)\nn
%   &=&   \int d^{4}\t\;\f{1}{4k}\,k\d^{ab}\;
%\LP. W^{\a\,a}W^{b}_{\a}\RP|_{\t\t}\nn
   &=&   \f{1}{4}\;\LP.W^{a\,\a}W^{a}_{\a}\RP|_{\t\t}.
\EEA
Here we have used the normalization (in the adjoint representation)
\BEA
    Tr\LP(T^{a}T^{b}\RP)  &=&  k\d^{ab}.
\EEA
Furthermore, eqs.~\r{W sup alpha}, \r{Spinor Relations prop 18},
\r{Spinor Relations prop 4}, \r{Spinor Relations prop 16}
and  \r{Spinor Relations prop 2}
yield
\BEA
\lefteqn{\f{1}{4}\;\LP.W^{a\,\a}W^{a}_{\a}\RP|_{\t\t}}\hspace{0.9cm} \nn
  &=&  \f{1}{4}\;\e^{\a\b}\LP\{
          \LP[i\,\la^{a}_{\b} - \LS{\t}{\b}\,D^{a}
            + \LP(\s^{\mu\nu}\RP)_{\b}^{\;\;\d}\LS{\t}{\d}\,V^{a}_{\mu\nu}
            - \t\t\,\s^{\mu}_{\b\dot{\b}}\LP(\P_{\mu}\bar{\la}^{\dot{\b}\,a}
        - g\,f^{abc}V^{b}_{\mu}\bar{\la}^{\dot{\b}\,c}\RP)\RP] \RP.
  \nmb \LP.\LP. \hspace{0.8cm} \times
       \LP[i\,\la^{a}_{\a} - \LS{\t}{\a}\,D^{a}
            + \LP(\s^{\rho\s}\RP)_{\a}^{\;\;\g}\LS{\t}{\g}\,V^{a}_{\rho\s}
            - \t\t\,\SA\LP(\P_{\rho}\bar{\la}^{\dot{\a}\,a}
             - g\,f^{ab'c'}V^{b'}_{\rho}\bar{\la}^{\dot{\a}\,c'}\RP)\RP]
             \RP\} \RP|_{\t\t} \nn
   &=&   -\f{i}{4}\;\la^{a}\s^{\mu}\LP(\P_{\mu}\bar{\la}^{a}
         - g\,f^{abc}V^{b}_{\mu}\bar{\la}^{c}\RP)
         +\f{1}{4}\;D^{a}D^{a}
   \nmb
         -\f{1}{4}\;\t \s^{\mu\nu} \t\;D^{a}V^{a}_{\mu\nu}
         +\f{1}{8}\;(\s^{\mu\nu})_{\b}^{\;\;\d}
     (\s^{\rho\s})_{\a}^{\;\;\g}\e^{\a\b}\e_{\d\g}V^{a}_{\mu\nu}V^{a}_{\rho\s}
   \nmb
         -\f{i}{4} \la^{a}\s^{\mu}
         \LP(\P_{\mu}\bar{\la}^{a} - g\,f^{abc}V^{b}_{\mu}\bar{\la}^{c}\RP)\nn
   &=&  -\f{i}{2}\la^{a}\s^{\mu}\LP(\P_{\mu}\bar{\la}^{a}
           - g\,f^{abc}V^{b}_{\mu}\bar{\la}^{c}\RP)
        +\f{1}{4}\;D^{a}D^{a}
%        +\f{1}{4}\;(\s^{\mu\nu})_{\a}^{\;\;\a}D^{a}V^{a}_{\mu\nu}
   \nmb
        +\f{1}{8}\;(\s^{\mu\nu})_{\b}^{\;\;\d}(\s^{\rho\s})_{\a}^{\;\;\g}
           \LP(\d^{\a}_{\;\;\g}\d^{\b}_{\;\;\d}-\d^{\a}_{\;\;\d}
             \d^{\b}_{\;\;\g}\RP)V^{a}_{\mu\nu}V^{a}_{\rho\s}.
            \label{WW prop 1}
\EEA
\newpage
By rewriting the last term of eq.~\r{WW prop 1} with use of
eqs.~\r{Pauli Matrix prop 11a} and \r{Pauli Matrix prop 11}, one obtains
\BEA
\lefteqn{\f{1}{8}\;(\s^{\mu\nu})_{\b}^{\;\;\d}(\s^{\rho\s})_{\a}^{\;\;\g}
          \LP(\d^{\a}_{\;\;\g}\d^{\b}_{\;\;\d}-\d^{\a}_{\;\;\d}
          \d^{\b}_{\;\;\g}\RP)V^{a}_{\mu\nu}V^{a}_{\rho\s}}\hspace{1.5cm} \nn
   &=& \f{1}{8}\LP((\s^{\mu\nu})_{\b}^{\;\;\b}(\s^{\rho\s})_{\a}^{\;\;\a}
  -(\s^{\mu\nu}\s^{\rho\s})_{\b}^{\;\;\b}\RP) V^{\a}_{\mu\nu}V^{a}_{\rho\s}\nn
   &=&  -\f{1}{16}\;\LP(g^{\mu\rho}g^{\nu\s}-
          g^{\mu\s}g^{\nu\rho}+i\e^{\mu\nu\rho\s}\RP)
           V^{a}_{\mu\nu}V^{a}_{\rho\s}\nn
   &=&  -\f{1}{8}\;V^{a\,\mu\nu}V^{a}_{\mu\nu}
        -\f{i}{16}\;\e^{\mu\nu\rho\s} V^{a}_{\mu\nu}V^{a}_{\rho\s},
\EEA
where we in the last transition have used the antisymmetry of
the conventional fieldstrength.

Thus one can conclude
\BEA
\lefteqn{\int d^{4}\t\;\f{1}{4k}\,Tr\LP(W^{\a}W_{\a}\RP)\,
           \d^{2}(\tb)}\hspace*{1.5cm} \nn
   &=&  -\f{i}{2}\la^{a}\s^{\mu}\LP(\P_{\mu}\bar{\la}^{a} -
            g\,f^{abc}V^{b}_{\mu}\bar{\la}^{c}\RP)
        +\f{1}{4}\;D^{a}D^{a}
%   \nmb
        -\f{1}{8}\;V^{a\,\mu\nu}V^{a}_{\mu\nu}
    \nmb
        -\f{i}{16}\;\e^{\mu\nu\rho\s} V^{a}_{\mu\nu}V^{a}_{\rho\s}.
          \label{WW prop 2}
\EEA
With eq.~\r{Spinor Relations prop 14} the first term of
the above equation can be rewritten as follows
\BEA
  \lefteqn{-\f{i}{2}\la^{a}\s^{\mu}\LP(\P_{\mu}\bar{\la}^{a}
            - g\,f^{abc}V^{b}_{\mu}\bar{\la}^{c}\RP)}\hspace{1.5cm}\nn
     &=& \f{i}{2}\LP(\P_{\mu}\bar{\la}^{a}-g
            f^{abc}V^{b}_{\mu}\bar{\la}^{c}\RP) \bar{\s}^{\mu}\la^{a} \nn
     &=& -\f{i}{2}\,\bar{\la}^{a}\bar{\s}^{\mu}
           \LP(\P_{\mu}\la^{a}-g f^{abc}V^{b}_{\mu}\la^{c}\RP)
           +\f{i}{2}\,\P_{\mu}\LP(\bar{\la}^{a}\bar{\s}^{\mu}\la^{a}\RP)\nn
     &=& -\f{i}{2}\,\bar{\la}^{a}\bar{\s}^{\mu}D_{\mu}\la^{a}
                +\f{i}{2}\,\P_{\mu}\LP(\bar{\la}^{a}\bar{\s}^{\mu}\la^{a}\RP).
\EEA
Here we have introduced the $SU(2)\times U(1)$-covariant derivative
\BEA
   D_{\mu} &=& \P_{\mu} + i g{\bf T}^{a} V^{a}_{\mu}+ig'\f{{\bf Y}}{2}
               V'_{\mu}, \hspace{1cm}
    a = 1,2,3,  \nonumber
\EEA
and when it operates on e.g. the gaugino $\la^{a}$,
which lay in the adjoint representation of the gauge group, i.e.
\BEA
       \LP[T^{c}_{adj}\RP]^{ab} &=& -if^{cab}, \nn
         Y_{adj}&=& 0, \nonumber
\EEA
we have
\BEA
   D_{\mu} \la^{a} &\equiv& \LP[D_{\mu}\la\RP]^{a}\nn
                    &=&  \LP[D_{\mu}\RP]^{ab}\la^{b}\nn
                    &=& \LP(\P_{\mu}\d^{ab} +ig\LP[T^{c}_{adj}\RP]^{ab}
                           V^{c}_{\mu}
                           +ig' \f{ Y_{adj}}{2}V'_{\mu}\RP)\la^{b}\nn
                    &=& \P_{\mu}\la^{a} - g f^{abc} V^{b}_{\mu}\la^{c}.
                        \label{app: Component Field Expansion prop 10asa1}
\EEA
This yields for eq.~\r{WW prop 2}
\BEA
\int d^{4}\t\;\f{1}{4k}\,Tr\LP(W^{\a}W_{\a}\RP)\,\d^{2}(\tb)
   &=&  -\f{i}{2}\bar{\la}^{a}\bar{\s}^{\mu} D_{\mu}\la^{a}
        +\f{1}{4}\;D^{a}D^{a}
%   \nmb
        -\f{1}{8}\;V^{a\,\mu\nu}V^{a}_{\mu\nu}
    \nmb
        -\f{i}{16}\;\e^{\mu\nu\rho\s} V^{a}_{\mu\nu}V^{a}_{\rho\s}
        +\f{i}{2}\P_{\mu}\LP( \bar{\la}^{a}\bar{\s}^{\mu}\la^{a}\RP).
          \label{WW prop 2aaa}
\EEA

Hence by hermitian conjugation (of eq.~\r{WW prop 2}), one has
\BEA
  \int d^{4}\t\;\f{1}{4k}\,Tr\LP(\bar{W}_{\dot{\a}}\bar{W}^{\dot{\a}}\RP)
            \,\d^{2}(\t)
   &=&
     \f{i}{2}\;D_{\mu}\la^{a} \s^{\mu}\bar{\la}^{a}
        +\f{1}{4}\;D^{a}D^{a}
%   \nmb
        -\f{1}{8}\;V^{a\,\mu\nu}V^{a}_{\mu\nu}
    \nmb
        +\f{i}{16}\;\e^{\mu\nu\rho\s} V^{a}_{\mu\nu}V^{a}_{\rho\s} \nn
%        -\f{i}{2}\,\P_{\mu}\LP(\bar{\la}^{a}\bar{\s}^{\mu}\la^{a}\RP)\nn
    &=&
       - \f{i}{2}\bar{\la}^{a}\bar{\s}^{\mu} D_{\mu}\la^{a}
         +\f{1}{4}\;D^{a}D^{a}
%   \nmb
        -\f{1}{8}\;V^{a\,\mu\nu}V^{a}_{\mu\nu}
    \nmb
        +\f{i}{16}\;\e^{\mu\nu\rho\s} V^{a}_{\mu\nu}V^{a}_{\rho\s}
%        -\f{i}{2}\,\P_{\mu}\LP(\bar{\la}^{a}\bar{\s}^{\mu}\la^{a}\RP),
          \label{WW prop 3}
\EEA
and by adding eqs.~\r{WW prop 2aaa} and \r{WW prop 3} we may conclude
\BEA
\lefteqn{\f{1}{4k}\,\int d^{4}\t\;\LP\{\,Tr\LP(W^{\a}W_{\a}\RP)\,\d^{2}(\tb)
           + Tr\LP(\bar{W}_{\dot{\a}}\bar{W}^{\dot{\a}}\RP)
          \,\d^{2}(\t) \RP\} }\hspace*{1.5cm} \nn
    &=&
       - i\,\bar{\la}^{a}\bar{\s}^{\mu} D_{\mu}\la^{a}
         +\f{1}{2}\;D^{a}D^{a}
%    \nmb
        -\f{1}{4}\;V^{a\,\mu\nu}V^{a}_{\mu\nu}
          +  \f{i}{2}\;\P_{\mu} \LP(\bar{\la}^{a}\bar{\s}^{\mu}\la^{a}\RP)
%        -\f{i}{2}\,\P_{\mu}\LP(\bar{\la}^{a}\bar{\s}^{\mu}\la^{a}\RP)
          \label{WW prop 4}.
\EEA
%If we like, eq.~\r{WW prop 4} can be siplified even further since

\section{Calculating $\int d^{4}\t\;\hat{\phi}^{\dagger}
        e^{2g\hat{V}+g'\hat{V}'} \hat{\phi} $.}
        \label{sect. left}

In this section we will derive the component form
of $\int d^{4}\t\;\hat{\phi}^{\dagger}e^{2g\hat{V}+g'\hat{V}'} \hat{\phi}$.
Here $\hat{\phi}(x,\t,\tb)$ is a chiral superfield
and $\hat{V}(x,\t,\tb)$ and $V'(x,\t,\tb)$
are gauge vector superfields for some non-Abelian group ${\cal G}$ and U(1)
respectively.

As usual we work in the WZ-gauge with Lie-algebra valued gauge superfields
of the form
\BEA
    \hat{V}(x,\t,\tb) &=& {\bf T}^{a}\hat{V}^{a}(x,\t,\tb),
          \label{LCEL prop 2} \SL
    \hat{V}'(x,\t,\tb) &=& {\bf Y}\hat{v}'(x,\t,\tb),
           \label{LCEL prop 3}
\EEA
and with the following component expansions for
the superfields
\BEA
  \hat{\phi}(x,\t,\tb)
             &=&    A(x)
                 +  i\;\TSTB \;\P_{\mu}A(x)
                 -  \f{1}{4}\;\t\t\;\tb\tb\;\P^{\mu}\P_{\mu}A(x)
             \nmb
                 +  \sqrt{2}\;\t \psi(x)
                 +  \f{i}{\sqrt{2}}\;\t\t\;\tb\bar{\s}^{\mu}\P_{\mu}\psi(x)
                 +  \t\t\;F(x),
           \label{LCEL prop 6}  \SL
  \hat{\phi}^{\dagger}(x,\t,\tb)
             &=&    A^{\dagger}(x)
                 -  i\;\TSTB \;\P_{\mu}A^{\dagger}(x)
                 -  \f{1}{4}\;\t\t\;\tb\tb\;\P^{\mu}\P_{\mu}A^{\dagger}(x)
             \nmb
                 +  \sqrt{2}\;\tb \bar{\psi}(x)
                 +  \f{i}{\sqrt{2}}\;\tb\tb\;\t\s^{\mu}\P_{\mu}\bar{\psi}(x)
                 +  \tb\tb\;F^{\dagger}(x), \SL
           \label{LCEL prop 7}
 \hat{V}^{a}(x,\t,\tb)
     &=& - \;\TSTB\;V^{a}_{\mu}(x)
         + i\;\t\t\;\tb\bar{\lambda}^{a}(x)
         - i\;\tb\tb\;\t\lambda^{a}(x)
         + \HA\;\t\t\,\tb\tb\;D^{a}(x),
           \label{LCEL prop 4} \SL
  \hat{v}'(x,\t,\tb)
     &=& - \;\TSTB\;V'_{\mu}(x)
         + i\;\t\t\;\tb\bar{\lambda}'(x)
         - i\;\tb\tb\;\t\lambda'(x)
         + \HA\;\t\t\,\tb\tb\;D'(x).
           \label{LCEL prop 5}
\EEA
Furthermore, $\hat{\phi}(x,\t,\tb)$ will be taken to lie in a representation
of the gauge group ${\cal G}~\times~U(1)$ described by the
matrix representation $T^{a}$ and the hypercharge quantum number Y.
Hence $\hat{\phi}(x,\t,\tb)$, and its component fields, are generally
matrix-valued. As our notation indicates, we  will work in the
$(x,\t,\tb)$-basis, and from now on this dependence will be suppressed.

Since the two gauge super-multiplets are commuting, i.e.
$[\hat{V},\hat{V}']=0$, and we are working in the WZ-gauge, we have
\BEA
 e^{2g\hat{V}+g'\hat{V}'} &=&
        \LP(    1
             +  2gT^{a}\hat{V}^{a}
             +  2g^{2}T^{a}T^{b}\hat{V}^{a}\hat{V}^{b}  \RP)
     \nmb
        \times
        \LP(    1
             +  g'Y\hat{v}'
             +  \HA g'^{2}Y^{2}\hat{v}'^{2}   \RP) \nn
     &=&      1
           +  g'Y\hat{v}'
           +  2gT^{a}\hat{V}^{a}
           +  \f{g'^{2}}{2}Y^{2}\hat{v}'^{2}
           +  2g^{2}T^{a}T^{b}\hat{V}^{a}\hat{V}^{b}
     \nmb
           +  2gg'YT^{a}\hat{V}^{a}\hat{v}.'
            \label{LCEL prop 11}
\EEA
Here in the last line we have used the fact that third powers of
vector superfields in the WZ-gauge always vanish.
Furtermore
%A straightforward generalization
%of eq.~\r{WZ-gauge prop 2} yields
\BEA
  \hat{V}^{a}\hat{V}^{b}  &=& \HA \;\t\t\;\tb\tb\;V^{a\;\mu}V^{b}_{\mu},
              \label{LCEL prop 12}\SL
  \hat{v}'^{2}            &=&  \HA \;\t\t\;\tb\tb\;V'^{\mu}V'_{\mu},
              \label{LCEL prop 13}\SL
  \hat{V}^{a}\hat{v}'    &=&  \HA \;\t\t\;\tb\tb\;V^{a\;\mu}V'_{\mu},
              \label{LCEL prop 14}
\EEA
and hence with eqs.~\r{LCEL prop 4}, \r{LCEL prop 5} and
\r{LCEL prop 12}--\r{LCEL prop 14}
substituted into eq.~\r{LCEL prop 11} one obtains
\BEA
    e^{2g\hat{V}+g'\hat{V}'}
     &=&     1
          -  \TSTB\;\LP[\;2gT^{a}V^{a}_{\mu}+g'YV'_{\mu}\;\RP]
     \nmb
          +  i\;\t\t\;\tb\;\LP[\;2gT^{a}\bar{\lambda}^{a}
                                 +g'Y\bar{\lambda}'\;\RP]
          -  i\;\tb\tb\,\t\;\LP[\;2gT^{a}\lambda^{a}
                                 +g'Y\lambda'\;\RP]
     \nmb
          +  \HA\;\t\t\;\tb\tb\;\LP[\;2gT^{a}D^{a}
                      + g'YD'
                      + 2g^{2}\,T^{a}T^{b}V^{a\;\mu}V_{\mu}^{b} \RP.
     \nmb  \hspace{2.3cm} \LP.
                      + \HA g'^{2}Y^{2}V'^{\mu}V'_{\mu}
                      + 2 gg'YT^{a} V^{a\;\mu}V'_{\mu} \;\RP].
\EEA

Postmultiplying the above expression with $\hat{\phi}$ yields
\BEA
  \lefteqn{e^{2g\hat{V}+g'\hat{V}'}\,\hat{\phi}}\hspace{1.5cm} \nn
     &=&     1
          -  \TSTB\;\LP[\;2gT^{a}V^{a}_{\mu}+g'YV'_{\mu}\;\RP]
     \nmb
          +  i\;\t\t\;\tb\;\LP[\;2gT^{a}\bar{\lambda}^{a}
                                 +g'Y\bar{\lambda}'\;\RP]
          -  i\;\tb\tb\,\t\;\LP[\;2gT^{a}\lambda^{a}
                                 +g'Y\lambda'\;\RP]
     \nmb
          +  \HA\;\t\t\;\tb\tb\;\LP[\;2gT^{a}D^{a}
                      + g'YD'
                      + 2g^{2}\,T^{a}T^{b}V^{a\;\mu}V_{\mu}^{b} \RP.
     \nmb  \hspace{2.3cm} \LP.
                      + \HA g'^{2}Y^{2}V'^{\mu}V'_{\mu}
                      + 2 gg'YT^{a} V^{a\;\mu}V'_{\mu} \;\RP] \nn
     &\;\;\times &
              \LP[       A
                 +  i\;\TSTB \;\P_{\mu}A
                 -  \f{1}{4}\;\t\t\;\tb\tb\;\P^{\mu}\P_{\mu}A \RP.
             \nmb \LP. \;\;\;
                 +  \sqrt{2}\;\t \psi
                 +  \f{i}{\sqrt{2}}\;\t\t\;\tb\bar{\s}^{\mu}\P_{\mu}\psi
                 +  \t\t\;F \RP] \nn
     &=&     A
          -  \TSTB\;\LP[\;2gT^{a}V^{a}_{\mu}
                        + g'YV'_{\mu}\;\RP]\;A
     \nmb
          +  i\;\t\t\;\tb\LP[\;2gT^{a}\bar{\lambda}^{a}
                                 +g'Y\bar{\lambda}'\;\RP]\;A
          -  i\;\tb\tb\,\t\LP[\;2gT^{a}\lambda^{a}
                                 +g'Y\lambda'\;\RP]\;A
     \nmb
          +  \HA\;\t\t\;\tb\tb\;\LP[\;2gT^{a}D^{a}
                      + g'YD'
                      + 2g^{2}\,T^{a}T^{b}V^{a\;\mu}V_{\mu}^{b} \RP.
     \nmb  \hspace{2.3cm} \LP.
                      + \HA g'^{2}Y^{2}V'^{\mu}V'_{\mu}
                      + 2 gg'YT^{a} V^{a\;\mu}V'_{\mu} \;\RP]
                      \;A
     \nmb
          +  i\;\TSTB\;\P_{\mu}A
          -  i\;\TSTB\;\t\s^{\nu}\tb\;\LP[\;2gT^{a}V^{a}_{\mu}
                                          + g'YV'_{\mu}\;\RP]\;
                                       \P_{\nu}A
     \nmb
          - \f{1}{4}\;\t\t\;\tb\tb\;\P^{\mu}\P_{\mu}A
          + \sqrt{2}\;\t \psi
          - \sqrt{2}\;\TSTB\;\LP[\;2gT^{a}V^{a}_{\mu}
                                   + g'YV'_{\mu}\;\RP]\;\t \psi
     \nmb
          - \sqrt{2}i\;\tb\tb\;\t\LP[\;2gT^{a}\lambda^{a}
                                     + g'Y\lambda'\;\RP]\;\t \psi
          + \f{i}{\sqrt{2}}\;\t\t\;\tb\bar{\s}^{\mu}\P_{\mu}\psi
          + \t\t\;F . \nonumber
\EEA
Now we rewrite the fourth and third  last term as follows:
\BEA
    - \sqrt{2}\;\TSTB\;\LP[\;2gT^{a}V^{a}_{\mu}
                                   + g'YV'_{\mu}\;\RP]\;\t \psi
    &=& \sqrt{2}\,\LP[\;2gT^{a}V^{a}_{\mu}+ g'YV'_{\mu}\;\RP]\;
           \US{\t}{\a}\US{\t}{\b}\SA \DUS{\t}{\a}\LS{\psi}{\b} \nn
    &=& - \sqrt{2}\,\LP[\;2gT^{a}V^{a}_{\mu}+ g'YV'_{\mu}\;\RP]\;
         \HA\,\e^{\a\b} \t\t \,\SA \DUS{\t}{\a}\LS{\psi}{\b} \nn
    &=& - \f{1}{\sqrt{2}}\,\t\t\,\tb\bar{\s}^{\mu}\psi\,
            \LP[\;2gT^{a}V^{a}_{\mu}+ g'YV'_{\mu}\;\RP], \NN
    - \sqrt{2}i\;\tb\tb\;\t\LP[\;2gT^{a}\lambda^{a}
                                     + g'Y\lambda'\;\RP]\;\t \psi
      &=& - \sqrt{2}i\;\tb\tb\;\US{\t}{\a}\LP[\;2gT^{a}\LS{\lambda}{\a}^{a}
               + g'Y\LS{\lambda}{\a}'\;\RP]
               \US{\t}{\b}\LS{\psi}{\b}   \nn
      &=& - \f{i}{\sqrt{2}}\;\t\t\;\tb\tb\LP[\;2gT^{a}\LS{\lambda}{\a}^{a}
               + g'Y\LS{\lambda}{\a}'\;\RP]
               \US{\psi}{\a} \nn
      &=&  \f{i}{\sqrt{2}}\;\t\t\;\tb\tb\LP[\;2gT^{a}\lambda^{a}
               + g'Y\lambda'\;\RP]\psi,
\EEA
and thus
\BEA
   \lefteqn{e^{2g\hat{V}+g'\hat{V}'}\,\hat{\phi}}\hspace{1.5cm} \nn
     &=&     A
          +  \sqrt{2}\;\t \psi
          +  \t\t\;F
          +  \TSTB\;\LP\{\;i\P_{\mu}A
                  -\LP[\;2gT^{a}V^{a}_{\mu}+g'YV'_{\mu}\;\RP]\;
                        A \;\RP\}
     \nmb
          +  \t\t\;\tb\LP\{\;i
                 \LP[\;2gT^{a}\bar{\lambda}^{a}
                           +g'Y\bar{\lambda}'\;\RP]\;A\RP.
     \nmb \LP. \hspace{1.4cm}
                     - \f{1}{\sqrt{2}}\;\LP[\;2gT^{a}V^{a}_{\mu}
                                            +g'YV'_{\mu}\;\RP]
                            \bar{\s}^{\mu}\psi
                     + \f{i}{\sqrt{2}}\;\bar{\s}^{\mu}\P_{\mu}\psi \RP\}
     \nmb
          -  i\;\tb\tb\;\t\LP[\;2gT^{a}\lambda^{a}
                              + g'Y\lambda'\;\RP]\;A
     \nmb
          +  \HA\;\t\t\;\tb\tb\;
                \LP\{\;\LP[\;2gT^{a}D^{a}
                            + g'YD'
                            + 2g^{2}\,T^{a}T^{b}V^{a\;\mu}V_{\mu}^{b} \RP.\RP.
    \nmb  \hspace{3cm} \LP.
                            + \HA g'^{2}Y^{2}V'^{\mu}V'_{\mu}
                            + 2 gg'YT^{a} V^{a\;\mu}V'_{\mu} \;\RP]
                               \;A
     \nmb  \hspace{2.2cm}
                    -i\;\LP[\;2gT^{a}V^{a}_{\mu}
                            + g'YV'_{\mu}\;\RP]\;\P^{\mu}A
                    - \HA\;\P^{\mu}\P_{\mu}A
     \nmb  \hspace{2.2cm} \LP.
                    + \sqrt{2}i\;\LP[\;2gT^{a}\lambda^{a}
                                     +g'Y\lambda'\;\RP]
                                     \;\psi
                 \RP\}.
\EEA

Finally we can address the main purpose of this section.
By premultiplying the above result by $\hat{\phi}^{\dagger}$
and projecting out the $\t\t\;\tb\tb$-component, equivalent to a Grassmann
integration, we obtain:
\BEA
  \lefteqn{ \int d^{4}\t \;\hat{\phi}^{\dagger}e^{2g\hat{V}+g'\hat{V}'}
           \hat{\phi}}\nn
   &=&\int d^{4}\t\;\LP[  A^{\dagger}
          -  i\;\TSTB \;\P_{\mu}A^{\dagger}
          -  \f{1}{4}\;\t\t\;\tb\tb\;\P^{\mu}\P_{\mu}A^{\dagger}\RP.
     \nmb \LP. \hspace{1.2cm}
          +  \sqrt{2}\;\tb \bar{\psi}
          +  \f{i}{\sqrt{2}}\;\tb\tb\;\t\s^{\mu}\P_{\mu}\bar{\psi}
          +  \tb\tb\;F^{\dagger} \RP] \nn
     & &  \hspace{0.8cm} \times   \LP[
             A
          +  \sqrt{2}\;\t \psi
          +  \t\t\;F
          +  \TSTB\;\LP\{\;i\P_{\mu}A
                  -\LP[\;2gT^{a}V^{a}_{\mu}+g'YV'_{\mu}\;\RP]\;
                        A \;\RP\} \RP.
     \nmb   \hspace{1.3cm}
          +  \t\t\;\tb\LP\{\;i
                 \LP[\;2gT^{a}\bar{\lambda}^{a}
                           +g'Y\bar{\lambda}'\;\RP]\;A\RP.
     \nmb \LP. \hspace{2.7cm}
                     - \f{1}{\sqrt{2}}\;\LP[\;2gT^{a}V^{a}_{\mu}
                                            +g'YV'_{\mu}\;\RP]
                            \bar{\s}^{\mu}\psi
                     + \f{i}{\sqrt{2}}\;\bar{\s}^{\mu}\P_{\mu}\psi \RP\}
     \nmb   \hspace{1.3cm}
          -  i\;\tb\tb\;\t\LP[\;2gT^{a}\lambda^{a}
                              + g'Y\lambda'\;\RP]\;A
     \nmb    \hspace{1.3cm}
          +  \HA\;\t\t\;\tb\tb\;
                \LP\{\;\LP[\;2gT^{a}D^{a}
                            + g'YD'
                            + 2g^{2}\,T^{a}T^{b}V^{a\;\mu}V_{\mu}^{b} \RP.\RP.
    \nmb  \hspace{4.3cm} \LP.
                            + \HA g'^{2}Y^{2}V'^{\mu}V'_{\mu}
                            + 2 gg'YT^{a} V^{a\;\mu}V'_{\mu} \;\RP]
                               \;A
     \nmb  \hspace{3.5cm}
                    -i\;\LP[\;2gT^{a}V^{a}_{\mu}
                            + g'YV'_{\mu}\;\RP]\;\P^{\mu}A
                    - \HA\;\P^{\mu}\P_{\mu}A
     \nmb  \hspace{3.5cm} \LP.\LP. \LP.
                    + \sqrt{2}i\;\LP[\;2gT^{a}\lambda^{a}
                                     +g'Y\lambda'\;\RP]
                                     \;\psi
                 \RP]\RP\}\RP]               \NN
     &=&  A^{\dagger}\;
             \LP[\;gT^{a}D^{a}
                   +\HA g'YD'
                   + g^{2}\,T^{a}T^{b}V^{a\;\mu}V_{\mu}^{b} \RP.
    \nmb  \hspace{1.5cm} \LP.
                   + \f{1}{4} g'^{2}Y^{2}V'^{\mu}V'_{\mu}
                   +  gg'YT^{a} V^{a\;\mu}V'_{\mu} \;\RP]
                     \;A
     \nmb
         -i\;A^{\dagger}\;
              \LP[\;gT^{a}V^{a}_{\mu}
                   + \HA g'YV'_{\mu}\;\RP]\;\P^{\mu}A
         - \f{1}{4}\;A^{\dagger}\P^{\mu}\P_{\mu}A
     \nmb
         + \sqrt{2}i\;A^{\dagger}\;\LP[\;gT^{a}\lambda^{a}
                             +\HA g'Y\lambda'\;\RP]\;\psi
     \nmb
         -i\;\t\s^{\nu}\tb\;\TSTB\;\P_{\nu}A^{\dagger}\;
             \LP.\LP\{\;i\P_{\mu}A
                    -2\LP[\;gT^{a}V^{a}_{\mu}+\HA\,g'YV'_{\mu}\;\RP]\;
                        A \;\RP\} \RP|_{\t\t\;\tb\tb}
         - \f{1}{4}\;\P^{\mu}\P_{\mu}A^{\dagger}A
     \nmb
         + \sqrt{2}\;\tb\bar{\psi}\;\t\t\;\tb
                \LP\{ \LP. \;2 i\,
                 \LP[\;gT^{a}\bar{\lambda}^{a}
                           +\HA \,g'Y\bar{\lambda}'\;\RP]
               \;A \RP\}\RP|_{\t\t\;\tb\tb}
     \nmb
         + \sqrt{2}\;\tb\bar{\psi}\;\t\t\;\tb
                \LP\{ \LP.
                     - \sqrt{2}\;\LP[\;gT^{a}V^{a}_{\mu}
                                      + \HA\;g'Y V'_{\mu}\;\RP]
                            \bar{\s}^{\mu}\psi
%     \nmb  \LP.\LP. \hspace{3cm}
             + \f{i}{\sqrt{2}}\;\bar{\s}^{\mu}\P_{\mu}\psi
                          \RP\}\RP|_{\t\t\;\tb\tb}
     \nmb
         +\LP. i\;\tb\tb\;\t\s^{\mu}\P_{\mu}\bar{\psi}\;\t
                   \psi\RP|_{\t\t\;\tb\tb}
         + F^{\dagger}F.
            \label{LCEL prop 15}
\EEA
With eq.~\r{Fierz Rearrangemant Formula prop 5} and the following results
\BEA
 \sqrt{2}\; \tb\bar{\psi}\;\t\t\;\DLS{\t}{\a}
           &=& \sqrt{2}\; \t\t\DLS{\t}{\a}\DLS{\t}{\b} \DUS{\psi}{\b}
           \;\;=\;\; -\f{1}{\sqrt{2}}\,\t\t\,\tb\tb\; \DLS{\psi}{\a},   \nn
  i\;\tb\tb\;\t\s^{\mu}\P_{\mu}\bar{\psi}\;\t \psi
           &=& -i\,\tb\tb\,\US{\t}{\a}\US{\t}{\b}\,
                \SA\,\P_{\mu}\DUS{\psi}{\a}\,\LS{\psi}{\b}
           \;\;=\;\; - \f{i}{2}\;\t\t\,\tb\tb\;
                 \psi\s^{\mu}\P_{\mu}\bar{\psi}, \nonumber
\EEA
eq.~\r{LCEL prop 15} becomes
\BEA
  \lefteqn{ \int d^{4}\t \;\hat{\phi}^{\dagger}e^{2g\hat{V}
            +g'\hat{V}'}\hat{\phi}}\hspace{1.5cm} \nn
      &=&  A^{\dagger}\;
             \LP[\;gT^{a}D^{a}
                   +\HA g'YD'
                   + g^{2}\,T^{a}T^{b}V^{a\;\mu}V_{\mu}^{b} \RP.
    \nmb  \hspace{1.5cm} \LP.
                   + \f{1}{4} g'^{2}Y^{2}V'^{\mu}V'_{\mu}
                   +  gg'YT^{a} V^{a\;\mu}V'_{\mu} \;\RP]
                     \;A
     \nmb
         -i\;A^{\dagger}\;
              \LP[\;gT^{a}V^{a}_{\mu}
                   + \HA g'YV'_{\mu}\;\RP]\;\P^{\mu}A
         - \f{1}{4}\;A^{\dagger}\P^{\mu}\P_{\mu}A
     \nmb
         + \sqrt{2}i\;A^{\dagger}\;\LP[\;gT^{a}\lambda^{a}
                             +\HA g'Y\lambda'\;\RP]\;\psi
     \nmb
         +\HA\,\P^{\mu}A^{\dagger}\P_{\mu}A
         + i\,\P^{\mu}A^{\dagger}\LP[\;gT^{a}V^{a}_{\mu}
             +\HA\,g'YV'_{\mu}\;\RP]\,A
         - \f{1}{4}\;\P^{\mu}\P_{\mu}A^{\dagger}A
     \nmb
          - \sqrt{2}i\,\bar{\psi}\LP[\;gT^{a}\bar{\lambda}^{a}
                           +\HA \,g'Y\bar{\lambda}'\;\RP]\;A
           +\bar{\psi}\LP[\;gT^{a}V^{a}_{\mu} + \HA\;g'Y V'_{\mu}\;\RP]
                            \bar{\s}^{\mu}\psi
     \nmb
         - \f{i}{2}\;\bar{\psi}\bar{\s}^{\mu}\P_{\mu}\psi
         - \f{i}{2}\,\psi \s^{\mu}\P_{\mu}\bar{\psi}
         + F^{\dagger}F.
            \label{LCEL prop 16}
\EEA

With the following identities
\BEA
  \psi\s^\mu\P_\mu\bar{\psi} &=&
     -\P_{\mu}\psi \s^\mu\bar{\psi}+ \P_{\mu}\LP(\psi\s^\mu\bar{\psi}\RP) \nn
                             &=& \bar{\psi}\bar{\s}^{\mu}\P_{\mu}\psi
       + \P_{\mu}\LP(\psi\s^\mu\bar{\psi}\RP), \nn
  \P^{\mu}\P_{\mu}A^{\dagger}A &=&
   -\P^{\mu}A^{\dagger}\P_{\mu}A + \P_{\mu}\LP(\P^{\mu}A^{\dagger}A\RP),\nn
  A^{\dagger}\P^{\mu}\P_{\mu}A &=&
-\P^{\mu}A^{\dagger}\P_{\mu}A + \P_{\mu}\LP(A^{\dagger}\P^{\mu}A\RP), \nonumber
\EEA
the final expression for the matter Lagrangian reads
\BEA
  \lefteqn{ \int d^{4}\t \;\hat{\phi}^{\dagger}
        e^{2g\hat{V}+g'\hat{V}'}\hat{\phi}}\hspace{1.5cm} \nn
      &=&  A^{\dagger}\;   \LP[\;gT^{a}D^{a} +\HA g'YD'\;\RP] A
         + \P^{\mu}A^{\dagger}\P_{\mu}A
      \nmb
        +   A^{\dagger}\; \LP[\;  g^{2}\,T^{a}T^{b} V^{a\;\mu} V^{b}_{\mu}
%    \nmb  \hspace{1.5cm} \LP.
                   + \f{1}{4} g'^{2}Y^{2}V'^{\mu}V'_{\mu}
                   +  g g'Y T^{a} V^{a\;\mu} V'_{\mu} \;\RP] A
    \nmb
         +i\;\P^{\mu}A^{\dagger}\LP[\;gT^{a}V^{a}_{\mu}
                      + \HA g'YV'_{\mu}\;\RP]A
         -i\;        A^{\dagger}\LP[\;gT^{a}V^{a}_{\mu}
                     + \HA g'YV'_{\mu}\;\RP]\P^{\mu}A
     \nmb
         + \sqrt{2}i\;A^{\dagger}\LP[\;gT^{a}\lambda^{a}
                     +\HA g'Y\lambda'\;\RP]\psi
         - \sqrt{2}i\;\bar{\psi}\LP[\;gT^{a}\bar{\lambda}^{a}
                     +\HA g'Y\bar{\lambda}'\;\RP] A
     \nmb
         + \bar{\psi}\bar{\s}^{\mu}\LP[\;gT^{a}V^{a}_{\mu}
                     + \HA g'YV'_{\mu}\;\RP] \psi
         - i\,\bar{\psi} \bar{\s}^{\mu}\P_{\mu}\psi
         + F^{\dagger}F + t.d.
            \label{LCEL prop 17}
\EEA
Here t.d. means a total derivative~(t.d.), and it can be neglected if we like.
This is so due to the four dimensional
 Gauss-theorem\footnote{The four dimensional
Gauss theorem states that
\BEA
    \int_{V} d^4x \,F(x) &=& \int_{S} d^3 S^\mu\,\P_\mu F(x)  \nonumber,
\EEA
where S is a 3-dimmentional surface enclosing the 4-dimensional volume
V. In our case, V denotes the total
4-space, and hence S is an surface at infinity.
Since the fields are assumed to vanish at infinity, the right hand side
vanish because F(x) is some function of quantum fields.}
which implies that the total derivative does not contribute to the action.

\subsection{Introducing the Covariant Derivative.}

Eq.~\r{LCEL prop 17} can be simplified even further if we introduce the
${\cal G}\times U(1)$-covariant derivative, defined by
\BEA
  D_{\mu} &=& \P_{\mu} +ig{\bf T}^{a}V_{\mu}^{a}+i g'\f{{\bf Y}}{2}V'_{\mu}.
     \label{LCEL prop 18}
\EEA
Here ${\bf T}^{a}$ and ${\bf Y}$ have the same meaning as in the previous
section.

With this definition one has
\BEA
   \lefteqn{\LP(D^{\mu} A\RP)^\dagger \LP(D_\mu A\RP)}\hspace{1.5cm}\nn
    &=& \LP(\P^{\mu}A +ig T^{a}V^{a\,\mu}A+i g'\f{ Y}{2}V^{'\mu}A\RP)^\dagger
         \LP(\P_{\mu}A +ig T^{b}V_{\mu}^{b}A+i g'\f{ Y}{2}V'_{\mu}A\RP)\nn
    &=&    \P^{\mu}A^{\dagger}\P_{\mu}A
    \nmb
         + i\,\P^{\mu}A^{\dagger}\LP(g T^{a}V_{\mu}^{a}+
                      g'\f{ Y}{2}V'_{\mu}\RP)A
         -i\,A^{\dagger}\LP(g T^{a}V_{\mu}^{a}+
                      g'\f{ Y}{2}V'_{\mu}\RP)\P^{\mu}A
    \nmb
        + A^{\dagger}\LP(g^2 T^a T^b V^{a\,\mu}V^b_{\mu}
            +g g'\, T^a Y V^{a\,\mu}V'_{\mu}
            +\f{g'^{2}}{4}  Y^{2} V'^\mu V'_{\mu}\RP)A,
               \label{LCEL prop 19}
\EEA
and substituting  eqs.~\r{LCEL prop 18} and \r{LCEL prop 19} into
eq.~\r{LCEL prop 17} yields
\BEA
\lefteqn{\int d^{4}\t \;\hat{\phi}^{\dagger}e^{2g\hat{V}
          +g'\hat{V}'}\hat{\phi}}\hspace{1.5cm} \nn
      &=&
         \LP(D^{\mu}A\RP)^{\dagger}\LP(D_{\mu}A\RP)
          -i\,\bar{\psi}\bar{\s}^{\mu}D_{\mu}\psi
     \nmb
          + A^{\dagger}\LP(\, gT^{a}D^{a}+g'\f{Y}{2}D'\RP)A
     \nmb
         + \sqrt{2}i\;A^{\dagger}\LP[\;gT^{a}\lambda^{a}
                 + g'\f{Y}{2}\lambda'\;\RP]\psi
         - \sqrt{2}i\;\bar{\psi}\LP[\;gT^{a}\bar{\lambda}^{a}
                 + g'\f{Y}{2}\bar{\lambda}'\;\RP] A
     \nmb
         + F^{\dagger}F + t.d.
            \label{LCEL prop 20}
\EEA
This concludes this section.

\section{Component Expansion of $\L_{SUSY}$.}

Now we will leave the general situation, and instead consider,
what is the purpose of this thesis, the electroweak $SU(2)\times U(1)$-theory.
{}From chapter~\ref{Chapter Supersymmetric Extension of QFD.}
we recall that the unbroken theory is
described by the Lagrangian
\BEA
   \L_{SUSY} &=&   \L_{Lepton}
                  + \L_{Gauge}
                  + \L_{Higgs},
                  \label{Component Expansion prop 1}
\EEA
where
\BEA
  \L_{Lepton}
     &=& \int d^{4}\t\;\LP[\,\hat{L}^{\dagger}e^{2g\hat{V}+g'\hat{V}'}
                           \hat{L} +
                           \hat{R}^{\dagger}e^{2g\hat{V}+g'\hat{V}'}
                           \hat{R}\,\RP],
                           \label{Component Expansion prop 2}\SL
  \L_{Gauge}
     &=&  \f{1}{4} \int  d^{4}\t\;
         \LP[\,
                \US{W}{a\,\a}\LS{W}{\a}^{a}
             +  \US{W}{'\,\a}\LS{W}{\a}' \,
         \RP]\d^{2}(\tb)   + h.c.\,,
         \label{Component Expansion prop 3}\SL
  \L_{Higgs}
     &=&  \int d^{4}\t\;\LP[\,\hat{H}_{1}^{\dagger}e^{2g\hat{V}+g'\hat{V}'}
                           \hat{H}_{1}
                         + \hat{H}_{2}^{\dagger}e^{2g\hat{V}+g'\hat{V}'}
                           \hat{H}_{2}
%                           \RP. \nn
%    & & \mbox{} \LP. \hspace{1.6cm}
               + W\,\d^{2}(\tb) + W^{\dagger}\,\d^{2}(\t) \RP].\hspace{0.5cm}
               \label{Component Expansion prop 4}
\EEA
Here the superpotential W is given by
\BEA
    W &=& W_{H} + W_{Y} \nn
      &=& \mu\; \e^{ij}\hat{H}_{1}^{i}\hat{H}_{2}^{j} +
          f\;\e^{ij}\hat{H}^{i}_{1}\hat{L}^{j}\hat{R}.
\EEA

{}From chapter~\ref{CHAPTER: Component Field Expansion} we recall the
component expansions of the various superfields of $\L_{SUSY}$.
They are
\BEA
  \hat{L}(x,\t,\tb)
             &=&    \tilde{L}(x)
                 +  i\;\TSTB \;\P_{\mu}\tilde{L}(x)
                 -  \f{1}{4}\;\t\t\;\tb\tb\;\P^{\mu}\P_{\mu}\tilde{L}(x) \nn
             & & \mbox{}
                 +  \sqrt{2}\;\t L^{(2)}(x)
                 +  \f{i}{\sqrt{2}}\;\t\t\;\tb\bar{\s}^{\mu}\P_{\mu}L^{(2)}(x)
                 +  \t\t\;F_{L}(x),
                 \label{Component Expansion prop 5}\SL
  \hat{R}(x,\t,\tb)
       &=&    \tilde{R}(x)
            + i\,\TSTB\;\P_{\mu} \tilde{R}(x)
            - \f{1}{4} \; \t\t\;\tb\tb\;\P^{\mu}\P_{\mu} \tilde{R}(x)\nn
       & & \mbox{}
            + \sqrt{2}\;\t R^{(2)}(x)
            + \f{i}{\sqrt{2}}\;\t\t\,\tb\bar{\s}^{\mu}\P_{\mu}R^{(2)}(x)
            + \t\t\;F_{R}(x),
            \label{Component Expansion prop 6}\SL
  \hat{H}_{1}(x,\t,\tb)
             &=&    H_{1}(x)
                 +  i\;\TSTB \;\P_{\mu}H_{1}(x)
                 -  \f{1}{4}\;\t\t\;\tb\tb\;\P^{\mu}\P_{\mu}H_{1}(x) \nn
             & & \mbox{}
                 +  \sqrt{2}\;\t \tilde{H}_{1}^{(2)}(x)
                 +  \f{i}{\sqrt{2}}\;\t\t\;\tb\bar{\s}^{\mu}
           \P_{\mu}\tilde{H}_{1}^{(2)}(x)
                 +  \t\t\;F_{1}(x),
                 \label{Component Expansion prop 7}
\EEA
and finally
\BEA
  \hat{H}_{2}(x,\t,\tb)
             &=&    H_{2}(x)
                 +  i\;\TSTB \;\P_{\mu}H_{2}(x)
                 -  \f{1}{4}\;\t\t\;\tb\tb\;\P^{\mu}\P_{\mu}H_{2}(x) \nn
             & & \mbox{}
                 +  \sqrt{2}\;\t \tilde{H}_{2}^{(2)}(x)
                 +  \f{i}{\sqrt{2}}\;\t\t\;
                     \tb\bar{\s}^{\mu}\P_{\mu}\tilde{H}_{2}^{(2)}(x)
                 +  \t\t\;F_{2}(x).
                 \label{Component Expansion prop 8}
\EEA
The various component fields are fully defined in the chapter
mentioned above and the quantum numbers are listed in
table~\ref{Table Superfields}.

With the results from the previous sections of this appendix together with
%eqs.~\r{Chiral Superfields prop 6a} and \r{Chiral Superfields prop 6b}
%i.e.
\BEA
   \Phi_{i}(y,\t)\Phi_{j}(y,\t) &=&
     A_{i}(y)A_{j}(y)
     +\sqrt{2}\t\LP[A_{i}(y)\psi_{j}(y)+\psi_{i}(y)A_{j}(y)\RP] \nn
     & & \mbox{}
     +\t\t\,\LP[A_{i}(y)F_{j}(y)+F_{i}(y)A_{j}(y)-\psi_{i}(y)\psi_{j}(y)\RP],
        \label{CE prop 5}
\EEA
and
\BEA
   \hspace{-0.5cm}
   \lefteqn{\Phi_{i}(y,\t)\Phi_{j}(y,\t)\Phi_{k}(y,\t)}\hspace{1.5cm}\nn
    &=&
     A_{i}(y)A_{j}(y)A_{k}(y) \nn
     & & \mbox{}
     +\sqrt{2}\t\LP[\psi_{i}(y)A_{j}(y)A_{k}(y)
                 +\psi_{j}(y)A_{k}(y)A_{i}(y)
                 +\psi_{k}(y)A_{i}(y)A_{j}(y)\RP] \nn
     & & \mbox{}
     +\t\t\LP[  F_{i}(y)A_{j}(y)A_{k}(y)
              + F_{j}(y)A_{k}(y)A_{i}(y)
              + F_{k}(y)A_{i}(y)A_{j}(y)\RP.  \nn
     & & \mbox{}\LP. \hspace{1cm}
              -\psi_{i}(y)\psi_{j}(y)A_{k}(y)
              -\psi_{j}(y)\psi_{k}(y)A_{i}(y)
              -\psi_{k}(y)\psi_{i}(y)A_{j}(y)\RP],
                      \label{CE prop 6}
\EEA
it is easy to calculate the expansion of $\L_{SUSY}$.
Note that the $\t\t$-component of eqs.~\r{CE prop 5}
and~\r{CE prop 6} is independent of basis.
We will now give the component expansions of the different terms
of eq.~\r{Component Expansion prop 1}.

\subsection{The Component Form of $\L_{Lepton}$.}
   \label{SUBSECT: The Component Form}

With eqs.~\r{LCEL prop 20}, \r{Component Expansion prop 5},
\r{Component Expansion prop 6}
and table~\ref{Table Superfields} one has
\BEA
    \L_{Lepton}
        &=& \int d^{4}\t\;\LP[\,\hat{L}^{\dagger}e^{2g\hat{V}+g'\hat{V}'}
                           \hat{L} +
                           \hat{R}^{\dagger}e^{2g\hat{V}+g'\hat{V}'}
                           \hat{R}\,\RP] \NN
         &=&
            \LP(D^{\mu}\tilde{L}\RP)^{\dagger}\LP(D_{\mu}\tilde{L}\RP)
      + \LP(D^{\mu}\tilde{R}\RP)^{\dagger}\LP(D_{\mu}\tilde{R}\RP)
      - i\;\bar{L}^{(2)}\bar{\s}^{\mu}D_{\mu}L^{(2)}
      - i\;\bar{R}^{(2)}\bar{\s}^{\mu}D_{\mu}R^{(2)}  \nn
  & & \mbox{}
      +\tilde{L}^{\dagger}\LP(gT^{a}D^{a}-\HA g'D'\RP)\tilde{L}
      +\tilde{R}^{\dagger}g'D'\tilde{R}  \nn
  & & \mbox{}
      + \sqrt{2}i\;\tilde{L}^{\dagger}\LP(gT^{a}\lambda^{a}
         -\HA g'\lambda'\RP)L^{(2)}
      - \sqrt{2}i\;\bar{L}^{(2)}\LP(gT^{a}\bar{\lambda}^{a}
          -\HA g'\bar{\lambda}'\RP)\tilde{L} \nn
  & & \mbox{}
      + \sqrt{2}i\;\tilde{R}^{\dagger}g'\lambda'R^{(2)}
      - \sqrt{2}i\;\bar{R}^{(2)}g'\bar{\lambda}'\tilde{R} \nn
  & & \mbox{}
      + F^{\dagger}_{L}F_{L}  +F^{\dagger}_{R}F_{R} + t.d.
\EEA
Here $D_{\mu}$ is the $SU(2)\times U(1)$-covariant derivative given
in complete agreement with eq.~\r{LCEL prop 18}.
Furthermore $T^{a}=\s^{a}/2$~($a=1,\ldots,3)$ and
this will be understood from now on.

\subsection{The Component Form of $\L_{Gauge}$.}

$\L_{Gauge}$ contains both an SU(2)- and an U(1)-piece.
The SU(2)-piece can be taken directly from eq.~\r{WW prop 4},
while the U(1)-piece is obtained by taking the non-Abelian limit
of the same equation.
Hence we may conclude
\BEA
  \L_{Gauge}
     &=&  \f{1}{4} \int  d^{4}\t\;
         \LP[\,
                \US{W}{a\,\a}\LS{W}{\a}^{a}
             +  \US{W}{'\,\a}\LS{W}{\a}' \,
         \RP]\d^{2}(\tb)   + h.c. \NN
     &=&
      - i\;\bar{\lambda}^{a}\bar{\s}^{\mu}\LP(
        \P_{\mu}\lambda^{a} -gf^{abc}V^{b}_{\mu}\la^{c}\RP)
      - i\;\bar{\lambda}'\bar{\s}^{\mu}\P_{\mu}\lambda' \nn
  & & \mbox{}
      - \f{1}{4}\;\LP(\;V^{a\;\mu\nu}V^{a}_{\mu\nu}
                       + V^{'\mu\nu}V'_{\mu\nu}\;\RP)
      + \HA\;\LP(\;D^{a}D^{a}+D'D'\;\RP) + t.d.
\EEA
Here $V^{a}_{\mu\nu}$ and $V'_{\mu\nu}$ are the (non-SUSY) fieldstrengths
for the SU(2)- and U(1)-gauge group respectively.

\subsection{The Component Form of $\L_{Higgs}$.}

The expansion of the kinetic terms of $\hat{H}_{1}$ and $\hat{H}_{2}$ are
obtained in a complete analogous way to what we did in the
subsect~\ref{SUBSECT: The Component Form}.

However, in order to give the full expression for $\L_{Higgs}$,
the component form of the superpotential piece has to be obtained.
This is done with eqs.~\r{CE prop 5} and \r{CE prop 6} and reads
\BEA
 \int d^{4}\t\; W\,\d^{2}(\tb)
   &=&   \int d^{4}\t\; \LP\{   \mu\; \e^{ij}\hat{H}_{1}^{i}\hat{H}_{2}^{j} +
          f\;\e^{ij}\hat{H}^{i}_{1}\hat{L}^{j}\hat{R}  \RP\} \d^{2}(\tb) \NN
   &=&
      \mu \e^{ij}\;
           \LP[\,H^{i}_{1}F_{2}^{j}
%              + H_{1}^{i\,\dagger}F_{2}^{j\,\dagger}
              + F_{1}^{i}H_{2}^{j}
%              + F_{1}^{i\,\dagger}H_{2}^{j\,\dagger}
              - \tilde{H}_{1}^{(2)\,i}\tilde{H}_{2}^{(2)\,j}\, \RP] \nn
%         - \bar{\tilde{H}_{1}^{(2)\,i}}\bar{\tilde{H}_{2}^{(2)\,j}}\;\RP] \nn
  & & \mbox{}
       + f\e^{ij}\;
           \LP[\;F_{1}^{i}\tilde{L}^{j}\tilde{R}
%              +  F_{1}^{i\,\dagger}\tilde{L}^{j\,\dagger}\tilde{R}^{\dagger}
              +  H_{1}^{i}F_{L}^{j}\tilde{R}
%           +  H_{1}^{i\,\dagger}F_{L}^{j\,\dagger}\tilde{R}^{\dagger}\RP. \nn
%  & & \mbox{} \hspace{1.8cm}
              +  H_{1}^{i}\tilde{L}^{j} F_{R} \RP.
   \nmb  \hspace{1.8cm} \LP.
%              +  H_{1}^{i\,\dagger}\tilde{L}^{j\,\dagger}F_{R}^{\dagger}
              -  \tilde{H}_{1}^{(2)\,i}L^{(2)\,j}\tilde{R}
%        -  \bar{\tilde{H}_{1}}^{(2)\,i}\bar{L}^{(2)\,j}\tilde{R}^{\dagger} \nn
%  & & \mbox{} \hspace{1.8cm} \LP.
              -  H_{1}^{i}L^{(2)\,j}R^{(2)}
%              -  H_{1}^{i\,\dagger}\bar{L}^{(2)\,j}\bar{R}^{(2)}
           -  R^{(2)}\tilde{H}_{1}^{(2)\,i}\tilde{L}^{j} \,\RP].\hspace{0.7cm}
%- \bar{R}^{(2)}\bar{\tilde{H}_{1}}^{(2)\,i}
%\tilde{L}^{j\,\dagger}\;\RP].\hspace{0.4cm}
                        \label{Komp higgs property 1}
\EEA
The corresponding expression for $W^{\dagger}$ is, of course, obtained by
hermitian conjugation.

Thus the expression for $\L_{Higgs}$ becomes
\BEA
  \L_{Higgs}
   &=&  \int d^{4}\t\;\LP[\,\hat{H}_{1}^{\dagger}e^{2g\hat{V}+g'\hat{V}'}
                           \hat{H}_{1}
                         + \hat{H}_{2}^{\dagger}e^{2g\hat{V}+g'\hat{V}'}
                           \hat{H}_{2}
               + W\,\d^{2}(\tb) + W^{\dagger}\,\d^{2}(\t)\, \RP] \NN
  &=&
    \LP(D^{\mu}H_{1}\RP)^{\dagger}\LP(D_{\mu}H_{1}\RP)
      + \LP(D^{\mu}H_{2}\RP)^{\dagger}\LP(D_{\mu}H_{2}\RP)
   \nmb
      - i\;\bar{\tilde{H}}_{1}^{(2)}\bar{\s}^{\mu}D_{\mu}\tilde{H}_{1}^{(2)}
      - i\;\bar{\tilde{H}}_{2}^{(2)}
           \bar{\s}^{\mu}D_{\mu}\tilde{H}_{2}^{(2)} \nn
  & & \mbox{}
      + H^{\dagger}_{1}\LP(gT^{a}D^{a}-\HA g'D'\RP)H_{1}
      + H^{\dagger}_{2}\LP(gT^{a}D^{a}+\HA g'D'\RP)H_{2} \nn
  & & \mbox{}
      + \sqrt{2}i\;H^{\dagger}_{1}\LP(gT^{a}\lambda^{a}-
              \HA g'\lambda'\RP) \tilde{H}_{1}^{(2)}
      - \sqrt{2}i\;\bar{\tilde{H}}_{1}^{(2)}\LP(gT^{a}\bar{\lambda}^{a}
             -\HA g'\bar{\lambda}'\RP) H_{1} \nn
  & & \mbox{}
      + \sqrt{2}i\;H^{\dagger}_{2}\LP(gT^{a}\lambda^{a}
            +\HA g'\lambda'\RP) \tilde{H}_{2}^{(2)}
      - \sqrt{2}i\;\bar{\tilde{H}}_{2}^{(2)}
          \LP(gT^{a}\bar{\lambda}^{a}+\HA g'\bar{\lambda}'\RP) H_{2} \nn
  & & \mbox{}
      + F^{\dagger}_{1}F_{1}  +F^{\dagger}_{2}F_{2} \nn
  & & \mbox{}
      + \mu \e^{ij}\;
           \LP[\;H^{i}_{1}F_{2}^{j}
              + H_{1}^{i\,\dagger}F_{2}^{j\,\dagger}
              + F_{1}^{i}H_{2}^{j}
              + F_{1}^{i\,\dagger}H_{2}^{j\,\dagger}
              - \tilde{H}_{1}^{(2)\,i}\tilde{H}_{2}^{(2)\,j}
              - \bar{\tilde{H}}_{1}^{(2)\,i}
                   \bar{\tilde{H}}_{2}^{(2)\,j}\;\RP] \nn
  & & \mbox{}
       + f\e^{ij}\;
           \LP[\;F_{1}^{i}\tilde{L}^{j}\tilde{R}
              +  F_{1}^{i\,\dagger}\tilde{L}^{j\,\dagger}\tilde{R}^{\dagger}
              +  H_{1}^{i}F_{L}^{j}\tilde{R}
              +  H_{1}^{i\,\dagger}F_{L}^{j\,\dagger}
                  \tilde{R}^{\dagger}\RP. \nn
  & & \mbox{} \hspace{1.8cm}
              +  H_{1}^{i}\tilde{L}^{j} F_{R}
              +  H_{1}^{i\,\dagger}\tilde{L}^{j\,\dagger}F_{R}^{\dagger}
              -  \tilde{H}_{1}^{(2)\,i}L^{(2)\,j}\tilde{R}
              -  \bar{\tilde{H}_{1}}^{(2)\,i}
                     \bar{L}^{(2)\,j}\tilde{R}^{\dagger} \nn
  & & \mbox{} \hspace{1.8cm} \LP.
              -  H_{1}^{i}L^{(2)\,j}R^{(2)}
              -  H_{1}^{i\,\dagger}\bar{L}^{(2)\,j}\bar{R}^{(2)}
              -  R^{(2)}\tilde{H}_{1}^{(2)\,i}\tilde{L}^{j}
              -  \bar{R}^{(2)}
            \bar{\tilde{H}_{1}}^{(2)\,i}\tilde{L}^{j\,\dagger}\;\RP]
  \nmb
     +   t.d.
\EEA

\subsection{Conclusion --- The Two-Component Form of $\L_{SUSY}$.}

By adding the results from the three previous subsections,
the expansion of $\L_{SUSY}$ is obtained.

Hence
\BEA
\L_{SUSY}
  &=&   \LP(D^{\mu}\tilde{L}\RP)^{\dagger}\LP(D_{\mu}\tilde{L}\RP)
      + \LP(D^{\mu}\tilde{R}\RP)^{\dagger}\LP(D_{\mu}\tilde{R}\RP)
%      \nn
%  & & \mbox{}
      - i\;\bar{L}^{(2)}\bar{\s}^{\mu}D_{\mu}L^{(2)}
      - i\;\bar{R}^{(2)}\bar{\s}^{\mu}D_{\mu}R^{(2)}  \nn
  & & \mbox{}
      +\tilde{L}^{\dagger}\LP(gT^{a}D^{a}-\HA g'D'\RP)\tilde{L}
      +\tilde{R}^{\dagger}g'D'\tilde{R}  \nn
  & & \mbox{}
      + \sqrt{2}i\;\tilde{L}^{\dagger}
          \LP(gT^{a}\lambda^{a}-\HA g'\lambda'\RP)L^{(2)}
      - \sqrt{2}i\;\bar{L}^{(2)}\LP(gT^{a}\bar{\lambda}^{a}
          -\HA g'\bar{\lambda}'\RP)\tilde{L} \nn
  & & \mbox{}
      + \sqrt{2}i\;\tilde{R}^{\dagger}g'\lambda'R^{(2)}
      - \sqrt{2}i\;\bar{R}^{(2)}g'\bar{\lambda}'\tilde{R} \nn
  & & \mbox{}
      + F^{\dagger}_{L}F_{L}  +F^{\dagger}_{R}F_{R} \nn
  & & \mbox{}
      - i\;\bar{\lambda}^{a}\bar{\s}^{\mu} D_{\mu}\la^{a}
      - i\;\bar{\lambda}'\bar{\s}^{\mu}D_{\mu}\lambda' \nn
  & & \mbox{}
      - \f{1}{4}\;\LP(\;V^{a\;\mu\nu}V^{a}_{\mu\nu}
                       + V^{'\mu\nu}V'_{\mu\nu}\;\RP)
      + \HA\;\LP(\;D^{a}D^{a}+D'D'\;\RP) \nn
  & & \mbox{}
      + \LP(D^{\mu}H_{1}\RP)^{\dagger}\LP(D_{\mu}H_{1}\RP)
      + \LP(D^{\mu}H_{2}\RP)^{\dagger}\LP(D_{\mu}H_{2}\RP) \nn
  & & \mbox{}
      - i\;\bar{\tilde{H}}_{1}^{(2)}\bar{\s}^{\mu}D_{\mu}\tilde{H}_{1}^{(2)}
      - i\;\bar{\tilde{H}}_{2}^{(2)}
            \bar{\s}^{\mu}D_{\mu}\tilde{H}_{2}^{(2)} \nn
  & & \mbox{}
      + H^{\dagger}_{1}\LP(gT^{a}D^{a}-\HA g'D'\RP)H_{1}
      + H^{\dagger}_{2}\LP(gT^{a}D^{a}+\HA g'D'\RP)H_{2} \nn
  & & \mbox{}
      + \sqrt{2}i\;H^{\dagger}_{1}\LP(gT^{a}\lambda^{a}
         -\HA g'\lambda'\RP) \tilde{H}_{1}^{(2)}
      - \sqrt{2}i\;\bar{\tilde{H}}_{1}^{(2)}\LP(gT^{a}\bar{\lambda}^{a}
         -\HA g'\bar{\lambda}'\RP) H_{1} \nn
  & & \mbox{}
      + \sqrt{2}i\;H^{\dagger}_{2}\LP(gT^{a}\lambda^{a}
        +\HA g'\lambda'\RP) \tilde{H}_{2}^{(2)}
      - \sqrt{2}i\;\bar{\tilde{H}}_{2}^{(2)}
          \LP(gT^{a}\bar{\lambda}^{a}+\HA g'\bar{\lambda}'\RP) H_{2} \nn
  & & \mbox{}
      + F^{\dagger}_{1}F_{1}  +F^{\dagger}_{2}F_{2} \nn
  & & \mbox{}
      + \mu \e^{ij}\;
           \LP[\;H^{i}_{1}F_{2}^{j}
              + H_{1}^{i\,\dagger}F_{2}^{j\,\dagger}
              + F_{1}^{i}H_{2}^{j}
              + F_{1}^{i\,\dagger}H_{2}^{j\,\dagger}
              - \tilde{H}_{1}^{(2)\,i}\tilde{H}_{2}^{(2)\,j}
              - \bar{\tilde{H}_{1}}^{(2)\,i}
                 \bar{\tilde{H}_{2}}^{(2)\,j}\;\RP] \nn
  & & \mbox{}
       + f\e^{ij}\;
           \LP[\;F_{1}^{i}\tilde{L}^{j}\tilde{R}
              +  F_{1}^{i\,\dagger}\tilde{L}^{j\,\dagger}\tilde{R}^{\dagger}
              +  H_{1}^{i}F_{L}^{j}\tilde{R}
              +  H_{1}^{i\,\dagger}F_{L}^{j\,\dagger}\tilde{R}^{\dagger}\RP.
                    \nn
  & & \mbox{} \hspace{1.8cm}
              +  H_{1}^{i}\tilde{L}^{j} F_{R}
              +  H_{1}^{i\,\dagger}\tilde{L}^{j\,\dagger}F_{R}^{\dagger}
              -  \tilde{H}_{1}^{(2)\,i}L^{(2)\,j}\tilde{R}
              -  \bar{\tilde{H}_{1}}^{(2)\,i}
                  \bar{L}^{(2)\,j}\tilde{R}^{\dagger} \nn
  & & \mbox{} \hspace{1.8cm} \LP.
              -  H_{1}^{i}L^{(2)\,j}R^{(2)}
              -  H_{1}^{i\,\dagger}\bar{L}^{(2)\,j}\bar{R}^{(2)}
              -  R^{(2)}\tilde{H}_{1}^{(2)\,i}\tilde{L}^{j}
              -  \bar{R}^{(2)}\bar{\tilde{H}_{1}}^{(2)\,i}
                  \tilde{L}^{j\,\dagger}\;\RP]
  \nmb
     +   t.d.
     \label{app comp L_sub_SUSY}
\EEA
and this appendix is concluded.

\cleardoublepage

\chapter{The Four Component-Form of the On-Shell Lagrangian $\L_{SUSY}$.}
  \label{App: 4 comp notation}

In this appendix the two component Lagrangian~\r{On-Shall L sub SUSY},
i.e.
\BEA
\L_{SUSY}
  &=&   \LP(D^{\mu}\tilde{L}\RP)^{\dagger}\LP(D_{\mu}\tilde{L}\RP)
      + \LP(D^{\mu}\tilde{R}\RP)^{\dagger}\LP(D_{\mu}\tilde{R}\RP)
      - i\;\bar{L}^{(2)}\bar{\s}^{\mu}D_{\mu}L^{(2)}
      - i\;\bar{R}^{(2)}\bar{\s}^{\mu}D_{\mu}R^{(2)}  \nn
  & & \mbox{}
      + \sqrt{2}i\;\tilde{L}^{\dagger}
            \LP(gT^{a}\lambda^{a}-\HA g'\lambda'\RP)L^{(2)}
      - \sqrt{2}i\;\bar{L}^{(2)}
           \LP(gT^{a}\bar{\lambda}^{a}-\HA g'\bar{\lambda}'\RP)\tilde{L} \nn
  & & \mbox{}
      + \sqrt{2}i\;\tilde{R}^{\dagger}g'\lambda'R^{(2)}
      - \sqrt{2}i\;\bar{R}^{(2)}g'\bar{\lambda}'\tilde{R} \nn
  & & \mbox{}
      - i\;\bar{\lambda}^{a}\bar{\s}^{\mu}
           \LP(\P_{\mu}\lambda^{a} - gf^{abc}V^{b}_{\mu}\la^{c}\RP)
      - i\;\bar{\lambda}'\bar{\s}^{\mu}\P_{\mu}\lambda'
      - \f{1}{4}\;\LP(\;V^{a\;\mu\nu}V^{a}_{\mu\nu}
                       + V^{'\mu\nu}V'_{\mu\nu}\;\RP) \nn
  & & \mbox{}
      + \LP(D^{\mu}H_{1}\RP)^{\dagger}\LP(D_{\mu}H_{1}\RP)
      + \LP(D^{\mu}H_{2}\RP)^{\dagger}\LP(D_{\mu}H_{2}\RP) \nn
  & & \mbox{}
      - i\;\bar{\tilde{H}}_{1}^{(2)}\bar{\s}^{\mu}D_{\mu}\tilde{H}_{1}^{(2)}
      - i\;\bar{\tilde{H}}_{2}^{(2)}
             \bar{\s}^{\mu}D_{\mu}\tilde{H}_{2}^{(2)} \nn
  & & \mbox{}
      + \sqrt{2}i\;H^{\dagger}_{1}\LP(gT^{a}\lambda^{a}
             -\HA g'\lambda'\RP) \tilde{H}_{1}^{(2)}
      - \sqrt{2}i\;\bar{\tilde{H}}_{1}^{(2)}\LP(gT^{a}\bar{\lambda}^{a}
              -\HA g'\bar{\lambda}'\RP) H_{1} \nn
  & & \mbox{}
      + \sqrt{2}i\;H^{\dagger}_{2}\LP(gT^{a}\lambda^{a}
               +\HA g'\lambda'\RP) \tilde{H}_{2}^{(2)}
      - \sqrt{2}i\;\bar{\tilde{H}}_{2}^{(2)}
            \LP(gT^{a}\bar{\lambda}^{a}+\HA g'\bar{\lambda}'\RP) H_{2} \nn
  & & \mbox{}
      -  \e^{ij}\;
           \LP[\;\mu\LP(\,
               \tilde{H}_{1}^{(2)\,i}\tilde{H}_{2}^{(2)\,j}
              + \bar{\tilde{H}}_{1}^{(2)\,i}\bar{\tilde{H}}_{2}^{(2)\,j}\,\RP)
       + f\LP(\,
                \tilde{H}^{(2)\,i}_{1} L^{(2)\,j}\tilde{R}
              +  \bar{\tilde{H}}_{1}^{(2)\,i}\bar{L}^{(2)\,j}
                   \tilde{R}^{\dagger}\,\RP) \RP. \nn
  & & \mbox{} \hspace{1.3cm} \LP.
      +  f\LP(
                H_{1}^{i}L^{(2)\,j}R^{(2)}
              +  H_{1}^{i\,\dagger}\bar{L}^{(2)\,j}\bar{R}^{(2)}
              +  R^{(2)}\tilde{H}_{1}^{(2)\,i}\tilde{L}^{j}
              +  \bar{R}^{(2)}\bar{\tilde{H}_{1}}^{(2)\,i}
                     \tilde{L}^{j\,\dagger}\,\RP)\RP] \nn
  & & \mbox{}
        -\mu^{2}\, H_{1}^{\dagger}H_{1}
        -\mu^{2}\, H_{2}^{\dagger}H_{2}
        -\mu f \LP[\, H_{2}^{\dagger}\tilde{L}\,\tilde{R}
                  +\tilde{L}^{\dagger}H_{2}\,\tilde{R}^{\dagger}\,\RP]\nn
   & & \mbox{}
     - f^{2}\LP[ \,
       \tilde{L}^{\dagger}\tilde{L}\,\tilde{R}^{\dagger}\tilde{R}
     + H_{1}^{\dagger}H_{1}\LP( \tilde{L}^{\dagger}\tilde{L}
                                + \tilde{R}^{\dagger}\tilde{R} \RP)
     - H_{1}^{\dagger}\tilde{L}\LP(H_{1}^{\dagger}\tilde{L}\RP)^{\dagger}\,\RP]
     \nmb
     -\f{g^2}{2}
         \LP(\,\tilde{L}^{\dagger}T^{a}\tilde{L}
                +H_{1}^{\dagger}T^{a}H_{1}+H_{2}^{\dagger}T^{a}H_{2}\,\RP)
         \LP(\,\tilde{L}^{\dagger}T^{a}\tilde{L}
                   +H_{1}^{\dagger}T^{a}H_{1}+H_{2}^{\dagger}T^{a}H_{2}\,\RP)
      \nmb
        - \f{g'^2}{8} \LP(\,\tilde{L}^{\dagger}\tilde{L}-2\tilde{R}^{\dagger}
              \tilde{R}+H_{1}^{\dagger}H_{1}-H_{2}^{\dagger}H_{2}\,\RP)^2
      + t.d.,
         \label{Four Component-Form prop 1}
\EEA
will be transformed into four-component notation~(i.e. to
introduce four-component spinors).

Our strategy will be as follows. First the following well known gauge boson
combinations
\BEA
   A_{\mu}(x) &=&   \cos\theta_{\mbox{w}}\, V'_{\mu}(x)
              + \sin\theta_{\mbox{w}}\,V^{3}_{\mu}(x),
              \label{Four Component-Form prop 2}  \SL
   Z_{\mu}(x)  &=&   - \sin\theta_{\mbox{w}}\,V'_{\mu}(x)
                  + \cos\theta_{\mbox{w}} \,V^{3}_{\mu}(x), \SL
   W^{\pm}_{\mu}(x) &=& \f{V^{1}_{\mu}(x) \mp i V^{2}_{\mu}(x)}{\sqrt{2}},
              \label{Four Component-Form prop 4}
\EEA
and corresponding relations for the spin-$1/2$ gauginos
\BEA
   \la_{A}(x)   &=&     \cos\theta_{\mbox{w}} \,\la'(x)
                   + \sin\theta_{\mbox{w}} \,\la^{3}(x),
                   \label{Four Component-Form prop 5}\SL
   \la_{Z}(x)   &=&   - \sin\theta_{\mbox{w}} \,\la'(x)
                   + \cos\theta_{\mbox{w}} \,\la^{3}(x), \SL
   \la^{\pm}(x) &=& \f{\la^{1}(x) \mp i \la^{2}(x)}{\sqrt{2}}
                    \label{Four Component-Form prop 7},
\EEA
will be introduced. Next, the two component spinors will be arranged in
various (four-component) Majorana- and Dirac-spinors.
As we will see, the S-QFD theory contains Photino-~($\tilde{A}$),
Zino-~($\tilde{Z}$) and two neutral
Higgsino-states~($\tilde{H}_{1}$, $\tilde{H}_{2}$)
defined in terms of two-component spinors as follows
\BEA
  \tilde{A}(x) &=& \LP( \BA{r}  -i \la_{A}(x) \\
                              i \bar{\la}_{A}(x)   \EA   \RP),
                              \label{Four Component-Form prop 8}\SL
  \tilde{Z}(x) &=& \LP( \BA{r}   -i \la_{Z}(x) \\
                               i \bar{\la}_{Z}(x)   \EA   \RP),
                               \label{Four Component-Form prop 9}\SL
  \tilde{H}_{1} &=& \LP( \BA{r}  \psi_{H_{1}}^{1} \\
                                 \bar{\psi}_{H{1}}^{1}   \EA   \RP) ,
                                 \label{Four Component-Form prop 10}\SL
  \tilde{H}_{2} &=& \LP( \BA{r}  \psi_{H_{2}}^{2} \\
                                 \bar{\psi}_{H{2}}^{2}   \EA   \RP).
                                 \label{Four Component-Form prop 11}
\EEA
These spinors are all of the Majorana type.

For the Dirac-spinors, we have the Winos~($\tilde{W}$) and
the charged Higgsinos~($\tilde{H}$) given by
\BEA
 & &  \tilde{W}(x) \;=\; \LP( \BA{r}   -i  \la^{+}(x) \\
                               i  \bar{\la}^{-}(x)   \EA   \RP),
                                \hspace{1.4cm}
  \tilde{W}^{c}(x) \;=\; \LP( \BA{r}   -i  \la^{-}(x) \\
                               i  \bar{\la}^{+}(x)   \EA   \RP),
                               \label{Four Component-Form prop 12}\\
 & &  \tilde{H}(x)  \;=\; \LP( \BA{r}  \psi_{H_{2}}^{1} \\
                              \bar{\psi}_{H_{1}}^{2}   \EA   \RP),
 \hspace{2.3cm}
  \tilde{H}^{c}(x)  \;=\; \LP( \BA{r}  \psi_{H_{1}}^{2} \\
                              \bar{\psi}_{H_{2}}^{1}   \EA   \RP)
                              \label{Four Component-Form prop 13}.
\EEA
Here the upper ``c" on $\tilde{W}^{c}$ and $\tilde{H}^{c}$ means
charge conjugation~(cf. eq.\r{Charge conjugation}).

Finally we have the leptons which
as usual are arranged in four-component Dirac-spinors
defined by
\BEA
  l &=&  \LP( \BA{r}  l^{(2)}_{L} \\
                      \bar{l}^{(2)}_{R}   \EA   \RP).
\EEA

After introducing all necessary notation, one is in position to
show attention to the main purpose of this appendix ---
the four-component formulation of the Lagrangian $\L_{SUSY}$.

The coming calculations rely heavily on the results of
subsect.~\ref{Subsect: Connection Between Two- and Four-Component Spinors.},
and in order to avoid clutter in our description, these results will be used
without any further reference.
Those readers not familiar with the connection between two- and
four-component spinors are guided to study this subsection most
carefully.

\section{Rewriting Kinetic Terms.}

\subsection{Slepton and Higgs Kinetic Terms.}

{}From eq.~\r{Four Component-Form prop 1} we see that the transcription of the
kinetic terms of sleptons and Higgses is completed once the
$SU(2)\times U(1)$-covariant derivative is written in terms of the new
field-combinations
eqs.~\r{Four Component-Form prop 2}--\r{Four Component-Form prop 4}.

Hence
\BEA
  D_{\mu} &=& \P_{\mu}+ig{\bf T}^{a}V^{a}_{\mu}+ig'\f{{\bf Y}}{2} V'_{\mu} \nn
    &=&    \P_{\mu}
        + \f{ig}{\sqrt{2}}\LP({\bf T}^{1}+i{\bf T}^{2}\RP)
             \LP(\f{V^{1}_{\mu}(x) - i V^{2}_{\mu}(x)}{\sqrt{2}}\RP)
          + \f{ig}{\sqrt{2}}\LP({\bf T}^{1}-i{\bf T}^{2}\RP)
             \LP(\f{V^{1}_{\mu}(x) + i V^{2}_{\mu}(x)}{\sqrt{2}}\RP)
    \nmb
           + i g {\bf T}^3\LP(\la_{A}\SWA+\la_{Z}\CWA\RP)
         + ig' \f{{\bf Y}}{2} \LP(\la_{A}\CWA -\la_{Z}\SWA\RP)\NN
   &=&    \P_{\mu}
          + \f{ig}{\sqrt{2}}  {\bf T}^{+} W^{+}_{\mu}
          + \f{ig}{\sqrt{2}}  {\bf T}^{-} W^{-}_{\mu} \nn
     & & \mbox{}
        + i\LP( g\sin\theta_{\mbox{w}}{\bf T}^{3}
              + g'\cos\theta_{\mbox{w}}\f{{\bf Y}}{2}\RP)A_{\mu}
         + i\LP( g\cos\theta_{\mbox{w}}{\bf T}^{3}
               - g'\sin\theta_{\mbox{w}}\f{{\bf Y}}{2}\RP)Z_{\mu}\NN
   &=&    \P_{\mu}
          + \f{ig}{\sqrt{2}}  {\bf T}^{+} W^{+}_{\mu}
          + \f{ig}{\sqrt{2}}  {\bf T}^{-} W^{-}_{\mu} \nn
     & & \mbox{}
          + i e \LP({\bf T}^{3}+\f{{\bf Y}}{2}\RP)A_{\mu}
          + \f{ig}{\cos\theta_{\mbox{w}}}
            \LP[\, {\bf T}^{3}-\LP({\bf T}^{3}+\f{{\bf Y}}{2}\RP)
                 \sin^{2}\theta_{\mbox{w}}\,\RP] Z_{\mu}\NN
   &=&    \P_{\mu}
          + \f{ig}{\sqrt{2}}  {\bf T}^{+} W^{+}_{\mu}
          + \f{ig}{\sqrt{2}}  {\bf T}^{-} W^{-}_{\mu}
          + i e {\bf Q} A_{\mu}
          + \f{ig}{\cos\theta_{\mbox{w}}}
            \LP[\, {\bf T}^{3}-{\bf Q}\sin^{2}\theta_{\mbox{w}}\,\RP] Z_{\mu},
              \label{Cov. deriv.}
\EEA
where we have used the SM-relations
\BEA
     e &=& g \sin \t_{\mbox{w}} \;\; = \;\; g'\cos \t_{\mbox{w}},
\EEA
and introduced the operators
\BEA
    {\bf T}^{\pm} &=& {\bf T}^{1} \pm i {\bf T}^{2} , \SL
    {\bf Q} &=& {\bf T}^{3} +\f{{\bf Q}}{2}.
\EEA
Here the operator ${\bf Q}$ is the charge operator, with
eigenvalues in units of the elementary charge e.

\subsection{Lepton and Higgsinos Kinetic Terms.}

After completing the rewriting of the covariant derivative in the
previous subsection, we have for the kinetic term of left-handed
leptons\footnote{Keep in mind that the covariant derivative
has $SU(2)\times U(1)$-indices, and that the neutrinos are assumed to
be completely left-handed.}
\BEA
-i\,\bar{L}^{(2)}\bar{\s}^{\mu}D_{\mu}L^{(2)}
   &=&  -i\,\bar{L}^{(2)\,i}\bar{\s}^{\mu}D_{\mu}^{ij} L^{(2)\,j},
               \hspace{1.5cm} i,j = 1,2, \nn
   &=&   -i\, \bar{\nu}^{(2)}_{l}\bar{\s}^{\mu}D_{\mu}^{11}\nu^{(2)}_{l}
         -i\, \bar{\nu}^{(2)}_{l}\bar{\s}^{\mu}D_{\mu}^{12} l^{(2)}_{L}
   \nmb
         -i\, \bar{l}^{(2)}_{L}\bar{\s}^{\mu}D_{\mu}^{21} \nu^{(2)}_{l}
         -i\, \bar{l}^{(2)}_{L}\bar{\s}^{\mu}D_{\mu}^{22} l^{(2)}_{L} \nn
%   &=&
%        -i\, \LP( \BA{cc} 0 & \bar{\nu}^{(2)}_{l} \EA \RP)
%            \LP( \BA{cc}  0               & \s^{\mu} \\
%                          \bar{\s}^{\mu}  & 0      \EA \RP)
%             D_{\mu}^{11}\LP(\BA{c} \nu^{(2)}_{l}\\  0 \EA \RP)
%   \nmb
%         -i\, \LP( \BA{cc} 0 & \bar{\nu}^{(2)}_{l} \EA \RP)
%            \LP( \BA{cc}  0               & \s^{\mu} \\
%                          \bar{\s}^{\mu}  & 0      \EA \RP)
%             D_{\mu}^{12} \LP(\BA{c} l^{(2)}_{L}\\  0 \EA \RP)
%   \nmb
%        -i\, \LP( \BA{cc} 0 & \bar{l}^{(2)}_{L} \EA \RP)
%            \LP( \BA{cc}  0               & \s^{\mu} \\
%                          \bar{\s}^{\mu}  & 0      \EA \RP)
%             D_{\mu}^{21} \LP(\BA{c} \nu^{(2)}_{l}\\  0 \EA \RP)
%   \nmb
%        -i\, \LP( \BA{cc} 0 & \bar{l}^{(2)}_{L} \EA \RP)
%            \LP( \BA{cc}  0               & \s^{\mu} \\
%                          \bar{\s}^{\mu}  & 0      \EA \RP)
%             D_{\mu}^{22} \LP(\BA{c} l^{(2)}_{L}\\  0 \EA \RP) \nn
    &=&
         -i\, \bar{\nu}_{l}\g^{\mu} D_{\mu}^{11} \nu_{l}
         -i\, \bar{\nu}_{l}\g^{\mu} D_{\mu}^{12} l_{L}
    \nmb
         -i\, \bar{l}_{L}\g^{\mu} D_{\mu}^{21} \nu_{l}
         -i\, \bar{l}_{L}\g^{\mu} D_{\mu}^{22} l_{L} \nn
    &=&
       -i \LP( \BA{cc} \bar{\nu}_{l} & \bar{l}_{L} \EA \RP)  \g^{\mu}
       \LP( \BA{cc}    D_{\mu}^{11}  &  D_{\mu}^{12} \\
                      D_{\mu}^{21}  &  D_{\mu}^{22}   \EA \RP)
       \LP( \BA{c} \nu_{l} \\ l_{L} \EA \RP)   \nn
    &=&
       -i\, \bar{L}\g^{\mu} D_{\mu} L.
       \label{Lepton and Higgsinos prop 1}
\EEA
Here $L=\LP( \BA{cc} \nu_{l} & l \EA \RP)_{L}^{T}$ is the SU(2)-doublet of
four-component Dirac-spinors, well known from the SM.

In a similar way, we can show that~($R = l_{R}$)
\BEA
  -i\,\bar{R}^{(2)}\bar{\s}^{\mu}D_{\mu}R^{(2)}
     &=& -i\,\bar{R}\bar{\s}^{\mu}D_{\mu}R,
            \label{Lepton and Higgsinos prop 2}
\EEA
for the right-handed leptons.

Furthermore, one has for the kinetic term of the
two-component Higgsino $\tilde{H}_{1}^{(2)}$
\BEA
-i\;\bar{\tilde{H}}^{(2)}_{1}\bar{\s}^{\mu}D_{\mu}\tilde{H}^{(2)}_{1}
  &=&
     -i\; \bar{\psi}^{1}_{H_{1}}\bar{\s}^{\mu}D_{\mu}^{11}\psi^{1}_{H_{1}}
     -i\; \bar{\psi}^{1}_{H_{1}}\bar{\s}^{\mu}D_{\mu}^{12}\psi^{2}_{H_{1}}
  \nmb
     -i\; \bar{\psi}^{2}_{H_{1}}\bar{\s}^{\mu}D_{\mu}^{21}\psi^{1}_{H_{1}}
     -i\; \bar{\psi}^{2}_{H_{1}}\bar{\s}^{\mu}D_{\mu}^{22}\psi^{2}_{H_{1}} \NN
  &=&
    - i\; \bar{\psi}^{1}_{H_{1}}\bar{\s}^{\mu}\P_{\mu}\psi^{1}_{H_{1}}
    + \f{g}{2\CWA}\; \bar{\psi}^{1}_{H_{1}}
            \bar{\s}^{\mu}\psi^{1}_{H_{1}}\;Z_{\mu}
  \nmb
    +\f{g}{\sqrt{2}}\; \bar{\psi}^{1}_{H_{1}}\bar{\s}^{\mu}
            \psi^{2}_{H_{1}}\;W^{+}_{\mu}
    +\f{g}{\sqrt{2}}\; \bar{\psi}^{2}_{H_{1}}\bar{\s}^{\mu}
             \psi^{1}_{H_{1}}\;W^{-}_{\mu}
  \nmb
    -i\;\bar{\psi}^{2}_{H_{1}}\bar{\s}^{\mu}\P_{\mu}\psi^{2}_{H_{1}}
    -e\;\bar{\psi}^{2}_{H_{1}}\bar{\s}^{\mu}\psi^{2}_{H_{1}}\;A_{\mu}
  \nmb
     -\f{g}{2\CWA}\LP(1-2\SWAS \RP)\bar{\psi}^{2}_{H_{1}}
             \bar{\s}^{\mu}\psi^{2}_{H_{1}}\;Z_{\mu}\NN
   &=&
      -\f{i}{2}\;\bar{\tilde{H}}_{1}\g^{\mu}\P_{\mu}\tilde{H}_{1}
      - \f{g}{4\CWA}\;\bar{\tilde{H}}_{1}\g^{\mu} \g_{5} \tilde{H}_{1}\;Z_{\mu}
   \nmb
      -\f{g}{\sqrt{2}}\;\bar{\tilde{H}}\g^{\mu}
                 P_{R}\tilde{H}_{1} \;W^{+}_{\mu}
      -\f{g}{\sqrt{2}}\;\bar{\tilde{H}}_{1}\g^{\mu}
                  P_{R}\tilde{H} \;W^{-}_{\mu}
   \nmb
      -i\; \bar{\tilde{H}}\g^{\mu} P_{R} \P_{\mu}\tilde{H}
      + e\;\bar{\tilde{H}}\g^{\mu} P_{R}\tilde{H}\;A_{\mu}
   \nmb
      +\f{g}{2\CWA}\LP(1-2\SWAS\RP)\bar{\tilde{H}}
                  \g^{\mu} P_{R}\tilde{H} \; Z_{\mu}
  %  \nmb
       + t.d.
       \label{Lepton and Higgsinos prop 3}
\EEA
Here the charge of the various (two-component) fields,
recapitulated in table~\ref{Table:  Total Component fields},
has been taken advantage of.

In a complete analogous way, we obtain for the kinetic term of
$\tilde{H}_{2}^{(2)}$
\BEA
-i\;\bar{\tilde{H}}^{(2)}_{2}\bar{\s}^{\mu}D_{\mu}\tilde{H}^{(2)}_{2}
 &=&
    -\f{i}{2}\;\bar{\tilde{H}}_{2}\g^{\mu}\P_{\mu}\tilde{H}_{2}
     + \f{g}{4\CWA}\;\bar{\tilde{H}}_{2}\g^{\mu} \g_{5} \tilde{H}_{2}\;Z_{\mu}
   \nmb
      +\f{g}{\sqrt{2}}\;\bar{\tilde{H}}\g^{\mu}
                 P_{L}\tilde{H}_{2} \;W^{+}_{\mu}
      +\f{g}{\sqrt{2}}\;\bar{\tilde{H}}_{2}\g^{\mu}
                 P_{L}\tilde{H} \;W^{-}_{\mu}
   \nmb
      -i\; \bar{\tilde{H}}\g^{\mu} P_{L} \P_{\mu}\tilde{H}
      + e\;\bar{\tilde{H}}\g^{\mu} P_{L}\tilde{H}\;A_{\mu}
   \nmb
      +\f{g}{2\CWA}\LP(1-2\SWAS\RP)\bar{\tilde{H}}
                 \g^{\mu} P_{L}\tilde{H} \; Z_{\mu}
  %  \nmb
       + t.d.
       \label{Lepton and Higgsinos prop 4}
\EEA
By adding eqs.~\r{Lepton and Higgsinos prop 3}
       and~\r{Lepton and Higgsinos prop 4},
and using eq.~\r{pl plus pr equiv one},
one may conclude
\BEA
\lefteqn{-i\;\bar{\tilde{H}}^{(2)}_{1}\bar{\s}^{\mu}D_{\mu}\tilde{H}^{(2)}_{1}
          -i\;\bar{\tilde{H}}^{(2)}_{2}\bar{\s}^{\mu}D_{\mu}
               \tilde{H}^{(2)}_{2}} \hspace{1cm}\NN
       &=&
      -i\; \bar{\tilde{H}}\g^{\mu}  \P_{\mu}\tilde{H}
      -\f{i}{2}\;\bar{\tilde{H}}_{1}\g^{\mu}\P_{\mu}\tilde{H}_{1}
      -\f{i}{2}\;\bar{\tilde{H}}_{2}\g^{\mu}\P_{\mu}\tilde{H}_{2}
   \nmb
      -\f{g}{\sqrt{2}}\LP[\,\,\LP(\bar{\tilde{H}}\g^{\mu} P_{R} \tilde{H}_{1}
                           - \bar{\tilde{H}}\g^{\mu} P_{L} \tilde{H}_{2} \RP)
                                W^{+}_{\mu}+ h.c.\RP]
      + e\;\bar{\tilde{H}}\g^{\mu} \tilde{H}\;A_{\mu}
   \nmb
      +\f{g}{2\CWA}\LP[\,\,\LP(1-2\SWAS\RP)\bar{\tilde{H}}\g^{\mu} \tilde{H}
          -\HA \LP(
           \bar{\tilde{H}}_{1}\g^{\mu}\g_{5}\tilde{H}_{1}
           -\bar{\tilde{H}}_{2}\g^{\mu}\g_{5}\tilde{H}_{2} \RP)\,\,\RP] Z_{\mu}
  \nmb
       + t.d.
       \label{Lepton and Higgsinos prop 5}
\EEA
%In the last line of the above equation, we have used
%that $\bar{\tilde{H}}_{i}\g^{\mu}\tilde{H}_{i}=0$, ($i=1,2$)
%according to eq.~\r{Haber A31} (since $\tilde{H}_{i}$ are
%Majorana spinors).

\subsection{Gaugino Kinetic Terms.}

With eqs.~\r{Component Field Expansion prop 10asa1} and
\r{Component Field Expansion prop 10asa2} we have
\BEA
   \lefteqn{-i\bar{\la}^{a}\bar{\s}^{\mu} D_{\mu}\la^{a}
             -i\bar{\la}'\bar{\s}^{\mu} D_{\mu}\la'}\hspace{1.5cm} \nn
          &=& -i\bar{\la}^{a}\bar{\s}^{\mu}\P_{\mu}\la^{a}
             -i\bar{\la}'\bar{\s}^{\mu} \P_{\mu}\la'
             +ig\,f^{abc}\bar{\la}^{a}\bar{\s}V^{b}_{\mu}\la^{c}.
               \label{RMFL prop 5}
\EEA

Using the inverse of the
transformations~\r{Four Component-Form prop 5}--\r{Four Component-Form prop 7}
yields for the two first terms of eq.~\r{RMFL prop 5}
\BEA
\lefteqn{-i\bar{\la}^{a}\bar{\s}^{\mu}\P_{\mu}\la^{a}
    -i\bar{\la}'\bar{\s}^{\mu}\P_{\mu}\la'}\hspace{1.5cm}\nn
     &=& -i\LP(\f{\bar{\la}^{-}+\bar{\la}^{+}}{\sqrt{2}}\RP)
       \bar{\s}^{\mu}\P_{\mu} \LP(\f{\la^{-}+\la^{+}}{\sqrt{2}}\RP)
     \nmb
         -i\LP(\f{\bar{\la}^{-}-\bar{\la}^{+}}{-i\sqrt{2}}\RP)
   \bar{\s}^{\mu}\P_{\mu} \LP(\f{\la^{-}-\la^{+}}{i\sqrt{2}}\RP)
     \nmb
         -i\LP(\bar{\la}_{A}\SWA+\bar{\la}_{Z}\CWA\RP)
   \bar{\s}^{\mu}\P_{\mu} \LP(\la_{A}\SWA+\la_{Z}\CWA\RP)
     \nmb
         -i\LP(\bar{\la}_{A}\CWA-\bar{\la}_{Z}\SWA\RP)
      \bar{\s}^{\mu}\P_{\mu} \LP(\la_{A}\CWA-\la_{Z}\SWA\RP) \NN
     &=& -i\,\bar{\la}^{+}\bar{\s}^{\mu}\P_{\mu}\la^{+}
         -i\,\bar{\la}^{-}\bar{\s}^{\mu}\P_{\mu}\la^{-}
         -i\,\bar{\la}_{A}\bar{\s}^{\mu}\P_{\mu}\la_{A}
         -i\,\bar{\la}_{Z}\bar{\s}^{\mu}\P_{\mu}\la_{Z}.
\EEA
For the last term of eq.~\r{RMFL prop 5} we have
\BEA
   \lefteqn{ig\, f^{abc}\bar{\la}^{a}\bar{\s}^{\mu}
           V^{b}_{\mu}\la^{c}}\hspace{1.5cm}\nn
      &=&
             ig\,f^{3ij} \bar{\la}^{3}\bar{\s}^{\mu}V^{i}_{\mu}\la^{j}
          +  ig\,f^{i3j} \bar{\la}^{i}\bar{\s}^{\mu}V^{3}_{\mu}\la^{j}
          +  ig\,f^{ij3} \bar{\la}^{i}\bar{\s}^{\mu}V^{j}_{\mu}\la^{3} \nn
      &=&
        ig\,\e^{ij}\LP[
            \bar{\la}^{3}\bar{\s}^{\mu}V^{i}_{\mu}\la^{j}
          - \bar{\la}^{i}\bar{\s}^{\mu}V^{3}_{\mu}\la^{j}
          + \bar{\la}^{i}\bar{\s}^{\mu}V^{j}_{\mu}\la^{3} \RP],
                 \hspace{0.5cm} i,j =1,2.
                 \label{hdskjhsdjkhkjs}
\EEA
Here we have used that
\BEA
    f^{ij3} &=& \e^{ij} \nonumber  ,
\EEA
where $\e^{ij}$ is the usual antisymmetric tensor defined by $\e^{12}=1$.

Now each term in square brackets of eq.~\r{hdskjhsdjkhkjs}
will be rewritten separately.
The results are:
\BEA
   \e^{ij}\, \bar{\la}^{3}\bar{\s}^{\mu}V^{i}_{\mu}\la^{j}
      &=& \bar{\la}^{3}\bar{\s}^{\mu}
             \LP( \la^{2}V^{1}_{\mu}-\la^{1}V^{2}_{\mu}\RP) \nn
      &=& i\LP(\bar{\la}_{A}\SWA+\bar{\la}_{Z}\CWA\RP)
           \bar{\s}^{\mu}\LP(\la^{+}W^{-}_{\mu}-\la^{-}W^{+}_{\mu}\RP),\NN
   -\e^{ij}\, \bar{\la}^{i}\bar{\s}^{\mu}V^{3}_{\mu}\la^{j}
      &=& \LP(\bar{\la}^{2}\bar{\s}^{\mu}
          \la^{1}-\bar{\la}^{1}\bar{\s}^{\mu}\la^{2}\RP)V^{3}_{\mu}\nn
      &=&  i \LP(\bar{\la}^{-}\bar{\s}^{\mu}\la^{-}-
      \bar{\la}^{+}\bar{\s}^{\mu}\la^{+}\RP)\LP(A_{\mu}\SWA+Z_{\mu}\CWA\RP),\NN
   \e^{ij}\, \bar{\la}^{i}\bar{\s}^{\mu}V^{j}_{\mu}\la^{3}
      &=&   \LP(\bar{\la}^{1}\bar{\s}^{\mu}V^{2}_{\mu}
           - \bar{\la}^{2}\bar{\s}^{\mu}V^{1}_{\mu}\RP)\la^{3} \nn
      &=&   i\LP(\bar{\la}^{+}\bar{\s}^{\mu}W^{+}_{\mu}
                   -\bar{\la}^{-}\bar{\s}^{\mu}W^{-}_{\mu}\RP)
                     \LP(\la_{A}\SWA+\la_{Z}\CWA\RP).\nonumber
\EEA

Hence, collecting terms yields
\BEA
   \lefteqn{ig\, f^{abc}\bar{\la}^{a}\bar{\s}^{\mu}
               V^{b}_{\mu}\la^{c}}\hspace{1.5cm}\nn
      &=&
            -g\LP(\bar{\la}_{A}\SWA+\bar{\la}_{Z}\CWA\RP)
              \bar{\s}^{\mu} \LP(\la^{+}W^{-}_{\mu}-\la^{-}W^{+}_{\mu}\RP)
      \nmb
          -g \LP(\bar{\la}^{-}\bar{\s}^{\mu}\la^{-}
         -\bar{\la}^{+}\bar{\s}^{\mu}\la^{+}\RP)\LP(A_{\mu}\SWA+Z_{\mu}\CWA\RP)
      \nmb
          -g \LP(\bar{\la}^{+}\bar{\s}^{\mu}W^{+}_{\mu}
 -\bar{\la}^{-}\bar{\s}^{\mu}W^{-}_{\mu}\RP)\LP(\la_{A}\SWA+\la_{Z}\CWA\RP)\NN
      &=&  g\CWA \LP[
                 \LP(\bar{\la}_{Z}\bar{\s}^{\mu}\la^{-}
      -\bar{\la}^{+}\bar{\s}^{\mu}\la_{Z}\RP)W^{+}_{\mu}
                -\LP(\bar{\la}_{Z}\bar{\s}^{\mu}\la^{+}
      -\bar{\la}^{-}\bar{\s}^{\mu}\la_{Z}\RP)W^{-}_{\mu}
                \RP.
     \nmb \hspace{2cm} \LP.
                + \LP(\bar{\la}^{+}\bar{\s}^{\mu}\la^{+}
      -\bar{\la}^{-}\bar{\s}^{\mu}\la^{-}\RP)Z_{\mu}\RP]
     \nmb
          + e \LP[
                  \LP(\bar{\la}_{A}\bar{\s}^{\mu}\la^{-}
       -\bar{\la}^{+}\bar{\s}^{\mu}\la_{A}\RP)W^{+}_{\mu}
               -   \LP(\bar{\la}_{A}\bar{\s}^{\mu}\la^{+}
       -\bar{\la}^{-}\bar{\s}^{\mu}\la_{A}\RP)W^{-}_{\mu}\RP.
     \nmb \hspace{0.8cm} \LP.
                +  \LP(\bar{\la}^{+}\bar{\s}^{\mu}\la^{+}
       -\bar{\la}^{-}\bar{\s}^{\mu}\la^{-}\RP)A_{\mu} \RP],
\EEA
and thus
\BEA
   \lefteqn{-i\bar{\la}^{a}\bar{\s}^{\mu} D_{\mu}\la^{a}
             -i\bar{\la}'\bar{\s}^{\mu} D_{\mu}\la'}\hspace{1.5cm} \nn
     &=& -i\,\bar{\la}^{+}\bar{\s}^{\mu}\P_{\mu}\la^{+}
         -i\,\bar{\la}^{-}\bar{\s}^{\mu}\P_{\mu}\la^{-}
         -i\,\bar{\la}_{A}\bar{\s}^{\mu}\P_{\mu}\la_{A}
         -i\,\bar{\la}_{Z}\bar{\s}^{\mu}\P_{\mu}\la_{Z}
     \nmb
         +  g\CWA \LP[
                 \LP(\bar{\la}_{Z}\bar{\s}^{\mu}\la^{-}
              -\bar{\la}^{+}\bar{\s}^{\mu}\la_{Z}\RP)W^{+}_{\mu}
                -\LP(\bar{\la}_{Z}\bar{\s}^{\mu}\la^{+}
                  -\bar{\la}^{-}\bar{\s}^{\mu}\la_{Z}\RP)W^{-}_{\mu}
                \RP.
     \nmb \hspace{2cm} \LP.
                + \LP(\bar{\la}^{+}\bar{\s}^{\mu}\la^{+}
                  -\bar{\la}^{-}\bar{\s}^{\mu}\la^{-}\RP)Z_{\mu}\RP]
     \nmb
          + e \LP[
                  \LP(\bar{\la}_{A}\bar{\s}^{\mu}\la^{-}
            -\bar{\la}^{+}\bar{\s}^{\mu}\la_{A}\RP)W^{+}_{\mu}
                 - \LP(\bar{\la}_{A}\bar{\s}^{\mu}\la^{+}
                -\bar{\la}^{-}\bar{\s}^{\mu}\la_{A}\RP)W^{-}_{\mu}\RP.
     \nmb \hspace{0.8cm} \LP.
                +  \LP(\bar{\la}^{+}\bar{\s}^{\mu}\la^{+}
               -\bar{\la}^{-}\bar{\s}^{\mu}\la^{-}\RP)A_{\mu} \RP].
                  \label{ingve er en proff}
\EEA

With eqs.~\r{Four Component-Form prop 8}, \r{Four Component-Form prop 9},
\r{Four Component-Form prop 12} and~\r{Spinor Relations prop 16},
the four-component form of~\r{ingve er en proff}  is easily obtained,
and it reads
\BEA
   \lefteqn{-i\bar{\la}^{a}\bar{\s}^{\mu} D_{\mu}\la^{a}
             -i\bar{\la}'\bar{\s}^{\mu} D_{\mu}\la'}\hspace{1.5cm} \nn
     &=&
           -i\,\bar{\tilde{W}}\g^{\mu}\P_{\mu}\tilde{W}
           -\f{i}{2}\,\bar{\tilde{A}}\g^{\mu}\P_{\mu}\tilde{A}
           -\f{i}{2}\,\bar{\tilde{Z}}\g^{\mu}\P_{\mu}\tilde{Z}
    \nmb
      -g\CWA \LP[ \; \bar{\tilde{Z}}\g^{\mu}\tilde{W}\;W^{-}_{\mu}
                 +  \bar{\tilde{W}}\g^{\mu}\tilde{Z}\;W^{+}_{\mu}
                 -  \bar{\tilde{W}}\g^{\mu}\tilde{W}\;Z_{\mu}  \RP]
   \nmb
     -e \LP[ \; \bar{\tilde{A}}\g^{\mu}\tilde{W}\;W^{-}_{\mu}
             +  \bar{\tilde{W}}\g^{\mu}\tilde{A}\;W^{+}_{\mu}
             -  \bar{\tilde{W}}\g^{\mu}\tilde{W}\;A_{\mu}  \RP]
             +   t.d.
             \label{GKT prop 3a}
\EEA

\subsection{Gauge-Boson Kinetic Terms.}

By introducing the practical ``scripted" quantities
\BEA
   {\cal A}_{\mu\nu}   &=& \CWA V'_{\mu\nu}+ \SWA V^{3}_{\mu\nu} ,\SL
   {\cal Z}_{\mu\nu}   &=& -\SWA V'_{\mu\nu}+ \CWA V^{3}_{\mu\nu} ,\SL
   {\cal W}^{\pm}_{\mu\nu}   &=&
       \f{V^{1}_{\mu\nu} \mp i V^{2}_{\mu\nu}}{\sqrt{2}},
\EEA
defined in complete analogy with the
eqs.~\r{Four Component-Form prop 2}--\r{Four Component-Form prop 4},
the kinetic terms of the gauge-bosons can be rewritten in a compact form
as we will see in a moment.
However, first the explicit form of these ``scripted" fieldstrengths will be
derived.
Hence
\BEA
  {\cal A}_{\mu\nu}
     &=&    V'_{\mu\nu}\CWA +V^{3}_{\mu\nu}\SWA \nn
     &=& \P_{\mu} A_{\nu}- \P_{\nu}A_{\mu}
           - g\SWA\,f^{312}\LP(V^{1}_{\mu}V^{2}_{\nu}
                 -V^{2}_{\mu} V^{1}_{\nu}\RP)\nn
     &=& A_{\mu\nu}
         + i e\LP(W^{+}_{\mu}W^{-}_{\nu}- W^{-}_{\mu}W^{+}_{\nu}\RP),\SL
           \label{KIN A}
  {\cal Z}_{\mu\nu}
     &=&  - V'_{\mu\nu}\SWA +V^{3}_{\mu\nu}\CWA \nn
     &=&  \P_{\mu}Z_{\nu}- \P_{\nu}Z_{\mu}
             - g\CWA\,f^{312}\LP(V^{1}_{\mu}V^{2}_{\nu}
            -V^{2}_{\mu} V^{1}_{\nu}\RP)\nn
     &=&  Z_{\mu\nu}
            + ig\CWA\LP(W^{+}_{\mu}W^{-}_{\nu}- W^{-}_{\mu}W^{+}_{\nu}\RP),
               \label{KIN Z}
\EEA
with
\BEA
   A_{\mu\nu} &=& \P_{\mu} A_{\nu}- \P_{\nu}A_{\mu}, \nn
   Z_{\mu\nu} &=& \P_{\mu}Z_{\nu}- \P_{\nu}Z_{\mu}. \nonumber
\EEA

Furthermore
\BEA
  {\cal W}^{\pm}_{\mu\nu}
     &=& \f{V^{1}_{\mu\nu}\mp i V^{2}_{\mu\nu}}{\sqrt{2}}   \NN
     &=& \P_{\mu}\LP( \f{V^{1}_{\nu} \mp i V^{2}_{\nu}}{\sqrt{2}} \RP)
         -\P_{\nu} \LP( \f{V^{1}_{\mu} \mp i V^{2}_{\mu}}{\sqrt{2}} \RP)
     \nmb
         -\f{g}{\sqrt{2}} \LP\{
              f^{123}\LP( V^{2}_{\mu}V^{3}_{\nu} -V^{3}_{\mu}V^{2}_{\nu} \RP)
            \mp i f^{231}\LP( V^{3}_{\mu}V^{1}_{\nu}
                -V^{1}_{\mu}V^{3}_{\nu} \RP) \RP\} \NN
     &=& \P_{\mu} W^{\pm}_{\nu} -\P_{\nu} W^{\pm}_{\mu}
      \nmb
         -\f{ig}{2}\LP[
            \LP(W^{+}_{\mu}-W^{-}_{\mu}\RP)\LP(A_{\nu}\SWA+Z_{\nu}\CWA\RP) \RP.
      \nmb \hspace{1.3cm} \LP.
           -\LP(A_{\mu}\SWA+Z_{\mu}\CWA\RP)\LP(W^{+}_{\nu}-W^{-}_{\nu}\RP) \RP.
      \nmb \hspace{1.3cm}
        \mp \LP\{
            \LP(A_{\mu}\SWA+Z_{\mu}\CWA\RP)\LP(W^{+}_{\nu}+W^{-}_{\nu}\RP) \RP.
      \nmb \hspace{1.3cm} \LP.\LP.
           -\LP(W^{+}_{\mu}+W^{-}_{\mu}\RP)
                 \LP(A_{\nu}\SWA+Z_{\nu}\CWA\RP) \RP\} \RP]\NN
     &=& \P_{\mu} W^{\pm}_{\nu} -\P_{\nu} W^{\pm}_{\mu}
      \nmb
        -\f{ig}{2} \LP[
             W^{+}_{\mu}\LP\{\LP(A_{\nu}\pm
                A_{\nu}\RP)\SWA +\LP(Z_{\nu}\pm Z_{\nu}\RP)\CWA\RP\}\RP.
       \nmb  \hspace{1.3cm}
            -W^{-}_{\mu}\LP\{\LP(A_{\nu}\mp
               A_{\nu}\RP)\SWA +\LP(Z_{\nu}\mp Z_{\nu}\RP)\CWA\RP\}
       \nmb  \hspace{1.3cm}
            -\LP\{\LP(A_{\mu}\pm A_{\mu}\RP)\SWA +
                     \LP(Z_{\mu}\pm Z_{\mu}\RP)\CWA\RP\}W^{+}_{\nu}
       \nmb  \hspace{1.3cm}\LP.
            +\LP\{\LP(A_{\mu}\mp A_{\mu}\RP)\SWA +
                \LP(Z_{\mu}\mp Z_{\mu}\RP)\CWA\RP\}W^{-}_{\nu}\RP],
               \label{jhjgadmdfkj}
\EEA
where
\BEA
   W^{\pm}_{\mu\nu} &=& \P_{\mu}W^{\pm}_{\nu}-\P_{\nu}W^{\pm}_{\mu},
       \label{KIN W1}
\EEA
is the ``normal" fieldstrength of the W-bosons.

Writing eq.~\r{jhjgadmdfkj} out in full yields~($e=g\SWA$)
\BEA
  {\cal W}^{+}_{\mu\nu}
     &=&  W^{+}_{\mu\nu}
           + i\,e\LP(A_{\mu}W^{+}_{\nu}-W^{+}_{\mu}A_{\nu}\RP)
      \nmb
           + i\,g\CWA \LP(Z_{\mu}W^{+}_{\nu}-W^{+}_{\mu}Z_{\nu} \RP),
             \label{KIN W2}
\EEA
and
\BEA
  {\cal W}^{-}_{\mu\nu}
     &=&  W^{-}_{\mu\nu}
           - i\,e\LP(A_{\mu}W^{-}_{\nu}-W^{-}_{\mu}A_{\nu}\RP)
      \nmb
           - i\,g\CWA \LP(Z_{\mu}W^{-}_{\nu}-W^{-}_{\mu}Z_{\nu} \RP).
\EEA
Note that ${\cal W}^{\pm}_{\mu\nu}$ contains
neither $W^{\mp}_{\mu}$ nor $W^{\mp}_{\nu}$ (reversed signs) as
we may have guessed in advance.

With the above relations established, we have for the kinetic terms of
gauge-bosons
\BEA
   \lefteqn{ -\f{1}{4}\LP( V^{a\,\mu\nu}V^{a}_{\mu\nu}
            +V'^{\mu\nu}V'_{\mu\nu}\RP)}\hspace{1.5cm}\nn
        &=&
            -\f{1}{4}\,{\cal W}^{+\,\mu\nu}{\cal W}^{-}_{\mu\nu}
            -\f{1}{4}\,{\cal W}^{-\,\mu\nu}{\cal W}^{+}_{\mu\nu}
            -\f{1}{4}\,{\cal Z}^{\mu\nu}{\cal Z}_{\mu\nu}
            -\f{1}{4}\,{\cal A}^{\mu\nu}{\cal A}_{\mu\nu}.
            \label{GBKP prop 1}
\EEA

This concludes this subsection.

\section{Rewriting Interaction terms.}

In this section the various interaction terms of
eq.~\r{Four Component-Form prop 1}
will be rewritten.

\subsection{Rewriting Interaction Terms Containing Gauginos.}

Before proceeding, a useful general calculation will be performed.
{}From the no-shall Lagrangian~\r{On-Shall L sub SUSY},
or equivalently from eq.~\r{LCEL prop 20}, we see that the transcription of
the matter field Lagrangian is completed once the expression (adopting
the general notation of sect.~\ref{sect. left})
\BEA
  \sqrt{2}i\;A^{\dagger}\LP[\;gT^{a}\lambda^{a} +\HA g'Y\lambda'\;\RP]\psi
         - \sqrt{2}i\;\bar{\psi}\LP[\;gT^{a}\bar{\lambda}^{a}
                +\HA g'Y\bar{\lambda}'\;\RP] A,
         \label{RMFL prop 3}
\EEA
is rewritten.
The first term of eq.~\r{RMFL prop 3},
in square brackets, can in analogy with the covariant derivative,
be written as
\BEA
  \lefteqn{gT^{a}\lambda^{a} +\HA g'Y\lambda'}\hspace{1.5cm} \nn
    &=&
    \f{g}{\sqrt{2}}\LP(T^{+}\la^{+}+T^{-}\la^{-}\RP)
          + e Q\,\la_{A}
         +\f{g}{\CWA} \LP[T^{3}-Q\SWAS\RP]\la_{Z}.
\EEA
Here  $T^{3}$ and Q are the representations of
${\bf T}^{3}$ and ${\bf Q}$ respectively.

By hermitian conjugation, one obtains for eq.~\r{RMFL prop 3}
\BEA
   \lefteqn{\sqrt{2}i\;A^{\dagger}\LP[\;gT^{a}\lambda^{a}
              +\HA g'Y\lambda'\;\RP]\psi
         - \sqrt{2}i\;\bar{\psi}\LP[\;gT^{a}\bar{\lambda}^{a}
                 +\HA g'Y\bar{\lambda}'\;\RP] A}\hspace{1.5cm} \nn
     &=&   ig \LP(A^\dagger T^+ \psi\;\la^+ -\bar{\la}^+\;\bar{\psi} T^- A \RP)
         + ig \LP(A^\dagger T^- \psi\;\la^- -\bar{\la}^-\;\bar{\psi} T^+ A \RP)
     \nmb
         +\sqrt{2}ie \LP(A^\dagger Q \psi\;\la_{A}-\bar{\la}_{A}\;\bar{\psi}
                      Q A \RP)
     \nmb
         +\f{\sqrt{2}ig}{\CWA}\LP(A^\dagger \LP[T^{3}-Q\SWAS\RP]
          \psi\;\la_{Z}-\bar{\la}_{Z}\;\bar{\psi} \LP[T^{3}-Q\SWAS\RP] A\RP)\NN
     &=&
           ig \LP(A^\dagger T^+ \psi\;\la^+ -\bar{\la}^+\;\bar{\psi} T^- A \RP)
         + ig \LP(A^\dagger T^- \psi\;\la^- -\bar{\la}^-\;\bar{\psi} T^+ A \RP)
     \nmb
         +\sqrt{2}ie Q_i \LP(A^{\dagger\,i}
           \psi^{i}\;\la_{A}-\bar{\la}_{A}\;\bar{\psi}^{i}  A^{i} \RP)
     \nmb
        + \f{\sqrt{2}ig}{\CWA}\LP[{\cal T}^{3}_{i}-Q_{i}
              \SWAS\RP]\LP(A^{\dagger\,i}  \psi^{i}\;\la_{Z}-
              \bar{\la}_{Z}\;\bar{\psi}^{i} A^{i}\RP)
            ,\hspace{0.5cm} i=1,2.
         \label{RMFL prop 4}
\EEA
Here ${\cal T}^{3}_{i}$ and $Q_{i}$ are the eigenvalues of $T^{3}$ and Q
respectively.

To introducing the new two-component spinors~($\la^{\pm},
 \la_{\mbox{A}}, \la_{\mbox{Z}}$)
in the various
interaction terms is thus straightforward in  view of
the general expression~\r{RMFL prop 4}.
Hence we have
\BEA
\lefteqn{ \sqrt{2}i\;\tilde{L}^{\dagger}
\LP(gT^{a}\lambda^{a}-\HA g'\lambda'\RP)L^{(2)}
      - \sqrt{2}i\;\bar{L}^{(2)}\LP(gT^{a}\bar{\lambda}^{a}
     -\HA g'\bar{\lambda}'\RP)\tilde{L}}\hspace{1cm} \nn
      &=&
         ig\LP( \tilde{L}^{\dagger\;1}\;L^{(2)\;2}\la^{+}
               - \bar{\la}^{+}\bar{L}^{(2)\;2}\;\tilde{L}^{1}\RP)
       + ig\LP( \tilde{L}^{\dagger\;2}\;L^{(2)\;1}\la^{-}
               - \bar{\la}^{-}\bar{L}^{(2)\;1}\;\tilde{L}^{2}\RP)
     \nmb
       - \sqrt{2}ie\LP( \tilde{L}^{\dagger\;2}\;L^{(2)\;2}\la_{A}
               - \bar{\la}_{A}\bar{L}^{(2)\;2}\;\tilde{L}^{2} \RP)
     \nmb
       +\f{\sqrt{2}ig}{\CWA} \LP({\cal T}^{3}_{i} -Q_{i}\SWAS\RP)
          \LP(  \tilde{L}^{\dagger\;i}\;L^{(2)\;i}\la_{Z}
               - \bar{\la}_{Z}\bar{L}^{(2)\;i}\;\tilde{L}^{i} \RP)\NN
     &=&
        -g \LP(\tilde{L}^{\dagger\;1}\;\bar{\tilde{W}}^{c}P_{L}L^{2}
               + \bar{L}^{2}P_{R}\tilde{W}^{c}\;\tilde{L}^{1}\RP)
        -g \LP(\tilde{L}^{\dagger\;2}\;\bar{\tilde{W}}P_{L}L^{1}
               + \bar{L}^{1}P_{R}\tilde{W}\;\tilde{L}^{2}\RP)
     \nmb
        +\sqrt{2}e \LP(\tilde{L}^{\dagger\;2}\;\bar{\tilde{A}}P_{L}L^{2}
               + \bar{L}^{2}P_{R}\tilde{A}\;\tilde{L}^{2}\RP)
      \nmb
        -\f{\sqrt{2}g}{\CWA} \LP( {\cal T}^{3}_{i}-Q_{i}\SWAS \RP)
             \LP(\tilde{L}^{\dagger\;i}\;\bar{\tilde{Z}}P_{L}L^{i}
               + \bar{L}^{i}P_{R}\tilde{Z}\;\tilde{L}^{i}\RP) \NN
      &=&
        -g \LP( \bar{L}^{2}\tilde{W}^{c}\;\tilde{L}^{1}
              + \bar{L}^{1}\tilde{W}\;\tilde{L}^{2} \RP)
        +\sqrt{2}e\;\bar{L}^{2}\tilde{A}\;\tilde{L}^{2}
       \nmb
        -\f{\sqrt{2}g}{\CWA} \LP( {\cal T}^{3}_{i}-Q_{i}\SWAS \RP)
                \bar{L}^{i}\tilde{Z}\;\tilde{L}^{i}
        + h.c.
           \label{RIT prop 2}
\EEA
Here in the last line we have utilized that $P_{L}L = L$.

The corresponding term for the right-handed leptons
is rewritten as follows
\BEA
\lefteqn{\sqrt{2}i\;\tilde{R}^{\dagger}g'\lambda'R^{(2)}
      - \sqrt{2}i\;\bar{R}^{(2)}g'\bar{\lambda}'\tilde{R}}\hspace{1.5cm} \NN
   &=&
      \sqrt{2}ig'\;\tilde{R}^{\dagger}
            \LP(\la_{A}\CWA-\la_{Z}\SWA\RP)R^{(2)}
   \nmb
      - \sqrt{2}ig'\;\bar{R}^{(2)}
             \LP(\bar{\la}_{A}\CWA-\bar{\la}_{Z}\SWA\RP)\tilde{R}\NN
   &=&
      \sqrt{2}i e \LP(\tilde{R}^{\dagger}
             \;R^{(2)}\la_{A}-\bar{\la}_{A}\bar{R}^{(2)}\;\tilde{R}\RP)
   \nmb
      -\sqrt{2}i g \f{\SWAS}{\CWA}
            \LP(\tilde{R}^{\dagger}\;R^{(2)}
         \la_{Z}-\bar{\la}_{Z}\bar{R}^{(2)}\;\tilde{R}\RP)   \NN
   &=&
      -\sqrt{2}e \LP(\tilde{R}^{\dagger}\;\bar{R}P_{L}\tilde{A}+
           \bar{\tilde{A}}P_{R}R\;\tilde{R} \RP)
   \nmb
       +\sqrt{2}g \f{\SWAS}{\CWA}
            \LP(\tilde{R}^{\dagger}\;\bar{R}P_{L}\tilde{Z}+
              \bar{\tilde{Z}}P_{R}R\;\tilde{R} \RP)  \NN
    &=&
       -\sqrt{2} e\; \bar{\tilde{A}}R\;\tilde{R}
       +\sqrt{2}g \f{\SWAS}{\CWA}  \;
          \bar{\tilde{Z}}R\;\tilde{R} + h.c.
       \label{RIT prop 3}
\EEA
Here we have used that $\hat{R}$ is a right-handed gauge-singlet
(and thus also the component fields),
and that $g'= g\tan{\t{\mbox{w}}}$.

Hence, adding eqs.~\r{RIT prop 2} and \r{RIT prop 3} yields
\BEA
\lefteqn{\hspace{-0.8cm} \sqrt{2}i\;\tilde{L}^{\dagger}
 \LP(gT^{a}\lambda^{a}-\HA g'\lambda'\RP)L^{(2)}
      - \sqrt{2}i\;\bar{L}^{(2)}\LP(gT^{a}\bar{\lambda}^{a}
        -\HA g'\bar{\lambda}'\RP)\tilde{L}}\nn
\lefteqn{\hspace{-0.8cm}+\sqrt{2}i\;\tilde{R}^{\dagger}g'\lambda'R^{(2)}
      - \sqrt{2}i\;\bar{R}^{(2)}g'\bar{\lambda}'\tilde{R}} \NN
   &=&
            -g\LP[\LP\{\bar{L}^{1}\tilde{W}\;\tilde{L}^{2}
               +  \bar{L}^{2}\tilde{W}^{c}\;\tilde{L}^{1}  \RP\}+h.c.\RP]
      +\sqrt{2} e
        \LP[\LP\{ \bar{L}^{2}\tilde{A}\;\tilde{L}^{2}
                - \bar{\tilde{A}}R \;\tilde{R} \RP\} + h.c. \RP]
   \nmb
         -\f{\sqrt{2}g}{\cos \theta_{\mbox{w}}}
           \LP[\LP\{
              \LP({\cal T}^{3}_{i}- Q_{i}\sin^{2}\theta_{\mbox{w}}\RP)
                    \bar{L}^{i}\tilde{Z}\;\tilde{L}^{i}
                  - \sin^{2}\theta_{\mbox{w}}\,
                     \bar{\tilde{Z}}R\;\tilde{R} \RP\}
                  + h.c.\RP].
            \label{RIT prop 4}
\EEA

With eq.~\r{RMFL prop 4} and the fact that ${\bf Q}H_{1}
     = \LP( \BA{cc} 0 & - H_{1}^{2} \EA\RP)^{T}$
we have
\BEA
\lefteqn{\sqrt{2}i\;H^{\dagger}_{1}\LP(gT^{a}\lambda^{a}
        -\HA g'\lambda'\RP) \tilde{H}_{1}^{(2)}
      - \sqrt{2}i\;\bar{\tilde{H}}_{1}^{(2)}
   \LP(gT^{a}\bar{\lambda}^{a}-\HA g'\bar{\lambda}'\RP) H_{1}} \hspace{1cm} \NN
   &=&
      ig\LP( H^{1\;\dagger}_{1}\;\psi^{2}_{H_{1}}\la^{+}
         - \bar{\la}^{+}\bar{\psi}^{2}_{H_{1}}\;H^{1}_{1}\RP)
     + ig\LP( H^{2\;\dagger}_{1}\;\psi^{1}_{H_{1}}\la^{-}
         - \bar{\la}^{-}\bar{\psi}^{1}_{H_{1}}\;H^{2}_{1}\RP)
   \nmb
     -\sqrt{2}i e
         \LP( H^{2\;\dagger}_{1}\;\psi^{2}_{H_{1}}\la_{A}
           - \bar{\la}_{A}\bar{\psi}^{2}_{H_{1}}\;H^{2}_{1}\RP)
   \nmb
     +\f{ig}{\sqrt{2}\CWA}
         \LP( H^{1\;\dagger}_{1}\;\psi^{1}_{H_{1}}\la_{Z}
                - \bar{\la}_{Z}\bar{\psi}^{1}_{H_{1}}\;H^{1}_{1}\RP)
   \nmb
     -\f{ig}{\sqrt{2}\CWA} \LP(1-2\SWAS\RP)
         \LP( H^{2\;\dagger}_{1}\;\psi^{2}_{H_{1}}\la_{Z}
               - \bar{\la}_{Z}\bar{\psi}^{2}_{H_{1}}\;H^{2}_{1}\RP) \NN
   &=&
       -g \LP( H^{1\;\dagger}_{1}\;\bar{\tilde{H}}P_{L}\tilde{W}
                     +\bar{\tilde{W}}P_{R}\tilde{H}\;H^{1}_{1}\RP)
       -g \LP( H^{2\;\dagger}_{1}\;\bar{\tilde{W}}P_{L}\tilde{H}_{1}
                     + \bar{\tilde{H}}_{1}P_{R}\tilde{W}\;H^{2}_{1}\RP)
   \nmb
      + \sqrt{2} e \LP( H^{2\;\dagger}_{1}\;\bar{\tilde{H}}P_{L}\tilde{A}
                     +\bar{\tilde{A}}P_{R}\tilde{H}\;H^{2}_{1}\RP)
   \nmb
      -\f{g}{\sqrt{2}\CWA}\LP( H^{1\;\dagger}_{1}\;\bar{\tilde{H}}_{1}P_{L}
         \tilde{Z}+\bar{\tilde{Z}}P_{R}\tilde{H}_{1}\;H^{1}_{1}\RP)
   \nmb
      +\f{g}{\sqrt{2}\CWA}\LP(1-2\SWAS\RP)\LP(H^{2\;\dagger}_{1}
               \;\bar{\tilde{H}}P_{L}\tilde{Z}+
                \bar{\tilde{Z}}P_{R}\tilde{H}\;H^{2}_{1}\RP) \NN
   &=&
      -g \LP(\bar{\tilde{W}}P_{R}\tilde{H}\;H^{1}_{1}
       + \bar{\tilde{H}}_{1}P_{R}\tilde{W}\;H^{2}_{1}\RP)
      + \sqrt{2} e \;\bar{\tilde{A}}P_{R}\tilde{H}\;H^{2}_{1}
   \nmb
      -\f{g}{\sqrt{2}\CWA}\;\bar{\tilde{Z}}P_{R}\tilde{H}_{1}\;H^{1}_{1}
      +\f{g}{\sqrt{2}\CWA}\LP(1-2\SWAS\RP)
                \bar{\tilde{Z}}P_{R}\tilde{H}\;H^{2}_{1}
   \nmb
      + h.c.
        \label{RIT prop 5}
\EEA

A corresponding calculation for the $H_{2}$-term yields
\BEA
\lefteqn{\sqrt{2}i\;H^{\dagger}_{2}
          \LP(gT^{a}\lambda^{a}+\HA g'\lambda'\RP) \tilde{H}_{2}^{(2)}
      - \sqrt{2}i\;\bar{\tilde{H}}_{2}^{(2)}
    \LP(gT^{a}\bar{\lambda}^{a}+\HA g'\bar{\lambda}'\RP) H_{2}}\hspace{1cm} \NN
   &=&
       -g \LP(\bar{\tilde{W}}P_{R}\tilde{H}_{2}\;H^{1}_{2}
          + \bar{\tilde{H}}P_{R}\tilde{W}\;H^{2}_{2}\RP)
      - \sqrt{2} e \;\bar{\tilde{H}}P_{R}\tilde{A}\;H^{1}_{2}
   \nmb
      -\f{g}{\sqrt{2}\CWA}\LP(1-2\SWAS\RP)
            \bar{\tilde{H}}P_{R}\tilde{Z}\;H^{1}_{2}
      +\f{g}{\sqrt{2}\CWA} \bar{\tilde{H}}_{2}P_{R}\tilde{Z}\;H^{2}_{2}
   \nmb
      + h.c.
              \label{RIT prop 6}
\EEA
Here we have used that for Majorana spinors
$\bar{\Psi}_1\Psi_2 = \bar{\Psi}_2\Psi_1$
and $\bar{\Psi}_1\g_{5}\Psi_2 = \bar{\Psi}_2\g_5\Psi_1$~(cf.
eqs.~\r{URBTFCS prop 0a} and \r{URBTFCS prop 0c})

Adding eqs.~\r{RIT prop 5} and \r{RIT prop 6} yields
\BEA
\lefteqn{\hspace{-0.8cm}\sqrt{2}i\;H^{\dagger}_{1}
      \LP(gT^{a}\lambda^{a}-\HA g'\lambda'\RP) \tilde{H}_{1}^{(2)}
      - \sqrt{2}i\;\bar{\tilde{H}}_{1}^{(2)}\LP(gT^{a}\bar{\lambda}^{a}
             -\HA g'\bar{\lambda}'\RP) H_{1}}\nn
\lefteqn{\hspace{-0.8cm}+\sqrt{2}i\;H^{\dagger}_{2}
        \LP(gT^{a}\lambda^{a}+\HA g'\lambda'\RP) \tilde{H}_{2}^{(2)}
      - \sqrt{2}i\;\bar{\tilde{H}}_{2}^{(2)}\LP(gT^{a}\bar{\lambda}^{a}
             +\HA g'\bar{\lambda}'\RP) H_{2}}\NN
    &=&
      -g \LP[\, \LP(\bar{\tilde{W}}P_{R}\tilde{H}\;H^{1}_{1}
                   + \bar{\tilde{H}}P_{R}\tilde{W}\;H^{2}_{2}
                   + \bar{\tilde{H}}_{1}P_{R}\tilde{W}\;H^{2}_{1}
                   + \bar{\tilde{W}}P_{R}\tilde{H}_{2}\;H^{1}_{2}\RP)+h.c.\RP]
    \nmb
      + \sqrt{2} e \LP[\,\LP(\bar{\tilde{A}}P_{R}\tilde{H}\;H^{2}_{1}
               -  \bar{\tilde{H}}P_{R}\tilde{A}\;H^{1}_{2}\RP)+h.c. \RP]
   \nmb
      -\f{g}{\sqrt{2}\CWA}\LP[\,\LP\{
           \bar{\tilde{Z}}P_{R}\tilde{H}_{1}\;H^{1}_{1}
          - \bar{\tilde{H}}_{2}P_{R}\tilde{Z}\;H^{2}_{2} \RP.\RP.
    \nmb \hspace{2.8cm} \LP.\LP.
       - \LP(1-2\SWAS\RP)\LP(
           \bar{\tilde{Z}}P_{R}\tilde{H}\;H^{2}_{1}
         - \bar{\tilde{H}}P_{R}\tilde{Z}\;H^{1}_{2}\RP) \RP\}+ h.c. \RP]
         \label{RIT prop 7}
\EEA

This completes this subsection.

\subsection{Rewriting the Cubic Interaction Terms.}

In the previous subsection, cubic interaction terms containing gauginos
were transcripted. The aim of the present subsection is to
perform a paraphrase of the remaining cubic interaction terms
of the Lagrangian~\r{Four Component-Form prop 1}.
The calculations go like this
\BEA
\lefteqn{\hspace{-0.8cm}- f\e^{ij} \LP(\,
   \tilde{H}^{(2)\,i}_{1}L^{(2)\,j}\tilde{R}
          +  \bar{\tilde{H}_{1}}^{(2)\,i}\bar{L}^{(2)\,j}\tilde{R}^{\dagger}
           +   H_{1}^{i}L^{(2)\,j}R^{(2)} \RP. } \nn
\lefteqn{\hspace{0.7cm} \LP. + H_{1}^{i\,\dagger}\bar{L}^{(2)\,j}\bar{R}^{(2)}
         +  R^{(2)}\tilde{H}_{1}^{(2)\,i}\tilde{L}^{j}
         +  \bar{R}^{(2)}\bar{\tilde{H}_{1}}^{(2)\,i}
                   \tilde{L}^{j\,\dagger}\RP)} \NN
    &=&
       -f\LP[\; \tilde{H}^{(2)\,1}_{1} L^{(2)\,2}\,\tilde{R}
              - \tilde{H}^{(2)\,2}_{1} L^{(2)\,1}\,\tilde{R}
              + \bar{\tilde{H}}^{(2)\,1}_{1} \bar{L}^{(2)\,2}\,
                      \tilde{R}^{\dagger}
              - \bar{\tilde{H}}^{(2)\,2}_{1} \bar{L}^{(2)\,1}\,
                    \tilde{R}^{\dagger} \RP.
   \nmb \hspace{0.8cm}
              + L^{(2)\,2}R^{(2)}\,H^{1}_{1}
              - L^{(2)\,1}R^{(2)}\,H^{2}_{1}
              + \bar{L}^{(2)\,2}\bar{R}^{(2)}\,H^{1\;\dagger}_{1}
              - \bar{L}^{(2)\,1}\bar{R}^{(2)}\,H^{2\;\dagger}_{1}
   \nmb \hspace{0.8cm}  \LP.
              + R^{(2)}\tilde{H}^{(2)\,1}_{1}\,\tilde{L}^{2}
              - R^{(2)}\tilde{H}^{(2)\,2}_{1}\,\tilde{L}^{1}
              + \bar{R}^{(2)}\bar{\tilde{H}}^{(2)\,1}_{1}\,
                       \tilde{L}^{2\;\dagger}
              - \bar{R}^{(2)}\bar{\tilde{H}}^{(2)\,2}_{1}\,
                   \tilde{L}^{1\;\dagger} \; \RP] \NN
    &=&
       f \LP[\;  \psi^{2}_{H_{1}}\nu^{(2)}_{l}\,\tilde{R}
               - \psi^{1}_{H_{1}} l^{(2)}_{L}\,\tilde{R}
               + \bar{\psi}^{2}_{H_{1}} \bar{\nu}^{(2)}_{l}\,
                     \tilde{R}^{\dagger}
               - \bar{\psi}^{1}_{H_{1}} \bar{l}^{(2)}_{L}\,
                      \tilde{R}^{\dagger} \RP.
    \nmb \hspace{0.4cm}
               + \nu^{(2)}_{l}l^{(2)}_{R}\,H^{2}_{1}
               - l^{(2)}_{L}l^{(2)}_{R}\,H^{1}_{1}
               + \bar{\nu}^{(2)}_{l}\bar{l}^{(2)}_{R}\,H^{2\,\dagger}_{1}
               - \bar{l}^{(2)}_{L}\bar{l}^{(2)}_{R}\,H^{1\,\dagger}_{1}
    \nmb  \hspace{0.4cm}\LP.
               + l^{(2)}_{R}\psi^{2}_{H_{1}}\,\tilde{L}^{1}
               - l^{(2)}_{R}\psi^{1}_{H_{1}}\,\tilde{L}^{2}
               + \bar{l}^{(2)}_{R}\bar{\psi}^{2}_{H_{1}}\,
                    \tilde{L}^{1\,\dagger}
               - \bar{l}^{(2)}_{R}\bar{\psi}^{1}_{H_{1}}\,
                     \tilde{L}^{2\,\dagger} \;\RP] \NN
     &=&
          f \LP[\;
                \bar{\tilde{H}}P_{L}\nu_{l}\,\tilde{R}
              + \bar{\nu}_{l}P_{R}\tilde{H}\,\tilde{R}^{\dagger}
              - \bar{\tilde{H}}_{1}P_{L}l_{L}\,\tilde{R}
              - \bar{l}_{L}P_{R}\tilde{H}_{1}\,\tilde{R}^{\dagger} \RP.
          \nmb \hspace{0.4cm}
              + \bar{l}_{R}P_{L}\nu_{l}\,H^{2}_{1}
                            + \bar{\nu}_{l}P_{R}l_{R}\,H^{2\,\dagger}_{1}
              - \bar{l}_{R}P_{L}l_{L}\,H^{1}_{1}
              - \bar{l}_{L}P_{R}l_{R}\,H^{1\,\dagger}_{1}
          \nmb \hspace{0.4cm} \LP.
              + \bar{l}_{R}P_{L}\tilde{H}^{c}\,\tilde{L}^{1}
                            + \bar{\tilde{H}}^{c} P_{R} l_{R}\,
                  \tilde{L}^{1\,\dagger}
              - \bar{l}_{R}P_{L}\tilde{H}_{1}\,\tilde{L}^{2}
              - \bar{\tilde{H}}_{1} P_{R} l_{R}\,
                   \tilde{L}^{2\,\dagger} \; \RP] \NN
     &=&
    f \LP[\; \LP\{
         \bar{\tilde{H}}L^{1}\;\tilde{R}
       - \bar{\tilde{H}}_{1}L^{2}\;\tilde{R}
       + \bar{R}L^{1}\;H_{1}^{2}
       - \bar{R}L^{2}\;H_{1}^{1}  \RP. \RP.
    \nmb \hspace{1cm} \LP.\LP.
       + \bar{R}\tilde{H}^{c}\;\tilde{L}^{1}
       - \bar{R}\tilde{H}_{1}\;\tilde{L}^{2} \RP\} +h.c.\; \RP].
       \label{RCI prop 1}
\EEA

%          f \LP[\;
%                \bar{\tilde{H}}L^{1}\,\tilde{R}
%              + \bar{L}^{1}\tilde{H}\,\tilde{R}^{\dagger}
%              - \bar{\tilde{H}}_{1}L^{2}\,\tilde{R}
%              - \bar{L}^{2}\tilde{H}_{1}\,\tilde{R}^{\dagger}
%              + \bar{R}L^{1}\,H^{2}_{1}
%              +\bar{
%\EEA
%

\subsection{Rewriting the Higgsino Mass Terms.}

In order to complete the rewriting of the
Lagrangian~\r{Four Component-Form prop 1}, one has to
transform the terms $\mu\e^{ij}\;\tilde{H}^{(2)\,i}_{1}\tilde{H}^{(2)\,j}_{2}$,
and their hermitian conjugated,
into four-component notation.
This is done like this
\BEA
\lefteqn{  -\mu\e^{ij}\LP[\;
   \tilde{H}^{(2)\,i}_{1}\tilde{H}^{(2)\,j}_{2}
   +\bar{\tilde{H}}^{(2)\,i}_{1}\bar{\tilde{H}}^{(2)\,j}_{2}\;\RP]}
           \hspace{1.5cm} \nn
  &=&
    \mu\LP[ \psi^{2}_{H_{1}}\psi^{1}_{H_{2}}
         +  \bar{\psi}^{2}_{H_{1}}\bar{\psi}^{1}_{H_{2}}
         -  \psi^{1}_{H_{1}}\psi^{2}_{H_{2}}
         -  \bar{\psi}^{1}_{H_{1}}\bar{\psi}^{2}_{H_{2}} \RP] \nn
   &=&
        \mu\,\bar{\tilde{H}}\tilde{H}
      - \f{\mu}{2}\,\bar{\tilde{H}}_{1}\tilde{H}_{2}
      - \f{\mu}{2}\,\bar{\tilde{H}}_{2}\tilde{H}_{1},
      \label{RMT prop 1}
\EEA
and finally the rewriting procedure is completed.

\section{Summation --- The On-Shell Lagrangian.}

In the two previous sections
the transcription from two- to  four-component notation
of the various terms of the Lagrangian~\r{Four Component-Form prop 1}
was completed.
In this section we will collect the results, and
with eqs.~\r{Cov. deriv.}, \r{Lepton and Higgsinos prop 1},
\r{Lepton and Higgsinos prop 2}, \r{Lepton and Higgsinos prop 5},
\r{GKT prop 3a}, \r{GBKP prop 1}, \r{RIT prop 4}, \r{RIT prop 7},
\r{RCI prop 1} and finally eq.~\r{RMT prop 1} we obtain
\BEA
  \L_{SUSY}
  &=&   \LP(D^{\mu}\tilde{L}\RP)^{\dagger}\LP(D_{\mu}\tilde{L}\RP)
      + \LP(D^{\mu}\tilde{R}\RP)^{\dagger}\LP(D_{\mu}\tilde{R}\RP)
      - i\;\bar{L}\g^{\mu}D_{\mu}L
      - i\;\bar{R}\g^{\mu}D_{\mu}R
  \nmb
       -g\LP[\LP\{\bar{L}^{1}\tilde{W}\;\tilde{L}^{2}
       +  \bar{L}^{2}\tilde{W}^{c}\;\tilde{L}^{1}  \RP\}+h.c.\RP]
       +\sqrt{2} e
        \LP[\LP\{ \bar{L}^{2}\tilde{A}\;\tilde{L}^{2}
                - \bar{\tilde{A}}R \;\tilde{R} \RP\} + h.c. \RP]
   \nmb
         -\f{\sqrt{2}g}{\cos \theta_{\mbox{w}}}
           \LP[\LP\{
              \LP({\cal T}^{3}_{i}- Q_{i}\sin^{2}\theta_{\mbox{w}}\RP)
                    \bar{L}^{i}\tilde{Z}\;\tilde{L}^{i}
                  - \sin^{2}\theta_{\mbox{w}}\,
                      \bar{\tilde{Z}}R\;\tilde{R} \RP\}
                  + h.c.\RP]
    \nmb
           -i\,\bar{\tilde{W}}\g^{\mu}\P_{\mu}\tilde{W}
           -\f{i}{2}\,\bar{\tilde{A}}\g^{\mu}\P_{\mu}\tilde{A}
           -\f{i}{2}\,\bar{\tilde{Z}}\g^{\mu}\P_{\mu}\tilde{Z}
    \nmb
      -g\CWA \LP[ \; \bar{\tilde{Z}}\g^{\mu}\tilde{W}\;W^{-}_{\mu}
                 +  \bar{\tilde{W}}\g^{\mu}\tilde{Z}\;W^{+}_{\mu}
                 -  \bar{\tilde{W}}\g^{\mu}\tilde{W}\;Z_{\mu}  \RP]
   \nmb
     -e \LP[ \; \bar{\tilde{A}}\g^{\mu}\tilde{W}\;W^{-}_{\mu}
             +  \bar{\tilde{W}}\g^{\mu}\tilde{A}\;W^{+}_{\mu}
             -  \bar{\tilde{W}}\g^{\mu}\tilde{W}\;A_{\mu}  \RP]
   \nmb
        - \f{1}{4} {\cal W}^{+\,\mu\nu} {\cal W}^{-}_{\mu\nu}
        - \f{1}{4} {\cal W}^{-\,\mu\nu} {\cal W}^{+}_{\mu\nu}
        - \f{1}{4} {\cal Z}^{\mu\nu} {\cal Z}_{\mu\nu}
        - \f{1}{4} {\cal A}^{\mu\nu} {\cal A}_{\mu\nu}
    \nmb
      + \LP( D^{\mu}H_{1}\RP)^{\dagger}\LP(D_{\mu}H_{1}\RP)
           - \mu^{2} H_{1}^{\dagger}H_{1}
      + \LP(D^{\mu}H_{2}\RP)^{\dagger}\LP(D_{\mu}H_{2}\RP)
        -\mu^{2} H_{2}^{\dagger}H_{2}
    \nmb
      - \bar{\tilde{H}}\LP(i\g^{\mu}\P_{\mu}-\mu\RP)\tilde{H}
      - \f{i}{2}\;\bar{\tilde{H}}_{1}\g^{\mu}\P_{\mu}\tilde{H}_{1}
      - \f{i}{2}\;\bar{\tilde{H}}_{2}\g^{\mu}\P_{\mu}\tilde{H}_{2}
      - \f{\mu}{2} \bar{\tilde{H}}_{1}\tilde{H}_{2}
      - \f{\mu}{2} \bar{\tilde{H}}_{2}\tilde{H}_{1}
    \nmb
      -\f{g}{\sqrt{2}}\LP[\,\,\LP(\bar{\tilde{H}}\g^{\mu} P_{R} \tilde{H}_{1}
                    - \bar{\tilde{H}}\g^{\mu} P_{L} \tilde{H}_{2} \RP)
                         W^{+}_{\mu}+ h.c.\RP]
      + e\;\bar{\tilde{H}}\g^{\mu} \tilde{H}\;A_{\mu}
   \nmb
      +\f{g}{2\CWA}\LP[\,\,\LP(1-2\SWAS\RP)\bar{\tilde{H}}\g^{\mu} \tilde{H}
          -\HA \LP(
           \bar{\tilde{H}}_{1}\g^{\mu}\g_{5}\tilde{H}_{1}
           -\bar{\tilde{H}}_{2}\g^{\mu}\g_{5}\tilde{H}_{2} \RP)\,\,\RP] Z_{\mu}
   \nmb
           -g \LP[\, \LP(\bar{\tilde{W}}P_{R}\tilde{H}\;H^{1}_{1}
                   + \bar{\tilde{H}}P_{R}\tilde{W}\;H^{2}_{2}
                   + \bar{\tilde{H}}_{1}P_{R}\tilde{W}\;H^{2}_{1}
                   + \bar{\tilde{W}}P_{R}\tilde{H}_{2}\;H^{1}_{2}\RP)+h.c.\RP]
    \nmb
      + \sqrt{2} e \LP[\,\LP(\bar{\tilde{A}}P_{R}\tilde{H}\;H^{2}_{1}
               -  \bar{\tilde{H}}P_{R}\tilde{A}\;H^{1}_{2}\RP)+h.c. \RP]
   \nmb
      -\f{g}{\sqrt{2}\CWA}\LP[\,\LP\{
           \bar{\tilde{Z}}P_{R}\tilde{H}_{1}\;H^{1}_{1}
          - \bar{\tilde{H}}_{2}P_{R}\tilde{Z}\;H^{2}_{2} \RP.\RP.
    \nmb \hspace{2.8cm} \LP.\LP.
       - \LP(1-2\SWAS\RP)\LP(
           \bar{\tilde{Z}}P_{R}\tilde{H}\;H^{2}_{1}
         - \bar{\tilde{H}}P_{R}\tilde{Z}\;H^{1}_{2}\RP) \RP\}+ h.c. \RP]
    \nmb
     + f \! \LP[ \LP\{
         \bar{\tilde{H}}L^{1}\,\tilde{R}
       - \bar{\tilde{H}}_{1}L^{2}\,\tilde{R}
       + \bar{R}L^{1}\,H_{1}^{2}
       - \bar{R}L^{2}\,H_{1}^{1}
       + \bar{R}\tilde{H}^{c}\,\tilde{L}^{1}
       - \bar{R}\tilde{H}_{1}\,\tilde{L}^{2} \RP\}\! +h.c. \RP]
    \nmb
     - \mu f\! \LP[H_{2}^{\dagger}\tilde{L}\,\tilde{R}
             + h.c. \RP]
%    \nmb
     - f^{2} \! \LP[
           \tilde{L}^{\dagger}\tilde{L}\,\tilde{R}^{\dagger}\tilde{R}
         + H_{1}^{\dagger}H_{1}\!\LP( \tilde{L}^{\dagger}\tilde{L}
                                + \tilde{R}^{\dagger}\tilde{R} \RP)
     - H_{1}^{\dagger}\tilde{L}\!\LP(H_{1}^{\dagger}\tilde{L}\RP)^{\dagger}\RP]
    \nmb
    -\f{g^2}{2}
         \LP(\,\tilde{L}^{\dagger}
       T^{a}\tilde{L}+H_{1}^{\dagger}T^{a}H_{1}+H_{2}^{\dagger}T^{a}H_{2}\,\RP)
         \LP(\,\tilde{L}^{\dagger}
       T^{a}\tilde{L}+H_{1}^{\dagger}T^{a}H_{1}+H_{2}^{\dagger}T^{a}H_{2}\,\RP)
     \nmb
        - \f{g^2 \tan^{2} \t_{\mbox{w}} }{8}
          \LP(\,\tilde{L}^{\dagger}\tilde{L}-2\tilde{R}^{\dagger}
           \tilde{R}+H_{1}^{\dagger}H_{1}-H_{2}^{\dagger}H_{2}\,\RP)^2
      + t.d.
         \label{app L sub SUSY 4-comp}
\EEA

Here $\tilde{W}^{c}$ is the charge
conjugated (defined in eq.~\r{Charge conjugation}) of the
spinor~\r{Introducing Four-Component Spinors prop 3} and
$P_{L}$ and $P_{R}$ are the left- and right-handed projection
operators given by  eqs.~\r{left proje} and \r{right proje}, i.e.
\BEA
     P_{L} &=& \HA\LP(1-\g_{5}\RP), \SL
     P_{R} &=& \HA\LP(1+\g_{5}\RP).
\EEA

\cleardoublepage

\chapter{The Two-Component Form of the On-Shell
           Lagrangian $\L_{SUSY}$.}
           \label{APP: On-shell lagrangian}

In this appendix, starting with the off-shell
Lagrangian~\r{Off-Shall L sub SUSY}, we will construct the
corresponding (two-component) on-shell Lagrangian, i.e.
we have to eliminate the auxiliary fields.

\section{The Auxiliary Fields.}

In sect.~\ref{sect. Elimination of the Auxiliary Fields}
we obtained, by using the Euler-Lagrange equations, the following relations
for the auxiliary fields
\BEA
    F_{L}^{j\,\dagger}   &=& -f\;\e^{ij}H_{1}^{i}\tilde{R},
       \label{aux prop 1}\SL
    F_{R}^{\dagger}      &=& -f\;\e^{ij}H_{1}^{i}\tilde{L}^{j}, \SL
    F_{1}^{i\,\dagger}   &=& -\mu\;\e^{ij}H_{2}^{j}
                             - f\;\e^{ij}\tilde{L}^{j}\tilde{R},\SL
    F_{2}^{j\,\dagger}   &=& -\mu\;\e^{ij}H_{1}^{i},
        \label{aux prop 4}
\EEA
and
\BEA
  D^{a} &=& -g\LP[\,
            \tilde{L}^{\dagger}T^{a}\tilde{L}
          + H_{1}^{\dagger}T^{a}H_{1}
          + H_{2}^{\dagger}T^{a}H_{2} \RP] ,
            \label{aux prop 5} \SL
  D' &=&   \f{g'}{2}\,\tilde{L}^{\dagger}\tilde{L}
         - g'\, \tilde{R}^{\dagger}\tilde{R}
         + \f{g'}{2}\, H_{1}^{\dagger}H_{1}
         - \f{g'}{2}\, H_{2}^{\dagger}H_{2}.
            \label{aux prop 6}
\EEA
In this appendix, the detailed calculations for the back-substitution
of these relations into $\L_{Aux}$,
given by eq.~\r{Elimination of the Auxiliary Fields prop 2},
will be performed,
and we start by eliminating the auxiliary F-fields.

\subsection{Auxiliary F-fields.}

With eqs.~\r{aux prop 1}--\r{aux prop 4} we have
\BEA
    \L_{Aux-F}  &=&
          F_{L}^{\dagger}F_{L}
       +  F_{R}^{\dagger}F_{R}
       +  F_{1}^{\dagger}F_{1}
       +  F_{2}^{\dagger}F_{2} \nn
   & & \mbox{}
      + \mu\;\e^{ij}\LP[\,
            H_{1}^{i}F_{2}^{j}
         +   H_{1}^{i\,\dagger}F_{2}^{j\,\dagger}
         +   F_{1}^{i}H_{2}^{j}
         +   F_{1}^{i\,\dagger}H_{2}^{j\,\dagger}\,\RP] \nn
    & & \mbox{}
      + f\;\e^{ij}\LP[\,
            F_{1}^{i}\tilde{L}^{j}\tilde{R}
         +  F_{1}^{i\,\dagger}\tilde{L}^{j\,\dagger}\tilde{R}^{\dagger}
         +  H_{1}^{i}F_{L}^{j}\tilde{R}
         +  H_{1}^{i\,\dagger}F_{L}^{j\,\dagger}\tilde{R}^{\dagger}\RP. \nn
     & & \mbox{}  \LP. \hspace{1.2cm}
         +  H_{1}^{i}\tilde{L}^{j}F_{R}
         +  H_{1}^{i\,\dagger}\tilde{L}^{j\,\dagger}F_{R}^{\dagger} \, \RP] \NN
     &=&
          \LP(-f\e^{ij}\,H_{1}^{i}\tilde{R}\RP)
              \LP(-f\e^{kj}\,H_{1}^{k\,\dagger}\tilde{R}^{\dagger}\RP)
       +  \LP(-f\e^{ij}\,H_{1}^{i}\tilde{L}^{j}\RP)
              \LP(-f\e^{kl}\,H_{1}^{k\,\dagger}\tilde{L}^{l\,\dagger}\RP)
    \nmb
       +  \LP(-\mu\e^{ij}\,H_{2}^{j}-f\e^{ij}\tilde{L}^{j}\tilde{R}\RP)
              \LP(-\mu\e^{ik}\,H_{2}^{k\,\dagger}
             -f\e^{ik}\tilde{L}^{k\,\dagger}\tilde{R}^{\dagger}\RP)
    \nmb
       +  \LP(-\mu\e^{ij}\,H_{1}^{i}\RP)
              \LP(-\mu\e^{kj}\,H_{1}^{k\,\dagger}\RP)
     \nmb
         + \mu\;\e^{ij}\, H_{1}^{i}\LP(-\mu\e^{kj}\,H_{1}^{k\,\dagger}\RP)
         + \mu\;\e^{ij}\,  H_{1}^{i\,\dagger}\LP(-\mu\e^{kj}\,H_{1}^{k}\RP)
     \nmb
         + \mu\;\e^{ij}\,  \LP(-\mu\e^{ik}\,H_{2}^{k\,\dagger}
         -f\e^{ik}\tilde{L}^{k\,\dagger}\tilde{R}^{\dagger}\RP) H_{2}^{j}
     \nmb
         + \mu\;\e^{ij}\,  \LP(-\mu\e^{ik}\,
          H_{2}^{k}-f\e^{ik}\tilde{L}^{k}\tilde{R}\RP)H_{2}^{j\,\dagger}\,
     \nmb
         + f\;\e^{ij}\,\LP(-\mu\e^{ik}\,H_{2}^{k\,\dagger}
   -f\e^{ik}\tilde{L}^{k\,\dagger}\tilde{R}^{\dagger}\RP)\tilde{L}^{j}\tilde{R}
     \nmb
               +  f\;\e^{ij}\,\LP(-\mu\e^{ik}\,H_{2}^{k}
              -f\e^{ik}\tilde{L}^{k}\tilde{R}\RP)
                     \tilde{L}^{j\,\dagger}\tilde{R}^{\dagger}
     \nmb
         +  f\;\e^{ij}\,H_{1}^{i}\LP(-f\e^{kj}
                   \,H_{1}^{k\,\dagger}\tilde{R}^{\dagger}\RP)\tilde{R}
         +  f\;\e^{ij}\,H_{1}^{i\,\dagger}\LP(-f\e^{kj}\,H_{1}^{k}\tilde{R}\RP)
                 \tilde{R}^{\dagger}
     \nmb
         +  f\;\e^{ij}\,H_{1}^{i}\tilde{L}^{j}\LP(-f\e^{kl}\,H_{1}^{k\,\dagger}
                 \tilde{L}^{l\,\dagger}\RP)
         +  f\;\e^{ij}\,H_{1}^{i\,\dagger}\tilde{L}^{j\,\dagger}\LP(-f\e^{kl}
                          \,H_{1}^{k}\tilde{L}^{l}\RP)  \NN
   &=&
       -\mu^{2}\, H_{1}^{\dagger}H_{1}
        -\mu^{2}\, H_{2}^{\dagger}H_{2}
        -\mu f \LP[\, H_{2}^{\dagger}\tilde{L}\,\tilde{R}
                  +\tilde{L}^{\dagger}H_{2}\,\tilde{R}^{\dagger}\,\RP]\nn
   & & \mbox{}
     - f^{2}\LP[ \,
       \tilde{L}^{\dagger}\tilde{L}\,\tilde{R}^{\dagger}\tilde{R}
     + H_{1}^{\dagger}H_{1}\LP( \tilde{L}^{\dagger}\tilde{L}
                                + \tilde{R}^{\dagger}\tilde{R} \RP)
     - H_{1}^{\dagger}\tilde{L}\LP(H_{1}^{\dagger}
                   \tilde{L}\RP)^{\dagger}\,\RP].
         \label{aux prop 6aaaaaa}
\EEA
Here in the last transition the following relations have been used:
\BEA
  \e^{ij}\e^{kj} &=& \d^{ik},\nonumber \SL
  \e^{ij}\e^{kl} &=& \d^{ik}\d^{jl}-\d^{il}\d^{jk}. \nonumber
\EEA

\subsection{Auxiliary D-fields.}

When one is going to rewrite $\L_{Aux-D}$, given by
\BEA
   \L_{Aux-D} &=&
          \HA\;\LP(\;D^{a}D^{a}+D'D'\;\RP) \nn
  & & \mbox{}
      +\tilde{L}^{\dagger}\LP(gT^{a}D^{a}-\HA g'D'\RP)\tilde{L}
      +\tilde{R}^{\dagger}g'D'\tilde{R}  \nn
  & & \mbox{}
      + H^{\dagger}_{1}\LP(gT^{a}D^{a}-\HA g'D'\RP)H_{1}
      + H^{\dagger}_{2}\LP(gT^{a}D^{a}+\HA g'D'\RP)H_{2},
                            \label{aux prop 8}
\EEA
it is practical to introduce the following temporary abbreviations
\BEA
   A &=& \tilde{L}^{\dagger}T^{a}\tilde{L},\nn
   B &=& H_1^\dagger T^a H_1,\nn
   C &=& H_2^\dagger T^a H_2, \nn
   D &=& \tilde{L}^\dagger\tilde{L}, \nn
   E &=& \tilde{R}^\dagger\tilde{R},\nn
   F &=& H_1^\dagger H_1,\nn
   G &=& H_2^\dagger H_2. \nonumber
\EEA
Here the SU(2)-index ``a"  has been suppressed for convenience.

With these abbreviations eqs.~\r{aux prop 5} and \r{aux prop 6} take on
the form
\BEA
   D^a &=& -g \LP[A+B+C \RP],\nn
   D'  &=&  g'\LP[\f{D}{2}-E+\f{F}{2}-\f{G}{2} \RP].  \nonumber
\EEA
We will now rewrite each term of eq.~\r{aux prop 8}.
Hence
\BEA
  \HA \, D^a D^a
      &=& \f{g^{2}}{2} \LP( A+B+C \RP)\LP( A+B+C \RP), \NN
  \HA\,D' D'
      &=& \f{g'^{2}}{2} \LP( \f{D}{2} - E + \f{F}{2} - \f{G}{2} \RP)
          \LP( \f{D}{2} - E + \f{F}{2} - \f{G}{2} \RP), \NN
   \lefteqn{\tilde{L}^{\dagger}\LP(\,gT^aD^a-\HA g'D'\,\RP)\tilde{L}}
                     \hspace{1.5cm} \nn
      &=& -g^2 A\LP[A+B+C \RP]
          -\HA g'^2 D\LP[\f{D}{2}-E+\f{F}{2}-\f{G}{2} \RP],\NN
   \tilde{R}^{\dagger}g'D'\tilde{R}
      &=& g'^2\; E \LP(\f{D}{2}-E+\f{F}{2}-\f{G}{2}\RP),\NN
   \lefteqn{H^{\dagger}_1\LP(\,gT^aD^a-\HA g'D'\,\RP)H_1}\hspace{1.5cm} \nn
      &=& -g^2 B\LP[A+B+C \RP]
          -\HA g'^2 F\LP[\f{D}{2}-E+\f{F}{2}-\f{G}{2} \RP], \NN
    \nn
   \lefteqn{H^{\dagger}_2\LP(\,gT^aD^a+\HA g'D'\,\RP)H_2}\hspace{1.5cm} \nn
      &=& -g^2 C\LP[A+B+C \RP]
          +\HA g'^2 G\LP[\f{D}{2}-E+\f{F}{2}-\f{G}{2} \RP].
          \nonumber
\EEA
For $\L_{Aux-D}$ this implies
\BEA
  \L_{Aux-D}
     &=& - \f{g^2}{2} \LP(A+B+C\RP)\LP(A+B+C\RP)
    \nmb
        - \f{g'^2}{2}
               \LP( \f{D}{2} - E + \f{F}{2} - \f{G}{2} \RP)
                \LP(\f{D}{2} - E + \f{F}{2} - \f{G}{2} \RP),
                 \nonumber
\EEA
or in terms of the S-QFD fields
\BEA
  \L_{Aux-D}
        &=& -\f{g^2}{2}
         \LP(\,\tilde{L}^{\dagger}T^{a}
          \tilde{L}+H_{1}^{\dagger}T^{a}H_{1}+H_{2}^{\dagger}T^{a}H_{2}\,\RP)
         \LP(\,\tilde{L}^{\dagger}T^{a}\tilde{L}
             +H_{1}^{\dagger}T^{a}H_{1}+H_{2}^{\dagger}T^{a}H_{2}\,\RP)
    \nmb
        -\f{g'^2}{8}
          \LP(\,\tilde{L}^{\dagger}\tilde{L}-2\tilde{R}^{\dagger}
           \tilde{R}+H_{1}^{\dagger}H_{1}-H_{2}^{\dagger}H_{2}\,\RP)^2.
                             \label{aux prop 9}
\EEA

\subsection{Conclusion.}

{}From the two previous subsections, we can conclude that the
expression for the ``auxiliary" Lagrangian is
\BEA
  \L_{Aux} &=& \L_{Aux-F}+\L_{Aux-D}\NN
         &=&
          -\mu^{2}\, H_{1}^{\dagger}H_{1}
        -\mu^{2}\, H_{2}^{\dagger}H_{2}
        -\mu f \LP[\, H_{2}^{\dagger}\tilde{L}\,\tilde{R}
                  +\tilde{L}^{\dagger}H_{2}\,\tilde{R}^{\dagger}\,\RP]\nn
   & & \mbox{}
     - f^{2}\LP[ \,
       \tilde{L}^{\dagger}\tilde{L}\,\tilde{R}^{\dagger}\tilde{R}
     + H_{1}^{\dagger}H_{1}\LP( \tilde{L}^{\dagger}\tilde{L}
                                + \tilde{R}^{\dagger}\tilde{R} \RP)
     - H_{1}^{\dagger}\tilde{L}\LP(H_{1}^{\dagger}\tilde{L}\RP)^{\dagger}\,\RP]
   \nmb
     - \f{g^2}{2} \LP(\,\tilde{L}^{\dagger}
     T^{a}\tilde{L}+H_{1}^{\dagger}T^{a}H_{1}+H_{2}^{\dagger}T^{a}H_{2}\,\RP)
         \LP(\,\tilde{L}^{\dagger}T^{a}\tilde{L}+H_{1}^{\dagger}
               T^{a}H_{1}+H_{2}^{\dagger}T^{a}H_{2}\,\RP)
    \nmb
        -\f{g'^2}{8}
          \LP(\,\tilde{L}^{\dagger}\tilde{L}-2\tilde{R}^{\dagger}
             \tilde{R}+H_{1}^{\dagger}H_{1}-H_{2}^{\dagger}H_{2}\,\RP)^2.
             \label{L sub auks}
\EEA
This concludes this section.

\section{The On-Shell Lagrangian.}

The on-shell Lagrangian $\L_{SUSY}$ is
with the results of the previous section, easily
obtained from the corresponding
off-shell Lagrangian~\r{Off-Shall L sub SUSY} by substituting for
eq.~\r{L sub auks}.

The result is:
\BEA
\L_{SUSY}
  &=&   \LP(D^{\mu}\tilde{L}\RP)^{\dagger}\LP(D_{\mu}\tilde{L}\RP)
      + \LP(D^{\mu}\tilde{R}\RP)^{\dagger}\LP(D_{\mu}\tilde{R}\RP)
      - i\;\bar{L}^{(2)}\bar{\s}^{\mu}D_{\mu}L^{(2)}
      - i\;\bar{R}^{(2)}\bar{\s}^{\mu}D_{\mu}R^{(2)}  \nn
  & & \mbox{}
      + \sqrt{2}i\;\tilde{L}^{\dagger}
          \LP(gT^{a}\lambda^{a}-\HA g'\lambda'\RP)L^{(2)}
      - \sqrt{2}i\;\bar{L}^{(2)}\LP(gT^{a}\bar{\lambda}^{a}
               -\HA g'\bar{\lambda}'\RP)\tilde{L} \nn
  & & \mbox{}
      + \sqrt{2}i\;\tilde{R}^{\dagger}g'\lambda'R^{(2)}
      - \sqrt{2}i\;\bar{R}^{(2)}g'\bar{\lambda}'\tilde{R} \nn
  & & \mbox{}
      - i\;\bar{\lambda}^{a}\bar{\s}^{\mu} D_{\mu}\lambda^{a}
      - i\;\bar{\lambda}'\bar{\s}^{\mu} D_{\mu}\lambda'
      - \f{1}{4}\;\LP(\;V^{a\;\mu\nu}V^{a}_{\mu\nu}
                       + V^{'\mu\nu}V'_{\mu\nu}\;\RP) \nn
  & & \mbox{}
      + \LP(D^{\mu}H_{1}\RP)^{\dagger}\LP(D_{\mu}H_{1}\RP)
      + \LP(D^{\mu}H_{2}\RP)^{\dagger}\LP(D_{\mu}H_{2}\RP) \nn
  & & \mbox{}
      - i\;\bar{\tilde{H}}_{1}^{(2)}\bar{\s}^{\mu}D_{\mu}\tilde{H}_{1}^{(2)}
      - i\;\bar{\tilde{H}}_{2}^{(2)}\bar{\s}^{\mu}D_{\mu}
                      \tilde{H}_{2}^{(2)} \nn
  & & \mbox{}
      + \sqrt{2}i\;H^{\dagger}_{1}\LP(gT^{a}\lambda^{a}
               -\HA g'\lambda'\RP) \tilde{H}_{1}^{(2)}
      - \sqrt{2}i\;\bar{\tilde{H}}_{1}^{(2)}\LP(gT^{a}
               \bar{\lambda}^{a}-\HA g'\bar{\lambda}'\RP) H_{1} \nn
  & & \mbox{}
      + \sqrt{2}i\;H^{\dagger}_{2}\LP(gT^{a}\lambda^{a}
                +\HA g'\lambda'\RP) \tilde{H}_{2}^{(2)}
      - \sqrt{2}i\;\bar{\tilde{H}}_{2}^{(2)}\LP(gT^{a}\bar{\lambda}^{a}
                +\HA g'\bar{\lambda}'\RP) H_{2} \nn
  & & \mbox{}
      -  \e^{ij}\;
           \LP[\;\mu\LP(\,
               \tilde{H}_{1}^{(2)\,i}\tilde{H}_{2}^{(2)\,j}
              + \bar{\tilde{H}}_{1}^{(2)\,i}\bar{\tilde{H}}_{2}^{(2)\,j}\,\RP)
       + f\LP(\,
                \tilde{H}^{(2)\,i}_{1} L^{(2)\,j}\tilde{R}
              +  \bar{\tilde{H}}_{1}^{(2)\,i}
                     \bar{L}^{(2)\,j}\tilde{R}^{\dagger}\,\RP) \RP. \nn
  & & \mbox{} \hspace{1.3cm} \LP.
      +  f\LP(
                H_{1}^{i}L^{(2)\,j}R^{(2)}
              +  H_{1}^{i\,\dagger}\bar{L}^{(2)\,j}\bar{R}^{(2)}
              +  R^{(2)}\tilde{H}_{1}^{(2)\,i}\tilde{L}^{j}
              +  \bar{R}^{(2)}\bar{\tilde{H}_{1}}^{(2)\,i}
                     \tilde{L}^{j\,\dagger}\,\RP)\RP] \nn
  & & \mbox{}
        -\mu^{2}\, H_{1}^{\dagger}H_{1}
        -\mu^{2}\, H_{2}^{\dagger}H_{2}
        -\mu f \LP[\, H_{2}^{\dagger}\tilde{L}\,\tilde{R}
                  +\tilde{L}^{\dagger}H_{2}\,\tilde{R}^{\dagger}\,\RP]\nn
   & & \mbox{}
     - f^{2}\LP[ \,
       \tilde{L}^{\dagger}\tilde{L}\,\tilde{R}^{\dagger}\tilde{R}
     + H_{1}^{\dagger}H_{1}\LP( \tilde{L}^{\dagger}\tilde{L}
                                + \tilde{R}^{\dagger}\tilde{R} \RP)
     - H_{1}^{\dagger}\tilde{L}\LP(H_{1}^{\dagger}\tilde{L}\RP)^{\dagger}\,\RP]
     \nmb
     -\f{g^2}{2}
         \LP(\,\tilde{L}^{\dagger}T^{a}\tilde{L}
                 +H_{1}^{\dagger}T^{a}H_{1}+H_{2}^{\dagger}T^{a}H_{2}\,\RP)
         \LP(\,\tilde{L}^{\dagger}T^{a}\tilde{L}+H_{1}^{\dagger}
                    T^{a}H_{1}+H_{2}^{\dagger}T^{a}H_{2}\,\RP)
      \nmb
        - \f{g'^2}{8} \LP(\,\tilde{L}^{\dagger}\tilde{L}-2\tilde{R}^{\dagger}
               \tilde{R}+H_{1}^{\dagger}H_{1}-H_{2}^{\dagger}H_{2}\,\RP)^2
      + t.d.
           \label{apendix On-Shall L sub SUSY}
\EEA
Hence this appendix is concluded.

\cleardoublepage

\chapter{Transcription of the Scalar Higgs Potential.}
   \label{APP: Transcription of the Scalar Higgs Potential.}

The aim of this appendix is to eliminate the SU(2)
representation matrices $T^{a}$ appearing in the scalar Higgs potential
given by eq.~\r{Radiative Breaking prop 1aaa}, i.e.
\BEA
    V_{Higgs} &=&   \LP(m_{1}^{2}+\mu^{2}\RP) H_{1}^{\dagger}H_{1}
          + \LP(m_{2}^{2}+\mu^{2}\RP) H_{2}^{\dagger}H_{2}
          - m_{3}^{2}\,\e^{ij}\LP(H_{1}^{i}H_{2}^{j} +h.c.\RP)
      \nmb
          +\f{g^{2}}{2} \LP( H_{1}^{\dagger}T^{a}H_{1}+H_{2}^{\dagger}
                      T^{a}H_{2}\RP)
            \LP( H_{1}^{\dagger}T^{a}H_{1}+H_{2}^{\dagger}T^{a}H_{2}\RP)
      \nmb
+ \f{g'^{2}}{8} \LP( H_{1}^{\dagger}H_{1}   -  H_{2}^{\dagger}H_{2}   \RP)^{2}.
              \label{Transcription of the Scalar Higgs Potential prop 1}
\EEA

Our starting point is the following general calculation
\BEA
   H^{\dagger}\!_{m}T^{a}H_{m}\,H^{\dagger}\!_{n}T^{a}H_{n}
     &=& \f{1}{4} \;
             H^{\dagger}\!_{m}\s^{a} H_{m} \, H^{\dagger}\!_{n} \s^{a} H_{n}
        \hspace{1.5cm}  m,n = 1,2\;\;\; \mbox{(no sum)}\NN
     &=& \f{1}{4}
         \LP[ H^{\dagger}\!_{m}\LP(\BA{rr} 0 & 1\\1 & 0 \EA \RP) H_{m}\;
              H^{\dagger}\!_{n}\LP(\BA{rr} 0 & 1\\1 & 0 \EA \RP) H_{n} \RP. \nn
         \nmb \hspace{0.4cm} +
              H^{\dagger}\!_{m}\LP(\BA{rr} 0 & -i\\i & 0 \EA \RP) H_{m}\;
              H^{\dagger}\!_{n}\LP(\BA{rr} 0 & -i\\i & 0 \EA \RP) H_{n} \nn
          \nmb \hspace{0.4cm}\LP. +
              H^{\dagger}\!_{m}\LP(\BA{rr} 1 & 0\\0 & -1 \EA \RP) H_{m}\;
              H^{\dagger}\!_{n}\LP(\BA{rr} 1 & 0\\0 & -1 \EA \RP) H_{n}
         \RP]\NN
     &=& \f{1}{4}  \LP[
        \LP(\LP. H_{m}^{1}\RP.^{\dagger}\! H_{m}^{2}
               + \LP. H_{m}^{2}\RP.^{\dagger}\! H_{m}^{1} \RP)
        \LP(\LP. H_{n}^{1}\RP.^{\dagger}\! H_{n}^{2}
               + \LP.H_{n}^{2}\RP.^{\dagger}\! H_{n}^{1} \RP) \RP.
    \nmb  \hspace{0.7cm}
        - \LP(\LP. H_{m}^{2}\RP.^{\dagger}\! H_{m}^{1}
               - \LP.H_{m}^{1}\RP.^{\dagger}\! H_{m}^{2} \RP)
          \LP(\LP. H_{n}^{2}\RP.^{\dagger}\! H_{n}^{1}
               - \LP.H_{n}^{1}\RP.^{\dagger}\! H_{n}^{2} \RP)
    \nmb  \hspace{0.7cm}\LP.
       +  \LP(\LP. H_{m}^{1}\RP.^{\dagger}\! H_{m}^{1}
               - \LP.H_{m}^{2}\RP.^{\dagger}\! H_{m}^{2} \RP)
          \LP(\LP. H_{n}^{1}\RP.^{\dagger}\! H_{n}^{1}
               - \LP.H_{n}^{2}\RP.^{\dagger}\! H_{n}^{2} \RP)\RP]\NN
    &=& \f{1}{4}
       \LP[  \;
              2\,\LP. H_{m}^{1}\RP.^{\dagger}\!
              H_{m}^{2}\,\LP. H_{n}^{2}\RP.^{\dagger}\! H_{n}^{1}
           +  2\,\LP. H_{m}^{2}\RP.^{\dagger}\!
               H_{m}^{1}\,\LP. H_{n}^{1}\RP.^{\dagger}\! H_{n}^{2} \RP.
   \nmb \hspace{0.6cm}
           + \LP. H_{m}^{1}\RP.^{\dagger}\!
               H_{m}^{1}\,\LP. H_{n}^{1}\RP.^{\dagger}\! H_{n}^{1}
           + \LP. H_{m}^{2}\RP.^{\dagger}\!
                H_{m}^{2}\,\LP. H_{n}^{2}\RP.^{\dagger}\! H_{n}^{2}
   \nmb \hspace{0.6cm} \LP.
           - \LP. H_{m}^{1}\RP.^{\dagger}\!
                H_{m}^{1}\,\LP. H_{n}^{2}\RP.^{\dagger}\! H_{n}^{2}
           - \LP. H_{m}^{2}\RP.^{\dagger}\!
          H_{m}^{2}\,\LP. H_{n}^{1}\RP.^{\dagger}\! H_{n}^{1}\RP]. \nonumber
\EEA
With this result we have
\BEA
\lefteqn{ \f{g^{2}}{2} \LP( H_{1}^{\dagger}T^{a}H_{1}
                 +H_{2}^{\dagger}T^{a}H_{2}\RP)
                        \LP( H_{1}^{\dagger}T^{a}H_{1}+H_{2}^{\dagger}
             T^{a}H_{2}\RP)}\hspace{2cm}\nn
          &=& \f{g^{2}}{8}\LP[\,
               \LP(\LP.H_{1}^{1}\RP.^{\dagger}\!H_{1}^{1}\RP)^{2}
             + \LP(\LP.H_{1}^{2}\RP.^{\dagger}\!H_{1}^{2}\RP)^{2}
             + \LP(\LP.H_{2}^{1}\RP.^{\dagger}\!H_{2}^{1}\RP)^{2}
             + \LP(\LP.H_{2}^{2}\RP.^{\dagger}\!H_{2}^{2}\RP)^{2} \RP.
        \nmb       \hspace{0.8cm}
             + 2\,\LP.H_{1}^{2}\RP.^{\dagger}\!
H_{1}^{1}\,\LP.H_{1}^{1}\RP.^{\dagger}\!H_{1}^{2}
             + 2\,\LP.H_{2}^{2}\RP.^{\dagger}\!
H_{2}^{1}\,\LP.H_{2}^{1}\RP.^{\dagger}\!H_{2}^{2}
        \nmb       \hspace{0.8cm}
             + 2\,\LP.H_{1}^{1}\RP.^{\dagger}\!
H_{1}^{1}\,\LP.H_{2}^{1}\RP.^{\dagger}\!H_{2}^{1}
             + 2\,\LP.H_{1}^{2}\RP.^{\dagger}\!
H_{1}^{2}\,\LP.H_{2}^{2}\RP.^{\dagger}\!H_{2}^{2}
        \nmb       \hspace{0.8cm}
             - 2\,\LP.H_{1}^{1}\RP.^{\dagger}\!
H_{1}^{1}\,\LP.H_{2}^{2}\RP.^{\dagger}\!H_{2}^{2}
             - 2\,\LP.H_{1}^{2}\RP.^{\dagger}\!
H_{1}^{2}\,\LP.H_{2}^{1}\RP.^{\dagger}\!H_{2}^{1}
        \nmb \LP.  \hspace{0.8cm}
             + 4\,\LP.H_{1}^{1}\RP.^{\dagger}\!
H_{1}^{2}\,\LP.H_{2}^{2}\RP.^{\dagger}\!H_{2}^{1}
             + 4\,\LP.H_{1}^{2}\RP.^{\dagger}\!
H_{1}^{1}\,\LP.H_{2}^{1}\RP.^{\dagger}\!H_{2}^{2}
             \;\RP] \NN
          &=& \f{g^{2}}{8}
              \LP[\; 2\ABS{\LP.H_{1}^{i}\RP.^{\dagger}\!H_{2}^{i}}^{2}
                   -2\ABS{\e^{ij}\,H_{1}^{i}H_{2}^{j}}^{2}\RP.
         \nmb \LP. \hspace{0.8cm}
                   + \LP( \LP.H_{1}^{i}\RP.^{\dagger}\!
H_{1}^{i} - \LP.H_{2}^{j}\RP.^{\dagger}\!H_{2}^{j} \RP)^{2}
                   + 2 \LP(\LP.H_{1}^{i}\RP.^{\dagger}\!
H_{1}^{i}\RP)\LP(\LP.H_{2}^{j}\RP.^{\dagger}\!H_{2}^{j}\RP)\RP]\NN
         &=&
            \f{g^{2}}{8} \LP[
               \LP( \LP.H_{1}^{i}\RP.^{\dagger}\!H_{1}^{i}
                        -  \LP.H_{2}^{j}\RP.^{\dagger}\!H_{2}^{j} \RP)^{2}
              + 4 \ABS{ \LP.H_{1}^{i}\RP.^{\dagger}\!
H_{2}^{i}}^{2} \RP]. \nonumber
\EEA
Here in the last transition we have used the identity
\BEA
   \ABS{ \LP.H_{1}^{i}\RP.^{\dagger}\!H_{2}^{j}}^{2}
   +\ABS{e^{ij}\,H_{1}^{i}H_{2}^{j}}^{2}
      &=& \LP( \LP.H_{1}^{i}\RP.^{\dagger}\! H_{1}^{i}\RP)
           \LP( \LP.H_{2}^{j}\RP.^{\dagger}\! H_{2}^{j}\RP),\nonumber
\EEA
which can be derived by straightforward calculations.

Hence the scalar Higgs
potential~\r{Transcription of the Scalar Higgs Potential prop 1}
reads
\BEA
    V_{Higgs} &=&   \LP(m_{1}^{2}+\mu^{2}\RP) H_{1}^{\dagger}H_{1}
          + \LP(m_{2}^{2}+\mu^{2}\RP) H_{2}^{\dagger}H_{2}
          - m_{3}^{2}\,\e^{ij}\LP(H_{1}^{i}H_{2}^{j} +h.c.\RP)
      \nmb
          + \f{1}{8}\LP(g^{2}+g'^{2}\RP) \LP( H_{1}^{\dagger}H_{1}
  -  H_{2}^{\dagger}H_{2}   \RP)^{2}
          +\f{g^{2}}{2}\ABS{ \LP.H_{1}\RP.^{\dagger}\!H_{2}}^{2},
              \label{Transcription of the Scalar Higgs Potential prop 2}
\EEA
and this concludes this appendix.

\cleardoublepage

\end{document}